\RequirePackage{fix-cm}
\documentclass[smallextended]{svjour3}       
\smartqed  
\usepackage{amsmath}
\usepackage{graphicx}
\usepackage{verbatim}
\usepackage{natbib}
\usepackage{xcolor}
\usepackage[varg]{txfonts}
\usepackage[printonlyused,withpage]{acronym}
\usepackage[colorlinks=true,citecolor=blue,urlcolor=blue,breaklinks]{hyperref}
\usepackage{comment}
\usepackage{siunitx}
\usepackage{nicefrac}
\usepackage{enumerate}
\newcommand{\msun}{\ensuremath{M_\odot}}
\newcommand{\rsun}{\ensuremath{R_\odot}}
\newcommand{\ace}{\ensuremath{\alpha_\mathrm{CE}}}
\newcommand*{\defeq}{\ensuremath{\mathrel{\vcenter{\baselineskip0.5ex \lineskiplimit0pt
                       \hbox{\scriptsize.}\hbox{\scriptsize.}}}%
                       =}}
\newcommand{\DD}{\ensuremath{\mathrm{D}}}

\renewcommand{\vec}[1]{\ensuremath\mbox{\boldmath{$#1$}}}
\def\aj{AJ\ }%
\def\araa{ARA\&A\ }%
\def\apj{ApJ\ }%
\def\apjl{ApJ\ }%
\def\apjs{ApJS\ }%
\def\aap{A\&A\ }%
\def\aapr{A\&A~Rev.\ }%
\def\aaps{A\&AS\ }%
\def\mnras{MNRAS\ }%
\def\prl{Phys.~Rev.~Lett.\ }%
\def\pasa{PASA}
\def\pasp{PASP\ }%
\def\sovast{Soviet~Ast.\ }%
\def\na{New Astronomy\ }%
\def\nar{New Astronomy Reviews\ }%
\def\nat{Nature\ }%
\def\aplett{Astrophys.~Lett.\ }%
\def\physrep{Phys.~Rep.\ }%
\DeclareSIUnit\year{yr}

\begin{document}

\title{Simulations of common-envelope evolution in binary stellar systems: physical models and numerical techniques}

\titlerunning{Simulations of common-envelope evolution in binary stellar systems}  

\author{Friedrich K.\ R{\"o}pke \and
        Orsola De Marco
}


\institute{F.\ K.\ R{\"o}pke \at
    Institut f{\"u}r Theoretische Astrophysik, Zentrum f{\"u}r
    Astronomie, Universit{\"a}t Heidelberg, Philosophenweg 12, 69120
    Heidelberg, Germany, and\\
    Heidelberger Institut f{\"u}r Theoretische Studien,
    Schloss-Wolfsbrunnenweg 35, 69118 Heidelberg, Germany
    \email{friedrich.roepke@h-its.org}
    \and
  O. De Marco \at
    School of Mathematical and Physical Sciences, Macquarie University, Sydney,
    NSW 2109, Australia, and\\
    Astronomy, Astrophysics and Astrophotonics Research Centre,
    Macquarie University, Sydney, NSW 2109, Australia
}

\date{Received: date / Accepted: date}

\maketitle

\begin{abstract}
When the primary star in a close binary system evolves into a giant
and engulfs its companion, its core and the companion temporarily
orbit each other inside a common envelope.  Drag forces transfer
orbital energy and angular momentum to the envelope
material. Depending on the efficiency of this process, the envelope
may be ejected leaving behind a tight remnant binary system of two
stellar cores, or the cores merge retaining part of the envelope
material. The exact outcome of common-envelope evolution is critical
for in the formation of X-ray binaries, supernova progenitors, the
progenitors of compact-object mergers that emit detectable
gravitational waves, and many other objects of fundamental
astrophysical relevance. The wide ranges of spatial and temporal
timescales that characterize common-envelope interactions and the lack
of spatial symmetries present a substantial challenge to generating
consistent models. Therefore, these critical phases are one of the
largest sources for uncertainty in classical treatments of binary
stellar evolution. Three-dimensional hydrodynamic simulations of at
least part of the common envelope interaction are the key to
gain predictive power in modeling common-envelope evolution. We review
the development of theoretical concepts and numerical approaches for
such three-dimensional hydrodynamic simulations. The inherent
multi-physics, multi-scale challenges have resulted in a wide variety
of approximations and numerical techniques to be exercised on the
problem. We summarize the simulations published to date and their main
results. Given the recent rapid progress, a sound understanding of the
physics of common-envelope interactions is within reach and thus there
is hope that one of the remaining fundamental problems of stellar
astrophysics may be solved before long.

\end{abstract}

\setcounter{tocdepth}{3}
\tableofcontents

\section{Introduction}
\label{intro}

The historic detection of gravitational waves from merging black holes
\citep{abbott2016a} and neutron stars \citep{abbott2017a} highlights
the importance of understanding the details of binary stellar
evolution. What are the physical processes that lead to mergers of
these ``double-compact objects'' \citep{kalogera2007a}? Reducing the
orbit such that a merger commences requires to drain orbital energy
and angular momentum from the system. In the last stage before the
actual merger, this proceeds due to the emission of gravitational
waves---a process that is highly inefficient. Orbital decay is
extremely slow. This final phase, driven by gravitational wave
emission, must therefore start with a very small orbital separation;
otherwise a merger could not occur within a Hubble time.

How is this possible?  For some compact objects, a capture of a close
companion in a dense stellar environment is conceivable. However, many
of the systems that eventually merge are thought to start out as
main-sequence star binaries.  This, of course, implies an initial
separation larger than the sum of the main-sequence star radii, so a
mechanism must exist to reduce the orbital separation of the stellar
cores\footnote{We use the terms ``stellar cores'' and ``core binary''
when referring to a system consisting of the core of a giant star and
a companion. We note, however, that the companion need not necessarily
be a core of a star. It can also be a main sequence star, a planet, or
a compact remnant, such as a neutron star or even a black hole.}
sufficiently that eventually they merge via the emission of
gravitational waves.

Similar evolutionary mechanisms must be responsible for the formation
of interacting binary systems involving one or two compact objects
(so-called single- or double-compact binaries). Prominent examples are
cataclysmic variables \citep[e.g.][]{meyer1979a,webbink1984a}, Type Ia
supernovae progenitors \citep[e.g.][]{wang2018a}, X-ray binaries
\citep[e.g.][]{tauris2014a}, short gamma ray burst progenitors, binary pulsars
\citep[e.g.][]{chen2014b}, and close white dwarf binaries
\citep[e.g.][]{zorotovic2010a, demarco2011a}.  Here, the same argument applies:
While still on the main sequence, the two stars in the progenitor
binary must be sufficiently separated to avoid interaction. The
observed phenomena, however, imply an orbital distance significantly
smaller than such a relatively large initial separation.

It is therefore clear that the progenitors of compact binaries were
much larger in the past and would not have fitted into the orbits of
the systems we observe today. The requirement of sufficient separation
of the stellar binary at birth, however, does not need to hold true
for the subsequent evolution, where strong interaction and mass
transfer is possible, in particular because the radii of stars
increase substantially for later evolutionary
stages. \citet{paczynski1976a} and
others\footnote{\citet{paczynski1976a}, often cited as the having
first formalized the idea of the common envelope interaction, credits
aspects of the concept to a number of previous works. As is often the
case, a number of separate steps are eventually formalized into a
clear hypothesis.} realized that this is not just another nuisance in
the description of the binary system, but rather the key to
understanding its evolution towards a close binary: They proposed a
phase in which the two stellar cores orbit each other sharing the
stellar envelope and coined the name \acfi{CE} phase for it. Such an
extreme case of binary stellar interaction is the ``magic trick'' to
explain the short orbital separation of single and double-compact
systems of various kinds: the orbit of the companion with the core of
the ``primary'' star inside the common envelope causes drag (whether
gravitational or hydrodynamical), which transfers orbital energy and
angular momentum to the envelope material unbinding it from the
system. The stellar cores approach each other, the envelope is ejected
and---voil\`a!---a close binary system emerges.

While critical for the formation of narrow binary systems, the \ac{CE}
phase is also ubiquitous in stellar evolution. A substantial fraction
of stars forms in multiples \citep[e.g.][]{moe2017a} and many of them
are bound to interact during their evolution. For example, practically
all massive stars are born in multiple systems and up to 70\%
experience binary interaction in their life \citep{sana2012a}. The
chances for interaction are not as high for low and intermediate-mass
stars, but the consequences are no less impactful for the evolution of
these systems \citep{demarco2017a}.

The implications of \ac{CE} interaction are far-reaching and
fundamental for answering questions such as:
\begin{itemize}
  \item What are the progenitor systems of Type Ia supernovae?
  \item What defines the rates and parameter distributions of novae
    and dwarf novae?
  \item Can planets survive when they are engulfed by their host
    star?
  \item How are the source systems of gravitational wave emission
    formed?
\end{itemize}
Answers these questions pivotally depend on understanding the post-\ac{CE}
separation distribution. This quantity cannot be determined reliably
at the moment because of shortcomings in modeling the inspiral of the
companion into the envelope of the primary, orbital tightening of the
core binary system, envelope ejection, and potentially other effects
in late stages of \ac{CEE}.

Theoretical modeling of \ac{CE} phases in binary stellar evolution has
a long history.  Early attempts used \ac{1D} parametrizations in the
framework of classical stellar evolution theory. There are, however,
reasonable doubts as to whether their results are physical. Such
parametrized models have certain advantages in covering processes that
act on longer timescales, but they struggle to capture what is
essentially a fast, dynamical phase with no symmetry
\citep{ivanova2016a}. Other models use parametrized prescriptions to
follow the evolution of large samples of binary systems or
populations. By calibrating the parameters on observational
constraints, some of the predictions reach a certain level of
confidence. Such population models are used, for instance, to estimate
the event rate in gravitational wave detectors
\citep[e.g.,][]{dominik2012a, belczynski2014a} and to distinguish
between different progenitor scenarios for Type Ia supernovae
\citep[e.g.,][]{ruiter2009a, toonen2012a}.

A more direct approach to understanding the physical processes
governing \ac{CEE} is the goal of numerical simulations that we focus
on in our review.  In addition to the listed examples of astrophysical
implications of \ac{CE} interaction, the process itself is fascinating
featuring strong nonlinear dynamical effects. Understanding the
underlying physical processes and making reliable predictions for the
outcome of \ac{CE} events cannot be accomplished without complex
\ac{3D} hydrodynamics models that cover phases of \ac{CEE}, because
these lack obvious spatial symmetries and proceed on dynamical
timescales. The aim of such simulations is to answer specific
questions about details of the inspiral and envelope ejection, which
then allow for tackling the higher-order questions listed above. Some
of these still unanswered, specific questions are:
\begin{itemize}
  \item How does the system enter \ac{CEE}?
  \item How are orbital energy and angular momentum of the core binary
    transferred to the envelope?
  \item What physical processes lead to the ejection of the envelope?
    Which energy reservoirs are tapped for accelerating the envelope
    material?
  \item What processes determine the final orbital separation of the
    remnant core binary?
  \item Which systems manage to eject the envelope completely?
  \item What are the conditions under which the process ``fails'',
    i.e. a merger of the core takes place without ejecting the
    envelope?
  \item What is the morphology of the ejected envelope in the case of
    success? Can it explain planetary nebulae?
  \item What is the role of magnetic fields in \ac{CEE}?
\end{itemize}      
The ultimate goal is to connect the initial conditions, i.e.\ the
parameters of the system before entering \ac{CE} interaction, with the
properties of the remnant system---most importantly its orbital
parameters in the case of successful envelope ejection---but also
other quantities that determine its future evolution, such as
left-over envelope material or a potentially forming circumbinary
disk.

\subsection{Observations of common-envelope evolution}
\label{sect:obs_constraints}

The concept of \ac{CE} was first envisaged by \citet{paczynski1976a} to
explain a specific star, V~471~Tau, a detached, compact binary
comprising a white dwarf and a K dwarf, rare at the time and presumed
to be a direct predecessor of cataclysmic variables. Only through a \ac{CE}
inspiral in the past evolution of the system, could today's parameters
be explained.

Over the years, many more low-mass systems were found that can only be
explained with a \ac{CE} interaction. Low and intermediate-mass star
binaries that have gone through a \ac{CE} interaction are observed as
post-red giant branch, sub-dwarf B (sdB) and white dwarf (WD) close binaries,
or as post-asymptotic giant branch WD binaries, with either
another WD or a main sequence companion. These are the immediate
progenitors of cataclysmic variables. In some cases even brown dwarfs
or giant planets have been deemed capable of ejecting the \ac{CE} and
surviving as binaries with the core of the original giant
\citep[e.g.,][]{geier2011a, schaffenroth2014a, schaffenroth2019a,
  vanderburg2020a}. Such lower mass, post-\ac{CE} systems are plentiful,
and their parameters can be measured with relative precision, thus
providing statistical information about the outcome of orbital
evolution in the \ac{CE} phase \citep{iaconi2019a}. The direct
consequences of envelope ejection can be seen in many \acp{PN}. At
least one in five \acp{PN} is an ejected \ac{CE} \citep{hillwig2016a,
  munday2020a, jacoby2021a}.

After the initial idea of a \ac{CE} inspiral was fleshed out, it became
clear how \ac{CE} phases can explain binaries comprising the remnant
of massive stars, neutron stars and black holes. Today, this group of
binaries is a prime target of multimessenger astronomy. The detection
statistics of gravitational-wave induced mergers of double-compact
systems is strongly linked to the parameters of the \ac{CE}
interaction. There are other massive systems that can be identified as
binaries and that must have emerged from a \ac{CE}, such as low and high
mass X-ray binaries, but unfortunately the \ac{CE} phase is not the only
event that determines the observed parameters of these systems, and
therefore they are not ideal to constrain \ac{CE} interaction
observationally.

Additionally, thanks to time-resolved astronomical surveys, an
increasing number of transient events has been discovered. Supported
by pre-outburst observations, several of them have been unequivocally
associated with \ac{CE} interactions and stellar mergers \citep[e.g.,
  V1309~Sco,][]{tylenda2011a}. Others have very likely undergone a \ac{CE}
inspiral even if there is no definitive proof (e.g., V838~Mon,
\citealp{bond2003a, goranskij2004a}, but see \citealp{tylenda2006a}
for a discussion of alternative models). By statistical necessity, the
majority of objects detected in these brief phases of interaction are
the most luminous events, which presumably derive from more massive
stars \citep{kochanek2014a}.  Examples include NGC~4490-2012OT1
\citep{smith2016a} and M101-2015OT1 \citep{blagorodnova2017a} that
have been detected in external, sometimes distant galaxies.

A new field of study has emerged from these discoveries, alongside
several competing classifications, nomenclatures and preliminary
explanations for the observed events. Such transients are commonly
known as \emph{gap transients}---observed in the gap between the
luminosities of novae and supernovae \citep{kasliwal2012b,
  pastorello2019a}---or \emph{intermediate-luminosity optical
transients} (ILOTs).  These include \emph{\aclp{ILRT}} (\acsp{ILRT}) \acused{ILRT}
\citep{botticella2009a} and \emph{\aclp{LRN}} (\acsp{LRN}) \acused{LRN}
\citep{munari2002a} distinguished by their lightcurves and spectral
behaviour. The latter are thought to be \ac{CE} interactions between
expanding, massive stars and companions that resulted in mergers,
although in most cases it is impossible to say for sure if the binary
has survived. Interestingly, while \acp{ILRT} are thought to be electron
capture supernovae from super-\ac{AGB} stars, there are some cases where
the classification is intermediate between \acp{ILRT} and \acp{LRN}
\citep{cai2019a}. Finally, a distinction between \acp{LRN} and \acp{RN} is
emerging, where the latter group are a less luminous and possibly less
massive counterpart of the former group \citep{pastorello2019a}.

In Sect.~\ref{sect:observational_constraints} we will return to these
observations and consider how they can constrain the simulations that
are the central topic of this review.

\subsection{Terminology of common-envelope modeling}
\label{ssec:terminology}

Many stars in the Universe form in binaries or higher-order multiples,
and a large fraction of these engages in some kind of interaction over
their evolution. The nature of such interaction can be manifold;
ranging from orbital perturbations mediated by gravitational
interaction, irradiation, and hydrodynamical mass-transfer processes
to the most violent case: a merger of two stellar objects. \ac{CE}
interaction is very close to this extreme case. In a sense, it
classifies as a merger of stars because at least for some period of
time two (or perhaps even more\footnote{While interaction in a stellar
binary system is perhaps the most relevant case, \ac{CEE} could also
proceed in higher-order stellar multiples and involve more than two
stellar cores. This case is in itself interesting but has received
less attention in simulations \citep[see][for an exception and references therein for implications of mass transfer in stellar triples]{glanz2021b}. The
``traditional'' case of a stellar binary is already challenging enough
and its processes remain sufficiently unclear to not open Pandora's
box and postpone such complications to future work. Our review will
therefore focus on modeling \ac{CE} interaction in binary systems.})
stellar objects join into a single envelope. Unlike a complete stellar
merger, however, this ``merged'' phase may be a transient
phenomenon. After the ejection of the \ac{CE}, separate stellar cores
emerge. Special conditions are required for this to work.

An obvious prerequisite is that one of the interacting stars, which we
refer to as ``primary'', has to exhibit a clearly separated
core--envelope structure.  Otherwise, it would not make sense to speak
of a \emph{common envelope} in the interaction phase and to define a
stellar core that emerges as part of the newly formed binary star
system after the envelope ejection. This implies that the primary star
has to reach an evolved, giant stage before engaging into the
interaction. For a time, the companion resides inside the envelope of
the giant primary star. While this companion can be a stellar object
of any kind, we encounter a special case if it is another giant
star. For this to happen, the initial mass ratio of the binary system
has to be very close to unity so that when the primary overfills its
Roche lobe as a giant, accretion onto the companion causes it to
expand and overfill its Roche lobe, too. A joint \ac{CE} establishes
that consists of material from both stars. Still, during the
interaction, two stellar cores orbit each other inside a \ac{CE}. This
case is sometimes (and confusingly) termed \emph{double core
evolution}, falling back to the term that was originally used for
general \ac{CE} phases. It is particularly interesting for massive
stars \citep{brown1995a, dewi2006a} and may contribute a substantial
fraction to the population of double neutron stars
\citep{vigna2018a}. Despite this ambiguity, a stellar merger with
evolved giant stars provides a useful definition of \ac{CE} interaction and
includes the classical case as well as the merger of two giant stars.

Naturally, as the evolution into giant stages is a continuous process,
the boundaries between \ac{CE} interaction and stellar mergers are
blurred. For massive stars in the giant stage (so-called
\emph{supergiants}), for instance, the separation between core and
envelope is less clear than the sharp interface in low-mass \ac{RG} or
\ac{AGB} stars. Another example is the system V1309~Sco
\citep{tylenda2011a}. It was modeled as a merger of a $1.5 \,\msun$
star expanding into the Hertzsprung gap with a low-mass (${\sim}\,
0.15 \, \msun$) main sequence companion. The primary is about 40 times
larger than the secondary and the system likely went through a brief
common envelope, shortly followed by the disruption of the companion
near the core of the primary \citep{nandez2014a}. As a Hertzsprung-gap
star, the primary would have had a less clear core-envelope
boundary. Therefore, the nature of the physical interaction in the
later inspiral is not quite the same as for a system with a more
distinguished core.

Another characteristic of \ac{CE} interaction is the aforementioned
interesting possibility that after envelope ejection a tight binary of
stellar cores emerges. This case is sometimes referred to as
\emph{successful} \ac{CEE} to distinguish it from the case where the
energy available to envelope ejection is insufficient and only parts
of it can be removed. The ultimate fate of such a system is a merger
of the stellar cores inside the \ac{CE}, which can also give rise to
interesting phenomena: for a high-mass primary star and neutron star
companion, such a scenario could lead to the formation of
Thorne--{\.Z}ytkov objects \citep{thorne1975a}. It has been also
suggested that because of the high angular momentum involved, a
neutrino-cooled accretion disk forms around a neutron star or a black
hole companion when merging with the core of the primary star. This
accretion liberates gravitational binding energy, which to a large
fraction is carried away by the neutrinos but some part may also be
released as mechanical energy---perhaps in form of material
accelerated in a disk wind or a jet-like outflow---feeding back on
the surrounding envelope material (a scenario originally proposed by
\citealp{fryer1998b}, and \citealp{zhang2001a}, to explain gamma-ray
bursts). The corresponding events have been suggested to resemble some
sub-classes of Type II supernovae \citep[such as Type IIn or Type II-P
  supernovae; e.g.][]{chevalier2012a, soker2018b, schroder2020a} and
are sometimes called ``common-envelope jets
supernovae''\footnote{Events powered by a jet due to accretion onto a
neutron star in the envelope instead of the core are termed
``common-envelope jets supernova impostors'' by \cite{gilkis2019a}}
\citep{soker2018b, soker2019b}. For low-mass stars involved in \ac{CE}
interactions with a white dwarf companion, a thermonuclear explosion
resulting from the merger of the degenerate cores has been suggested
to explain some peculiar Type Ia supernovae \citep{sparks1974a,
  livio2003a}. This mechanism is also called the``core degenerate
scenario'', e.g.][]{kashi2011a,ilkov2012a,soker2013a}. While
  technically for all these events one may speak of a mergers, the term \emph{failed}
  \ac{CEE} is commonly used for it, or sometimes, as a compromise,
  \emph{\ac{CE} merger}. It highlights the fact that a giant star was
  involved in the interaction. Also from a practical point of view it
  makes sense to discuss these cases in conjunction with successful
  \ac{CE} interactions because they constitute an interesting limit in
  the parameter space of such systems.
 
That said, it may be useful to give examples of cases that do not
qualify as \ac{CE} interaction. A merger of two similarly sized (but
not giant) stars, for instance between two white dwarf dwarfs
(corresponding \ac{3D} simulations were pioneered by
\citealp{benz1990a}; see, e.g., \citealp{dan2009a},
\citealp{dan2012a}, \citealp{zhu2015a} for more recent simulations),
two neutron stars (e.g. \citealp{rasio1992a}
\citealp{ruffert1996b}, and \citealp{rosswog1999a}; see also
\citealp{faber2012a}, \citealp{rosswog2015a}, \citealp{baiotti2017a},
\citealp{shibata2019a}, and \citealp{radice2020a} for reviews), or
two main sequence stars \citep[e.g.,][]{benz1987a, schneider2019a}. Such
interactions are distinct from \ac{CEE}. The initial mass transfer,
that invariably commences the interaction, leads to orbital decrease
and soon thereafter a tidal disruption of one or both of the stars. We
would be hard pressed to state which star is in which star's envelope.

\subsection{Phases of common-envelope evolution}
\label{sect:phases}

To set the stage for our later discussion, we review some ideas on the
temporal evolution of \ac{CE} interaction proceeds. A classification
of stages through which \ac{CEE} proceeds was given by
\citet{ivanova2013a}. This classification was largely inspired by
concepts of classical stellar evolutionary theory. Here, we focus on
the computational modeling of \ac{CE} interaction.  In this context,
it is more useful to think of \ac{CEE} as a three-part interaction,
paying attention to the vastly different timescales (to be defined
later in our discussion) on which the stages proceed and consequently
our ability to model them in numerical simulations:
\begin{enumerate}[(i)]
  \item
    The first part starts with mass-transfer triggered by a star
    growing into its Roche lobe, or from the synchronized orbit
    becoming Darwin-unstable and leading to a loss of co-rotation.
    The companion starts to approach the primary. This phase ends when
    the mass-transfer process assumes a runaway nature.
  \item
    The second part is the inspiral of the companion into the envelope
    of the primary. It leads to the typical structure in which two
    stellar cores orbit each other inside a \ac{CE}. This phase is
    essentially dynamical.
  \item
    The third phase is a combination of the self-regulating phase
    (phase III of \citet{ivanova2013a}), its termination (their phase IV)
    and what comes after (their phase V). This phase starts with the
    companion--core distance stabilizing and the envelope adjusting
    thermally to a new equilibrium. 
\end{enumerate}
Our classification intentionally leaves out the long-term evolution
of the binary system up to \ac{CE} interaction and the fate of the
system after completion of the \ac{CE} phase. The former is important
because it determines the global parameters of the interacting system
and sets the initial conditions of the \ac{CE} interaction. However,
we take these as given and focus on the \ac{CE} phase itself in this
review. Likewise, however interesting from an astrophysical point of
view, the long-term evolution of the system resulting from \ac{CEE} is
not the subject of our discussion.

There are a lot of subtleties to this picture and to an extent it is a
matter of taste how we decide to break down the interaction into
different parts. It is also clear that the last ten years of
development have refined some of the ideas behind the phases listed by
\citet{ivanova2013a}. For example, the pre-inspiral phase is more
complex than a simple unstable mass-transfer starting out at the time
of Roche lobe overflow. If the mass ratio is in the proximity of the
analytically-derived threshold for instability \citep{tout1991a}, one
may expect a longer phase of semi-stable mass-transfer, that only
eventually leads to the formation of a \ac{CE} system.  Similarly, the
post-inspiral phase may not or not only consist of a self-regulating
phase as described in the literature. It is still unclear whether in
some setups all envelope material is ejected in the dynamic inspiral
phase so that a subsequent self-regulated phase does not
establish. The slowing down of the inspiral observed in hydrodynamic
simulations is more due to the companion bringing local gas into
co-rotation, than an evacuation of the orbit \citep[at least for lower
  mass stars;][]{reichardt2019a}. And the evolution of the orbit after
the end of the dynamical inspiral may have more to do with the
distribution and evolution of angular momentum of the gas as the inner
envelope contracts. It is also possible that the evolution in late \ac{CE}
phases involves phases of mass transfer between the remaining cores or
that a circum-binary disk forms.

It is likely not fruitful to complicate the nomenclature while a
clearer picture has not yet emerged. So for now we elect to think
about these three phases in lose terms as
\begin{enumerate}[(i)]
  \item the \emph{pre-\ac{CE} phase,}
  \item the \emph{inspiral phase,} and
  \item the \emph{post-inspiral phase.}
\end{enumerate}
While the first two proceed largely on dynamical timescales
(potentially longer in the pre-\ac{CE} phase), the post-inspiral phase may
include dynamical effects as well as processes that take place on
longer, thermal timescales. We use this simple subdivisions to provide
a reference frame for the discussion that follows. While classical
\ac{1D} models of stellar evolution are suitable for addressing
processes on longer thermal timescales, the decisive part of \ac{CE}
is part (ii) of our classification. Some aspects of parts (i) and
(iii) proceed on dynamical timescales, which need to be addressed with
the hydrodynamical simulations we focus on in our review.

\subsection{Structure of this review}
\label{ssec:structure_of_this_review}

Our review concentrates on advances in the understanding of the common
envelope interaction based on multidimensional hydrodynamic
simulations. In Sect.~\ref{sect:understanding} we set this discussion
into context by reporting the current stance on the common-envelope
interaction. It derives from years of modelling using distinct
approaches: parametric formalisms
(Sect.~\ref{sec:parametrized_models}) use primarily energy and angular
momentum considerations to connect an initial stellar binary system to
the outcome of the interaction. One-dimensional ``mechanical'' models
as described in Sect.~\ref{sec:one_dimensional_implicit_approaches}
try to determine the drag force experienced by the companion when
spiralling into the envelope of the primary star. Integrating a simple
equation of motion results in a model for orbital evolution that can
be coupled energetically to the envelope whose evolution is then
followed with a classical one-dimensional stellar evolution code. Such
\ac{CE} models are of interest not only from a historical
perspective. They may help to interpret the general behaviour of more
complicated hydrodynamic models and to derive a predictive model from
them to be used in binary stellar evolution theory. The development of
\ac{3D} hydrodynamic models is only briefly mentioned in
Sect.~\ref{sec:multi_dimensional_hydro_models}, because they are at
the core of the remainder of this review.

In Sect.~\ref{sect:challenges_physical}, we discuss the physics that
governs \ac{CEE} and its modeling. This section provides the physical
basis of numerical implementations but also points out the modeling
challenges caused by physical effects. After introducing the
fundamental (but probably insufficient) gravo-hydrodynamic model that
describes \ac{CE} interaction as arising exclusively from hydrodynamic and
gravitational interaction in Sect.~\ref{sect:gravohydro}, we explain
why multidimensional simulations of \ac{CE} phases are required to properly
represent the relevant physical processes in
Sect.~\ref{sect:whymultid}. The question of why in such simulations it
remains difficult to decide on the most pressing problems---envelope
ejection and final orbital separation of the remnant binary
system---is discussed in
Sect.~\ref{ssec:why_is_CE_ejection_hard_to_achieve}. Physical
phenomena that should be included in \ac{CE} simulations are mentioned in
Sect.~\ref{sect:whymultiphysics}. We conclude
Sect.~\ref{sect:challenges_physical} with pointing out why the dynamic
inspiral phase---although demanding to model and decisive for the
outcome of \ac{CE} interaction---is not the full story and the phases
preceding and following it have to be taken into account in order to
establish a comprehensive model of \ac{CEE}.

Approaches to numerical implementations of \ac{CE} models are reviewed
in Sect.~\ref{sect:challenges_numerical} with a focus on computational
fluid dynamics (Sect~\ref{sect:cfd}), self-gravity
(Sect.~\ref{sect:gravity}) and problems of energy and angular momentum
conservation (Sect.~\ref{sect:conservation}). Different discretization
approaches have been used to model \ac{CEE} and we try to point out
advantages and drawbacks of specific methods. Apart from some specific
drawbacks of certain schemes, there are general problems arising from
the multi-physics and multi-scale nature of the problem that have
posed challenges to virtually all \ac{CE} simulations to date, and we
discuss them, too.

A related topic is the setup of \ac{3D} hydrodynamic \ac{CE}
simulations. It is sufficiently important, but also delicate in its
treatment, to deserve the separate Sect.~\ref{sect:setups}.

The actual \ac{3D} hydrodynamic \ac{CE} simulations published so far
are reviewed in Sect.~\ref{sect:simulations}. After discussing
simulations pertaining to the pre-\ac{CE} phase (i) of our classification
in Sect.~\ref{ssec:simulations_of_the_pre-dynamical_inspiral_phase},
we give some account of simulations that focus on resolving the
details of the interaction between the companion and the gas of the
\ac{CE} in Sect.~\ref{ssec:windtunnel}. While in these---as a
necessary compromise for reaching high spatial resolution---only some
part of the system is included in the simulated domain, global \ac{CE}
simulations try to follow the evolution of the entire system through
the inspiral phase (ii). As discussed in Sect.~\ref{sect:sim_global},
these have been developed over the past decades, and we give some
account for the historical development of ideas and techniques. At the
moment, a rapid increase in the quality of such hydrodynamic
simulations is observed that originates from a combination of
increasing computational power and new numerical modeling techniques.

In Sect.~\ref{sec:future_perspective} we draw conclusions
(Sect.~\ref{sect:current}) and give future perspectives for \ac{CE}
modeling. We discuss physical modeling aspects (such as missing
physical effects, Sect.~\ref{ssec:missing_phisical_effects}) as well
as numerical issues (e.g. Sect.~\ref{sect:unresolved_num} on
unresolved numerical problems). The post-inspiral phase (iii) is
difficult to capture in \ac{3D} hydrodynamic approaches. Therefore,
such simulations have to be augmented with \ac{1D} classical
treatments of stellar evolution as discussed in
Sect.~\ref{ssec:interfacing_3d_global_simulations_with_1d_simulations_of_the_ce_remnant}. An
ultimate goal of \ac{CE} modeling is to connect directly to binary
evolution calculations
(Sect.~\ref{ssec:extracting_1d_parametrization_from_multiD_simulations})
and to astronomical observations
(Sect.~\ref{sect:observational_constraints}).

\section{Current theoretical understanding of common envelope interaction}
\label{sect:understanding}

In the previous section, we have introduced the \ac{CE} phase as a
necessary model to explain a number of observed systems, and we have
emphasized the importance of this phase to explain and connect a
number of new observations, such as gravitational wave detections. In
this section, we report on the progress that has been made to
elucidate the details of this interaction since the days when it first
was proposed in the late 1970s.  Attempts to model \ac{CEE} can be
classified into three categories: parametric formalisms,
one-dimensional (\ac{1D}) ``mechanical'' models and \ac{3D}
hydrodynamic simulations. In Sects.~\ref{sec:parametrized_models} and
\ref{sec:one_dimensional_implicit_approaches} we discuss the former
two categories. This is meant to summarize approaches that can be seen
as complimentary to the \ac{3D} hydrodynamic simulations. While
potentially useful for the interpretation and---potentially---for
casting the results of such simulations into a simpler framework
applicable in classical stellar evolution approaches, they lack
predictive power and do not require sophisticated computational
methods. Arriving at \ac{3D} hydrodynamic simulations of \ac{CEE} in
in Sect.~\ref{sec:multi_dimensional_hydro_models} sets the stage for
an in-depth discussion of these models in the following sections of
our review.

It has to be pointed out, however, that none of the three approaches
has, to-date, provided a satisfactory picture of the interaction, nor
have they enabled a predictive model of \ac{CEE}. Nevertheless, they
have laid a solid foundation for the modern efforts, which will
hopefully lead to the next wave of breakthroughs.

\subsection{Parametric models: the energy ($\alpha_{\rm CE}$) and angular momentum ($\gamma$) formalisms}
\label{sec:parametrized_models}

Parametric formalisms for modeling \ac{CE} phases are based on general
considerations of the energy or angular momentum budget in the
interacting systems. They determine the orbital energy and angular
momentum required to achieve envelope ejection. This simple estimate
allows to determine whether successful \ac{CEE} is possible or a
\ac{CE} merger is more likely. From the energy or angular momentum
budgets, the formalisms allow to derive the orbital parameters of the
surviving binary system. This way, they provide the end stage of
\ac{CEE} for a given initial system without attempting to model the
dynamical processes linking them. Therefore, such models give little
insight into the physics of \ac{CE} phases. Their predictions seem
questionable, but they allow us to construct population models to
explain observations and they provide guidance for multidimensional
simulations of \ac{CEE}.

The \emph{energy formalism} \citep{vandenheuvel1976a, webbink1984a} is
the coarsest parametrization of \ac{CEE}. It equates the binding
energy of the envelope, $E_\mathrm{bind}$, with the loss of orbital
energy, $\mathrm{\Delta} E_\mathrm{orb}$, during the inspiral
\begin{equation}
  \label{eq:alpha_CE}
  E_\mathrm{bind} = \alpha_\mathrm{CE} \mathrm{\Delta} E_\mathrm{orb},
\end{equation}
where the efficiency parameter \ace\ accounts for the fact that only
some of the deposited orbital energy may be needed to unbind the
entire envelope with the remainder being expended to accelerate all or
part of the envelope to speeds in excess of the escape velocity, or
being radiated or convected away. This would imply values of \ace\ in
the range from $0$ to $1$. Accounting for the fact that other energy
reservoirs than the orbital energy of the companion may be tapped to
unbind the envelope, $\ace > 1$ is formally possible.

\begin{figure}
    \centering
    \includegraphics[width=\textwidth]{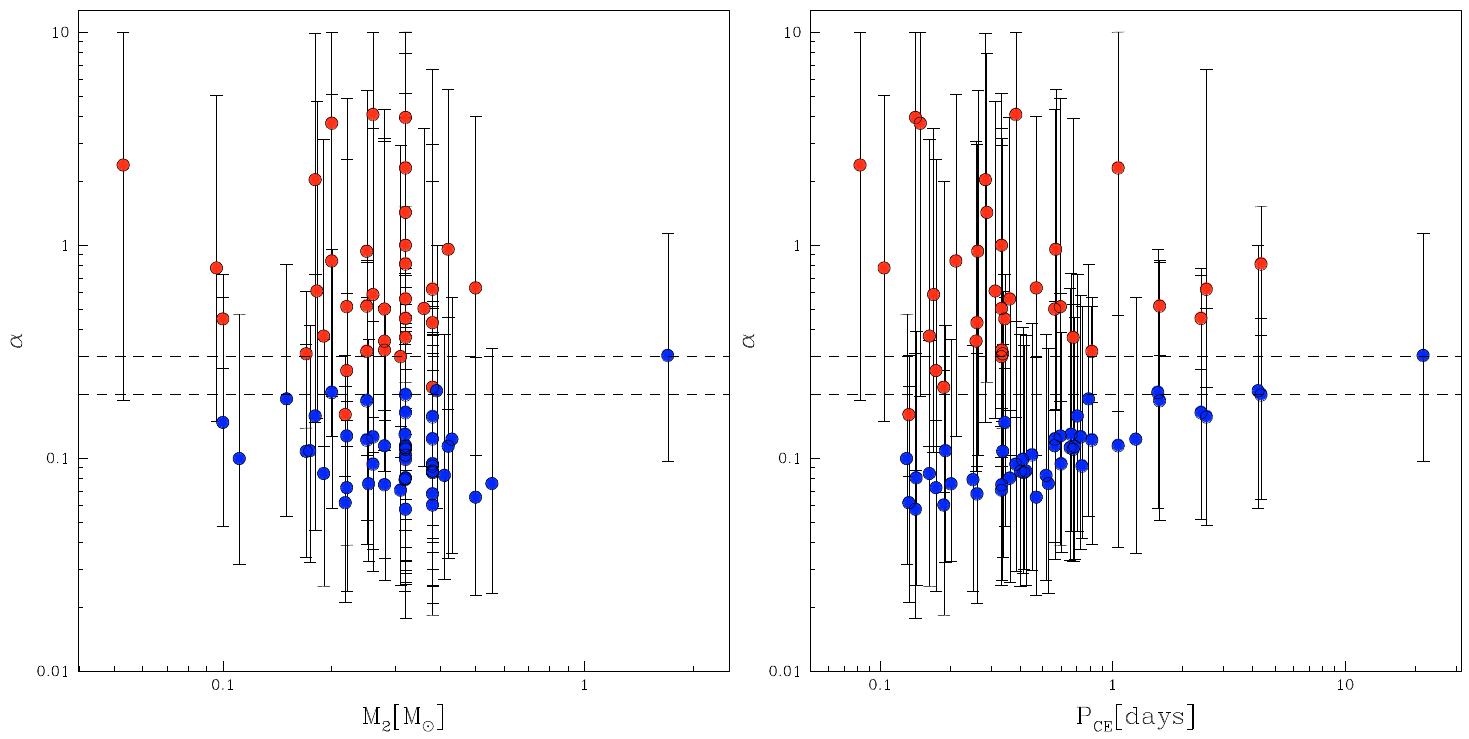}
    \caption{Values of $\ace$ derived by reconstructing or measuring
      the binding energy, the initial and the final separations of
      observed compact single degenerate post-\ac{CE} binaries. Colors
      indicate the nature of the primary star in the progenitor
      system: red dots mark \ac{RG} stars and blue dots \ac{AGB}
      stars. The results are plotted as a function of companion mass
      $M_2$ (left panel) or of the period after the \ac{CE}
      interaction $P_\mathrm{CE}$ (right panel). Figure adapted from
      \citet{zorotovic2010a}.}
    \label{fig:zorotovic+10}
\end{figure}

A reliable determination of \ace\ is required to turn the energy
formalism into a predictive model, but this is difficult to reach.  In
fact, many questions have been raised about the effectiveness of using
this pasteurization to describe the event. What exactly is the value
of \ace? Should be a function of parameters, or of time?
\citet{ivanova2013a} point out that $\alpha_\mathrm{CE}$ being exactly
equal to one is an extreme fine-tuning problem.  It could be a
function of stellar and binary parameters, but has traditionally been
set to a constant value for use in population synthesis models that
need to predict the final separation of systems given their envelope
binding energy \citep[e.g.,][]{toonen2018a}. Some population studies
tried to use a functional value of \ace\ \citep{politano2007a}, but
this has not enlightened the discussion. Observational studies based
on close binaries involving white dwarf stars (particularly
single-degenerate systems, where the companion is a main sequence
star) have attempted a calibration of \ace. Some of these studies
proposed a particular constant value for \ace\ \citep[][see
  Fig.~\ref{fig:zorotovic+10}]{zorotovic2010a} while others advocated
a variable value \citep{demarco2011a}. In the end, all these
approaches are plagued by the large uncertainties introduced in the
many steps required to reconstruct the parameters of the white dwarf
progenitor star.

The energy formalism involves uncertainties that go beyond the tunable
efficiency parameter.  They relate to the definition of the binding
energy of the envelope. Even if a non-parametric expression can be
used, for example by calculating the binding energy from \ac{1D} stellar
structure models, according to
\begin{equation}\label{eq:ebind}
  E_\mathrm{bind} = \int_{M_\mathrm{core}}^{M_\star}\left[ - G \frac{M(r)}{r}
    + \alpha_\mathrm{th} u \right] \mathrm{d} m
\end{equation}
\citep{dewi2000a}, there is always the issue of the mass coordinate
$M_\mathrm{core}$ at which the core of the primary star (of total mass
$M_\star$) ends and its envelope (of mass $M_\mathrm{env}$) begins
\citep[e.g.][]{dewi2000a, demarco2011a, kruckow2016a}. The choice of this
parameter greatly alters the result. For the outcome of \ac{CE}
interaction, the mass coordinate down to which the envelope has to be
removed so that the remaining object does not re-expand---sometimes
called \emph{bifurcation point}---is of particular interest. It sets
the minimum orbital separation that the system has to reach in order
to terminate \ac{CEE} \citep{tauris2001a, deloye2010a, ge2010a,
  marchant2021a, vigna2022a}. 

The first term in the integral of Eq.~(\ref{eq:ebind}) is clear: it is
the gravitational binding energy of the envelope gas with $G$ denoting
the gravitational constant. The specific\footnote{\emph{Specific} here
means per unit mass.} internal energy $u$ in the second term, however,
is a bit ambiguous. It should contain all other energies of the
envelope that potentially decide on its binding status, such as the
thermal energy of the gas, the energy of the radiation field
(expressed via the Stefan-Boltzmann law), the ionization energy of the
atoms, the dissociation energy of molecules \citep{dewi2000a}. The
problem with including these energies is that in order to support
envelope ejection, they must be converted into kinetic energy of the
gas during \ac{CE} interaction. Because it is unclear for which of these
energies and to which extent this is the case, a fudge factor
$\alpha_\mathrm{th}$ was introduced \citep{han1995a,
  dewi2000a}. Unfortunately, the introduction of two tunable
parameters, confusingly both called $\alpha$ but with partially
different meanings\footnote{To add to the confusion the index ``th''
seems to refer to thermal energy of the envelope gas although it
parametrizes also other contributions to its internal energy.}, leads
to degeneracies in the formalism. The variation captured in
$\alpha_\mathrm{th}$ could as well enter the global $\ace$. In order
to express the binding energy as a function of the structure of the
envelope in a simple way, the structural parameter $\lambda$ was
introduced\footnote{A similar formalism can be found in
\citet{webbink1984a}.} by \citet{dekool1987b} via
\begin{equation}
  E_\mathrm{bind} = -G \frac{M_\mathrm{env} (M_\mathrm{env} +
    M_\mathrm{core})}{\lambda R},
    \label{eq:ebind_para}
\end{equation}
with the stellar radius $R$. The structural parameter $\lambda$ scales
the radius of the star to represent its degree of internal
concentration. This parameter can be determined by fitting to \ac{1D}
stellar structure models, or it can remain a free parameter (often
used in the combination $\alpha \lambda$, that summarizes both the
binding energy and the envelope ejection efficiency). Obviously, the
value of $\lambda$ for a given stellar structure depends on how the
binding energy of the envelope is defined---and thus on the
uncertainties captured in $\alpha_\mathrm{th}$ in
Eq.~(\ref{eq:ebind}). For example, using the $\lambda$ values from the
table of \citet{loveridge2011a} requires to adopt their analytical
representation for the binding energy, which differs from that of
other authors \citep[see the discussion in][]{iaconi2019a}.

Returning to the question of which parts of internal energy can
support envelope unbinding (which would justify their inclusion in the
second term in the integral of Eq.~(\ref{eq:ebind}) as counteracting
the first gravitational binding energy term), \emph{ionization energy}
of the envelope gas seems to be particularly relevant. It has first
been argued by \citet{han1995a} that as material in the expanding
envelope falls below the ionization threshold, the released
recombination energy can be used to expand or even eject the envelope.
Its inclusion in the energy budget effectively implies that the
binding energy of the envelope is less negative, or indeed even
positive, in which case \ace\ is larger than unity.  The exact
fraction of recombination energy that is used to do work is in
question at the moment because recombined hydrogen is far more
transparent to photons than ionized hydrogen, implying that a fraction
of the hydrogen recombination energy could escape as radiation instead
of being thermalized. Helium recombination energy, in contrast, may be
used more efficiently as its release happens deeper in the star, in
layers that stay optically thick for longer. Although several
treatments of this phenomenon have been presented
\citep[e.g.,][]{ivanova2018a,soker2018a}, the answer appears to be
outside the ability of analytical theory, and gives another reason to
recur to simulations that include radiation transport. Additional
processes that may be relevant but are not accounted for in the simple
energy formalism include nuclear energy generation, tidal heating,
energy released by the accretion of material onto the companion star,
radiation losses and magnetic field generation \citep{ivanova2013a}.

Instead of considering energy balance, \citet{nelemans2000a} used the
angular momentum balance to relate the pre- and post-\ac{CE} states of the
system. Assuming a linear reduction of angular momentum
$\mathrm{\Delta} J$ with mass loss $\mathrm{\Delta} M$ (which could be
equated to the envelope mass for full envelope ejection), this gives
\begin{equation}
  \frac{\mathrm{\Delta} J}{J} = \gamma \frac{\mathrm{\Delta} M}{M},
\end{equation}
again introducing a tunable parameter $\gamma$. The motivation of this
\emph{angular momentum}, or \emph{gamma-formalism} was to model the
evolution of double helium white dwarfs, which requires $\gamma
\approx 1.75$ \citep{nelemans2001a, nelemans2001b, nelemans2005b}. The
authors argued that explaining these systems with the energy formalism
fails, although \citet{webbink2008a} remarks that the reason is that
the first of the two mass transfer phases in the systems discussed by
\citet{nelemans2005b} was not a common envelope, but a
quasi-conservative mass transfer phase. While this explanation is not
perfectly satisfying, as \citet{webbink2008a} also discuss, the main
problem with the gamma-formalism stems from its lack of predictive
power. The value of $\gamma$ is in fact constrained to lie in the
narrow range between $\nicefrac{5}{3}$ and $\nicefrac{5}{8}$, within
which it predicts a wide range of values for the final orbital
separation of the post-\ac{CE} core system.

Parametric models provide a shortcut through the complex \ac{CE}
interaction for population synthesis studies.  The validity of the
energy and angular momentum formalisms is still to be confirmed with
simulations that capture the details of the actual physical
interaction. It is therefore important to point out the fundamental
differences and the relations between these formalisms and detailed
hydrodynamic simulations of \ac{CE} interaction. The simplified
descriptions are based on the global budgets of conserved quantities,
but they do not specify where energy is injected and
what fraction of it is used for unbinding the envelope. They assume
certain initial and end stages without modeling the dynamical
processes linking them and therefore give little insight into the
physics of \ac{CEE}. But even within the global view of parametric
models, questions arise: Are the efficiency parameters universal? If
not, how do they depend on the parameters of the initial models, such
as the mass ratio of the stellar components, the evolutionary state of
the primary star, or even the plunge-in time?  Consequently,
parameters such as \ace\ or $\gamma$ by themselves cannot shed much
light on the physical processes and they are of little
use to constrain \ac{3D} hydrodynamic models. Conversely, a
goal of more advanced simulations of \ac{CEE} is to fix or predict
parameters like $\ace$ to enable their further use in parametric
studies. A full treatment with detailed numerical simulations is
ultimately required to validate such parametrizations and---if
possible---generate prescriptions for parametrized approaches that
supply them with predictive power.

\subsection{One-dimensional inspiral models}
\label{sec:one_dimensional_implicit_approaches}

A more detailed description of \ac{CE} phases than that provided by
the simple parametric formalisms discussed in
Sect.~\ref{sec:parametrized_models} is obtained from \ac{1D} stellar
structure models that integrate the change of orbital separation
during the inspiral of the companion into the envelope of the primary
star as obtained from a parametrized approach to the inspiral
\citep[e.g.][]{alexander1976a, fragos2019a}. These models do not
provide a self-consistent description of the drag forces acting on the
stellar cores. With an assumed drag force model, they integrate a
\ac{1D} equation of motion for the inspiral of the companion into the
envelope of the primary star. A coarser approach summarizes the
inspiral by a bulk mass-loss rate that simply removes the envelope
\citep{xiong2017a, clayton2017a}. Before we can give details of these
models, we need to discuss analytical approximations of the
gravitational drag force.

\subsubsection{Analytical approximation of the gravitational drag}
\label{sssec:drag}

The main energy reservoir driving the \ac{CE} interaction is the ``orbit"
of the initial binary system. The mechanism of the transfer of orbital
energy to the envelope gas of the primary star is mainly due to the
interaction between the companion and the envelope as well as the core
of the giant and the envelope.

Quantifying the drag around stellar cores in \ac{CEE} is a difficult
task. Several effects contribute to the total force
\citep{livio1984a}: dynamical gravitational braking due to deflection
of surrounding gas that transfers momentum onto the companion, drag
due to accretion (or evaporation) onto (from) the companion, viscous
hydrodynamical or turbulent friction, and tidal forces.  In the case
of light companions such as massive planets and brown dwarfs, it is
easy to see analytically that hydrodynamic friction onto the companion
easily competes with gravitational friction, complicating the inspiral
\citep{staff2016a}, but for most other situations, gravitational drag
dominates.

\begin{figure}
    \centering
    \includegraphics[width=0.7\textwidth]{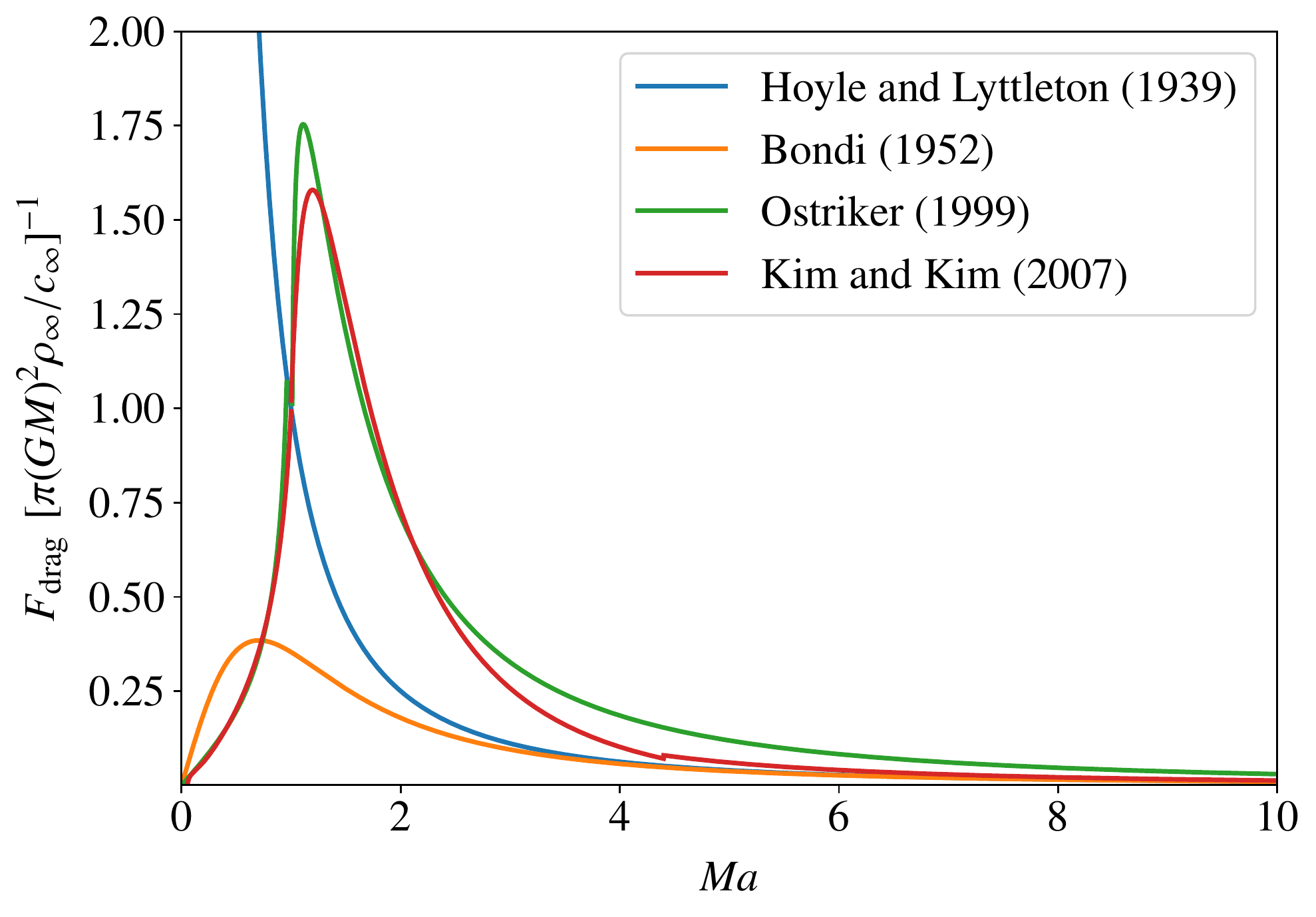}
    \caption{Drag force as a function of Mach number according to
      Eqs.~(\ref{eq:hoyle-lyttleton}), (\ref{eq:BH}),
      (\ref{eq:drag_coulomb} -- \ref{eq:I_supersonic}) assuming $r_\mathrm{max}/r_\mathrm{min} = 20$ and the
      expression given by \citet{kim2007a} with $r_\mathrm{max}/r_\mathrm{min} = 10$. Figure modified from L.~Thielbeer, Bachelor thesis, Heidelberg 2021.}
    \label{fig:drag}
\end{figure}

When a body moves through gas, gravitational interaction leads to a
deflection of the surrounding material into a wake behind the
body. Some part of the gas is accreted onto the gravitating object
\cite[see][for a detailed and pedagogical
  derivation]{edgar2004a}. This accretion flow was first studied by
\citet{hoyle1939a}, who, under the assumption of a sufficiently
rarefied gas, considered the problem in terms of a collisionless
system. Gas particles around the moving object---modeled as a point
mass---are deflected by its gravity and approach the axis of its
motion (in this case the symmetry axis of the problem) downstream of
it in a ballistic trajectory. If gas particles are gravitationally
bound to the object (i.e.\ in absolute value their kinetic energy is
smaller their gravitational potential energy) they will be accreted
with a rate that depends on the distance from the object at which they
reach the collimation axis. This, in turn, specifies a critical impact
parameter $R_\mathrm{a}$ around the moving object inside of which the
gravitational deflection directs particles to a point on the axis
where they are bound. From this simple picture, an accretion rate is
derived:
\begin{equation}
  \label{eq:hoyle-lyttleton}
  \dot{M}_\mathrm{HL} = \pi R_\mathrm{a}^2 \rho_\infty v_\infty,
\end{equation}
where $\rho_\infty$ and $v_\infty$ denote the density and velocity
upstream of the gas \citep{hoyle1939a, edgar2004a}. The subscript
`$\infty$' indicates that the values are measured sufficiently far
away from the moving object so that they are not affected by its
gravitational attraction. The critical impact parameter is usually
called the \emph{accretion radius} and it is given by
$$
R_\mathrm{a} = \frac{2 G M}{v_\infty^2}.
$$
Here, $G$ is the gravitational constant and $M$ is the mass of the
accreting object. \citet{bondi1944a} extended the analytic treatment
of \citet{hoyle1939a} by relaxing the assumption of a collisionless
system. Collimation of the flow to the downstream axis of motion by
gravitational deflection increases the density in a collisionless
system to infinity, which is unphysical. Instead, a hydrodynamic
pressure builds up and a wake of finite density forms behind the
object, which \citet{bondi1944a} called an \emph{accretion column}. The
finite density of the wake reduces the mass accretion rate to
\citep{bondi1944a, edgar2004a}
\begin{equation}
  \label{eq:bondi-hoyle}
  \dot{M} = \alpha \pi R_\mathrm{a}^2 \rho_\infty v_\infty,
\end{equation}
where, according to \citet{bondi1944a}, the factor $\alpha$ ranges
between $\nicefrac{1}{2}$ and $1$.  The setup considered by
\citet{hoyle1939a} applies to highly supersonic velocities of the
gravitating object with respect to the gas, but it fails in cases of slow motion. A
different setup that avoids dynamical effects was considered as a
limiting case by \citet{bondi1952a}: a static gravitating object
inside a gas that is initially at rest and from which it accretes in a
spherically symmetric way. This led \citet{bondi1952a} to introduce a
characteristic accretion length scale,
$$
  R_\mathrm{B}  = \frac{G M}{c^2_\infty},
$$ where $c_\infty$ denotes the sound speed far away from the object.
  Outside of this \emph{Bondi radius} the accretion flow is subsonic
  and the density is uniform; inside of it the gas becomes supersonic
  and accretes onto the object in free fall \citep{edgar2004a}. The
  nature of the flow---subsonic or supersonic---is characterized by
  the \emph{Mach number,}
\begin{equation}
  \mathit{Ma} = \frac{v}{c}.
\end{equation}
The resulting accretion rate onto the static gravitating object is given by \citep{bondi1952a}
\begin{equation}
  \label{eq:bondi}
  \dot{M} = 4 \pi \lambda R_\mathrm{B}^2 \rho_\infty c_\infty,
\end{equation}
with an order unity non-dimensional parameter $\lambda$. This
parameter\footnote{The notation of the parameters $\alpha$ and
$\lambda$ follows the original references. They are not to be confused
with the stellar structure and efficiency parameters of
Sect.~\ref{sec:parametrized_models}.} plays a role analogous to
$\alpha$ in Eq.~(\ref{eq:bondi-hoyle}). The similarity of
Eq.~(\ref{eq:bondi}), valid for the limiting case of a Mach number
$\mathit{Ma}_\infty = 0$, with Eq.~(\ref{eq:bondi-hoyle}), that
applies to the case of $\mathit{Ma}_\infty \gg 1$, led
\citet{bondi1952a} to propose the interpolation
\begin{equation}
  \label{eq:BH}
  \dot{M}_\mathrm{BH} \sim \frac{\pi (G M)^2
    \rho_\infty}{\left(c^2_\infty + v^2_\infty\right)^{3/2}},
\end{equation}
often called to as \emph{Bondi--Hoyle accretion rate}. Following
\citet{edgar2004a} we will refer to the general framework as the \ac{BHL}
formalism. The considered setups cannot be steady: material is
accreted onto the gravitating object, which not only increases its
mass, but also leads to a transfer of momentum. The momentum transfer
corresponds to a drag force. Dimensional considerations very
approximately  suggest the relation \citep{edgar2004a}
\begin{equation}\label{eq:dragforce}
  F_\mathrm{drag}  = M \dot{v}_\infty \sim \dot{M} v_\infty, 
\end{equation}
so that the accretion rates discussed above can be translated into
corresponding drag forces that act on a gravitating object moving
though gas. It is important to note that the origin of this drag force
is not a direct friction of the gas exerted on the moving object---in
the model the object is a point mass and does not have a surface---but
rather an effect of the material deflected into and accreted from its
wake \citep{edgar2004a}.  While, according to
Eq.~(\ref{eq:hoyle-lyttleton}), the drag force in the Hoyle-Lyttleton
formalism decreases monotonically with the Mach number of the relative
motion, the interpolation of Bondi [Eq.~(\ref{eq:BH})] predicts a peak
in it at $\mathit{Ma}_\infty = \sqrt{\nicefrac12}$ (see
Fig.~\ref{fig:drag}).

This, however, has an additional effect: more mass is present on the
downstream side of the moving object than upstream of it and this
causes a backward-directed gravitational pull on it---an effect that
is usually referred to as \emph{dynamical friction} and was first
studied by \citet{chandrasekhar1943a} under the assumption of a
collisionless system. \citet{dokuchaev1964a} extend the discussion to
a collisional fluid dynamical system.  For the resulting drag force in
the hypersonic case, \citet{ruderman1971a} proposed:
\begin{equation}
  \label{eq:drag_dyn}
  F_\mathrm{drag} = \dot{M} v_\infty \ln \left( \frac{r_\mathrm{max}}{r_\mathrm{min}}\right).
\end{equation}
Because of its similarity to the formalism describing Coulomb
collisions in plasma physics, the logarithm in this expression is
usually called the \emph{Coulomb logarithm.} It introduces two length
scales, $r_\mathrm{min}$ and $r_\mathrm{max}$. They result from
integration boundaries that represent the minimum and maximum radius
of gravitational interaction of the moving object with the
gas. Unfortunately, there is no clear definition of what these scales
should be and their choice remains somewhat arbitrary. In the
hypersonic case, the density asymmetry is a maximum because the
material is collimated in the region trailing the moving object. For
lower velocities, this is no longer the case and in the subsonic
regime the dynamical drag should be smaller. Therefore, the Coulomb
logarithm should involve a quantity that characterizes the
flow. \citet{rephaeli1980a} suggested a form for the supersonic case
that depends on the Mach number of the flow, $\mathit{Ma}$, but they
argued that in the subsonic case the dynamical drag should
vanish. Based on a linear perturbation analysis, \citet{ostriker1999a}
derived an expression for the drag force
\begin{equation}
  \label{eq:drag_coulomb}
   F_\mathrm{drag} = \frac{\pi R_\mathrm{a}^2 \rho_\infty}{v_\infty^2}
   I
\end{equation}
with the Coulomb logarithm
\begin{equation}
  I_\mathrm{subsonic} = \frac{1}{2} \ln \left( \frac{1 +
    \mathit{Ma}}{1 - \mathit{Ma}} \right) - \mathit{Ma}
  \label{eq:I_subsonic}
\end{equation}
  for subsonic relative velocities and
\begin{equation}
  I_\mathrm{supersonic} = \frac{1}{2} \ln \left( 1 -
  \frac{1}{\mathit{Ma}^2} \right) + \ln\left( \frac{v_\infty
    t}{r_\mathrm{min}}\right)
  \label{eq:I_supersonic}
\end{equation} 
for supersonic velocities, where $r_\mathrm{min}$ is the effective
size of the moving object. The resulting drag force is plotted in
Fig.~\ref{fig:drag} as a function of the Mach number.

Still, the specification of the upper integration limit for the
gravitational interaction in the supersonic case remains problematic
for the application of this formalism to
\ac{CEE}. \citet{ostriker1999a} considered a perturber moving through
a uniform gas on a straight line over some finite period of time $t$
and integrate over the path it travels. \citet{kim2007a} investigated
the drag force on a perturber moving on a circular orbit in a uniform
gaseous medium. The main effect of this change in geometry is that a
trailing tail forms behind the perturber in which the density
distribution is asymmetric. Therefore, the drag force has components
in lateral as well as radial direction, although \citet{kim2007a} found
that the latter contributes little to the orbital decay. Based on a
semi-analytic approach, \citet{kim2007a} provided fits for $I$ in
Eq.~(\ref{eq:drag_coulomb}) in different flow
regimes with a functional dependency on the Mach number that resembles
the form of \citet{ostriker1999a}, see also Fig.~\ref{fig:drag}.

In addition to shrinking the orbital separation between the stellar
cores and transferring angular momentum and orbital energy to the
envelope so that it is ejected, drag also affects the eccentricity of
the core binary system. \citet{szolgyen2022a} argue that while the
hydrodynamic drag acting on the companion tends to
circularize the orbit, gravitational drag can both enhance or decrease
eccentricity, depending on the density structure of the envelope.

The analytic expressions for the drag force discussed here are
important to understand the underlying effects and the functional
dependencies on characteristic quantities, chiefly the Mach number of
the relative velocity between perturber and gas. They are, however,
not fully self-consistent because the derived relations are
proportional to the drag force up to multiplicative factors, that, at
best, are constants. Whether these constants are universal or whether
they depend on details of the envelope structure and the inspiral of
the companion has to be determined in \ac{3D} hydrodynamic
simulations. In Sect.~\ref{ssec:windtunnel} we discuss further
attempts at modelling the gravitational drag using hydrodynamic
simulations that do not necessarily comprise the entire \ac{CE} system
but rather simulate objects in ``wind tunnels''. In the future, such
simulations may join forces with global \ac{CE} simulations and
\ac{1D} mechanical models to build a more complete understanding of
the entire interaction.

\subsubsection{Parametrized 1D models of dynamical common-envelope inspiral}
\label{sssec:1d_stellar_structure_models_of_the_dynamicsl_ce_inspiral}

Among the first computational models of \ac{CE} interaction were
approaches that follow the inspiral of the companion into the envelope
of the primary and, in most cases, the associated energy deposition
into the envelope gas with \ac{1D} stellar evolution codes
\citep[see][]{taam1978a, meyer1979a}. These approaches determine drag
forces and energy (and/or angular momentum injection) according to
local conditions as encountered in a \ac{1D} structure model of the
primary star's envelope. This formulation allows for a simple \ac{1D}
``mechanical'' model of the inspiral phase, i.e.\ Phase (ii) of our
classification in Sect.~\ref{sect:phases}. It is based on an equation
of motion for the companion under the action of a gravitational force
and a drag force,
\begin{equation}\label{eq:motion}
  \ddot{\vec{r}} = - \frac{G \left(M_\star^* + M_2\right)}{r^3} \vec{r} +
  \frac{1}{M_2} \vec{F}_\mathrm{drag},
\end{equation}
where $\vec{r}$ is the radius vector pointing from the center of the
primary star to the companion of mass $M_2$
\citep{alexander1976a}. The mass of the primary star inside the radius
$\vec{r}$ is denoted as $M_\star^*$.

Obviously, such a model is a gross simplification of the physical
processes at work in \ac{CEE}. It assumes a symmetry of the setup that
diverges from reality, where the envelope is deformed by the
gravitational interaction. Moreover, for the determination of the
orbital shrinkage and the associated energy (and angular momentum)
deposition due to drag, it considers only the motion of the companion
in the envelope of the primary, whose core is assumed to be fixed in
space. This setup thus misses the drag force that acts on the core of
the primary star when it orbits the center of mass of the system and
thus also moves relative to the envelope gas.

Nonetheless, \ac{1D} ``mechanical'' models extend beyond the simple
arguments of energy or angular momentum conservation used in the
formalisms discussed in Sect.~\ref{sec:parametrized_models}. They
attempt a physical description of the dynamical interaction between
companion and envelope, albeit in a parametric way: The inspiral of
the companion into the envelope is followed according to a drag force
acting on it. This model raises questions: (i) What exactly is the
nature of the interaction between the companion and the envelope gas
and how can the drag force $F_\mathrm{drag}$ perturbing the orbital
motion in Eq.~(\ref{eq:motion}) be modeled?  (ii) How does the
envelope react to the injection of energy and angular momentum and
what are the consequences of structural changes for the drag force?
 
As discussed in Sect.~\ref{sssec:drag}, Question (i) has been
addressed in analytic treatments in the framework of linear theory,
resulting in scaling relations \citep[e.g.][]{dokuchaev1964a,
  ostriker1999a}. The semi-analytical model of \citet{kim2007a}
extends the original setup of linear motion to the more appropriate
case of circular orbits. The omission of nonlinear effects and density
gradients in the envelope is accounted for by fudge factors that have
to be determined from numerical simulations. We discuss \ac{3D}
simulations which attempt to quantify the gravitational drag force in
Sect.~\ref{ssec:windtunnel}. They have been calculated either in
wind-tunnel type models (where high spatial resolution can be achieved
around the point mass that acts as the inspiralling companion, see
\citealp{macleod2015a} \citealp{macleod2017a}, and \citealp{de2020a}),
for massive perturbers on circular orbits in a gaseous medium
\citep{kim2010a}, or in a full \ac{CE} simulation where the reaction of the
envelope to the inspiral can be gauged, but the resolution is inferior
\citep{reichardt2019a, chamandy2019a}. Parametric approaches to the
description of the drag force have been used relatively rarely in
\ac{1D} ``mechanical'' models. Perhaps the most modern and
comprehensive approach was that of \citet{fragos2019a}, who used the
parametrization of the gravitational drag determined by the \ac{3D}
hydrodynamic models of \citet{macleod2015a}.

Question (ii) concerns the expansion and change of structure of the
envelope. In the simplest approach, the envelope is treated as static
and unperturbed by the inspiral, as assumed, for instance, in the
study of \citet{macleod2015b}. In principle, however, the co-evolution
of orbit and envelope can be followed in \ac{1D} hydrodynamic stellar
evolution models. The idea is to inject the orbital energy and/or
angular momentum released by the drag force into the stellar envelope
and to model its response in a \ac{1D} hydrodynamic stellar evolution
code. Different groups have developed recipes for this: Some inject
released orbital energy into the thermal energy of the envelope, while
others model angular momentum transfer to envelope material that also
increases its kinetic energy. The employed \ac{1D} parametrizations
cannot self-consistently determine what fractions of energy and
angular momentum are to be injected into each of these
reservoirs. Consequently, the implementations vary, but the general
approach has been followed in a number of early \ac{CE} studies
\citep{taam1978a, meyer1979a, taam1979a, delgado1980a,livio1984a,
  livio1984b, soker1984a}. In the late 1980s and the 1990s, \ac{1D}
``mechanical'' models were abandoned because of the asymmetries
observed in \ac{3D} hydrodynamic simulations that called the
assumption of spherical symmetry into question. The work of
\citet{podsiadlowski2001a}, however, revived the method.

The list of deficiencies of \ac{1D} ``mechanical'' \ac{CE} models' is long: The
assumed spherical symmetry in the parametrized models is a gross
simplification. It implies a local transfer of orbital energy and
angular momentum and injects them into a spherical shell. In reality,
however, the gravitational perturbation of the envelope is non-local
and the energy is not deposited into spherical shells. In fact, the
plunge-in of the companion quickly destroys any sphericity of the
primary star's envelope. Moreover, the orbit in the \ac{1D}
parametrized models remains circular by necessity, with concomitant
non-conservation of angular momentum.

Despite all these shortcomings, however, such models have their value
and hold promise for capturing effects that would be very difficult to
include in \ac{3D} hydrodynamic simulations of \ac{CEE}. Their recent
application is therefore motivated by the need to bridge scale gaps
that are hard to overcome in \ac{3D} hydrodynamic models.

An example is the inspiral of neutron stars into a massive red
supergiant primary star, which pose a strong spatial scale
challenge. This was addressed with a \ac{1D} parametrized model by
\citet{fragos2019a} generating information about the final product: a
binary comprising a $2.6 \,\msun$ helium star and a neutron star at a
separation of $3.3$ to $5.7 \,\rsun$.

\citet{clayton2017a} explore the evolution of \ac{CE} systems after the
inspiral phase (ii) when entering post-inspiral evolution [Phase (iii)
  of our classification in Sect.~\ref{sect:phases}], which poses a
time-scale challenge. Generally, during the slower phases, several
physical processes such as radiation transfer, convective energy
transport, and nuclear burning, can have a substantial effect on the
interaction and need to be modelled. These longer phases are
themselves needed to understand the interaction as a whole, so the
role of \ac{1D} parametrized models could be critical and was
advocated as the most important step to further progress by
\citet{ivanova2013a}.

\subsection{Three-dimensional hydrodynamic models}
\label{sec:multi_dimensional_hydro_models}

The concept of \ac{CE} interaction is deceivingly simple, but the
details are complex. As discussed above, finding a trustworthy way to
parametrize actual \ac{CEE} turns out to be extremely
challenging. Given a set of initial binary parameters, parametric
formalisms and \ac{1D} models are still unable to reliably predict the
outcome of the interaction.

This is why the first \ac{3D} hydrodynamics attempts started as early as
1987 \citep{dekool1987a}. After early works \citep[see
  also][]{livio1988a, terman1994a}, efforts with modern codes are
those of \citet[][]{rasio1996a}, who used a \ac{SPH} technique and of
\citet[][]{sandquist1998a}, who used a static nested-grid technique
(see Sect.~\ref{sect:sim_global} for details). These simulations
generated a number of useful pieces of information, but they must be
considered as being at the toy-model level from today's perspective,
mostly because the vast ranges of spatial and temporal scales to be
resolved was unmatched by early computational resources.

A second wave of models involved a new suite of codes, both using
Eulerian grid and SPH techniques
\citep{ricker2008a, ricker2012a, passy2012a}. These modern efforts were
fortunately soon followed by a number of additional simulations, which
somewhat expanded the considered parameter space, and involved yet
more codes and techniques
\citep[e.g.,][]{nandez2014a, iaconi2016a, ohlmann2016a, ohlmann2016b,
  chamandy2018a, prust2019a, reichardt2019a}.

The main appeal of \ac{3D} hydrodynamic simulations is that they can
in principle model the interaction between the stellar cores and the
envelope gas in a self-consistent way, although scale problems and
resolution issues seriously hinder this prospect. Moreover, \ac{3D}
simulations of the fast inspiral revealed just how non-spherical the
CE interaction is, with outflows relatively close to the equatorial
plane, and they revealed the complex flow structures involved in Phase
(ii) of \ac{CE} interaction. They also clearly showed that ejecting
the envelope using only orbital energy is very difficult, if not
impossible, and paved the way to a deeper understanding of the energy
exchange, including the impact of recombination energy on the
expanding envelope. While---contrary to initial, perhaps na\"ive,
hopes---\ac{3D} hydrodynamic \ac{CE} simulations have not provided a
convincing value of $\ace$ for the energy formalism discussed in
Sect.~\ref{sec:parametrized_models}, they do provide a far deeper
understanding of the parameter itself. These and other discoveries,
which will be described in detail in Sect.~\ref{sect:sim_global}, have
convinced many groups to pursue this investigation avenue. Computers
and numerical techniques are, after all, advancing rapidly and enable
ever more complex simulations. Yet, this enthusiasm also gives way to
the realization that, in the end, it might be a clever combination of
techniques that will take the podium.

\section{Physical modeling and challenges}
\label{sect:challenges_physical}

The system to be modeled consists of (at least) two stellar objects, a
giant star and a companion (which can be a star but also a compact
object or even a planet). This implies that the objects themselves can
be described with a combination of fluid dynamics, thermodynamics, and
gravity. The main problem for simulating the evolution of \ac{CE}
systems arises from the discrepancy of scales---most obvious in the
required distinction of an extended envelope from a small core in the
structure of the giant primary stars (see also
Sect.~\ref{ssec:terminology}).

Thus, before turning to the numerical implementation, the physical
model for \ac{CEE} has to be set out. We review it here together with
the challenges that arise from the \emph{physical modeling}
itself---\emph{numerical} implementation problems often resulting from
the chosen physical model will be discussed in
Sect.~\ref{sect:challenges_numerical}.

\subsection{The gravo-hydrodynamic model}
\label{sect:gravohydro}

Astrophysical models are commonly plagued by the different nature the
various physical processes at play and the wide ranges of spatial and
temporal scales on which they act. Models of \ac{CE} phases are no
exception. In fact, \ac{CEE} is driven by a complex interaction of
physical phenomena combining fluid dynamics, thermodynamics, atomic
physics, magnetic fields, radiation, accretion physics, and perhaps
other effects.  While some of the approaches discussed in
Sect.~\ref{sect:understanding} are based on coarse approximations, we
are interested here in a modeling basis that rests on fundamental
physical principles and avoids approximations.

The most basic model of \ac{CEE}---which we will refer to as
\emph{gravo-hydrodynamic model} in the following---implements the
equations of fluid dynamics, most commonly the Euler
equations\footnote{A more general approach could use the
Navier--Stokes equations instead, that explicitly account for
viscosity. However, because numerical viscosity dominates in
implementations, physical viscosity can usually be neglected.},
combined with a gravitational source term. The first of them is a
scalar equation,
\begin{equation} \label{eq:mass_cons}
  \frac{\partial \rho}{\partial t} + \nabla \cdot (\rho \vec{v}) = 0,
\end{equation}
with $\rho$, $t$, and $\vec{v}$ denoting mass density, time, and
velocity, respectively. It expresses \emph{mass conservation.} The
second (vector) equation is the \emph{momentum balance,}
\begin{equation} \label{eq:momentum_cons}
  \rho \frac{\partial \vec{v}}{\partial t} + \rho \left(\vec{v} \cdot
  \nabla\right) \vec{v} + \nabla P = - \rho \nabla \mathrm{\Phi},
\end{equation}
which accounts for pressure forces $\nabla P$ in the momentum flux and
includes gravity (with the gravitational potential $\mathrm{\Phi}$) as a source
term. Finally, \emph{energy balance} is expressed in terms of the
specific total energy density,
\begin{equation}
  e_\mathrm{tot} \defeq \frac{1}{2} |\vec{v}|^2 +
  \frac{\epsilon}{\rho},
\end{equation} 
where $\epsilon$ denotes the density of the internal energy. This
gives the last (again scalar) equation
\begin{equation} \label{eq:energy_cons}
  \frac{\partial \rho e_\mathrm{tot}}{\partial t} + \nabla \cdot (\rho
  e_\mathrm{tot} \vec{v}) + \nabla \cdot (P\vec{v}) = - \rho \vec{v}
  \cdot \nabla \mathrm{\Phi}.
\end{equation}

Equations (\ref{eq:mass_cons}), (\ref{eq:momentum_cons}), and
(\ref{eq:energy_cons}) are formulated for a fixed frame of
reference. This is called \emph{Eulerian specification}. An
alternative is obtained from transforming the system into a frame of
reference co-moving with the fluid. This is achieved by introducing
the \emph{substantial} (or \emph{Lagrangian}) derivative
\begin{equation}
\frac{\DD}{\DD t} \defeq \frac{\partial}{\partial t} + \vec{v} \cdot \nabla.
\end{equation}
After this transformation, the Euler equations in \emph{Lagrangian specification} read
\begin{eqnarray}
  \frac{\DD \rho}{\DD t} + \rho \nabla \cdot \vec{v} &=& 0,\label{eq:lagrange_mass}\\
  \rho \frac{\DD \vec{v}}{\DD t} + \nabla P &=& - \rho \nabla \mathrm{\Phi},\label{eq:lagrange_momentum}\\
  \rho \frac{\DD e_\mathrm{tot}}{\DD t} +  \nabla\cdot P \vec{v} &=& -
  \rho \vec{v} \cdot \nabla \mathrm{\Phi}.\label{eq:lagrange_energy}
\end{eqnarray}
A special case is encountered for situations where gravity is exactly
balanced by the pressure gradient. In this case, the acceleration in
Eq.~\ref{eq:lagrange_momentum} vanishes. This mechanical equilibrium
is an excellent approximation for the stratification of stellar
envelopes in long-lasting stages of stellar evolution. It is called
\acfi{HSE}:
\begin{equation}\label{eq:hse}
\frac{\nabla P}{\rho} = - \nabla \mathrm{\Phi}.
\end{equation}

The gravitational potential is determined by Poisson's equation:
\begin{equation}\label{eq:poisson}
  \mathrm{\Delta} \mathrm{\Phi} = 4 \pi G \rho,
\end{equation}
where $G$ is Newton's gravitational constant.

The Euler equations form a system of hyperbolic partial differential
equations and thus pose an initial value problem. For their solution,
initial conditions must be supplied. We devote Sect.~\ref{sect:setups}
on how to set up \ac{CE} simulations. To close the system of Equations
(\ref{eq:mass_cons}), (\ref{eq:momentum_cons}), and
(\ref{eq:energy_cons}), an equation of state is needed that relates
pressure, density and internal energy.  The equation of state is an
essential part of our model system. It accounts for micro-physical
interaction between the constituents of matter in the effective
macroscopic framework of thermodynamics. In the simplest case, an
ideal gas is assumed as a model for stellar material, but radiation
pressure can be added via the Stefan--Boltzmann law in a
straightforward way.

Obviously, the physical model described by this set of equations is
not very specific for the problem at hand---the same set of equations
forms the basis for the theoretical treatment of many astrophysical
objects ranging from planets to the large-scale structures in the
Universe\footnote{These equations also form the foundation of
classical stellar modeling. Under the simplifying assumptions of
spherical symmetry and hydrostatic equilibrium, the first two of the
stellar structure equations can directly be derived from them. Energy
balance with suitable source terms gives the third of the stellar
structure equation. In addition, a prescription of energy transport in
stellar material is needed that constitute the fourth equation of the
set.}. Despite its generic nature, however, the gravo-hydrodynamic
model captures the gravitational and hydrodynamic interaction between
the orbiting cores and the envelope. Setting up a \ac{CE} model in
this framework, gravitational (and also hydrodynamic) drag is
naturally included. The model thus self-consistently accounts for the
transfer of orbital energy and angular momentum from the core binary
to the envelope gas---one of the most important processes in
\ac{CEE}. In addition, the model also captures convection if it occurs
in the stellar envelope.

From a physical point of view, the gravo-hydrodynamic model does not
pose particular challenges, but analytic solutions of the system of
coupled nonlinear hyperbolic partial differential equations are
virtually impossible for complex setups. Therefore, numerical
approaches are needed and---as discussed in
Sect.~\ref{sect:challenges_numerical}---neither fluid dynamics nor
gravitational interaction are easy to deal with computationally.

\subsection{Why do we need multi-D hydrodynamical simulations?}
\label{sect:whymultid}

\begin{figure}
  \centering
  \includegraphics[width=\textwidth]{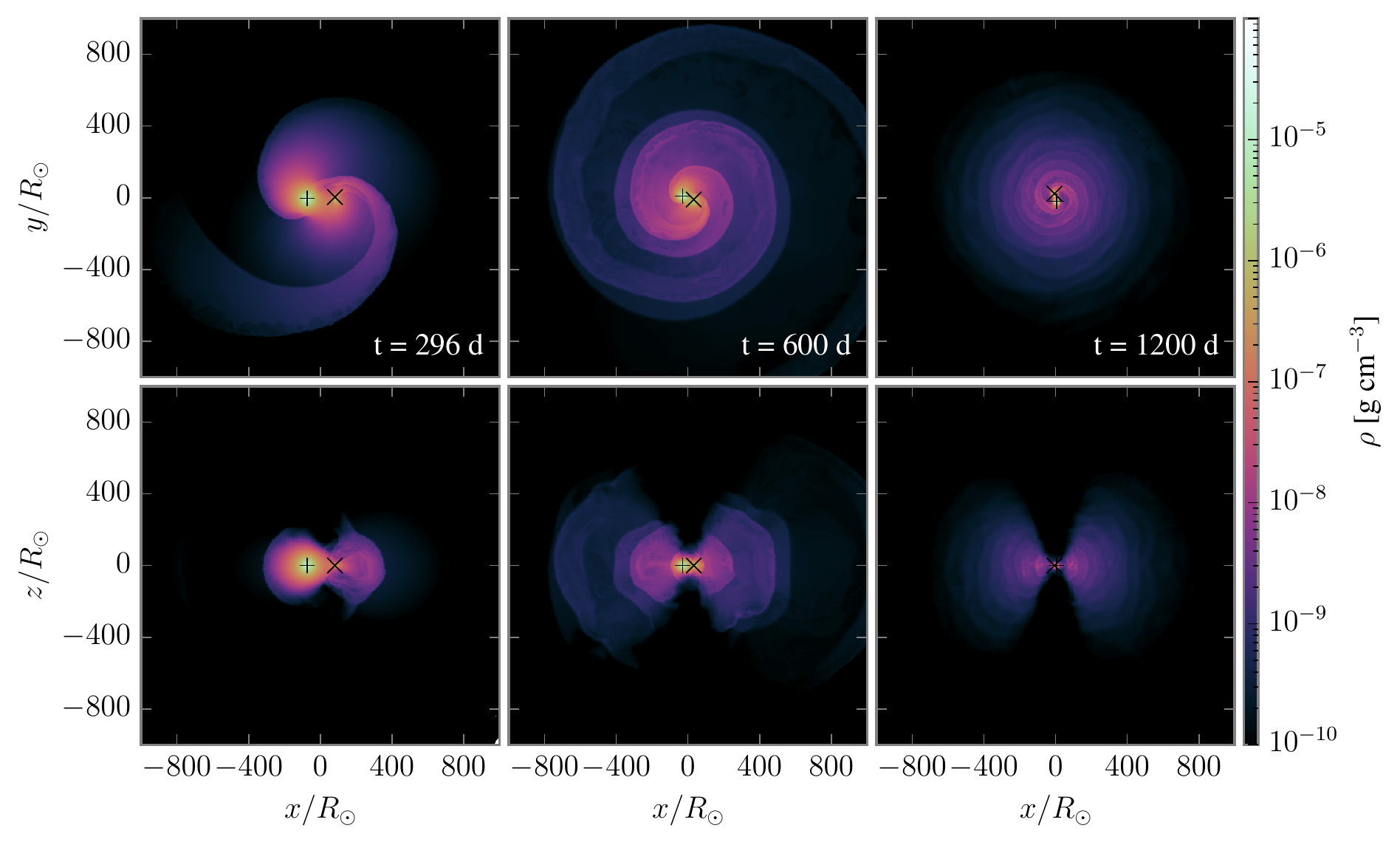}
  \caption{Density on slices through the orbital plane (top row) and
    perpendicular to it (bottom row) in a \ac{3D} hydrodynamic
    simulation of \ac{CE} inspiral of a $0.49 \, \msun$ companion
    (marked with a $\times$ symbol) into a $0.97 \, \msun$ early
    \ac{AGB} star \citep[core marked with a $+$ symbol; Model O.50
      of][courtesy of Christian Sand]{sand2020a}. The times of the
    snapshots after initiation of the \ac{CE} interaction are
    indicated in the respective upper panels.}
  \label{fig:ce_3D}
\end{figure}

In Sects.~\ref{sec:parametrized_models} and
\ref{sec:one_dimensional_implicit_approaches} we have listed the
shortcomings of parametric and \ac{1D} mechanical models and in
Sect.~\ref{sec:multi_dimensional_hydro_models} we have explained that
combining these models with \ac{3D} hydrodynamics models can fill in the
glaring gaps in our understanding of \ac{CEE}.  This in particular
applies to the inspiral of the stellar cores into the \ac{CE},
i.e.\ Phase (ii) of our classification in Sect.~\ref{sect:phases}:
Only \ac{3D} hydrodynamic approaches capture the physics of the
interaction between the stellar cores and the envelope material
consistently. They resolve the flow around the cores and the changes
in the structure and morphology of the envelope. Adopting the
gravo-hydrodynamical model described in Sect.~\ref{sect:gravohydro},
\ac{3D} simulations are able to tell where and how the transfer of
orbital energy and angular momentum from the core binary onto the
envelope material takes place. In this sense, \ac{3D} hydrodynamic
simulations are the key to understanding the elusive physics of \ac{CE}
interaction.

As \ac{CEE} progresses, the envelope will strongly diverge from
spherical symmetry. In an edge-on projection, the density structure of
the envelope assumes a toroidal shape (Fig.~\ref{fig:ce_3D}, bottom
row). It is immediately clear that perturbations induced to the
envelope by the inspiralling core binary system cannot be captured by
spherically symmetric models. In addition to the toroidal deformation
of the envelope, there is a pronounced spiral shock structure issuing
from the core binary that dominates the density structure
(Fig~\ref{fig:ce_3D}, top row). In later phases of the evolution,
shear-induced hydrodynamic instabilities become visible
(Fig.~\ref{fig:ce_3D}, snapshots at $600\, \mathrm{d}$ and at
$\num{1200} \, \mathrm{d}$).  While all this does not necessarily mean
that the basic results of the interaction such as envelope ejection
and orbital decay of the core binary system cannot be represented in
\ac{1D} models or parametric formalisms, it seems unlikely that the
complex hydrodynamic flows observed in \ac{CEE} would give rise to
universal parameters and thus self-consistent predictive prescriptions
in such approaches. Full \ac{3D} hydrodynamic models remain
indispensable and may be needed to calibrate the simplified models in
very fragmented ranges of parameter space.

Moreover, as discussed in Sect.~\ref{sssec:drag}, determining the drag
around an obstacle is an extremely challenging problem of \ac{3D}
hydrodynamics. A reliable description of accretion and deflection of
material around the companion star in \ac{CEE} and the formation of a
turbulent wake behind it are of fundamental importance to predict the
envelope ejection efficiency and the orbital shrinkage. In complex
setups with non-uniform background densities and circular orbits,
\ac{3D} hydrodynamic simulations are required to capture the effects
in a quantitative way.

Clearly, \ac{3D} simulations of the inspiral phase of \ac{CEE} are a
fundamental building block for developing a comprehensive model of
this important phase of binary stellar evolution and they are the
focus of our review. However, despite their obvious advantages, it is
unlikely that they are sufficient.  Adopting a fully \ac{3D} geometry
requires to take a huge leap in the methods and in the computational
resources spent on the modeling. Even pushing techniques, algorithm
efficiency and computational resources to the limit, there are some
inherent difficulties that make it impossible to carry out full
\ac{3D} hydrodynamic simulations for all the relevant timescales
involved in the problem in the foreseeable future---in particular if
pre-\ac{CE} and post-inspiral phases [Phases (i) and (iii) defined in
  Sect.~\ref{sect:phases}] are to be part of the model. So, while in
principle the \ac{3D} hydro-gravitational model takes into account the
interactions between envelope gas and core/companion system,
limitations such as resolution may curtail the initial
advantages; see also Sect.~\ref{sect:unresolved_num}.

\subsection{CE ejection and the end of the dynamical inspiral}
\label{ssec:why_is_CE_ejection_hard_to_achieve}

Simulating envelope ejection is a main goal of the numerical modeling
of \ac{CEE}, but it is also one of the most elusive aspects of the
interaction. In models, complete envelope removal seems hard to
achieve. What are the physical aspects that determine the process and
why is it challenging to incorporate them into numerical simulations?
The question of envelope ejection is fundamentally related to the
question of the final orbital separation of the stellar cores after
completing \ac{CEE}, although, as discussed in
Sect.~\ref{ssec:interfacing_3d_global_simulations_with_1d_simulations_of_the_ce_remnant},
it may not be the only process determining it.

\begin{figure}
    \centering
    \includegraphics[width=0.75\textwidth]{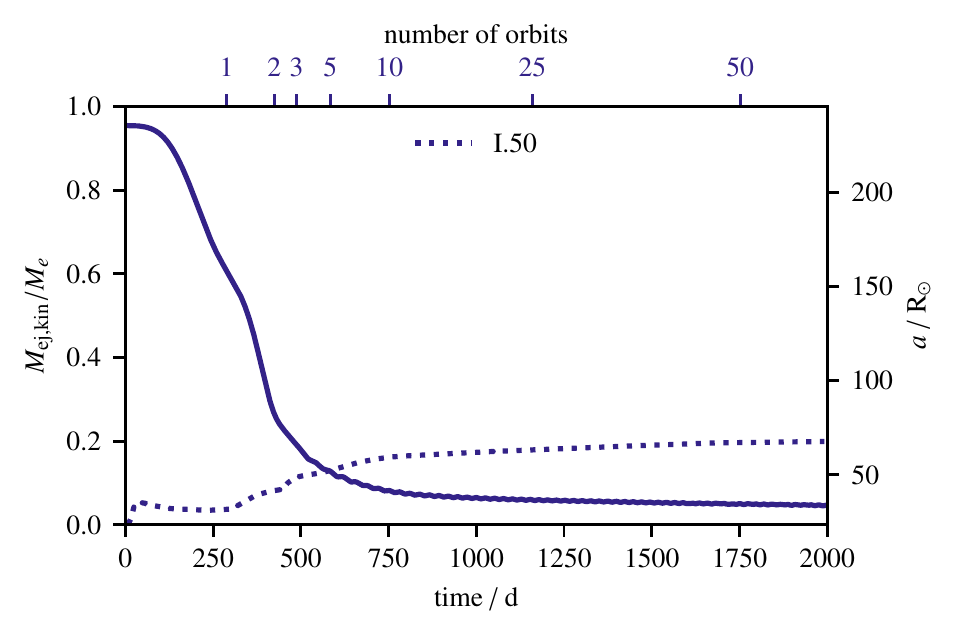}
    \caption{Simulation of \ac{CEE} in the gravo-hydrodynamic model
      for a system with a $1.0 \, \msun$ early-\ac{AGB} primary star
      with a companion of mass ratio $0.5$. An ideal-gas equation of
      state was assumed in envelope material (Simulation I.50 of
      \citet{sand2020a}, see Sect.~\ref{sect:sim_movingmesh}). Shown
      are the orbital evolution (solid line, right axis) and the
      fraction of unbound envelope mass (dotted line, left axis)
      according to the criterion of counting gas in a cell as unbound
      if its kinetic energy is larger than its gravitational binding
      energy. Figure courtesy of Christian Sand.
    \label{fig:sand_01}}
\end{figure}

The gravo-hydrodynamic model implements the basic mechanism of
\ac{CEE}: orbital shrinkage and unbinding of envelope gas due to
transfer of energy and angular momentum from the core binary to the
envelope. In its simplest version, it describes the stellar envelope
material with an ideal gas equation of state, in some cases adding
radiation pressure.  This model, however, fails in all \ac{3D} simulations
that have been carried out so far. The orbital shrinkage stalls or
retards when a significant fraction of the envelope is still bound
(see Fig.~\ref{fig:sand_01} for an example). This leads to the
question of how to measure the efficiency of envelope unbinding in a
numerical simulation.  In a common approach, material is counted as
bound if its total energy, determined as the sum of its gravitational
potential energy and its kinetic energy is negative. In
Sect.~\ref{sect:ionization}, we discuss other forms of energy that
could enter this estimate, but here we note that generally arguments
based on energy criteria may overestimate the amount of unbound
material: Whether or not material is ejected depends not only on its
energy but also on its location within the complex perturbed envelope
where the total energy is positive. If surrounded by bound material,
an unbound fluid element may dissipate its energy without being
ejected.

\begin{figure}
    \centering
    \includegraphics[width=\textwidth]{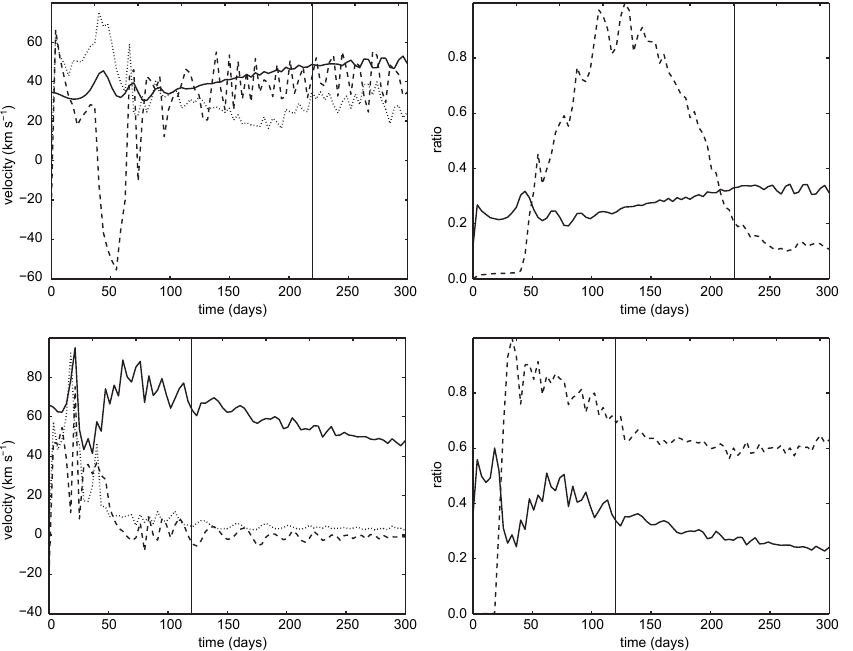}
    \caption{Parameters playing a role in the gravitational drag in
      simulations of \ac{CE} interaction of a $0.88 \, \msun$, $90 \,
      \rsun$ \ac{RGB} primary star (top row) and a $2.0 \, \msun$, $66
      \, \rsun$ \ac{RGB} primary star (bottom row) with a $0.6 \,
      \msun$ companion. The vertical lines approximately represent the
      points where the rapid inspiral terminates. Left column:
      companion velocity (solid line), local average gas velocity
      projected on the direction of the companion velocity (dashed
      line), and local average gas velocity perpendicular to the
      direction of the companion velocity (dotted line).  Right
      column: companion Mach number (solid line) and normalized
      average gas density in the companion’s proximity (dashed
      line). Figure adapted from \citet{iaconi2018a}.}
    \label{fig:Iaconi+18-Fig12}
\end{figure}

The failure of the basic gravo-hydrodynamic \ac{CE} model to eject the
envelope shows that at some point in the evolution the energy transfer
from the cores to the gas becomes inefficient. The fact that the
orbital decay ceases (or becomes very slow) indicates a weaker drag
acting on the cores.  There are several effects that can potentially
contribute to this phenomenon:
\begin{itemize}
  \item The inspiral of the companion and the associated transfer of
    orbital energy from the core binary to the surrounding gas perturb
    the \ac{HSE} of the envelope such that it expands globally. In the
    vicinity of the cores, the bow shocks caused by the supersonic
    orbital motion heat the gas while at the same time the gas is
    dragged along by the companion and spun up. This injection of
    energy and angular momentum causes envelope expansion.  As a
    result, the density in the vicinity of the cores decreases with an
    overall reduction of the drag force.  In the analytic treatment of
    Eqs.~(\ref{eq:BH}) and (\ref{eq:drag_dyn}), the effect corresponds
    to $\rho_\infty \to 0$. Thus, orbital energy release has two
    counteracting implications: it drives expansion of the envelope
    and makes its ejection possible in the first place, but, at the
    same time, dilution of the gas around the stellar cores reduces
    the efficiency of energy conversion. This makes it hard to predict
    whether envelope ejection will succeed in the gravo-hydrodynamic
    model.
  \item The perturbation of the envelope injecting angular momentum as
    described above may bring the gas into co-rotation with the core
    binary. For a vanishing relative velocity between core and gas,
    $v_\infty \to 0$ or, equivalently $\mathit{Ma} \to 0$, the drag
    force vanishes, see Eqs.~(\ref{eq:BH}), (\ref{eq:dragforce}),
    (\ref{eq:drag_dyn}), and (\ref{eq:drag_coulomb}). For example,
    \citet{reichardt2019a} found that establishing co-rotation stalls
    their inspiral of a $0.6\, \msun$ point mass companion into the
    envelope of a $0.88\, \msun$ \ac{RG} star. Similarly,
    \citet{iaconi2018a} analyzed the reason for the halting of the
    inspiral in two simulations with giants of different masses (0.88
    and 2.0~\msun) and a range of companions (from 0.1 to
    0.9~\msun). They found that only the heavier companions can
    evacuate the orbit and bring gas into co-rotation. The lighter
    companions struggle to do either (see
    Fig.~\ref{fig:Iaconi+18-Fig12}). They also found that for lighter
    envelopes (lighter primaries) it takes a more massive companion to
    evacuate the orbit and/or bring the gas into co-rotation.
  \item The Mach number of the orbital motion may decrease because the
    local sound speed increases when the companion reaches the inner,
    high-temperature parts of the envelope. Except for the
    Hoyle--Lyttleton formalism [Eq.~(\ref{eq:hoyle-lyttleton})], all
    analytic treatments discussed in Sect.~\ref{sssec:drag} include
    the subsonic regime, where the drag force decreases. In these
    models, the drag acquires a maximum value at some finite Mach
    number and vanishes for $\mathit{Ma} \to 0$, see
    Fig.~\ref{fig:drag}. The more elaborate treatments of
    \citet{ostriker1999a} and \citet{kim2007a} predict a maximum of
    the drag force in the range $1 < \mathit{Ma} <
    2$. \citet{staff2016a} showed that a transition through the sonic
    point may be responsible for the reduction of the the
    gravitational drag in their simulation. This was not observed in
    the simulations of \citet{iaconi2018a}, possibly due to the fact
    that even their lightest companions were chosen substantially more
    massive than the planets studied by \citet{staff2016a}.
  \item Another explanation of why the orbital separation stabilizes,
    is that the gravitational force between the cores increases as the
    cores approach each other. Therefore, one needs a larger drag
    force to noticeably perturb the orbital separation. Thus when the
    cores approach each other, maintaining the same rate of inspiral
    requires a larger and larger force, just when in fact orbit
    evacuation and a reduction of the velocity contrast is reducing
    the drag.
\end{itemize}

It is conceivable that all of these effects in
combination play a role and conspire to halt the orbital decay. For a
better physical understanding, an analysis of the relative
contribution and relevance of the processes remains essential and it
has to be based on \ac{3D} hydrodynamic simulations. This should be
possible, because all listed effects are in principle accounted for in
the context of the gravo-hydrodynamic base model of \ac{CEE}.

It is possible, indeed likely, that the dynamics of the envelope would
be different if physical effects beyond the simple gravo-hydrodynamic
model were accounted for. As discussed in
Sect.~\ref{sect:understanding}, an important phenomenon is the change
in ionization as the envelope expands and cools below the ionization
temperatures of hydrogen and helium. The released ionization energy
can lead to a substantial increase in envelope unbinding and---for
some systems---can even achieve complete envelope ejection
\citep{nandez2016a, ohlmann_phd, reichardt2020a, prust2019a, sand2020a,
  ondratschek2022a}.

While the debate continues on how much of the recombination energy can
actually contribute to envelope unbinding
\citep{ivanova2018a, soker2018a}, we question how the end of the inspiral
actually takes place and the extent to which it can be simulated by
explicit simulations. It is likely that even if the envelope is not
entirely ejected, material may be ``lifted'' sufficiently to stall the
inspiral. Some of this material could, however, fall back rapidly,
even dynamically/ballistically and the inspiral could resume. 

As the
non-ejected envelope contracts, a ``self-regulated inspiral'' could be
established \citep{meyer1979a, ivanova2013a}: the envelope adjusts
such that orbital energy is released at the same rate as convection in
the remaining envelope can transport it to the surface where it is
radiated away. If the heating becomes too weak, slow contraction
revives the interaction of cores and gas, while too strong an
interaction leads again to an expansion and dilution of the gas. In
this sense, an equilibrium between orbital energy release and energy
transport to the stellar surface is established and it is kept stable
by self-regulation. This is certainly a slow process and the orbital
evolution in the self-regulated inspiral regime is difficult or
impossible to follow in \ac{3D} hydrodynamic simulations. Another
complication is that the in-falling gas would have substantially more
angular momentum, making the final stable configuration difficult to
predict. It is possible that instead of a self-regulated envelope, a
circumbinary disk could forms that interacts with the orbit of the
stellar cores and modifies its parameters. Clearly, understanding the
end of the inspiral extends beyond the applicability of the basic
gravo-hydrodynamic model of \ac{CEE}.

\subsection{Why are common-envelope phases a multi-physics problem?}
\label{sect:whymultiphysics}

The basic gravo-hydrodynamic description of \ac{CEE} certainly does
not capture all effects associated with this phase of binary stellar
evolution. A model extension to account for additional physics must be
decided according to its relevance for specific aspects of \ac{CEE}:
some effects may pertain to the dynamical inspiral [Phase (ii)]
itself, others to the pre-\ac{CE} or post-inspiral phases [Phases (i) and
  (iii)], or even to later phases such as the shaping planetary nebulae
expected to form in connection with envelope ejection of some
stars. Including additional effects requires to modify the set of
model equations (\ref{eq:mass_cons}), (\ref{eq:momentum_cons}), and
(\ref{eq:energy_cons}).

Even without diverging very far from the basic gravo-hydrodynamic
model as discussed in Sect.~\ref{sect:gravohydro}, the results of
simulated \ac{CEE} can change dramatically.  Modeling the envelope of
giant stars and their \ac{CE} as ideal gas is reasonable and commonly
used in simulations for lower mass stars, while for massive stars
radiation pressure needs to be taken into account. This can
be done with an additive term in the equation of state.

In the following, we discuss changes in the ionization structure of
the envelope gas and magnetic fields as important ingredients for
model extensions. Other phenomena that are not part of the basic
gravo-hydrodynamic model of \ac{CEE} but still may be relevant include
nuclear reactions that may occur in dense accreted material, neutrino
cooling in the vicinity of compact object companions, accretion of
material, jet formation, evaporation of the companion, and perhaps
even relativistic effects. A rigorous assessment of the relevance of
these effects for \ac{CEE} is still missing, but some aspects of
accretion onto the stellar cores, the potential formation of a disk
and jet-like outflows have been explored in more detail
\cite[e.g.][]{macleod2015a, macleod2015b, chamandy2018a, soker2016b,
  moreno2017a, lopez2019a, shiber2019a, lopez2020a, lopez2022a, hillel2022a,
  zou2022a, schreier2019a, schreier2021a, schreier2023a}.

These examples illustrate that \ac{CEE} is a pronounced multi-physics
problem. Progress in understanding the dynamical evolution of the
system cannot only achieved by improving the numerical methods for
simulating it. The extension of the gravo-hydro\-dynamic base model
with additional effects is essential to reach a reliable physical
description.

\subsubsection{Ionization and recombination effects}
\label{sect:ionization}

\begin{figure}
    \centering
    \includegraphics[width= 0.7\textwidth]{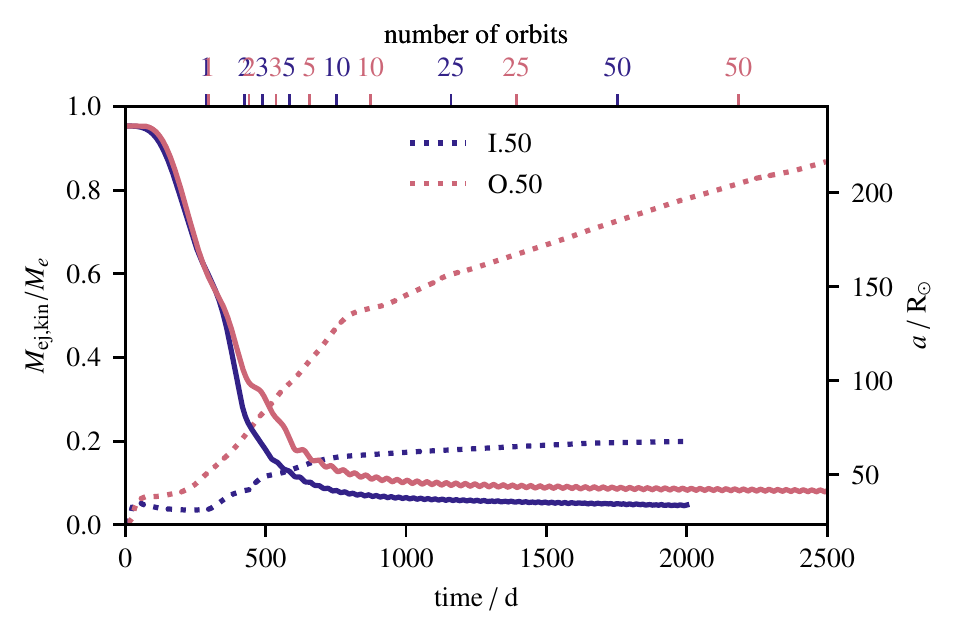}
    \caption{Fraction of unbound envelope mass (dotted lines, left axis) and
      orbital evolution of the cores (solid lines, right axis) for Simulation I.50 of
      \citet{sand2020a} as shown in Fig.~\ref{fig:sand_01} (blue
      color). In addition, in red color we plot the same quantities,
      but this time for Simulation O.50 of \citet{sand2020a}, that
      accounts for changes in the ionization of envelope
      material. Figure courtesy of Christian Sand.}
    \label{fig:sand2020}
\end{figure}

Over the past decades it has been realized that ionization effects
play an important role in \ac{CEE} \citep{ivanova2015a}. As the
envelope is lifted and starts to expand due to the perturbation by the
inspiralling companion, its outer layers may gradually fall below the
ionization thresholds of hydrogen and helium and previously ionized
material recombines.

This is a micro-physical matter effect leading to
a source term in the energy equation of fluid dynamics
(\ref{eq:energy_cons}). A simple model assuming local thermodynamic
equilibrium is given by the corresponding Saha equation, but a general
treatment would be achieved by implementing a detailed network
accounting for changes in the ionization states and releasing the
corresponding recombination energies. As a consequence, temperature
and pressure of the gas change and this may cause a transfer into
kinetic energy finally supporting envelope ejection. Instead of
solving the equations explicitly, the ionization of matter as a
micro-physical effect can be absorbed in a suitable equation of state,
for instance the \textsc{opal} equation of state of
\citet{rogers1996a} and \citet{rogers2002a} or its derivative in the
\textsc{mesa} code \citep{paxton2011a}, see
Sect.~\ref{sect:eos}. Thus, while formally sticking to the basic
gravo-hydrodynamic \ac{CE} model, important physical effects can be
incorporated via the equation of state.

The consequence of including
ionization effects in \ac{CE} simulations is shown in
Fig.~\ref{fig:sand2020}. Compared with the basic gravo-hydrodynamic
model, envelope ejection increases dramatically. Whether or not the
released recombination energy can---at least in parts---support
envelope ejection depends on its ability to thermalize. Instead of
adding to the thermal energy, it may be transported outwards by
radiation (\citealp{grichener2018a}, but see \citealp{ivanova2018a})
or convection \citep{sabach2017a} and lost from surface of the
envelope. \citet{wilson2019a} point out that convection may also remove energy released by the orbital decay directly so that deeper inspiral of the companion is necessary to eject the envelope. While convection is in principle part of the basic
gravo-hydrodynamic model (but may not be sufficiently resolved in \ac{3D} hydrodynamic simulaitons of \ac{CEE}), radiation transport constitutes an extension
of the model. Because all simulations that successfully expel the
\ac{CE} rely on the contribution of recombination energy to the
unbinding of material, it is urgent to include radiation transport
into the modeling.  Without radiation transport the use of a tabulated
equation of state that includes the release of recombination energy
may severely overestimate the envelope unbinding, because a (large)
fraction of that energy might instead leak out of the star.

This illustrates the ambiguity of defining the unbound fraction of
envelope material. Should thermal and ionization energy be counted in
into the energy budget? Certainly, the criterion comparing the
gravitational binding energy with the kinetic energy of the gas only
is the most conservative choice, although it may still overestimate
the effect of released recombination energy if no radiation transfer
is included. Counting in the thermal energy of the gas is a widely
accepted approximation, while including the ionization energy in
addition marks the most optimistic case. The approximation relying on
kinetic energy only is illustrated in
Fig.~\ref{fig:sand2020}.

Recombination energy, if it can be fully used to expand the envelope,
acts in the immediate post-inspiral phases when the envelope is
expanding substantially. During these phases it has been demonstrated
that the conditions in the envelope are ideal for dust formation
\citep{glanz2018a, reichardt2020a, iaconi2020a} which would likely
have an impact on the dynamics of the envelope by the increase in
opacity. Dust-driven winds are extremely effective in reducing the
envelope mass of \ac{AGB} stars over short timescales with mass-loss rates
as high as $10^{-4}$ to $10^{-3} \,\msun \, \mathrm{yr}^{-1}$
\citet{kwok1978a} and these stars are not dissimilar in mass and size
to those typical simulated in the context of \ac{CEE}.

\subsubsection{Magnetic fields}

\begin{figure}
  \centering
    \includegraphics[width=0.6\textwidth]{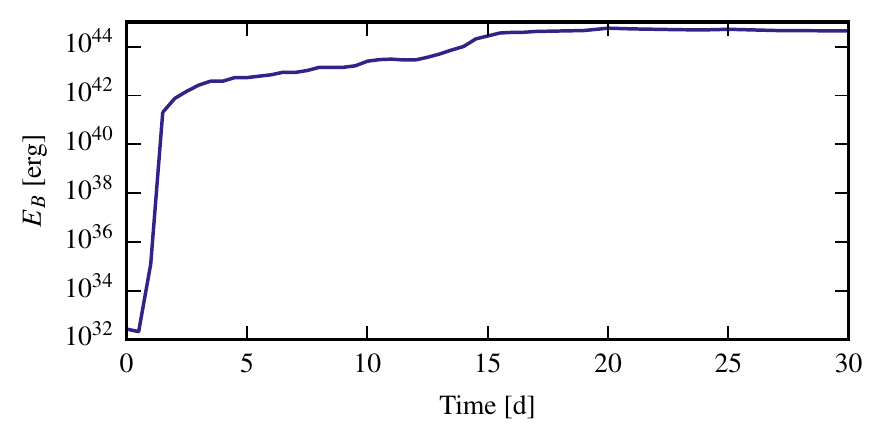}
    \includegraphics[width=\textwidth]{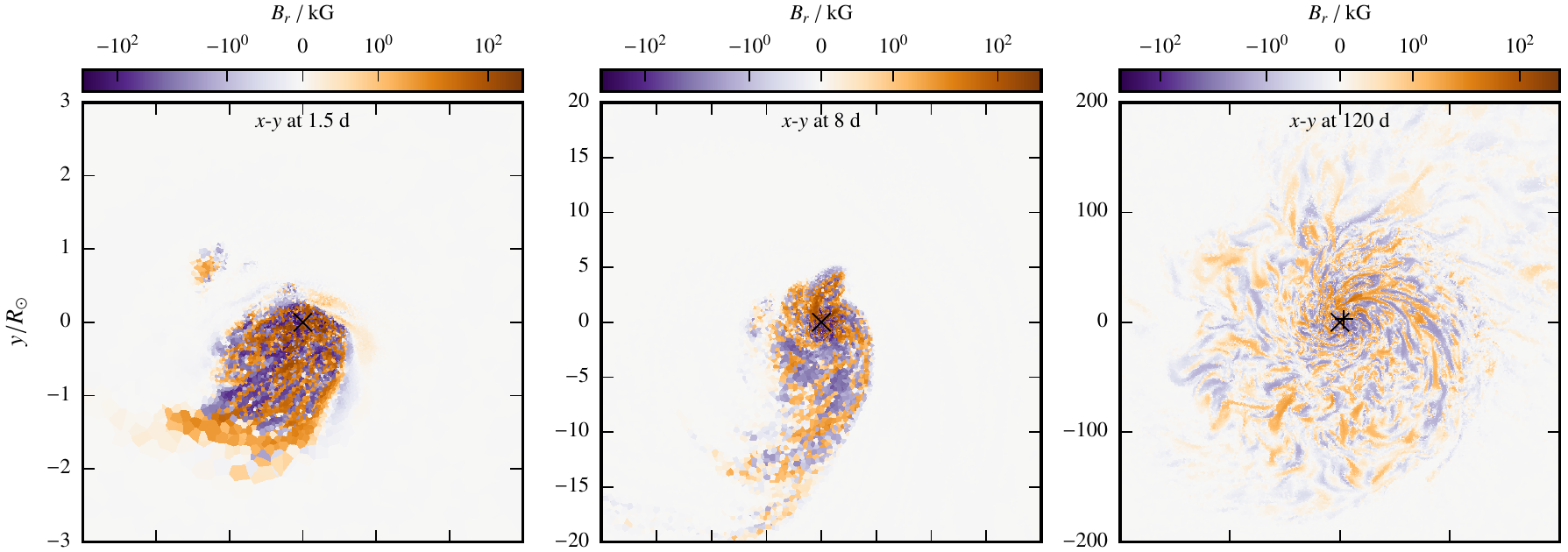}
    \caption{Magnetic field energy over time (top) and radial magnetic
      field strength $B_r$ in the orbital plane (bottom) of the
      high-resolution simulation of \citet{ohlmann2016b} with a seed
      field of $10^{-6} \, \mathrm{G}$. The times at which the
      snapshots were taken are indicated in the panels. Note that the
      left and middle panel show the vicinity of the companion (marked
      with a $\times$ symbol). The position of the core of the
      \ac{RGB} primary star is marked by a $+$ symbol and visible in
      the right panel. Top figure courtesy of Sebastian
      Ohlmann. Bottom figures from \citet{ohlmann2016b}.}
    \label{fig:Bfields-sim}
\end{figure}

A second prominent example of a physical phenomenon that should be
considered, are magnetic fields. These can be accounted for by
extending to the equations of \ac{MHD}; see, e.g.,
\citet{campbell2018a} for a detailed treatment of \ac{MHD} in binary
stars. Apart from the very challenging numerical implementation of
\ac{MHD}, the presence of magnetic fields makes the range of physical
phenomena extremely rich and complex. In addition to hydrodynamic flow
instabilities, \ac{MHD} instabilities can play a role. It has been
discussed whether the action of magnetic fields can help with envelope
ejection. While magnetic fields do not introduce a new energy
source---they can only be amplified at the expense of kinetic energy
of the gas---they can redistribute energy and angular momentum in the
envelope. Based on analytic considerations, \citet{regos1995a} argued
that magnetic fields should be strongly amplified in \ac{CE} inspirals,
while \citet{tocknell2014a} used jet-like features observed in post-\ac{CE}
planetary nebulae as an indication that strong fields must have been
present in the late \ac{CE} stages. The first \ac{MHD} simulation of \ac{CEE}
was presented by \citet{ohlmann2016b}, see also
Sect.~\ref{sect:sim_movingmesh}. It showed that the accretion flow
around the companion star leads to strong amplification of magnetic
fields (see left and middle panels of Fig.~\ref{fig:Bfields-sim}). The
reason is proposed to be the magnetorotational instability
\citep[MRI;][]{balbus1991a}. In late phases, the amplified magnetic
field disperses over the entire \ac{CE} (Fig.~\ref{fig:Bfields-sim},
right panel).  Despite the strong amplification, however, magnetic
fields are not dynamically relevant in the situation modeled by
\citet{ohlmann2016b}. Still, they may play a role in transport
processes and in shaping the ejected envelope material as shown in the
\ac{MHD} simulations of \citet{ondratschek2022a}.

\subsection{Why are common-envelope phases a multi-scale problem?}
\label{sect:multiscale}

Like many other problems in astrophysics, \ac{CEE} is plagued by
severe multi-scale challenges. These can be divided into two
categories: a spatial and a \emph{temporal} scale challenge.

The \emph{spatial scale problem} is rather obvious and caused by the
discrepancy in sizes between the core of the primary star and the
companion on the one side and the total radius of the primary star on
the other side. Although typical values vary somewhat, for \acf{RGB}
and \ac{AGB} stars core radii, $R_\mathrm{core}$, are around
$0.03\,\rsun$ while the size of the stars, $R_\star$, are between
${\sim} \, 20 \, \rsun$ and ${\sim} \, 600\, \rsun$. This gives a
range of scales extending over up to four orders of magnitude, which
is a lot but not overly impressive in comparison with other
astrophysical scale challenges. For red supergiants, the cores have a
radius of $\sim \, 0.07 \,\rsun$ and stellar radii are hundreds to
thousands of solar radii.  Companions can be of various nature, but
cases of interest include objects as small as planets, white dwarfs,
and neutron stars, thus extending the scale range to eight orders of
magnitude. A special case are systems with black hole companions.  The
orbital separation in the pre-\ac{CE} stage and the expansion of the
envelope in the \ac{CE} interaciton extend the range of relevant
spatial scales involved in the problem even more.

The \emph{temporal} scale challenge is usually discussed in terms of
the relevant characteristic timescales of the involved objects.  One
of these is given by the mechanical structure of the envelope. It can be
defined as the \emph{free-fall timescale,}
\begin{equation}
\tau_\mathrm{ff} \sim \frac{\pi}{2}
\frac{R_\star^{3/2}}
     {\left[ 2 G (M_\mathrm{core} + M_\mathrm{com}) \right]^{1/2}},
\end{equation}
with the gravitational constant, $G$, the core mass of the primary
star, $M_\mathrm{core}$ and the mass of the companion,
$M_\mathrm{com}$. Note that $\tau_\mathrm{ff}$ differs from the
orbital period of the companion at a distance of $R_\star$,
\begin{equation}
  \tau_\mathrm{orbit} = 2\pi
  \frac{R_\star^{3/2}}{\left[G\,(M_\mathrm{core} +
      M_\mathrm{com})\right]^{1/2}},
\end{equation}
only by a factor of $2^{-5/2}$. Both are due to the gravitational
force and estimate the timescale of changes to the system driven by
gravitational forces, that has to be captured in simulations of
\ac{CEE}. The \emph{dynamical timescale} $\tau_\mathrm{dyn}$, in
contrast, estimates the timescale of changes induced by hydrodynamic
pressure effects. It is given by the sound crossing time over the
system,
\begin{equation}
  \label{eq:tdyn}
  \tau_\mathrm{dyn} \sim \frac{R}{c_\mathrm{sound}},
\end{equation}
with the sound speed, $c_\mathrm{sound}$. Close to hydrostatic
equilibrium, pressure effects should compensate gravity and thus
$\tau_\mathrm{dyn} \,{\sim}\, \tau_\mathrm{ff} \,{\sim}\,
\tau_\mathrm{orbit}$.

The dynamical timescale of the envelope of \ac{RGB} and \ac{AGB} stars
can be estimated to be $\tau_\mathrm{dyn, envelope} \, {\sim} \, 50 \,
\mathrm{d}$ by adopting $R = R_\star$ and a typical sound speed of
${\sim}\, 10 \, \mathrm{km} \, \mathrm{s}^{-1}$. The cores of \ac{RGB}
or \ac{AGB} stars and their envelopes have starkly different
thermodynamic properties. The temperature in the cores is about four
orders of magnitude higher than in the envelopes and their densities
exceed the envelope densities by ${\sim} \, 10$ orders of magnitude.
Therefore, the sound speed exceeds $10^3 \, \mathrm{km} \,
\mathrm{s}^{-1}$ in the core of such stars, and the dynamical
timescale is as short as $\tau_\mathrm{dyn, core} \, {\approx} \, 20 \,
\mathrm{s}$.

Some late episodes of \ac{CEE} may rely on thermal readjustment of the
envelope which is associated with the much longer \emph{thermal}
(Kelvin--Helmholtz) timescale,
\begin{equation}
  \tau_\mathrm{KH} \sim \frac{G M_\star^2}{R_\star L_\star},
\end{equation}
where $L_\star$ is the luminosity of the star. For \ac{RGB} and
\ac{AGB} stars with $L_\star \, {\sim} \, 3000 \, L_\odot$, the
thermal timescale $\tau_\mathrm{KH} \, {\sim}\, 10^4 \mathrm{yr}$,
leading to a timescale discrepancy compared with $\tau_\mathrm{dyn,
  envelope}$ of $10^5$. For late \ac{CEE} stages, however, the
relevant thermal timescale is that of the remaining envelope when part
of the material has already been removed, which may differ
significantly from our estimate for unperturbed stars.

\subsection{The need to simulate the pre-common envelope and post-inspiral phases}
\label{sect:pre-dynamical_phase}

We conclude our summary of the physical challenges to simulating
\ac{CEE} by discussing the phase that leads to the dynamical
inspiral---Phase (i), pre-\ac{CE} evolution---and the phase that follows
it---Phase (iii), post-\ac{CE} evolution. Both of them are likely to take
place on thermal, rather than dynamical timescales. This fact alone
makes them difficult to treat numerically. The \emph{need} to simulate
both of these phases stems from the suspicion that what precedes the
inspiral may change the playing field for the inspiral itself and that
what follows the inspiral ultimately settles the final orbital
separation for the surviving binary or even determines whether a
binary survives at all. Here, we quickly set the stage for the
physical challenge presented by these phases.

Before engaging into the inspiral phase of \ac{CEE}, binary
interactions may substantially affect the parameters of the system.
Pre-CE evolution is likely to involve mass loss from the system, and
this effect can be critical in determining whether or not the system
enters \ac{CEE} in the first place
\citep[e.g.][]{ivanova2013a}. Moreover, mass loss can change the
angular momentum budget of the system and the binding energy of the
primary's envelope. An extreme case of binary interaction preventing
\ac{CEE} is proposed by \citet{soker2015a}, who argues that if the
companion grazes the envelope of the giant primary star and accretes
mass at a sufficiently high rate, it can launch jets that remove
material and angular momentum. Additionally, mass and angular momentum
can be lost through the second Lagrange point. This way, the envelope
is ablated as the companion approaches the core of the primary star
without ever entering a \ac{CE} phase. The orbital decay is very slow and if existing, it is driven by
tidal interaction although the companion is never deeply immersed in
the envelope. \citet{soker2015a} call this alternative to \ac{CEE}
``grazing envelope evolution'' (GEE) and \citet{sabach2015a} discuss
the scenario in the context of triple star systems.  Mass loss is also
thought to lead to observable features, some of which have been
connected to the \ac{LRN} class of optical transients (see
\citealp{soker2006a, pejcha2014a, soker2015a, soker2016c, macleod2017a} and
Sect.~\ref{sect:observational_constraints}).
  
Once the primary star overfills its Roche lobe, mass transfer sets
in. Whether or not this leads to the formation of a \ac{CE} depends on
the dynamical stability of this mass transfer, which is determined by
the relative readjustments of the Roche lobe to mass transfer and the
donor (primary) star's structure to mass loss: if the donor expands
relative to its Roche lobe, mass loss becomes unstable and a runaway
in the mass transfer rate sets in. The details involve several
complications---for a discussion we refer to
\citet{ivanova2013a}---but stability analyses usually determine a
critical mass ratio of donor and accretor (i.e.\ companion)
$q_\mathrm{crit}$ for stability, which itself depends sensitively on
the type of donor \citep{hjellming1987a}.

This, however, is not the only criterion determining the events
preceding the fast inspiral. Tides dominate the early phases of
interaction, even before Roche lobe contact. Giant stars expand and
when their companions are located within a distance of some stellar
radii \citep[][the exact number is uncertain and can
  vary]{soker1994a, mustill2012a, madappatt2016a}, 
interaction between the
two tends to synchronize and circularize the orbit. While the
timescales for these processes are reasonably well understood
\citep[see][]{zahn1977a}, the concomitant stellar evolution alters the
playing field. In particular, an instability occurs when the angular
momentum of the giant is larger than a certain fraction of the orbital
angular momentum \citep{darwin1879a}. This tends to destabilize a
tidally relaxed system leading to a reduction of its orbital
separation. All in all, a complex interplay between tidal instability,
mass-transfer instability, and stellar evolution set the initial
conditions of the subsequent fast inspiral. While not all of these
processes take place on the Kelvin--Helmholtz timescale characterizing
changes in the thermal structure of the star---tidal interaction for
instance is a dynamical process---the timescales of these phenomena
are likely still too long to be modelled by global \ac{3D}
simulations.
  
After the dynamical inspiral stalls, a number of physical processes
\citep[see Table 1 of][for an overview of some of these]{soker2017a}
will take hold that operate on lengthening timescales.  While some of
the current \ac{3D} hydrodynamic simulations indicate successful
envelope ejection, a general problem seems to be that the orbital
separation reached over the simulation time is too large compared with
observations of post-\ac{CE} binaries \citep[e.g.][]{iaconi2017a,
  kramer2020a}. This may be associated with shortcomings in the
physical modeling or numerical deficiencies in the simulations, but
there is a fair chance that the system parameters are affected by
processes that act on timescales longer than the simulation
period. Self-regulated inspiral (see
Sect.~\ref{ssec:why_is_CE_ejection_hard_to_achieve}) is one
possibility, but even if most of the envelope was unbound, some
returning gas could interfere with the orbit once more, or form a disk
around the binary which could alter the orbital parameters. If most of
the envelope is still bound, the binary will presumably go through a
renewed phase of inspiral. If the new delivery of orbital energy is
still not sufficient to unbind the remaining envelope, then a merger
of the stellar cores may be the final outcome of the interaction. The
newly formed star would then relax on thermal timescales as it
radiates the excess energy. However, the orbital angular momentum
would have to find a final resting place in the envelope,
redistributing itself in some way. Another possibility is
that---instead of merging---the companion is destroyed during deep
inspiral into the \ac{CE}. This is in particular possible for low-mass
companions such as brown dwarfs or giant planets \citep{livio1984a,
  soker1998a, nelemans1998a, jia2018a, kramer2020a}
  
Modelling the immediate pre- and post-inspiral, i.e.\ the parts of
Phases (i) and (iii) that are closest to Phase (ii), with global
\ac{3D} simulations is not impossible, as we will discuss in
Sects.~\ref{ssec:simulations_of_the_pre-dynamical_inspiral_phase} and
\ref{sect:current}, but some simplifications must be incorporated,
which may or may not alter the results. The longer timescales likely
will need different modelling approaches and, ideally, these will be
interfaced with global simulations that best model the dynamical phase
(see
Sect.~\ref{ssec:interfacing_3d_global_simulations_with_1d_simulations_of_the_ce_remnant}).
After discussing the \emph{physical} challenges we face in formulating
a full description of \ac{CEE}, we turn to modelling problems that
arise from \emph{numerical} obstacles in the following section.

\section{Numerical approaches and limitations}
\label{sect:challenges_numerical}

Once the physical model is fixed---which, as discussed above, is
highly non-trivial---a valid numerical discretization of the
underlying equations is bound to converge to the correct result of
that model. In the ideal world of infinite numerical resolution, any
consistent and stable numerical scheme should give the same
answer\footnote{For linear finite difference equations, this is the
Lax--Richtmyer equivalence theorem \citep{lax1956a}.}. This, however,
is far from reality: the discrepancy of relevant spatial and temporal
scales implies that such a convergence cannot be reached easily.
While some effects are treated accurately by numerical approaches,
others can only be approximated. Gas flows in the envelope are usually
resolved, but effects closer to the cores are increasingly difficult
to represent correctly. The challenge is to find appropriate
approximations. Various choices are made which may interfere with the
original physical modeling so that differences in the numerical
results are to be expected. The goal, however, is to reach convergence
to the correct results in some key quantities.

Implementing the basic gravo-hydrodynamic model of \ac{CEE} as defined
in Sect.~\ref{sect:gravohydro} requires two ingredients: (i) a solver
for the Euler equations of fluid dynamics (\ref{eq:mass_cons},
\ref{eq:momentum_cons}, \ref{eq:energy_cons}), and (ii) a self-gravity
solver that accounts for the Poisson equation (\ref{eq:poisson}). Both
are in the standard repertoire of computational astrophysics, but also
mark some of the more challenging aspects of it. We review methods to
obtain solutions to the hydrodynamics in the presence of gravity in
Sects.~\ref{sect:cfd} and \ref{sect:gravity}, but omit implementation
details; these can be found in the literature. Instead, we focus on
some of the more specific aspects arising from the application of
standard techniques to \ac{CEE} simulations.

It is likely that physical effects beyond the basic gravo-hydrodynamic
model become more commonly included in future simulations of \ac{CE}
interaction. These would deserve a dedicated discussion in this
section. A prominent example is radiation transfer, which, however,
has not yet been established as a standard in the simulations and we
therefore refrain from including it at the moment. The same applies to
magnetohydrodynamic simulations of \ac{CEE}, that thus far have only
been carried out by a single group \citep{ohlmann2016a} with one
specific way of treating magnetic field. While we mention these
simulations in Sect.~\ref{sect:sim_movingmesh}, a review of different
approaches is not yet called for.

\subsection{Computational fluid dynamics}
\label{sect:cfd}

There are numerous approaches to discretize the Euler equations of
fluid dynamics. A fundamental distinction is the frame of reference in
which the basic equations are formulated. One possibility, where a
fixed lab frame of reference is chosen, is given in
Eqs.~(\ref{eq:mass_cons}, \ref{eq:momentum_cons}, and
\ref{eq:energy_cons}). Discretizing this formulation gives rise to the
so-called \emph{Eulerian schemes} for solving the equations of fluid
dynamics. Alternatively, discretization of
Eqs.~(\ref{eq:lagrange_mass}, \ref{eq:lagrange_momentum}, and
\ref{eq:lagrange_energy}) in Lagrangian specification with a co-moving
frame of reference results in \emph{Lagrangian schemes} of
computational fluid dynamics. Lagrangian approaches are particularly
well-suited to simulations of interactions between astrophysical
objects---not only in the context of \ac{CEE}, but in various other
setups reaching from the simulation of mergers of compact objects
\citep[see][for a review]{rosswog2015a} to mergers and evolution of
galaxies in cosmological contexts \citep[e.g.,][]{liu2016b, croft2009a,
  stinson2013a}.

Both the Eulerian and the Lagrangian approaches have their particular
strengths and weaknesses. Whether these play out in specific
applications depends on the context, the setup, and the research
questions to be answered. \ac{CEE} is in some sense a prototypical
``Lagrangian problem'': two main objects orbiting each other and a
rotating stellar envelope embedded in vacuum make methods following
the fluid flow attractive. The main advantage of Lagrangian methods is
that the geometry is adapted to the problem at hand. Following the
motion of the fluid, advection errors are minimized. This reduces
common problems with angular momentum conservation (see
Sect.~\ref{sect:SPH}) and diffusion over the surfaces of the stellar
objects. This said, it is not \textit{a priori} clear that Lagrangian
schemes perform best in all respects. This is related to the problem
that in Lagrangian specification a grid-based discretization approach
leads to complicated grid geometries and the danger of tangling if the
grid is to be advected along with complex \ac{3D} hydrodynamic
flows. Therefore, a grid-free Lagrangian approach is commonly chosen
(see Sect.~\ref{sect:SPH}), which, however, lacks the precision of
grid-based computational fluid dynamics.  Eulerian approaches do not
face this problem. Their grid-based discretizations are unproblematic
and they benefit straightforwardly from the accuracy of Riemann-based
hydrodynamics solvers (see Sect.~\ref{sect:eulerian}). Therefore, no
general statement is justified giving preference to one or the other
method and we discuss their application in the context of particular
simulations in Sect.~\ref{sect:simulations}. \ac{CE} simulations with
moving-mesh codes (see Sect.~\ref{sssec:moving_mesh_approaches})
combine the advantages of a (nearly) Lagrangian scheme with the
accuracy of a grid-based discretization of the equations of fluid
dynamics.

\subsubsection{Eulerian, grid-based methods}
\label{sect:eulerian}

There are various general strategies to discretize and solve the Euler
equations in Eulerian specification (Equations~\ref{eq:mass_cons},
\ref{eq:momentum_cons}, \ref{eq:energy_cons}), once initial conditions
are specified. Overviews are given in standard textbooks, such as
\citet{ferziger2020a}.

In astrophysical applications, finite-volume methods have defined the
quasi-standard for the past decades. They discretize space into
computational grid cells and assign cell-averages of the conserved
hydrodynamic quantities to their centers. Reconstructing these
averaged conserved quantities to the cell edges introduces jumps over
the interfaces between neighboring cells. Such initial conditions with
one discontinuity define \emph{Riemann problems.} In \emph{Godunov
schemes} \citep{godunov1959a}, they are solved to determine the
numerical fluxes, which are used to update the cell-averaged conserved
quantities.

This discretization of the equations of fluid dynamics in space has to
be augmented with a discretization in time. To this end, two main
classes of approaches exist: \emph{explicit} and \emph{implicit} time
discretization schemes. The former determine the state at the new
timestep from the state at the current timestep while the latter
solves for the state at the new timestep from the given current and
the new state, which requires a numerically expensive matrix
inversion. While implicit time discretization can be unconditionally
stable, explicit time stepping is restricted by the so-called \acfi{CFL}
condition \citep{courant1928a}. It states that for stability of the
scheme, the time step has to be chosen smaller than the crossing time
of the fastest wave emerging from the Riemann problem over a grid
cell. A rough estimate for this would be the sound crossing time over
a cell. So far, all hydrodynamic schemes employed for simulations of
\ac{CE} interaction use explicit time discretization and their time
stepping is therefore subject to the \ac{CFL} condition.

Different methods exist to ensure higher-order accuracy with specific
reconstruction and time integration schemes and for enhancing the
efficiency with approximate Riemann solvers. For the details of such
methods, we refer to textbooks such as that by \citet{toro2009a}. The
resulting high-resolution shock capturing methods are advantageous for
following the spiral shocks in the \ac{CE}. Fundamentally, this
property results from the fact that finite-volume methods derive from
the \emph{integral form} of the equations of fluid dynamics that in
contrast to the differential form as in
Eqs.~(\ref{eq:mass_cons}--\ref{eq:energy_cons}) allows for
discontinuous (so-called \emph{weak}) solutions. Shocks are possible
in the solutions of the Riemann problems at the cell
interfaces. Depending on the quality of the employed approximate
Riemann solver and the numerical dissipation of the scheme, however,
shocks are usually smeared out over a few grid cells. In addition to
the shock-capturing feature, Godunov methods are known to reproduce
phenomena such as hydrodynamic instabilities and turbulence very well.

The orbital motion of the cores and the rotation of the envelope,
however, imply that matter will be dragged over the spatially fixed
discretization grid. Transforming into a rotating frame of reference
can alleviate the problem, but because of the strongly differential
motion of the involved matter the problem cannot be completely
eliminated. This situation leads to advection errors that can only be
mitigated with high spatial resolution, which renders Eulerian
grid-based approaches to simulate \ac{CEE} a computationally expensive
choice.

Covering the required large domains with equally-sized grid cells for
discretization would either lead to low resolution of important parts
of the system in case of large grid cells or excessive computational
cost when trying to resolve the relevant spatial scales. For this
reason, non-uniform grids concentrating resolution in the center
around the core binary system or, even better, \ac{AMR} techniques
\citep[e.g.][]{berger1984a, berger1989a} are commonly employed in
\ac{CE} simulations. \ac{AMR} methods subdivide grid cells once
pre-defined refinement criteria are met and they de-refine when they
are no longer fulfilled. Multiple levels of refinement are possible so
that very high spatial resolution can be reached in some patches of
the domain. In \ac{CE} simulations, mesh refinement is successfully
applied around the cores to reduce the advection errors. While not as
excessive as global fine resolution, \ac{AMR} still comes at the price
of the higher computational effort associated with larger cell numbers
and shorter numerical timesteps in smaller cells required by the
\ac{CFL} condition plus a significant overhead for the bookkeeping of
the complex \ac{AMR} grid structure.

Nonetheless, once the resolution is high enough to ensure reasonable
conservation of angular momentum, Eulerian grid-based approaches show
excellent reproduction of hydrodynamical effects such as shock waves
and instabilities. Moreover, their methodology offers more flexibility
for extending the physical modeling base. Magnetohydrodynamics, for
instance, is best implemented in grid-based techniques and the
formulation of subgrid-scale models for unresolved physical effects is
an attractive perspective for the future development of \ac{CE}
simulations.

\subsubsection{Lagrangian, smooth particle hydrodynamics}
\label{sect:SPH}

The problem of growing complexity of a computational grid with the
evolution of the fluid-dynamical system in Lagrangian approaches is
avoided in \ac{SPH} methods (\citealp{lucy1977a},
\citealp{gingold1977a}; for reviews see \citealp{monaghan1992a},
\citealp{springel2010c}, \citealp{price2012a}, and
\citealp{rosswog2015a}).  A particle-based discretization of the mass
distribution is applied instead of a computational grid. The pressure
gradient appearing in Eq.~(\ref{eq:lagrange_momentum}), however,
raises the question of evaluating the gradients of quantities in a
particle representation of matter. To address this problem, the
particles are not understood as point masses in \ac{SPH} schemes, but
they rather mark the centers of smoothed-out distributions of the
conserved hydrodynamic quantities. The amount of smoothing is
quantified by the choice of a \emph{smoothing kernel}, $W(r, h)$,
which assigns quantities to a location at a distance of $r = | \vec{x}
- \vec{x}_i |$ from the \ac{SPH} particle at point $\vec{x}_i$ and is
characterized by a certain \emph{smoothing length}, $h$. The mass
density at point $\vec{x}$, for instance, is obtained from summing
over all $N$ \ac{SPH} particles:
\begin{equation}\label{eq:dens_sph}
\rho (\vec{x}) = \sum\limits_{i = 1}^{N} m_i W(|\vec{x} - \vec{x}_i|, h).
  \end{equation}
The mathematical form of the smoothing kernel is arbitrary, but has to
obey the normalization conditions of giving $1$ when integrated over
all space and degenerating to Dirac's delta function for $h \to 0$. An
efficient choice are kernels with compact support. They restrict the
overlap to nearby particles and thus reduce the terms that have to be
evaluated for the sum in Eq.~(\ref{eq:dens_sph}). A popular form is the spline
kernel of \citet{monaghan1985a},
\begin{equation}\label{eq:smoothing_kernel}
  W(r, h) = \frac{1}{\pi h^3}
  \begin{cases}
    1 - \frac{3}{2} \frac{r}{h}^2 + \frac{3}{4} \frac{r}{h}^3 &
    \mbox{if } 0 \le \frac{r}{h} \le 1;\\
    \frac{1}{4} (2 - \frac{r}{h})^3 & \mbox{if } 1 \le \frac{r}{h} \le
    2;\\
    0 & \mbox{otherwise,}
  \end{cases}
\end{equation}
but there are various alternatives to this choice
\citep[e.g.][]{dehnen2012a}. At any given point in space, the
hydrodynamic quantities thus defined can be determined by summing over
contributions with overlapping smoothing kernels. Therefore
characteristic size of the kernel determines the spatial resolution in
\ac{SPH} simulations. For all practically useful kernel functions, the
standard deviation of the kernel is very similar to the chosen
smoothing length. Efficient \ac{SPH} implementations employ variable
smoothing lengths assigning an $h_i$ to each particle such that
quantities are determined by summing over similar numbers of
neighboring particles independently of the local density. Hence, $h_i$
is an approximate measure of the linear resolution at the location of
particle $i$.

This particle-based discretization leads to a set of ordinary
differential equations, the equations of motion of \ac{SPH}
\citep{lucy1977a, gingold1977a, monaghan1992a}, that have to be solved
numerically. Stability of explicit time integration is again
determined by the \ac{CFL} condition, requiring timesteps smaller than
the sound crossing time over a smoothing length \citep{lucy1977a,
  gingold1977a}.

\ac{SPH} methods allow to exploit the advantages of Lagrangian schemes
in \ac{CE} simulations and they have been employed extensively for
this purpose. Their main benefit is that they avoid advection
errors. This leads to exact conservation of angular momentum (see also
Sect.~\ref{sect:conservation}). Reaching an acceptable degree of
angular momentum conservation with Eulerian grid-based approaches
(which is fundamental for \ac{CE} simulations) requires high spatial
resolution of the rotating objects which often can only be reached
with \ac{AMR} techniques. This makes these schemes computationally
more expensive than \ac{SPH}-based simulations.

Another feature of Lagrangian schemes is Galilean invariance---the
solution is not affected by the presence of bulk velocities. Violation
of Galilean invariance may suppress hydrodynamic instabilities when
altering the background velocity and keeping the spatial resolution
fixed. Again, Galilean invariance in Lagrangian schemes results from
avoiding advection errors and distinguishes \ac{SPH} from grid-based
approaches.

Conceptually, \ac{SPH} is very similar to numerical schemes for
simulating the evolution of collisionless $N$-body systems under the
action of gravity. The only differences in the underlying equations are
the terms involving pressure. For such systems, efficient methods have
been developed to calculate self-gravity (see
Sect.~\ref{sect:gravity}) and these are exploited in the
\ac{SPH}-based simulations of \ac{CEE}, where this effect is
fundamental for the dynamics.

Another advantage is that no pseudo-vacuum has to be introduced for
embedding the system of interest. \ac{SPH} particles are placed where
the matter of the object of interest is located. The computational
domain is arbitrary, free boundary conditions are natural to the
scheme and without excessive computational cost the computational
domain can be large enough so that no material leaves and conservation
of global quantities can be checked. The placement of particles
exclusively inside the objects of interest in combination with summing
up the contributions to conserved quantities from neighboring
particles, however, introduces an imbalance that causes an artificial
surface tension in \ac{SPH}.

Despite these advantages, there is a potential problem with \ac{SPH}
schemes. They are based on a discretization of
Eqs.~(\ref{eq:lagrange_mass}--\ref{eq:lagrange_energy})---a
\emph{differential form} of the equations of fluid dynamics---and
therefore allow only for differentiable (so-called \emph{strong})
solutions and not for discontinuities such as shocks. In order to
treat shocks in \ac{SPH}, artificial viscosity has to be introduced to
broaden the shock to a numerically resolvable continuous structure
\citep{monaghan1983a}. This at the same time allows for the necessary
dissipation of energy in the shock. A careful choice of the form of
the artificial viscosity is required to avoid undesired effects such
as dissipation in non-shocked regions or violation of angular momentum
conservation. A correct representation of pre- and post-shock states
is possible within the \ac{SPH} framework, but high resolution
(i.e.\ a local high density of particles) is required to keep the
shock structure reasonably thin. In contrast, no artificial viscosity
is needed in finite-volume Godunov schemes to represent shocks.

The \ac{SPH} discretization into particles naturally and necessarily
follows the mass distribution of the considered system. As opposed to
other astrophysical application examples of \ac{SPH}, this is a clear
disadvantage in \ac{CE} simulations, where the features in the gas
dynamics in a rather dilute envelope may need to be followed.

Still, with limited computational resources, \ac{SPH} simulations of
\ac{CEE} published so far show better performance than their
grid-based Eulerian counterparts. They are able to follow the gross
dynamics of the system with a fairly low number of particles without
sacrificing angular momentum conservation. However, reaching fine
resolution in the fluid flows and resolving instabilities requires a
drastic increase in the number of particles so that there seems to be
some break-even point where grid-based Eulerian techniques become more
efficient.

\subsubsection{Moving-mesh approaches}
\label{sssec:moving_mesh_approaches}

A rather recent development is the use of moving-mesh hydrodynamics
codes for \ac{CE} simulations. The first of such simulations---based
on the \textsc{arepo} code developed by \citet{springel2010a} for
cosmological simulations with the modifications of \citet{pakmor2016a}
to ensure second-order accuracy of the scheme---was presented by
\citet{ohlmann2016a}. A number of follow-up studies expanded on this
including additional physical effects and testing parameters of the
setup model \citep{ohlmann2016b, sand2020a, kramer2020a}. An
independent implementation of the moving-mesh technique in the
\textsc{manga} code was used for \ac{CE} simulations by
\citet{prust2019a}.

\begin{figure}
    \centering
    \includegraphics[width=0.6\textwidth]{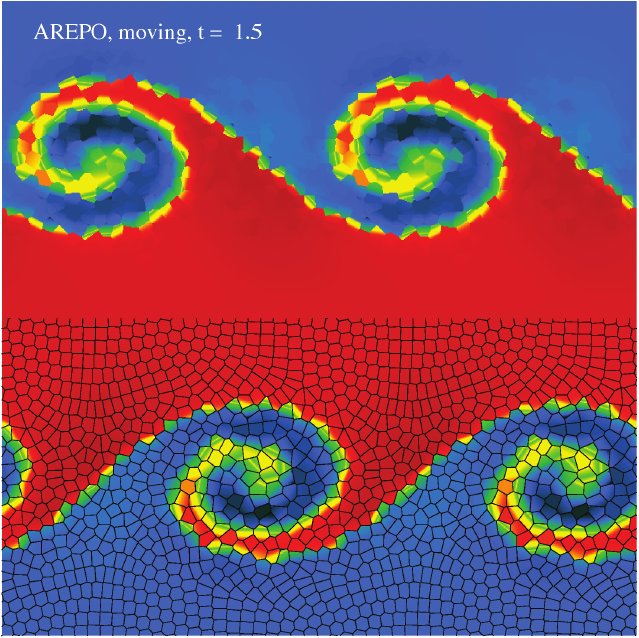}
    \caption{Simulation of a Kelvin-Helmholtz instability between two
      shearing fluids with the code \textsc{arepo} in a low-resolution
      (50 × 50) to illustrate the Voronoi mesh overlaid in black in
      the lower half of the box. Figure from \citet{springel2010a}.}
    \label{fig:Springel2010-Fig32d}
\end{figure}

The moving-mesh method can be seen as combining advantages of the
Lagrangian particle-based \ac{SPH} technique and Eulerian grid-based
approaches. A Voronoi tessellation of space based on a set of discrete
points is used to construct a complex unstructured mesh of polyhedral
cells, see Fig.~\ref{fig:Springel2010-Fig32d}. As in conventional,
finite-volume grid-based codes, numerical flux functions across the
thus defined cell boundaries are computed in a Godunov-type approach
using Riemann solvers. These provide a scheme to update the
cell-averaged values of conserved hydrodynamical quantities. The
mesh-generating points can in principle be moved arbitrarily or even
kept fixed---in which case the scheme recovers an Eulerian grid-based
approach to numerical fluid dynamics. Advecting the points with the
fluid flow, however, yields a nearly Lagrangian method. Its advantages
of reducing advection errors, conserving angular momentum, and
guaranteeing Galilean invariance\footnote{The failure of Eulerian
schemes to obey Galilean invariance is impressively demonstrated in
Figure~10 of \citet{springel2010a}, who shows that a background
velocity suppresses the development of Kelvin--Helmholtz instabilities
on a fixed grid. \citet{robertson2010a}, however, argue that Eulerian methods converge may converge to the correct Galilean-invariant solution.}  are combined with the superior shock capturing
capabilities and the improved reproduction of hydrodynamic
instabilities of grid-based Godunov-type schemes. For a detailed
discussion of the advantages of the moving mesh scheme, we refer to
the original publication of the \textsc{arepo} code
\citep{springel2010a}.

While moving-mesh methods overcome the notorious problems of grid
tangling in grid-based Lagrangian schemes, some measures of correcting
stretched cell shapes that would result from complex flows are still
required to guarantee numerical benignity
\citep{springel2010a}. Therefore, the resulting moving-mesh schemes
are not exactly Lagrangian but very close to it\footnote{For this
reason, advection errors are minimized but not eliminated, and angular
momentum is conserved with high accuracy, but not exactly.}.  The
Godunov-type finite-volume scheme for updating the cell values is
based on the integral form of the Euler equations of fluid
dynamics. It eliminates the need of an artificial viscosity as in \ac{SPH}
schemes. Moreover, it avoids the inherent sampling noise errors in the
gradient estimates of \ac{SPH} \citep{pakmor2013a}.

Based on a finite-volume discretization, moving-mesh schemes retain
the versatility of Eulerian approaches with respect to the
implementation of additional physical effects. Magnetic fields, for
example, can be accounted for \citep{pakmor2011d, pakmor2013b}. The
first \ac{3D} \ac{MHD} simulations of dynamical \ac{CE} interaction
were performed by \citet{ohlmann2016b} with the moving-mesh code
\textsc{arepo}. Unresolved effects can in principle be included via
subgrid-scale models. While utilized extensively in cosmological
simulations with \textsc{arepo} \citep[e.g.][]{vogelsberger2013a},
this capability has not been exploited yet in simulations of \ac{CEE}.

The flexibility of mesh construction can be used to implement \ac{AMR}
capabilities that, in contrast to \ac{SPH} approaches, are not limited to
improving the resolution of high-density regions. This is of
particular interest in \ac{CE} simulations because here the focus is
on resolving effects in the rather dilute envelope gas. By adding grid
generation points, cells can be split up and the resolution is
increased locally; removal of grid generation points reverses this
procedure. This spatial adaptivity can be combined with adaptive
individual time stepping (as implemented in \textsc{arepo} and
\textsc{manga}), which significantly enhances the computational
efficiency because it pays attention to the temporal multi-scale
nature of the problem as discussed in
Sect.~\ref{sect:challenges_physical}.  A moving, adaptively refined
mesh as implemented in \textsc{arepo} and \textsc{manga} is a powerful
approach to simulate \ac{CEE}, where certain parts of the envelope,
particularly the regions around the cores, benefit from higher spatial
resolution, while other parts of the envelope can be represented with
less computational effort.

Another advantage of the currently employed moving-mesh codes is that
they were derived from powerful particle-based codes, that had been
written for simulations of self-gravitating systems. The efficient
parallel implementation of the tree-based gravity solver of
\textsc{gadget-2} \citep{springel2005a} carries over to
\textsc{arepo}, and that of \textsc{ChaNGa} \citep{menon2015a} to
\textsc{manga}.

The origin of the moving-mesh codes from tools to simulate galaxy
evolution in a cosmological context required some
modifications. Methods for capturing the relevant conditions and
effects in stellar matter had to be implemented \citep{pakmor2013a,
  ohlmann2017a, prust2019a}, see Sects.~\ref{sect:challenges_setups} and \ref{sect:sim_movingmesh}.

\subsection{Implementing the equation of state}
\label{sect:eos}

As mentioned in Sect.~\ref{sect:gravohydro}, an equation of state
relating pressure to density and internal energy (and possibly other
quantities such as chemical composition and ionization state) has to
be provided in order to close the fluid dynamics equations set.

Realistic treatments of matter require the inclusion of various
effects. Already when augmenting an ideal-gas equation of state with
radiation pressure, the computational effort increases
significantly. The reason is that the equation of state derived from
statistical physics and thermodynamics usually provides pressure and
internal energy as a function of density and temperature. A
fluid-dynamical model, however, follows the conserved hydrodynamic
quantities, i.e.\ density and total energy, from which the kinetic
energy can be subtracted to obtain the internal energy of the
gas. While an ideal-gas equation of state can still be inverted
analytically to give pressure as a function of density and internal
energy, this is no longer possible when including radiation pressure
and therefore an iteration is required. Since the equation of state is
called at least in every hydrodynamic update of each computational
cell, the computational expense of this iteration affects the total
demand of the simulation significantly.

The situation becomes even more critical when more complex physical
effects are included such as degeneracy or ionization. In such cases,
modern implementations revert to tabulated data. The look-up of table
values and interpolation between them can add substantial
computational cost to the hydrodynamics scheme.

As discussed in Sect.~\ref{sect:ionization}, in the context of \ac{CE}
simulations, ionization effects are of particular interest. When
envelope gas expands and cools down below the ionization thresholds of
hydrogen and helium, the corresponding recombination energies are
released into thermal energy. Assuming local thermal equilibrium, the
effect can be included into the equation of state as done by virtually
all \ac{3D} hydrodynamic \ac{CE} simulations that accounted for these
effects. This approach does not provide the detailed ionization
structure of the gas, which would require the solution of the Saha
equation or a detailed ionization network; however, for \ac{CE}
simulations this is not of immediate interest. As an input, the
equation of state is given the local density and the internal energy
of the gas (that corresponds to the sum of thermal and ionization
energy). From this, temperature and pressure are determined.  This can
result in a local over-pressure when recombination energy is released,
giving rise to expansion and acceleration of material. If no losses
due to cooling or radiation are included, all released recombination
energy is transferred to kinetic energy of the envelope gas. The
processes determining the equation of state are complex and therefore
the native implementation of the \textsc{opal} equation of state
\citep{rogers1996a, rogers2002a} or its derivative in \textsc{mesa}
\citep{paxton2011a} rely on pre-computed tabulated data.

As a side remark, we mention that applying such more elaborate
equations of state requires to use suitable exact Riemann solvers
\citep[see, e.g.,][]{colella1985a, chen2019a} or approximate Riemann solvers for
Eulerian grid-based methods. This is accounted for in most
implementations of modern astrophysical hydrodynamics codes.

\subsection{Modelling self-gravity}
\label{sect:gravity}

Gravitational forces define the \ac{CE} interaction. They mediate (1)
the binding of the stellar components and the binding of the
progenitor binary system, (2) the interaction and orbiting motion of
the stellar cores, and (3) the interaction of the cores and the
envelope gas giving rise to drag and energy transfer. This highlights
the importance of treating gravity accurately and efficiently in
\ac{CE} simulations. As a long-range force, gravity enters the
modeling as a source term in the momentum
[Eqs.~(\ref{eq:momentum_cons}) and (\ref{eq:lagrange_momentum})] and
energy equations [Eqs.~(\ref{eq:energy_cons}) and
  )\ref{eq:lagrange_energy})]. Source terms are often discretized
separately from the equations of fluid dynamics in an operator
splitting approach. For the case of gravity, there are two basic
choices: the gravitational potential can be determined by solving the
Poisson equation (\ref{eq:poisson}) in the continuum model of matter,
or the gravitational forces are computed directly from Newton's law of
gravity between point masses in particle-based discretizations of
matter. We note that this choice is independent of the discretization
of the equations of fluid dynamics. While \ac{SPH} codes preferably
use a particle-based discretization for modeling gravity, mapping to
grids in the particle-mesh method is also employed
occasionally. Grid-based Poisson solvers seem a natural choice for
grid-based discretizations of the equations of fluid
dynamics. However, since stellar cores are often modelled as point
masses, grid-based schemes are sometimes coupled with particle-based
gravity solvers instead.

Approximations, such as a gravitational potential that is constant in
time or one that is derived from a low-order multipole expansion of
the overall gravitational potential, are inappropriate for capturing
\ac{CE} dynamics---a full treatment of self-gravity of the involved
envelope gas and stellar cores is inevitable. The Poisson equation of
Newtonian gravity (\ref{eq:poisson}) poses an elliptic problem and is
numerically very challenging. A direct summation of the gravitational
forces gives rise to algorithms that scale with the square of the
number of cells or particles $N$ quickly leading to prohibitive costs
in highly-resolved simulations. In the following, we discuss various
approaches employed in \ac{CE} simulations to reduce this
$\mathcal{O}(N^2)$ operation count to a more modest algorithm
complexity, such as $\mathcal{O}(N \log N)$.

\subsubsection{Tree methods to calculate self-gravity}
\label{sect:tree}

In particle-based discretizations, a preferred method for calculating
self-gravity of astronomical objects is a tree-based approximation
\citep{barnes1986a}. The method is based on the fact that the
individual contribution of distant particles to the force of a target
particle is negligible, so that it is sufficient to treat the
collective effect of distant particles in an approximate way, i.e.\ as
a low-order multipole of the gravitational potential. The interaction
with close particles, in contrast, is treated accurately. To
distinguish close from distant particles, a tessellation of space in a
tree hierarchy is constructed. For details, we refer to the literature
\citep[e.g.][]{appel1985a, barnes1986a}. This method reduces the
complexity of the algorithm to the desired $\mathcal{O}(N \log N)$.

Tree-based methods are commonly employed in \ac{CE} simulations. They
are a natural choice when particle-based, Lagrangian schemes
(\ac{SPH}) are used to solve the equations of fluid dynamics, because
they can also help to efficiently search for neighbor particles as
required for kernels with compact support (see
Sect.~\ref{sect:SPH}). But also in grid-based approaches to fluid
dynamics, a particle-based discretization may be used to model the
gravitational interaction. This is popular in particular in the
context moving-mesh methods. A problem with tree-based methods,
however, is that they introduce asymmetries. The approximate treatment
of the gravitational interaction of a target particle with distant
particles violates Newton's third law (\emph{actio} equals
\emph{reactio}). This is in conflict with global momentum conservation
and introduces spurious motions. Indeed, many \ac{CE} simulations
performed with this methods exhibit a slow drift of the center of mass
of the system. There are ways to alleviate this problem. The important
notion is that random force errors would cancel during the
simulation. The overall drift and momentum conservation violation
results from \emph{correlations} between the errors
\citep{springel2021a}. These correlations can be broken, and for this purpose
\cite{springel2021a} propose to randomize the relative location of the
particles with respect to the computational box.

\subsubsection{Alternative methods to calculate self-gravity}
\label{sect:grav_alt}

An approach that appears as a natural choice in the context of
grid-based discretizations, but is also used with a particle-based
schemes \citep[see][]{taam2006a}, are \emph{Fourier methods.} Solving
the Poisson equation (\ref{eq:poisson}) with a Green's function gives
\begin{equation}
  \mathrm{\Phi} (\vec{r}, t) = - G \int \frac{\rho(\vec{r}', t)}{|
    \vec{r} - \vec{r}' |} \mathrm{d}^3 r',
\label{eq:grav_alt}
\end{equation}
which corresponds to a convolution of the Green's function for the
Laplacian with the inhomogeneity of the equation, i.e.\ the
density. In Fourier space, this reduces to a simple multiplication and
the required discrete Fourier transforms can efficiently carried out
with the Fast Fourier Methods (FFM). Similar to the tree-based
methods, this reduces operation count for calculating gravity in an
$N$-body system to $\mathcal{O}(N \log N)$.

A fast grid-based approach that reduces the effort to at least
$\mathcal{O}(N \log N)$ is the \emph{multigrid technique}
\citep{brandt1977a}. It accelerates iterative relaxation methods for
solving the Poisson equation \citep{press2007a}. Changing the elliptic
original problem into a pseudo-time dependent expression, a diffusion
equation is obtained that---when relaxed to equilibrium where the
pseudo-time derivative vanishes---yields the solution of the original
problem. Starting from an initial guess for the gravitational
potential, the relaxation is achieved in an iteration. The error in
the initial guess can be seen as being distributed over the domain
with various modes of different wavelength. Since standard relaxation
schemes couple only neighboring cells, short-wavelength errors are
quickly eliminated, but damping the long-wavelength error components
takes many iteration steps. This is avoided by mapping to a hierarchy
of subsequently coarser grids where the original long-wavelength
components are short compared to the grid spacing and can be damped
out in few relaxation steps. Subsequent projection back to the
original fine grid combined with further relaxation steps and cycling
through this hierarchy of grids efficiently produces the desired
equilibrium solution. The multigrid method is has been used in \ac{CE}
simulations by \citet{ricker2012a}.

\subsubsection{Gravitational softening}
\label{sect:gravsoft}

If the Poisson equation (\ref{eq:poisson}) is discretized on a grid
together with the equations of fluid dynamics, the grid spacing limits
the spatial resolution for both the hydrodynamic quantities and the
gravitational potential. As discussed above, however, self-gravity can
also be solved in a particle-based discretization, independently of
the discretization choice for the hydrodynamic part of the model; and
even in grid-based approaches the stellar cores are often modeled as
point masses. Particles can get arbitrarily close and one may expect
an arbitrarily fine spatial resolution in the discretization of
gravitational interaction---a potential advantage when combined with
a particle-based or moving-mesh discretization of fluid dynamics. As
particles approach each other, however, the gravitational forces
between them grow without limits requiring the scheme to take ever
smaller timesteps. This produces undesirable time sinks in the
numerical scheme.

Therefore, a fix is commonly employed: the gravitational potential is
unphysically altered to flatten for short distances, so that the
force approaches zero instead of scaling with the inverse of the
squared distance. Different functional forms are employed in practice,
but the simplest \emph{softened} gravitational potential---sometimes
called the Plummer form of softening \citep[e.g.,][]{dehnen2001a,
  price2007a}---reads
\begin{equation}\label{eq:gravsoft}
  \mathrm{\Phi}_\mathrm{soft} (r) \defeq - \frac{G m}{\sqrt{r^2 +
      r_\mathrm{soft}^2}},
\end{equation}
where $r$ denotes the distance to the particle of mass $m$, and
$r_\mathrm{soft}$ is the parameter altering the physical law---the
so-called \emph{softening length}.

In grid-based discretizations of fluid dynamics combined with
particle-based gravity solvers, the size of the grid cells $\delta$
provides an obvious choice for the scale of the gravitational
softening. However, the softening parameter does not necessarily have
to be a constant but it may depend on the distance $r$.
\citet{ruffert1993a} suggested the form
\begin{equation}\label{eq:ruffert}
  \mathrm{\Phi}_\mathrm{Ruffert} (r) \defeq - \frac{G m}{\sqrt{r^2 +
      \epsilon^2 \delta^2 \exp [-r^2/(\epsilon \delta)^2]}},
\end{equation}
with $\epsilon = 1.5$.

In \ac{SPH} approaches, the only sensible scale for the softening
length is the \ac{SPH} smoothing length. This can be implemented by using a
general formulation for force softening \citep{dehnen2001a,
  price2007a} with the gravitational potential due to all particles
$i = 1 \ldots N$,
\begin{equation}\label{eq:phi_general}
  \mathrm{\Phi} (\vec{r}) = - G \sum\limits_{i=1}^N m_b \phi(|\vec{r}
  - \vec{r}_i|, h).
\end{equation}
The \emph{softening potential kernel} $\phi$ is related to the
\ac{SPH} smoothing kernel (\ref{eq:smoothing_kernel}) via the Poisson
equation
 \begin{equation}\label{eq:phisoft}
W(r, h) = - \frac{1}{4 \pi r^2} \frac{\partial}{\partial r} \left( r^2
\frac{\partial \phi}{\partial r}\right)
\end{equation}
\citep{price2007a}.  As a consequence, the gravitational potential is
softened with a spline function. Such a choice seems preferable
because, in contrast to the expressions (\ref{eq:gravsoft}) and
(\ref{eq:ruffert}), splines reproduce the exact gravitational
potential outside a finite radius. 

\begin{figure}
    \centering
    \includegraphics[scale=0.75]{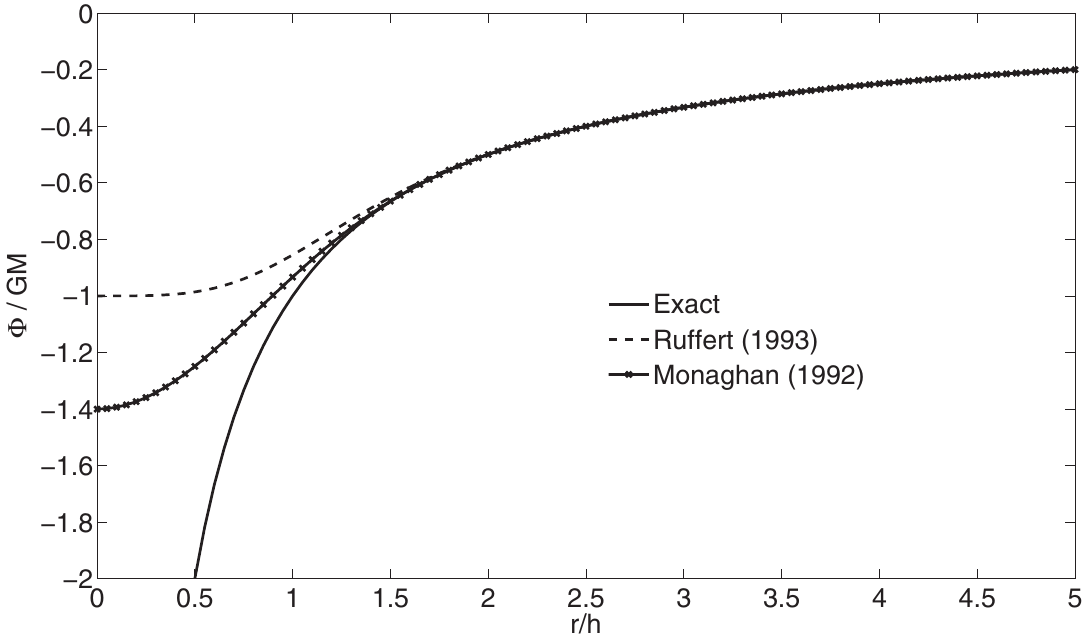}
    \caption{Exact
      gravitational potential compared with the softened potential
      according to Eq.~(\ref{eq:ruffert}), labeled as ``Ruffert
      (1993)'' and
      Eqs.~(\ref{eq:phi_general},\ref{eq:phisoft},\ref{eq:smoothing_kernel})
      labeled as ``Monaghan (1992)'' in arbitrary units with $h =
      \epsilon \delta = 1$. Figure from \citet{passy2012a}.}
    \label{fig:passy12}
\end{figure}

While alleviating the time sink problem, a consequence of
gravitational softening is that inside $r_\mathrm{soft}$ the simulated
dynamics is altered and less trustworthy. The physical law of
gravitational interaction is modified, most strongly at distances
shorter than the softening length. This is no fundamental problem for
simulating self-gravitating fluids in \ac{SPH} approaches or
grid-based methods combined with particle-based discretization of
gravity. The scale down to which fluid dynamics is resolved usually
corresponds to the scale of the softening; therefore, the
gravitational interactions are sufficiently well modeled.

In the context of \ac{CE} simulations, however, the stellar cores are
usually represented as point particles (in the case of \ac{SPH} approaches,
they are not to be confused with the \ac{SPH} particles; see
Sect.~\ref{sect:representing_cores}). These core particles are
distinct from the envelope gas. They interact via gravitational forces
but not via hydrodynamical effects. Since the core particles comprise
a substantial fraction of the total mass of the system in two points,
the associated gravitational potentials fall off very steeply close to
them. Gravitational softening must be applied to the core particles in
order to avoid fatal time sinks. For this, a softening length has to
be set. Again, depending on the discretization approach of the
hydrodynamics in the envelope gas, assigning $r_\mathrm{soft}$ to
either the grid spacing scale or the smoothing length would be
desirable. This, however, requires to follow the very steep gradient
of the potential close to the core particle, which makes the numerical
representation of mechanical equilibrium configurations (such as
required in the setup of \ac{CE} simulations, see
Sect.~\ref{sect:challenges_setups}) challenging. The pressure gradient
counteracting gravity has to be resolved in the hydrodynamic part of
the simulation and the required fine resolution leads to unacceptably
small time steps. In this situation, a softening according to
Eq.~(\ref{eq:ruffert}), has a practical advantage over the physically
more accurate spline softening: at close distances it is shallower
(see Fig.~\ref{fig:passy12}) and a shallower gradient can be resolved
with fewer discretization elements.

Whatever choice is made for the functional form of the softening,
however, setting the gravitational softening to the scale of the
hydrodynamic discretization elements does not ensure sufficiently
large time steps for currently affordable simulations. Therefore,
\ac{CE} simulations usually shy away from a safe choice of
gravitational softening parameters---from a computational point of
view, the softening length should be large to reduce the computational
cost of the time integration. But large $r_\mathrm{soft}$ implies a
large region around the core particles in which the
gravo-hydrodynamical interaction is suspicious. The dynamics around
the cores are the driver for \ac{CEE} and should be modeled as
accurately as possible. In fact, \citet{reichardt2020a} found that
smaller softening lengths in their simulations lead to faster
interaction and earlier unbinding of the envelope
material. \citet{staff2016b} carried out a similar test using an
Eulerian grid-based hydrodynamics code and the interaction seems to be
strengthened for smaller softening lengths leading to a slight
increase in the final orbital separations of the cores and slightly
more unbound mass. In virtually all \ac{CE} simulations softening length
for the core particles is set according to computational efficiency
considerations at the expense of physical accuracy. Therefore, careful
convergence tests are necessary.

\subsection{Advantages and drawbacks of hydrodynamic simulation methods}
\label{sect:simulations_comparison}

In this section, we summarize some advantages and drawbacks of each of
the three main techniques used in \ac{3D} hydrodynamic \ac{CE}
simulations: \ac{SPH}, Eulerian grid-based schemes, and moving-mesh
approaches. Only few direct comparisons have been carried out
\citep[most notably][who compare \ac{SPH} and Eulerian grid-based
  simulations]{passy2012a}, and the available results do not point to
major systematic discrepancies between simulations employing different
numerical methods. It is far more likely that different results are
obtained because of individual choices within each method. This, of
course, is not a universally true statement but merely an assessment of
the current state-of-the-art. As long as the physical models and
implementation details of the methods dominate, the specific features
of the fundamental schemes may remain hidden.

For future developments, however, it is important to be aware of the
different capabilities of the methods. The current status of \ac{CE}
simulations, discussed in Sect.~\ref{sect:current}, merely scratches
the surface of the problems involved in this complex phase of binary
stellar evolution. The basic question of whether envelope ejection is
possible under certain modeling assumptions can be addressed with
different numerical schemes. For the current simulations, this choice
affects mostly the efficiency of \ac{CE} simulations: The spatial
resolution required to capture certain physical processes depends on
the discretization method and it differs with its choice. As a
consequence, this affects the fidelity of the simulations that can be
carried out with given computational resources.

This situation may be regarded as an indication of the immaturity of
the field of \ac{CE} simulations: A detailed understanding of \ac{CEE}
requires improved simulations and the different features of methods
will play out more in the future.  Each technique has particular
drawbacks that may compromise the fidelity of the simulated parameters
of the \ac{CE} interaction. Generally, the important aspects to
consider are accuracy, efficiency and parallelizability, boundary
conditions and flexibility with the geometry of the setups,
conservation properties, and general versatility when it comes to
include additional physics.

\subsubsection{Boundary conditions and setup geometries}
\label{sect:boundaries}

The system to be modeled consists of two stars orbiting each other and
engaging into a dynamical interaction. This setup is far from a
spherically-symmetric configuration.  Because of the lack of spatial
symmetries in \ac{CEE}, special grid geometries (such as a spherical
grid) do not provide any advantage and there are no obvious boundaries
of the domain of interest such as stellar surfaces. Eulerian
grid-based \ac{CE} simulations are usually conducted on Cartesian
meshes in cubical boxes. This leads to conflicts with the fact that
the considered \ac{CE} systems are to a good approximation embedded in
vacuum\footnote{Pre-CE stages, however, may lead to mass loss and some
circum-binary material may be present in reality.}. Empty grid cells
cause numerical problems and therefore ``true'' vacuum cannot be
represented straightforwardly in grid-based approaches---a problem for
both Eulerian and moving-mesh discretizations. Therefore, cells that
are located outside of the system are usually filled with a
low-density \emph{pseudo-vacuum} in the hope that it does not affect
the dynamics of the simulated \ac{CE} interaction. If hot enough,
however, this pseudo-vacuum may have a welcome effect on the initial
setup: it can reduce the steep pressure gradient at the surface of the
primary star and therefore stabilize the stellar model against the
dispersion observed in the case of insufficient spatial resolution of
the gradient. But this comes at the price that observables cannot be
meaningfully be extracted from such simulations \citep[see][for a
  detailed discussion]{galaviz2017a}. Moreover, if too hot,
pseudo-vacuum cells may reduce the \ac{CFL} timestep. Therefore, it is
advisable to keep the pseudo-vacuum as cold as possible. With certain
setup procedures (see Sect.~\ref{sect:challenges_setups}), this is
possible.

In grid-based \ac{CE} simulations, the simulated domain has to be
finite in its spatial extent and therefore outer boundary conditions
at the edges of the computational domain have to be implemented in a
way such that they do not interfere with the modeled
system. Certainly, no material should flow into the simulation box,
but, in principle, outflow of material would be acceptable. A loss of
material, however, makes checking for conservation of mass, energy,
angular and linear momentum inaccurate and thus the simulations cannot
be fully validated using these global criteria. For this reason, the
domain sizes are usually chosen so large that even with successful
envelope ejection no material leaves the domain. This requires to fill
large volumes outside of the interacting system with pseudo-vacuum,
which can have some unwanted dynamical effects on the simulation
\citep[for a discussion see][]{chamandy2019a, iaconi2018a}.  To
minimize such artifacts, the pseudo-vacuum should contain as little
mass as possible---i.e.\ for the desired large domain sizes it should
have very low densities.

Representing \ac{CE} systems on a static grid as in the Eulerian
approaches causes strong advection errors, implying violation of
conservation properties. The only cure to this is an increase in
spatial resolution---a common numerical approach to alleviate
problems. Because of the spatial scale problem
(Sect.~\ref{sect:multiscale}) and in the light of the desire to make
the simulation domain as large as possible to avoid interaction with
the boundaries, this is only promising if high spatial resolution is
only realized certain, temporarily changing parts of the
domain. Therefore, as discussed in Sect.~\ref{sect:eulerian}, \ac{AMR}
techniques are widely used in \ac{CE} simulations.

\ac{SPH} methods do not require sophisticated outer boundary
conditions. Empty spaces between the objects of interest are possible
and no pseudo-vacuum is needed. The artificial surface tension
mentioned in Sect.~\ref{sect:SPH} supports the surface of the primary
star model and no advection errors disperse the stellar
material. These features make \ac{SPH} codes a natural choice for
simulations of \ac{CEE} simulations. However, as discussed in
Sect.~\ref{sect:SPH}, these have drawbacks when it comes to resolving
low-density envelope material and hydrodynamic processes in it.

For this reason, moving-mesh approaches present a favorable
compromise. They combine the accuracy of grid-based discretizations
with the flexibility of a Lagrangian method. With their spatial
discretization based on grid-generating points, adaptive mesh
refinement is straightforward and sufficiently large computational
domains pose no problem. The need of grid-based methods to fill voids
with pseudo-vacuum, however, persists.

\subsubsection{Accuracy, efficiency, and versatility}

A satisfying flow representation can in principle be achieved with any
converging scheme. The key question, however, is the numerical
resolution required to reach a certain level of accuracy---and this is
where the schemes differ fundamentally. In this context, it has to be
pointed out that an often used comparison between the number of
particles in an \ac{SPH} simulation and the number of cells in a
Godunov-like grid-based simulation is misleading. Whereas \ac{SPH}
methods average over the particles within the smoothing kernel,
Godunov schemes compute the fluxes of hydrodynamic quantities between
cells and reach an effective resolution defied by the grid scale. The
effective spatial resolution of \ac{SPH} simulations with the same
number of particles is usually lower by a factor of a few.

Generally, finite-volume grid-based discretization techniques are
superior over \ac{SPH} methods in resolving hydrodynamic flow features
accurately. However, at low resolution, \ac{SPH} methods often have to
be given preference. They yield robust results and, in the context of
\ac{CE} simulations, are capable of determining global properties of
the evolution with rather little computational cost. While grid-based
simulations have to cover vast amounts of pseudo-vacuum with the same
effort as the actual stars, \ac{SPH} focuses the resources on modeling
the physical objects of interest.

While \ac{AMR} techniques alleviate the problems with grid-based
\ac{CE} simulations, they introduce a significant computational
overhead and require some implementation effort. Still, Eulerian
grid-based simulations reach an accuracy that is very hard to match
with \ac{SPH} techniques. These tend to under-resolve low-density
regions. Moreover, implementing additional physics such as magnetic
fields and other effects seems more natural on a given grid than in
the \ac{SPH} approach, although it is not impossible there either
\citep{price2012a}. Subgrid-scale models representing unresolved
physical effects require a well-defined grid-scale to connect to.

As mentioned before, moving-mesh techniques combine the best of both
worlds: the accuracy and versatility of grid-based finite-volume
methods with the geometric flexibility of a Lagrangian method. A
drawback is the complex grid-generation procedure that tessellates
space in every timestep. This overhead renders moving-mesh approaches
computationally more expensive than \ac{SPH} but they are still
superior over Eulerian grid-based approaches in terms of efficiency.

\subsubsection{Energy and angular momentum conservation}
\label{sect:conservation}

Many hydrodynamics codes guarantee conservation of total
(i.e.\ kinetic plus internal) energy and linear momentum. For
finite-volume schemes, cell-averaged conserved quantities are updated
via fluxes over the cell edges. What leaves one cell enters its
neighbor. This is the reason why finite-volume approaches conserve
momentum and energy globally (but not necessarily locally)---inside
the domain, net loss or gain in conserved quantities is impossible. In
the case of \ac{SPH}, mass and energy are directly ascribed to the
discretizing particles and as long as no particle is lost, these
quantities are conserved in this scheme, too. Conservation of total
energy and linear momentum, however, is not sufficient for reliable
\ac{CE} simulations for two main reasons.

The first reason is that for the problem at hand, \emph{angular}
momentum conservation is critical. In contrast to linear momentum and
energy balance, the Euler equations of fluid
dynamics---(\ref{eq:mass_cons}) to (\ref{eq:energy_cons}) or
(\ref{eq:lagrange_mass}) to (\ref{eq:lagrange_energy})---do not
explicitly include angular momentum balance, but only imply
it. Whether or not these quantities are conserved in numerical
simulations depends critically on the discretization approach. In
grid-based Eulerian schemes, angular momentum conservation is not
guaranteed due to the inherent advection errors. These can be
decreased by adapting the geometry of the spatial discretization to
the expected flow, but this is neither universal nor always possible
in complex astrophysical situations. As emphasized above, the flows in
\ac{CEE} certainly lack clear symmetries that could be
exploited. Therefore, the only way to improve angular momentum
conservation in these approaches is a costly increase of numerical
resolution. As discussed in Sect~\ref{sect:SPH}, Lagrangian schemes
are superior in conserving angular momentum because they avoid
advection errors by construction. In fact, a variational derivation of
the \ac{SPH} method \citep[e.g.][]{monaghan2001a, springel2002a}
proves this feature: Within the framework of Lagrangian mechanics, the
equations of fluid dynamics can be derived from a Lagrangian
\citep{eckart1960a} via a variational principle.  Instead of
discretizing the resulting partial differential equations, however,
one can discretize the Lagrangian in a particle-based approach and
then derive the (discretized) \ac{SPH} equations directly. Symmetries
of the original and discretized Lagrangians give rise to conservation
properties: the absence of explicit time dependence implies energy
conservation, their translational invariance gives linear momentum
conservation, and---most importantly---their rotational invariance
leads to angular momentum conservation, not only at the level of the
partial differential equations, but also for the discretized
equations.

The second problem is that while total energy and linear momentum
conservation is---for conservative schemes---guaranteed for the
numerical implementation of the equations of fluid dynamics without
source terms, they do not generally hold in cases where external
forces are discretized in different approaches than the hydrodynamic
quantities. This is the case for gravity---which, as for many
astrophysical problems, is of fundamental importance for \ac{CEE}. The
basic gravo-hydrodynamic model relies on an exact treatment of the
conversion of gravitational binding energy and orbital energy to
internal energy and acceleration of the envelope gas. Mismatches
between the discretizations of the hydrodynamic variables and the
gravitational potential can lead to problems with maintaining the
expected mechanical equilibrium in stars (see
Sect.~\ref{sect:challenges_setups}), but they may also lead to
accumulating errors in the conserved quantities over long simulation
periods modeling the \ac{CE} interaction.  Apart from these more
subtle issues, the very treatment of self-gravity already poses
challenges to numerical schemes. As argued in
Sect.~\ref{sect:gravity}, an exact direct summation over all
contributions of $N$ discrete elements (cells in grid-based
approaches, or particles in the corresponding schemes) falls into the
rather unacceptable $\mathcal{O}(N^2)$ time complexity
class. Therefore, approximate algorithms, such as discussed in
Sects.~\ref{sect:tree} and \ref{sect:grav_alt}, are preferred. The
employed approximations induce errors that potentially can accumulate
over the duration of the simulations---not only in the energy but also
in linear angular momentum. The center-of-mass of the considered setup
can start to drift. Apart from these numerically and algorithmically
caused errors, the softening of the gravitational potential
(Sect.~\ref{sect:gravsoft}) alters the interactions at the fundamental
physical modeling level. Still being a central force field,
conservation of energy should not be affected, but changes due to
adaptive gravitational softening lengths may cause problems because
they artificially modify the shape of the gravitational potential in
time.
\begin{figure}
    \centering
    \includegraphics[width = 0.48 \textwidth]{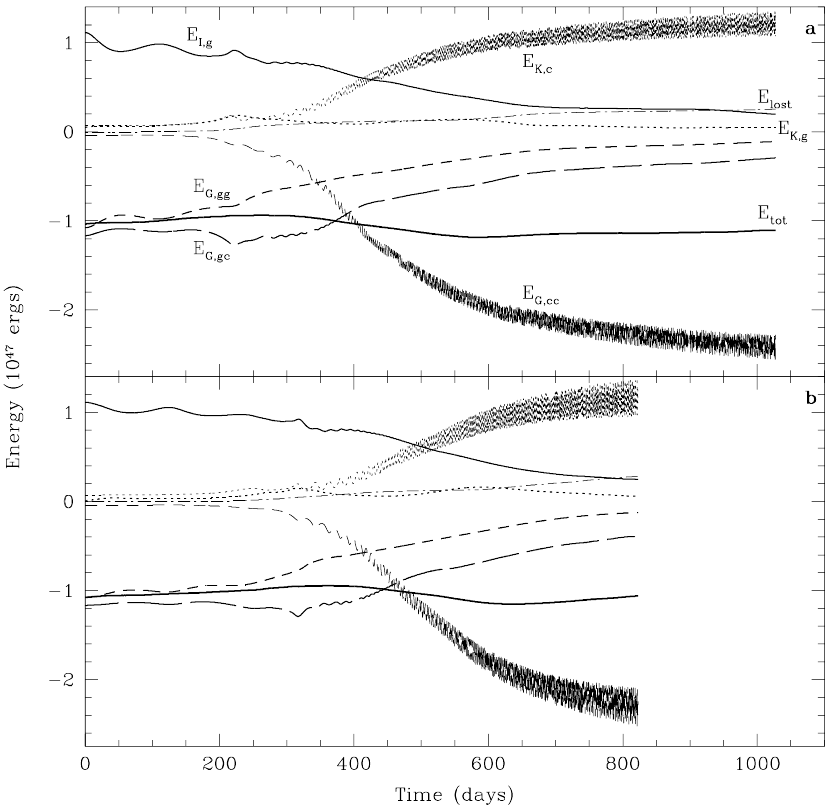}\hfill
    \includegraphics[width = 0.48 \textwidth]{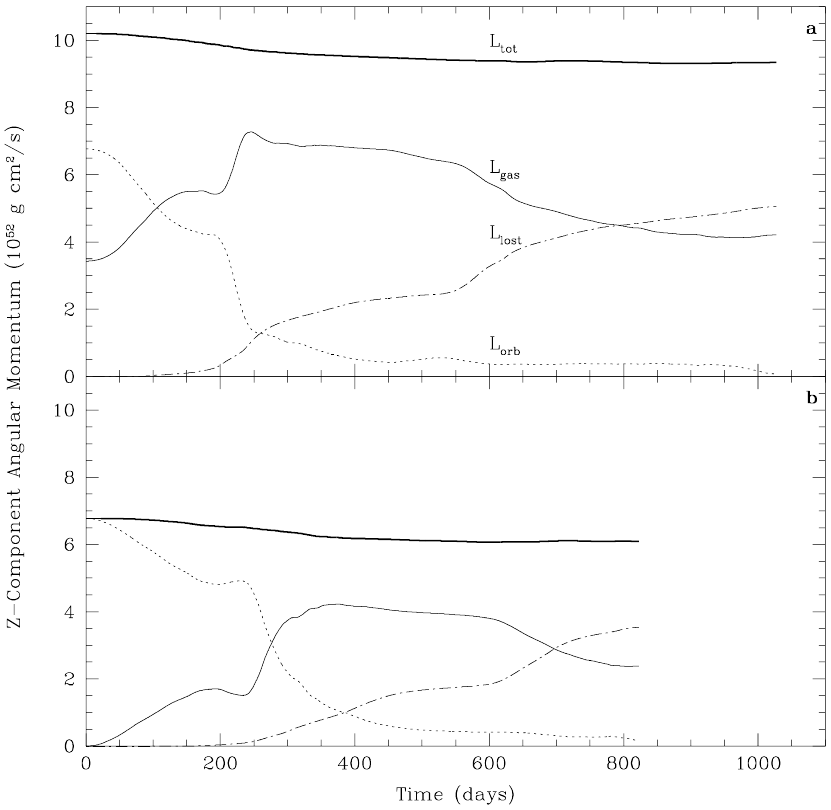}
    \caption{Evolution of energy and angular momentum in two
      simulations (panels a and b) of \citet{sandquist1998a} with
      different initial parameters. Shown are different components:
      the kinetic energy of the core particles and the gas
      ($E_\mathrm{K,c}$ and $E_\mathrm{K,g}$, respectively), the
      internal energy of the gas ($E_\mathrm{I,g}$), energy lost from
      the simulated domain ($E_\mathrm{lost}$), potential energies
      from gravitational interaction between gas and gas
      ($E_\mathrm{G,gg}$), gas and core particles ($E_\mathrm{G,gc}$),
      and between the core particles ($E_\mathrm{G,cc}$), the
      $z$-components of the total angular momentum ($L_\mathrm{tot}$),
      the angular momentum of the gas ($L_\mathrm{gas}$), and the core
      system ($L_\mathrm{orb}$), and the angular momentum lost from
      the domain ($L_\mathrm{lost}$). By the end of the simulations,
      conservation of total energy and angular momentum was violated
      at the $\sim$10\% level.  Figures from \citet{sandquist1998a}.}
    \label{fig:sandquist1998_conservation}
\end{figure}

Numerical errors in the conservation of angular momentum and energy
are an unavoidable compromise when striving for efficient
simulations. It is, however, essential to monitor them. Energy errors
are particularly critical in simulations of \ac{CEE}, because one of
the goals is to determine the unbinding of the envelope, which is
rather loosely bound in the giant primaries anyway. It has to be
ensured that any unbinding of material is physical and not caused by
numerical inaccuracies. The question of what level of non-conservation
can be tolerated is an important one. \citet{sandquist1998a} published
their conservation of energy and angular momentum showing that over
their entire simulations they tolerated a change of approximately
10~\%, see Fig.~\ref{fig:sandquist1998_conservation}. There is no
strict quantification of an acceptable level and it might affect
different output quantities differently. Clearly, if at the end of a
numerical simulation the binding energy of remaining envelope gas is
of the same order of magnitude as the accumulated energy error, little
can be said about the success of envelope ejection. It is therefore
mandatory to track the conservation of energy and angular momentum and
the results should be published alongside with the simulations.

Generally, more work is called for to identify and localize the
numerical problems that violate energy and momentum conservation. This
most likely requires extensive resolution studies.

\subsubsection{Code speed and scalability}
\label{ssec:speed-and-scalability}

Performing \ac{3D} hydrodynamic simulations of \ac{CEE} is
computationally very challenging. The system has to be followed over
long periods of time with high accuracy, i.e.\ many numerical time
integration steps are to be performed and the spatial resolution has
to be sufficiently high. Clearly, such simulations require high
performance computing (HPC) resources and are only possible when
employing parallelization techniques. Therefore, scalability to large
numbers of computational cores to be used in parallel is an important
criterion when choosing a code for \ac{CE} simulations.

Parallelization in the spatial domain is standard for the codes
currently used for \ac{CE} simulations. In finite-volume approaches,
the hydrodynamics solvers can usually be parallelized on
distributed-memory machines with some acceptable communication
overhead. Distributed-memory parallelization is more challenging for
particle-based discretization approaches as used for modeling
self-gravity (see Sect.~\ref{sect:tree}) or the hydrodynamics itself
(see Sect.~\ref{sect:SPH}). For this reason, some of the \ac{SPH}
codes employ shared-memory parallelization \citep[e.g., {\sc
    phantom};][]{price2018a}, which restricts the simulations to a
moderate number of computational cores, but there are also \ac{SPH}
implementations with distributed-memory parallelization,
e.g. \textsc{gadget-2} \citep{springel2005a} and \textsc{snsph}
\citep{fryer2006a}. In moving-mesh approaches, the grid generation
poses an additional challenge to efficient parallelization.

For an approximately fixed spatial resolution and a given numerical
algorithm, the speed-up that can be reached with parallelization is
limited by Amdahl's law \citep{amdahl1967a} or similar
considerations. Beyond this limit, an increase of the number of
employed compute cores no longer reduces the execution time
proportionally. Therefore, it is not only scalability of the employed
code that determines the feasibility of a certain \ac{CE} simulation,
but also its overall efficiency, i.e.\ the required wall-clock time to
perform a time integration step for a given number of discretization
elements in a fixed parallel setup. A common experience in code
development is that these aspects---scalability and overall
efficiency---are competing. A higher efficiency often causes
load-imbalances and thus comes at the expense of parallel
scalability. This does not necessarily imply a disadvantage of the
respective code for performing \ac{CE} simulations. Some of the codes
employed for such simulations have been developed originally for
extensive cosmological simulations and are highly optimized for
efficiency; for example the \textsc{arepo} code of
\citet{springel2010a} has been used for the groundbreaking
\emph{Illustris} project \citep{vogelsberger2014a}.

Ideally each simulation would report the total number of CPU-hours,
the number of nodes and cores it was executed on, wall-clock times,
and the computer specifications. It is assumed that at least a
fraction of the publications would be presenting simulations at the
limit of what is feasible. For example the 1 million particle
simulation presented by \citet{iaconi2017a} was at the limit of
feasibility for the OpenMP, shared memory code \textsc{phantom}.

\subsection{CE phases accessible to 3D hydrodynamic simulations}
\label{sect:3d_accessible}

The numerical methods for \ac{3D} hydrodynamic simulations discussed here
can be used to follow the dynamical part of \ac{CE} interaction. This
includes the initiation of \ac{CEE} (perhaps only the last part of
Phase (i) of our classification in Sect.~\ref{sect:phases}, but not
the complete pre-\ac{CE} evolution) and the plunge in and inspiral of the
companion, classified as Phase (ii).

Phase (iii), however, is characterized by different timescales
involving the long $\tau_\mathrm{KH}$, which, as discussed in
Sect.~\ref{sect:multiscale}, differs from the dynamical timescale of
the envelope of \ac{RGB} and \ac{AGB} stars by five orders of
magnitude.  The discrepancies in relevant timescales are a serious
challenge to hydrodynamic simulations and it seems unlikely that the
long-term \ac{CEE} can be followed in \ac{3D} hydrodynamic
simulations. A reasonable resolution of the objects of interest
requires about $10^2$ to $10^3$ discretization elements (grid cells or
softening lengths) per spatial dimension. Taking the dynamical
timescale as a proxy for the \ac{CFL} time step restriction in these
discretization elements shows that covering timescales relevant for
modeling Phase (iii) requires at least $10^7$ to $10^8$ time
integration steps with hydrodynamic schemes, which makes such
simulations unrealistic.

A solution to this problem would be to couple \ac{3D} hydrodynamic
simulations of the inspiral phase (ii) with classical \ac{1D} stellar
evolution models to describe the subsequent evolution. While the
\ac{3D} simulations provide the initial conditions, mapping them in a
useful way to \ac{1D} is highly non-trivial. We discuss this problem
further in
Sect.~\ref{ssec:extracting_1d_parametrization_from_multiD_simulations}.

\section{Initial conditions and setups of common-envelope simulations}
\label{sect:setups}

We now turn to the question of what hydrodynamic simulations of
\ac{CEE} are possible based on the physical foundations discussed in
Sect.~\ref{sect:challenges_physical} and the numerical approaches
described in Sects.~\ref{sect:cfd} and \ref{sect:gravity} with
available computational resources.  We start with the construction of
a sensible initial model for global \ac{3D} hydrodynamic simulations
of Phase (ii) of \ac{CEE} according to our classification in
Sect.~\ref{sect:phases}. This is not only difficult from a technical
point of view, but we also have to ensure that such a model represents a
reasonable end point of the evolution \emph{preceding} the \ac{CE}
interaction, i.e.\ at the end of Phase (i).

The high densities in the cores of giant stars are the reason why
degeneracy effects play a role in them, which cannot be described by a
classical ideal-gas equation of state. Core material must be treated
instead as an ideal quantum gas. This is no fundamental problem. The
hard limit of what can be accommodated in simulations is set by the
scale problems discussed in Sect.~\ref{sect:multiscale}.

\subsection{Can the stellar cores be resolved?}
\label{sect:comp_feasible}

At the first glance, the \emph{spatial scale discrepancy} between the
radius of the core, $R_\mathrm{core}$ and the outer envelope of a
giant star, $R_\star$, of about four orders of magnitude (see
Sect.~\ref{sect:multiscale})---and for some special cases of
early-stage giants perhaps even less---seems challenging but not
impossible to represent in numerical treatments. However, a closer
look quickly reveals that simulations in which these components are
fully resolved are quite unrealistic. A representation of the core
structures would require to resolve spatial scales at least a factor
of 10 to 100 below $R_\mathrm{core}$. Thus, a reasonable resolution of
the core leads to a spatial scale difference of about $10^5$ between
the integral scale of the primary star and the sizes of discrete
elements relevant for the \ac{CFL} condition. Still, this is extremely
challenging but not completely hopeless to be resolved in future
simulations.

A computational domain that allows to capture the expansion of the
envelope during \ac{CEE} without losing material over its boundaries
(which, as discussed in Sect.~\ref{sect:boundaries} would make it hard
to track the conservation properties of the simulation) has to extend
by several orders of magnitude beyond the radii of the giant
stars. Covering such domains uniformly with the finest required
resolution requires computational resources far beyond those of the
foreseeable future. Adaptive resolution refinement as inherent to
\ac{SPH} methods or included in grid-based approaches via \ac{AMR}
techniques can alleviate the problem somewhat but not sufficiently to
render fully resolved simulations a realistic option.

There is yet another, more restrictive reason why \ac{CE} simulations
with resolved cores are unrealistic. It arises from the \emph{timescale
problem} mentioned in Sect.~\ref{sect:multiscale}. The dynamical
timescale defined in Eq.~(\ref{eq:tdyn}) is the relevant quantity for
estimating the computational cost of the simulation. Generally,
hydrodynamic simulations aim to follow the evolution of the modeled
system over a duration close to $\tau_\mathrm{dyn}$, which therefore
sets a coarse order-of-magnitude estimate for the required number of
time integration steps. This argument, however, can be formalized
further. As discussed in Sects.~\ref{sect:eulerian} and
\ref{sect:SPH}, the \ac{CFL} condition states that numerical stability
of hydrodynamic schemes in explicit time discretization necessitates
discrete time integration steps smaller than the sound crossing
time\footnote{The sound crossing time is sufficient for our estimate,
but the actual \ac{CFL} condition is more specific and refers to the
hydrodynamic signal speeds.}  over a grid cell in grid-based
discretizations, or a smoothing length in \ac{SPH}. As estimated in
Sect.~\ref{sect:multiscale}, the dynamical timescale of the envelopes
of \ac{RGB} or \ac{AGB} stars is about $50$ days. In the cores of such
stars, $\tau_\mathrm{dyn, core}$ is only ${\approx} \, 20 \, \mathrm{s}$
(see Sect.~\ref{sect:multiscale}), i.e.\ it is lower than
$\tau_\mathrm{dyn, envelope}$ by five orders of magnitude. When
covering the radius of the core with $100$ resolution elements (grid
cells or \ac{SPH} smoothing lengths), the \ac{CFL} condition dictates
timesteps that are seven orders of magnitude smaller than the
dynamical timescale of the envelope. Global \ac{CE} simulations
striving to resolve the core of the primary star while also following
the evolution for a dynamical timescale of the envelope therefore face
a temporal scale discrepancy of $10^7$, i.e.\ ten million numerical
time integration steps are necessary to follow the system even over
only one $\tau_\mathrm{dyn, envelope}$. Even in the most optimistic
cases with a low spatial resolution of the stellar core, the timescale
difference does not reduce below five orders of magnitude. Therefore,
at best for some very favorable configurations fully resolved \ac{3D}
hydrodynamic simulations of \ac{CEE} can possibly attempted with
future computational hardware, but this will not become a standard
approach anytime soon.

Local time stepping focuses the computational resources on the stellar
core---meaning that not the entire domain is evolved in each small
time step. Nonetheless, the effort needed to follow the evolution of
the envelope remains prohibitive. The \ac{CFL} condition can be relaxed for
unconditionally stable implicit time integration schemes, but such
schemes are computationally costly and difficult to implement, so that
they have not yet been applied to simulations of \ac{CEE}. Moreover,
although stability is not a concern in time-implicit schemes, accuracy
of the solution is still important. If hydrodynamic flows inside the
core are to be resolved, a \ac{CFL}-like condition applies that replaces
the sound speed with the flow speed and a substantial gain is only
achieved for strongly subsonic flows.

\subsection{Representing the cores of the primaries and the companions}
\label{sect:representing_cores}

As discussed in the previous section, the question of resolving the
core of the giant star is not primarily a spatial scale problem; the
limiting factor is the timescale problem introduced by the core of the
primary star. Depending on its nature, one may hope to resolve the
companion. While compact objects such as white dwarfs or neutron stars
are too small to achieve sufficient resolution, a main sequence
companion may come into reach for future simulations. Still, for a
reasonably resolved sun-like companion star, the timescale discrepancy
of ${\sim}\, 10^5$ remains extremely challenging.

When \ac{CE} interactions take place between a primary star that has
just initiated its expansion, and a main sequence companion,
simulations may be able to represent both stars in their entirety. An
example are simulations of the transient V1309~Sco, whose progenitor
system is thought to have comprised a slightly enlarged (${\sim} \, 4
\,\rsun$) $1.5 \, \msun$ post-main sequence primary and a $0.15 \,
\msun$ main sequence companion
\citep{tylenda2011a}. \citet{nandez2014a} carried out a number of
simulations of this system using, for some of them, a full
representation of the two stars. These simulations were conducted with
an \ac{SPH} code and extremely low particle numbers of $\num{100000}$ to
$\num{200000}$ for the primary star and $\num{5000}$ to $\num{20000}$
for the companion. Lacking a credible convergence test, caution is
required in the interpretation of the results. However, in principle
one may attempt such simulations with higher resolution. This setup,
however, is already at the rather blurred boundary between \ac{CEE}
and stellar mergers and may, because of the lack of a clear
core-envelope structure, not completely qualify as \ac{CE} interaction
(see our discussion in Sect.~\ref{ssec:terminology}).  For
completeness we should mention again that stellar interaction
simulations with both stars fully represented are also carried out for
compact objects such as white dwarfs and neutron stars, as well as for
massive main sequence stars, where the two stars have comparable sizes
(see Sect.~\ref{ssec:terminology} for examples). However, these
interactions, which result in mergers, do not follow the same path as
the \ac{CEE} described here in that one of the stars is usually tidally
shredded rather than moving into the envelope of the other.

For more standard \ac{CE} setups, the timescale problem arising from
the extremely compact nature of the core of the primary giant star
remains an insurmountable obstacle to global, fully resolved
simulations of well-defined \ac{CE} interactions in the foreseeable
future. There are two ways of dealing with this fundamental
problem.

When \ac{CEE} results in the survival of the binary and the ejection
of the envelope, simulations should aim to accurately reproduce the
core--envelope interaction, the timing and mechanism of envelope
ejection and the post-ejection orbital separation of the cores. In
this first case, the stellar cores are largely unaffected by the
inspiral. Resolving their internal structure is therefore less
important than a precise representation of the envelope of the primary
star that later becomes the common envelope. It is therefore common
practice to replace the cores by a simple model while trying to
capture the physics of the envelope on the simulation domain. This is
usually achieved by substituting the core of the primary star and also
the companion with \emph{point particles} that interact mutually and
with the envelope gas only via gravity. In this context, such point
masses are often referred to as \emph{core particles}. The main
drawback of this approach is that only the gravitational interaction
of the cores with the envelope is modeled and any hydrodynamic
interaction, such as accretion, expansion, evaporation, and jet
formation is ignored. Yet such phenomena may strongly affect the
envelope gas. Therefore, subgrid-scale models have been employed to
represent them in global \ac{CE} simulations
\citep[e.g.][]{chamandy2018a, zou2022a}.

If, however, the evolution of the large envelope is not
the primary interest of the simulation, a second option is to follow
inner parts of the system with resolved cores. This may apply to
systems where additional simulations capturing the envelope or other
considerations lead to the conclusion that envelope ejection fails and
a merger of the cores or an evaporation of the companion is the focus.

A clear core--envelope structure of the primary star is a prerequisite
for replacing its core by a point particle. But an exact definition of
a sharp core--envelope boundary is difficult to give. Depending on the
evolutionary stage of the giant and the mass of the star, the boundary
may be set according to different parameters. Supergiant stars usually
have a less clearly defined core--envelope structure than \ac{RGB} or
\ac{AGB} stars. From a stellar evolution point of view, chemical
composition may provide a useful discrimination: the core may have
undergone nuclear burning and the mass fractions of the corresponding
``ashes'' could be used to indicate the core--envelope boundary. Shell
burning, however, leads to ambiguities. For the \ac{CE} dynamics, it
would be more useful to determine the point down to which the envelope
material has to be removed so that the remaining bound objects does
not re-expand, i.e.\ the bifurcation point (see
Sect.~\ref{sec:parametrized_models}).

The strategy of a simplified representation of the cores has been
implemented in various ways. In \ac{SPH} simulations, a definition of
a core particle is rather straightforward because the envelope
material is also discretized as particles.

Grid-based approaches, in contrast, require somewhat more thought. One
option here is to plainly ignore the problem when mapping the stellar
structure from the initial \ac{1D} model to the hydrodynamic
computational domain. Such setups do not use core particles.  The core
of the primary star is simply taken to be part of the central grid
cell that comprises the stellar core and some envelope material. In
principle, mass flows into and out of this central cell are
possible. The density in the central cell containing the core is bound
to be high, but its value depends on the spatial resolution. This
rather ambiguous and resolution-dependent representation of the
stellar core makes convergence studies difficult and must therefore be
discarded in modern \ac{3D} hydrodynamic simulations of \ac{CEE}.

The introduction of a core particle alone, however, does not solve the
problem satisfactorily straight away. It introduces two ways of
treating stellar material---a particle modeling the core and a
grid-based discretization of the envelope. These need to be defined
and coupled in a physically consistent way. A na\"ive approach that
simply assigns the mass of the stellar core to the particle leaves a
resolution-dependent amount of envelope material in the central grid
cell. This high-density material is usually not sufficiently resolved
to guarantee a correct treatment of core-envelope interaction. Leaving
the envelope profile unaltered, as was common practice in many
simulations of dynamical inspiral in \ac{CEE}, leads to perturbations
of the \ac{HSE} in the envelope structure of the primary star and
generates spurious velocities. As a resolution-dependent model,
numerical convergence of the results of \ac{CE} simulations is hard to
establish with such a setup.

\begin{figure}
    \centering
    \includegraphics[width=\textwidth]{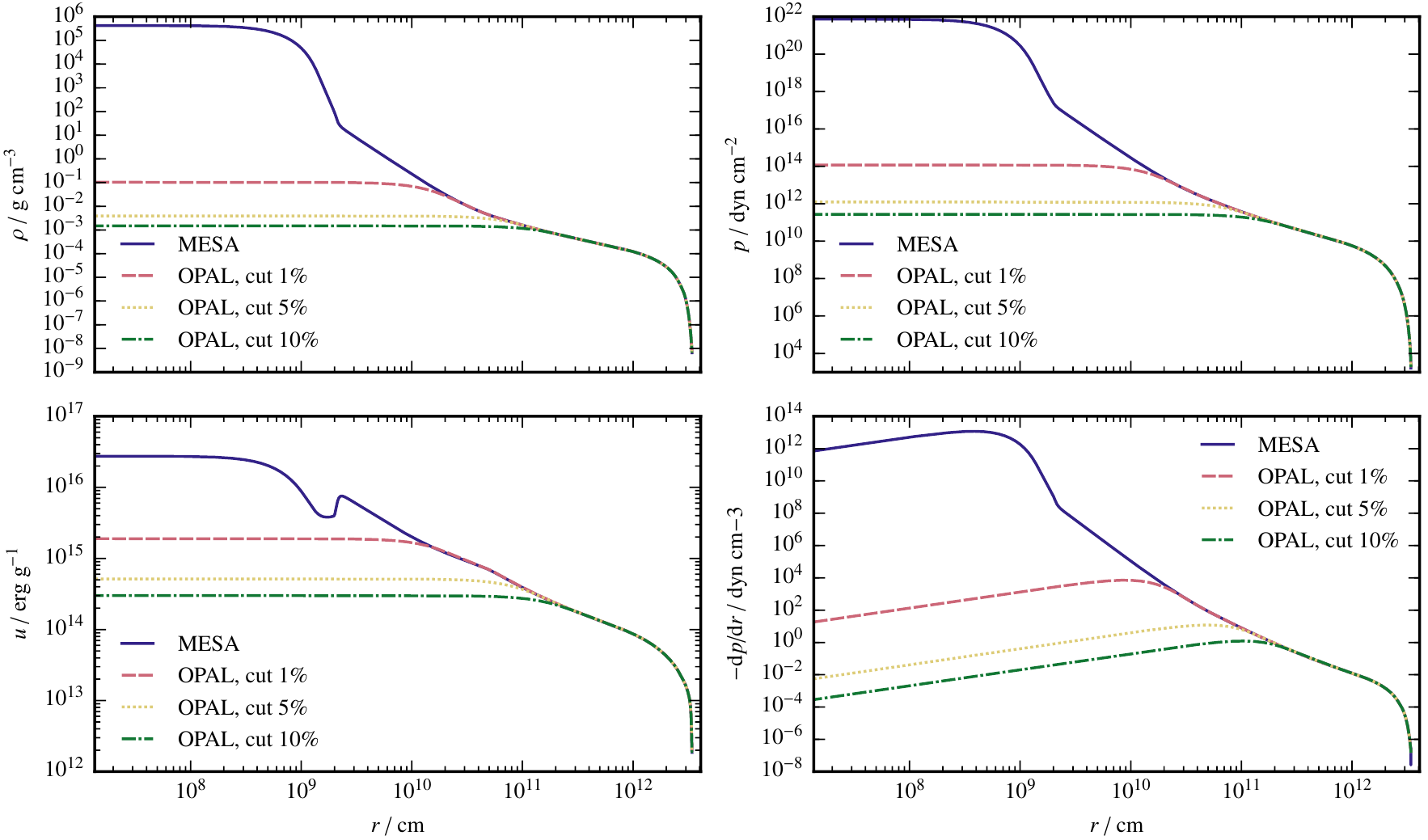}
    \caption{Radial profiles of density (top left), internal energy
      (bottom left), pressure (top right), and derivative of pressure
      (bottom right) of an original \textsc{mesa} model for a
      $2\,\msun$ \ac{RGB} star with reconstructed profiles
      (\textsc{``opal''}), in which the cores were cut out at the
      indicated percentage of the total stellar radius. Figure from
      \citet{ohlmann2017a}.}
    \label{fig:ohlmann17}
\end{figure}

For this reason, \cite{ohlmann2017a} proposed a procedure to cut the
original stellar profile (that in practical simulations often results
from preceding \ac{1D} stellar evolution calculations) at a certain
chosen radius into a thus defined ``core'' and an ``envelope''. The
``core'' is then represented by the core particle and the ``envelope''
is constructed on the grid in a way that preserves its expected
\ac{HSE} according to Eq.~(\ref{eq:hse}) before the onset of \ac{CE}
interaction.  The hydrostatic stratification of the gas discretized on
the grid is reconstructed with a modified Lane--Emden equation
\citep{ohlmann2017a}. In contrast to the original Lane--Emden equation
for polytropic stellar models \citep[see, e.g.,][]{kippenhahn2012a},
it includes the gravitational force $g_\mathrm{core}(r)$ due to the
core particle as a function of radius $r$ in addition to self-gravity
of the envelope gas:
\begin{equation}
\label{eq:le_mod}
  \frac{1}{r^2}\frac{\mathrm{d}}{\mathrm{d} r} \left[ \frac{r^2}{4 \pi
      G \rho} \frac{\mathrm{d} P}{\mathrm{d} r} + \frac{r^2
      g_\mathrm{core}(r)}{4 \pi G}\right] + \rho = 0
\end{equation}

This procedure provides a reliable and resolution-independent
reconstruction of the envelope profiles (see
Fig.~\ref{fig:ohlmann17}). The important parameter of this setup model
is the cut radius below which material is assigned to the core and
represented by a particle. This radius is chosen according to physical
considerations and numerical necessities. As discussed in
\ref{sect:comp_feasible}, a resolved physical core would act as a
time sink in numerical simulations and this is also true for the
higher-density regions around the core. Resolving the bifurcation
point on the grid would be desirable, but it often is located in
high-density material. Extending the correct envelope gas
representation to this point can result in unacceptable computational
costs and compromises are made. According to Eq.~(\ref{eq:le_mod}),
the core particle is still embedded in gas within the cut radius, but
its density profile does not provide a realistic representation of the
envelope close to the core. In fact, depending on the choice of the
cut radius, some envelope material is absorbed in the core
particle. Despite these deficiencies, the setup procedure of
\cite{ohlmann2017a} provides a well-controlled setup with a
clearly-defined, if not entirely astrophysical core--envelope
boundary. This has not been the case in many other setups employed
before.

\ac{SPH} based setups for \ac{CE} simulations face a similar
problem. The envelope gas represented by \ac{SPH} particles in the
vicinity of the distinct (gravitation-only) core particles and within
their gravitational softening radii is not necessarily in \ac{HSE}
when they are placed randomly. To ensure equilibrium, a distribution
of \ac{SPH} particles in this region can be chosen that represents a
density profile according to Eq.~(\ref{eq:le_mod}).  Whether or not
such a structure captures a realistic core--envelope transition has to
be tested and may depend on the number of \ac{SPH} particles as well
as on the value of the softening length \citep{reichardt2020a}.

Although the fluid-dynamical treatment of the core material is
discarded for efficiency reasons, it is important to capture the
gravitational effects of the cores as precisely as possible. In a
fully particle-based implementation in the framework of \ac{SPH} this
is relatively simple; the numerical treatment of self-gravity derives
from $N$-body simulations and is well-suited to deal with the problem.
When combining a grid-based discretization of the envelope gas with a
particle representation of the cores, however, the treatment of
self-gravity is less obvious. One option is to model the gravitational
interaction of the gas cells in a particle-based approach, such as
using the tree method (see Sect.~\ref{sect:tree}). To alleviate the
inherent asymmetries of tree methods, a direct summation of the
contributions to gravitational forces from the core particles is
possible. Calculating the gravitational interaction between the core
particles and between core particles and gas cells exactly adds a
tolerable computational overhead only as long as the gravitational
interaction of the many gas cells is treated in an approximate
$\mathcal{O} (N \log N)$ approach. This is implemented for instance in
the moving-mesh \ac{CE} simulations of \cite{ohlmann2016a}.

A full, grid-based treatment of gravity employing Fast Fourier
techniques or multigrid methods (see Sect.~\ref{sect:grav_alt}), in
turn, requires to map the masses of core particles onto the grid, for
instance with a cloud-in-cell algorithm \citep{taam2006a}. To avoid
force anisotropies, a large number of particles moving rigidly
together can be chosen to represent the cores \citep[][use $10^5$
  particles per core]{ricker2008a}.

The approach of replacing the cores by point masses naturally removes
the timescale problem. It seems, however, likely that future
generations \ac{3D} hydrodynamic \ac{CEE} simulations will strive to
improve the representation of the cores. Even if resolving them
spatially remains out of reach, subgrid-scale models may help to
refine the implementation of physical properties of the cores and
associated effects such as hydrodynamic friction, mass accretion, or
evaporation.

\subsection{Setting up the primary star in hydrostatic equilibrium}
\label{sect:challenges_setups}

The first step in simulating \ac{CEE} is to set up a model for the
primary star. This step requires some diligence. Although reaching
envelope ejection in \ac{CE} simulations remains a challenge, primary
stars are giants and have rather loosely bound envelopes. It is far
from trivial to keep them stable over many dynamical timescales in
\ac{3D} hydrodynamic simulations.

Usually, the initial model for the primary star is obtained from a
\ac{1D} stellar evolution calculation. Such calculations and their
results differ in many aspects from the representations of stars in
multidimensional hydrodynamic treatments. Two of them are of
particular significance when mapping models of primary stars into
\ac{CE} simulations. (1) Due to their \ac{1D} nature, stellar
evolution calculations can afford much higher resolutions in spatial
or mass coordinates. This is additionally supported by the common
choice of a Lagrangian frame of reference. (2) Classical \ac{1D}
models are constructed under the assumption of a mechanical
equilibrium in the stellar structure.

Aspect (1) directly links to the question of how to geometrically
place the discretization elements in the different \ac{3D}
hydrodynamic approaches. Adapting the geometry of the discretization
scheme to the shape of the simulated system becomes increasingly important
with lower spatial resolution. While observing the overall spherical
symmetry of a stellar model, spherical grids are plagued by their
axial and central coordinate singularities and are ill-suited for the
later simulation of the binary system evolution. In contrast, the
geometry of Cartesian meshes is not adapted to the spherical symmetry
of the stellar model and their spatial resolution is anisotropic with
preferred axis directions. Adaptive mesh refinement techniques can
alleviate the geometry problems somewhat.

Particle-based discretizations follow the mass distribution inside the
envelope by construction and may therefore provide poor spatial
resolution at certain locations (most obviously near the stellar
surface). Still, the mesh-free approach of \ac{SPH} and moving-mesh
discretizations of hydrodynamics automatically adapt to the geometric
properties of the simulated object as it evolves. This advantage,
however, comes at a price to pay at the initial setup: unstructured
moving meshes and random placements of particles in \ac{SPH} lack any
symmetry. Mapping a spherical stellar model onto them introduces noise
resulting in spurious velocities in the material. \citet{ohlmann2017a}
tested a variety of grid structures for placing the discretization
elements in the initial setup---a regular cubic grid, a nested cubic
grid, an adaptive cubic grid, and a grid based on a HEALPix
tessellation \citep{gorski2005a, pakmor2012b}. For reasons discussed
below, however, they found that the initial grid geometry has little
effect on the simulation results.

Aspect (2) is related to the problem of ensuring \ac{HSE} in
hydrodynamic simulations.  Except for the explosive termination
phases, stellar structures remain very close to \ac{HSE} throughout
the evolution of the star.  This, to good approximation, should hold
true for the envelope of the primary star at the onset of
\ac{CEE}. Simulations of the dynamic inspiral phase aim to determine
the response of the \ac{CE} to the perturbations caused by the
orbiting stellar cores. A key requirement for this is that the
employed numerical model is able to maintain \ac{HSE} in the
unperturbed envelope---any change in the envelope structure should
originate from the core inspiral and not from numerical errors. This
is challenging because, in contrast to the classical stellar structure
calculations providing the initial model for the primary star,
\ac{HSE} is not explicit at the level of the fundamental equations in
a hydrodynamic model. It rather constitutes a special solution and
there is no guarantee that the delicate balance between pressure
gradient and gravity in the loosely bound envelope of a giant star is
maintained over many dynamical timescales by the approximate numerical
schemes. Moreover, the goal is not to force the setup into some
arbitrary \ac{HSE} configuration but to reproduce the original stellar
model as exactly as possible in the multidimensional hydrodynamic
simulation.

To enable the correct \ac{HSE} in the numerical model, the
gravitational force as well as counteracting pressure gradient should
be reproduced accurately. Since both effects are discretized
independently (see Sects.~\ref{sect:cfd} and \ref{sect:gravity}),
there is little hope for errors to cancel reliably\footnote{There is a
special class of numerical schemes---so-called \emph{well-balancing
methods}---that aim at ensuring \ac{HSE} in hydrodynamic codes
\citep[see, e.g.,][]{edelmann2021a}, but these have not yet been
employed to \ac{CE} simulations.}.

A minimal requirement for global stability of the stellar model is a
sufficiently precise numerical representation of the pressure
gradient. To this end, a pressure scale height should be resolved by a
certain number of discrete elements (grid cells or
particles). Depending on the discretization approach, this can be a
difficult task. For finite-volume hydrodynamic schemes, it is possible
to estimate the resolution required in order to keep the Mach number
of spurious velocities below a certain threshold
\citep{zingale2002a}. The result depends on the chosen scheme and the
order of reconstruction, but generally a resolution of several cells
per pressure scale height is required for reasonably low velocity
perturbations. For the solver implemented in the \textsc{arepo} code,
\citet{ohlmann2017a} estimated the Mach number of velocity fluctuations
when resolving a pressure scale height $H_p$ with a grid spacing
$\mathrm{\Delta}$ to be
\begin{equation}\label{eq:minmach}
  \mathit{Ma} = \frac{C_\mathrm{CFL}}{12 \gamma} \left(
  \frac{\mathrm{\Delta}}{H_p} \right)^3 + \mathcal{O} \left[ \left(
    \frac{\mathrm{\Delta}}{H_p} \right)^4
  \right],
\end{equation}
where $C_\mathrm{CFL} < 1$ is the fudge factor introduced into the
necessary but not sufficient \ac{CFL} criterion to stabilize the scheme and
$\gamma$ is the adiabatic index of the gas. To stabilize the
hydrostatic stratification with this scheme assuming $\gamma =
\nicefrac{5}{3}$ and $C_\mathrm{CFL} = 0.4$ down to a Mach number of
$10^{-2}$, one pressure scale height must be resolved with more than
three cells \citep{ohlmann2017a}.

Particular problems arise near the core of the giant star and at its
surface, where the pressure gradients are steep. Here, adaptive mesh
refinement techniques can help to alleviate the resulting
computational costs. \citet{ohlmann2016a} found that \ac{HSE} can be
maintained only when retaining a certain number of grid cells inside
the softening length around the core particle of the primary star
model. The rather steep density gradient has to be sufficiently
resolved. A universal criterion for the necessary resolution, however,
is difficult to establish as it is expected to not only depend on the
parameters of the star but also on the numerical methods and their
implementation (e.g.\ the order of spatial reconstruction).
\citet{sand2020a}, for example, found that a higher resolution is
required in the case of \ac{AGB} primary stars as compared with the
\ac{RG} models of \cite{ohlmann2016a}.

Reaching a sufficient resolution of the pressure gradient near the
surface of the star is a harder task because of the even steeper
pressure gradient encountered here. It turns out, however, that this
is less critical for the overall stability of the modeled stellar
envelope than resolving the region around the core. Small departures
from equilibrium are often tolerable because they only lead to a
slight expansion of the very outer stellar layers that contain little
material. Insufficient resolution of the outer layers, and a
correspondingly large mass loss over the surface, however, may cause a
progressive dilution of the inner parts of the star. Moreover, similar
to the problems noted by \citet{galaviz2017a}, the lost material may
affect the prediction of the color and the lightcurve of the simulated
\ac{CE} event.

In low-resolution \ac{CE} simulations with grid codes, a practical way
to stabilize the surface of the primary star is to impose a high
temperature in the pseudo-vacuum embedding the star such that it is in
pressure equilibrium with the surface layers
\citep[e.g.,][]{sandquist1998a, passy2012a, chamandy2018a}. This,
however, may lead to problems with predicting observables from \ac{CE}
simulations, because the artificially hot pseudo-vacuum prevents a
useful estimate of a photospheric temperature \citep{galaviz2017a}. By
construction, \ac{SPH} codes have low resolution in the low density
regions of the surface. However, they tend to retain the stellar shape
without the addition of a hot pseudo-vacuum. In \ac{SPH} schemes,
hydrodynamic quantities and their gradients are computed by summing
over contributions from a certain set of neighboring particles. At the
surface of an object, this leads to a one-sided evaluation of
properties that can cause an artificial ``surface tension''
\citep[e.g.][]{springel2010c}.  On the one hand, this may help to
prevent the slow leakage of low-density material. On the other hand,
however, the artificial surface tension may affect the modeling of the
phase directly preceding the plunge-in of the companion into the
envelope.

\begin{figure}
    \centering
    \includegraphics[width=\textwidth]{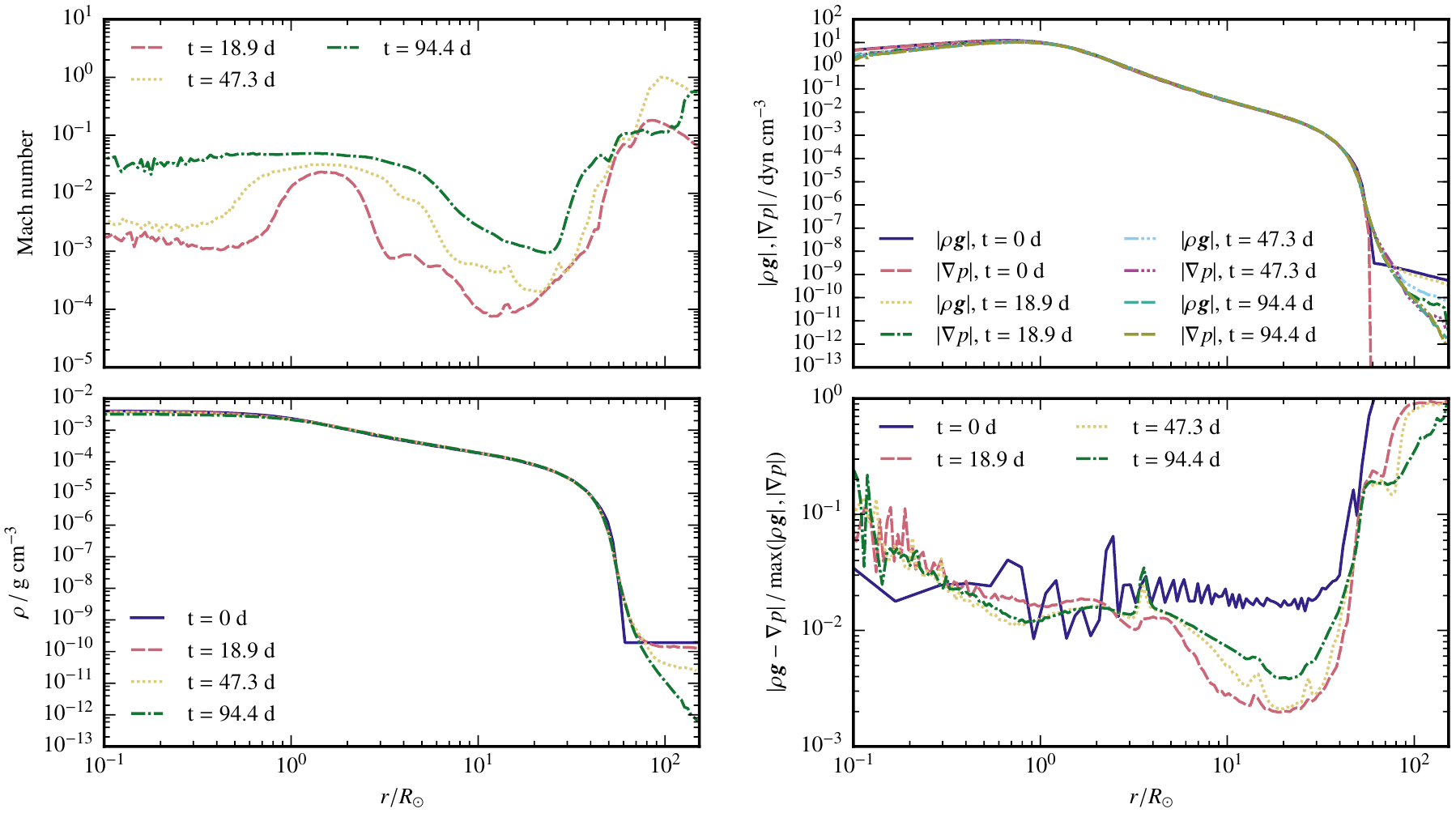}
    \caption{Relaxation of the setup model shown in
      Fig.~\ref{fig:ohlmann17}. Shown are the radial profiles of the
      Mach number (top left), the density (bottom left), the two terms
      in the equation of \ac{HSE} (\ref{eq:hse}) (top right) and the
      relative error in \ac{HSE} for different times during the
      relaxation procedure. The density profile is well-preserved and
      the deviation from \ac{HSE} is at the percent level within the
      stellar structure (but not in the pseudo-vacuum embedding
      it). The Mach number increases but stays reasonably low inside
      the stellar envelope. Figures from
      \citet{ohlmann2017a}.}
      \label{fig:ohlmann_relax}
\end{figure}

A careful preparation of the \ac{3D} stellar model involves not only
the mapping from the original \ac{1D} structure but in addition it
requires a damping to ensure initial \ac{HSE} in the discretized
structure. In particular, the arbitrary initial placement of the
grid-generating points in moving-mesh hydrodynamics schemes and of the
particles in \ac{SPH} requires care to not disturb local \ac{HSE}
conditions, which would lead to spurious velocities in the envelope of
the primary star. But also Eulerian grid-based simulations should be
damped to \ac{HSE} because of the change in resolution in the mapping
process. The damping can be implemented as an artificial friction term
\citep{rosswog2004a, pakmor2012b} proportional to
\begin{equation}
  \dot{\vec{v}}  = - \frac{1}{\tau} \vec{v},
  \end{equation}
where $\tau$ is the timescale over which the velocities are
damped. Following \citet{pakmor2012b}, \citet{ohlmann2017a} perform
the damping with a constant $\tau$ over several dynamical timescales
of the object, reduce the damping over a certain period and then
follow the undamped evolution over some dynamical timescales to ensure
stability of the obtained configurations, see
Fig.~\ref{fig:ohlmann_relax}. The necessary damping step is the reason
why the choice of the initial grid geometry is unimportant as long as
it provides a reasonable representation of the structure: The adaptive
moving mesh or the positions of the \ac{SPH} particles adjust
themselves in the damping process and attain a structure that is
largely independent of the initial choice. The purpose of the
relaxation step is to approach the lower limit for the Mach numbers of
spurious motions implied by criteria such as
Eq.~(\ref{eq:minmach}). Similar damping procedures have been applied
successfully for many years, starting with \citep{ricker2008a} and
they are used time and again from \citet{passy2012a} to
\cite{ohlmann2017a}, \citet{chamandy2018a}, and \citet{prust2019a}.

An additional problem arises when a correct representation of the
thermal structure of the \ac{1D} stellar model in the hydrodynamic
simulation setup is intended. This usually implies the need to use
exactly the same equation of state in both the stellar evolution code
providing the \ac{1D} model and in the hydrodynamic code. With minimal
changes in the hydrodynamical structure, however, it is possible to
preserve the correct convective properties of the envelope when using
the setup procedure proposed by \citet{ohlmann2017a} based on
\citet{edelmann2017a}.

\citet{ohlmann2017a} list five quality criteria for the representation
of primary stars in \ac{3D} hydrodynamic \ac{CE} simulations:
\begin{enumerate}
  \item Correct representation of the mechanical structure: deviations
    from the original \ac{1D} model in the pressure and density
    profiles should be small.
  \item Correct representation of the \ac{HSE}: the difference between
    both sides of Eq.~(\ref{eq:hse}) should be small.
  \item The Mach numbers in the envelope of the primary star model
    should be small compared to the Mach numbers expected in the later
    \ac{CE} evolution.
  \item Correct thermal structure of the envelope: if the convective
    structure of the primary star is retained in the mapping and
    relaxation procedures, a stationary state of convection should be
    reached in the model.
    \item Stability against large-scale oscillations: If the original
      primary star is not pulsating, the potential energy of the model
      should be constant.
\end{enumerate}
These criteria should be checked carefully before adding a companion
and starting the actual \ac{CE} simulation. Again, we advocate to
perform such tests as a standard procedure when performing \ac{CE}
simulations and to document the results in the related publications.

\subsection{Setting up the binary system}
\label{sect:setup_binary}

After setting up a primary star model and checking its stability, a
companion is added. This requires to choose the parameters of the
progenitor system, such as the initial orbital separation, the spin of
the two stars, and the eccentricity of the orbit. Physically, these
parameters are determined by the evolution of the progenitor binary
system and set by choosing the time at which the simulation of this
system is started. From this point of view, initializing the actual
\ac{CE} simulation as early as possible seems advisable in order to
capture the formation of the \ac{CE} correctly. The pre-\ac{CE} evolution,
however, takes place on timescales that are not easily accessible to
\ac{3D} hydrodynamic models (see Sects.~\ref{sect:pre-dynamical_phase}
and \ref{sect:3d_accessible}). Therefore, an initial condition for the
binary evolution must be constructed that reproduces an expected
structure at a certain stage of the pre-\ac{CE} evolution. Starting at the
onset of Roche lobe overflow appears to be a natural choice: The
initial system is set up at the time when the primary, with radius
$R_\star$, just fills its Roche lobe, whose radius,
$r^\mathrm{Roche}_\star$, can be approximated by
\begin{equation}
  R_\star = r^\mathrm{Roche}_\star = a \, \frac{0.49 \, q^{2/3}}{0.6
    \, q^{2/3} + \ln (1+ q^{1/3})},
\end{equation}
where $q = M_\star / M_\mathrm{comp}$ is the mass ratio of the primary
star to the companion \citep{eggleton1983a}. This determines the
initial orbital separation, $a$.

We could ask whether starting the simulation at onset of Roche lobe
overflow, as, e.g., in \citet{rasio1996a} and \citet{nandez2014a}, is
actually necessary. After all, we do not know whether the simulations
capture correctly the transition between stable Roche lobe overflow
and the onset of instability and inspiral. Following this argument,
the choice for the starting point of \ac{3D} hydrodynamic \ac{CE}
simulations is often made according to practical considerations. A too
large initial separation makes the simulations very expensive. Many
orbits of the binary have to be followed from before the onset of mass
transfer until the actual plunge-in of the companion into the envelope
of the primary. This is possible provided that the employed
hydrodynamics scheme preserves angular momentum well, which, of
course, is desirable. With this, we face the paradoxical situation
that the better the codes reproduce physical conservation laws, the
more expensive the simulations become. As a compromise, many
simulations position the companion near the surface of the primary
\citep[e.g.,][]{sandquist1998a, ohlmann2016a}. The companion could
even be initially placed inside the envelope---a choice motivated with
the argument that there is very little binding energy in the outer
layers \citep{law-smith2020a}. The danger with this approach is that a
simple binding energy argument is insufficient. The inspiral is a
dynamical process driven by the drag force. Therefore, an extreme
choice may select an orbital separation of the initial core binary
system that seems energetically plausible but is never reached in the
actual dynamical evolution because the envelope is lifted earlier and
the drag force vanishes.

Placing the companion close to the primary star---be it at Roche lobe
filling distance or closer---raises the question of how to define a
consistent setup configuration. It implies that early tidal effects
are missed. For a late initiation of the simulation, the primary star
cannot be assumed to be spherical, but many simulations start out with
exactly this assumption. From a practical point of view, a spherical
setup of the primary star has the advantage that sufficient stability
of its envelope can be guaranteed, e.g. by the procedures described in
Sect.~\ref{sect:challenges_setups}.  This allows to separate physical
effects during the \ac{CE} evolution from numerical artifacts, which would
be more difficult to ensure in a binary setup.  Therefore, the
assumption of a spherical primary star gives rise to numerically
well-defined initial conditions for \ac{CE} simulations. Still,
concerns remain as to whether this is a physically valid approach.

There are methods that allow to initiate the binary with a separation
larger than Roche lobe contact and to follow the evolution in a
numerically efficient way. After creating a stable model of a
spherical primary star as in Sect.~\ref{sect:challenges_setups},
another step is taken before simulating the plunge-in of the
companion. This intermediate step aims at relaxing the envelope of the
primary star into the potential of the binary system close to the
onset of mass transfer. For white dwarf merger simulations, such a
procedure has been suggested by \citet{dan2011a}. The idea is to place
the spherical stellar models in orbit with a companion at an orbital
separation large enough that the tidal interaction is negligible. The
computational cost of following the slow inspiral due to gravitational
wave emission is prohibitive. To speed up the orbital evolution, the
distance is artificially decreased on a timescale that is
substantially longer than the dynamical timescale of the stars. This
allows the stellar structures to adapt to the changing potential. Such
initial models, however, are not common in \ac{CE} simulations
yet. Retaining force balance in the loosely bound envelopes of giant
stars is more challenging than in the compact structures of white
dwarfs.

Even agreeing to start all simulations at the time of Roche lobe
overflow, we next would need to decide on the rotation of the
interacting stellar objects. Since the companion and the core of the
primary star have to be treated as point particles (see
Sect.~\ref{sect:representing_cores}), this question boils down to the
rotation of the envelope of the primary.  The general idea of the
onset of \ac{CEE} is a loss of co-rotation between the orbit of the
binary companion and the envelope of the primary. This triggers the
inspiral, because for the drag force to act, a non-vanishing relative
velocity between the companion and the envelope material is
necessary. It remains unclear, however, how much co-rotation is to be
expected in the epoch the simulation is initialized (provided, of
course, that this initialization reflects a physical evolution stage
in the first place). Moreover, not all progenitor systems may reach
full co-rotation before entering \ac{CEE}. In simulations, anything
between full co-rotation and a non-rotating primary star has been
used. It may be surprising that even full co-rotation leads to
inspiral, but this is due to non-relaxed initial conditions in the
binary potential which quickly re-arrange the density distributions so
that tidal forces can act.

The issue of primary-star rotation and the extent of primary--orbit
co-rotation goes hand in hand with the issue of stabilization of the
binary in a certain rotating configuration. While the primary is
always stabilized as a single star in the inertial frame (see
Sect.~\ref{sect:challenges_setups}), some simulations also stabilize
the binary in the rotating frame
\citep{rasio1996a}. \citet{nandez2014a} and following papers stabilize
the binary in the corotating frame using the method of
\citet{lombardi2011a}, but such a method is not universally adopted by
all simulations. \citet{reichardt2019a} evolved a Roche lobe contact
system in the rotating frame, but like \citet{sandquist1998a}, they
did not stabilize the binary in that frame, effectively spinning up a
star that was in equilibrium only when non-rotating.

An additional, dynamically important but poorly constrained parameter
is the initial eccentricity of the system. Many simulations assume
circularization of the orbit prior to the onset of \ac{CEE}, and this
is also supported by simulations \citep{macleod2022a}, but cases with
eccentric initial orbits have also been considered in numerical
simulations \citep{staff2016b,glanz2021a}. Eccentric post-\ac{CE} binaries
are observed and may result from pre-existing eccentricity or
generation of eccentricity during the \ac{CE} interaction. In
simulations with the companion initialized in a spherical orbit close
to the edge of the primary star, the generation of substantial
eccentricity in the plunge-in is indeed observed.

Our discussion in the previous sections indicates that the setup of
\ac{CE} simulations is a very delicate issue. Unlike the inspiral
phase of \ac{CE} interaction, the prospects for \ac{3D} hydrodynamic
simulations of large parts of the pre-\ac{CE} evolution are discouraging,
leaving us with rather arbitrary choices of initial system
parameters. This unsatisfactory situation calls for careful testing
of the impact these parameters have on the results. The studies
carried out thus far indicate some, but surprisingly little impact of
the initial separation and even the rotation of the primary star on
the final orbital separation \citep{sandquist1998a, iaconi2017a,
  prust2019a}. This, however, may depend on secondary parameters, such
as the mass and evolutionary stage of the giant star and the mass of
the companion. More extended parameter studies are required to settle
the impact of the choice of initial parameters on the final orbital
separation, but also on the mass and morphology of ejected envelope
material.

\section{Multidimensional hydrodynamic simulations}
\label{sect:simulations}

For reasons given in Sects.~\ref{sect:challenges_physical},
\ref{sect:challenges_numerical} and \ref{sect:setups}, only certain
stages of \ac{CEE} are accessible to \ac{3D} hydrodynamic
modeling. Although all phases are methodologically challenging, the
dynamical inspiral, Phase (ii) of our classification in
Sect.~\ref{sect:phases}, is out of reach for classical approaches to
stellar evolution modeling. Yet, it is perhaps the most critical phase
for the fate of the system. Progress in binary stellar evolution
fundamentally depends on a detailed understanding of this phase. We
argue that only \ac{3D} hydrodynamic simulations provide the required
realism and predictive power to solve this problem.

In this section, we review simulations pertaining to parts of the
pre-\ac{CE} evolution of the progenitor system, the dynamical \ac{CE}
interaction in the plunge-in phase, and a few tens to hundreds of
orbits afterwards. These have been carried out over the past decades
by several groups. In contrast to the previous sections, where we
discussed general aspects of \ac{CE} simulations and expressed our own
view on the approaches taken, we here summarize the findings reported
by the respective authors of the studies. The interpretations and
conclusions drawn are often complementary, but occasionally also
contradictory. This is an unavoidable consequence of the historical
development of the field, which we intend to convey from a neutral
perspective. We therefore try to hold back our own views.

\subsection{Simulations of the pre-dynamical inspiral phase and the onset of CE interaction}
\label{ssec:simulations_of_the_pre-dynamical_inspiral_phase}

Until recently, work on the pre-\ac{CE} evolution was mostly carried out
based on analytical models, stellar evolution calculations or binary
population synthesis tools. We have briefly discussed some important
aspects of the pre-\ac{CE} interaction in
Sect.~\ref{sect:pre-dynamical_phase} and refer the reader to Chapter~6
of \citet{ivanova2013a} for a more comprehensive review.  A broader
scope of modeling binary interactions, both with \ac{1D} stellar
evolution approaches and with multidimensional hydrodynamic methods,
is covered in the review of \cite{demarco2017a}. Here, we restrict our
discussion to the few available multidimensional, hydrodynamic models
that pertain to the the onset of \ac{CEE}.

There is no doubt that the physical parameters of the binary system
together with the nature and evolutionary stage of the primary star
and the companion determine its fate in and after the \ac{CE} phase. But
still we lack a comprehensive understanding of which systems achieve
successful envelope ejections leaving behind a compact binary and
which of them end up in a merger of the cores. It is expected that
this is determined by the physical initial conditions and the many
resulting processes that lead to the onset of \ac{CEE} discussed in
Sect.~\ref{sect:pre-dynamical_phase}. As mentioned there, there is
little hope to capture this complex phase comprehensively in \ac{3D}
hydrodynamic simulations, but a few attempts exist.

\citet{pejcha2016b} and \citet{pejcha2016a} carried out \ac{3D},
\ac{SPH}-based radiation-hydrodynamic simulations aimed at
understanding the pre-merger phase of the transient V1309~Sco
\citep{tylenda2011a}. This is now established to have been the merger
of a slightly enlarged, ${\sim} 1.5 \, \msun$ main sequence or
sub-giant primary star and its much less massive main sequence
companion. As discussed in Sect.~\ref{ssec:terminology}, this
configuration does not qualify as a show-case example of a \ac{CE}
system, but is rather at the blurry interface between \ac{CEE} and
classical stellar mergers. Nonetheless, the similarity of this system
and the less extreme spatial scale challenges render it a suitable
example to test some models with numerical simulations and comparisons
to observations. The simulations of \citet{pejcha2016b} and
\citet{pejcha2016a} modelled a very specific process: the mass loss of
the interacting binary system from the outer Lagrange point $L_2$ and
its observational consequences.

\ac{SPH} was coupled to a simplified treatment of radiation that
splits the process into flux-limited diffusion of radiation energy,
radiative cooling in the direction perpendicular to the orbital plane,
and---only in \citet{pejcha2016b}---irradiation of the gas from the
central binary. A suitable equation of state was used to account for the
ionization structure of the gas. The setup of \cite{pejcha2016b}
consisted of a central binary with a mass ratio between $0.06$ and
$0.8$, modeled only including the mutual gravitational force of the
stars, but with no back-reaction of the gas flow on the binary
orbit. The \ac{SPH} particles representing the gas component were
injected directly at the $L_2$ Lagrange point of the binary system
allowing one to follow the structure of the outflow, which was not
self-consistently modeled in the hydrodynamic simulation but assumed
in the setup. The spiral arms generated by the outflow overlapped and
merged at a certain distance from the binary, which is of interest for
observing the corresponding system. Shock waves thermalized and about
10\% to 20\% of the radial kinetic energy of the outflow was converted
to thermal energy. Depending on the system parameters, some of this
energy was radiated away, diffused or generated adiabatic expansion of
the gas and dominated the resulting luminosity. Conditions for dust
formation were found to prevail in the ejected material. The
temperatures and luminosities detected in the simulations of
\citet{pejcha2016b} matched observations of \ac{LRN} in order of
magnitude, but because the model did not follow the dynamical
evolution of the binary itself, lightcurves could not be reproduced.
 
\citet{pejcha2016a} extended the study to other mass ratios of the
binary system and found outflows of unbound material from $L_2$ if the
thermal content of the ejecta was sufficiently high, i.e., the ratio
between the sound speed and the orbital velocity at the injection
point was large. On the other hand, cold outflows outside the mass
ratio range determined in \citet{pejcha2016b} remained marginally
bound. A rich phenomenology was observed in the simulations. Again,
the expected optical emission from the implied processes was found to
explain prolonged activity of systems due to $L_2$ mass loss before
the peak of the emission was reached in the actual merger
event. Informed by the radiation-SPH simulations of
\citet{pejcha2016a, pejcha2016b}, \cite{metzger2017a} proposed that
the interaction of material ejected in the dynamical stellar merger or
\ac{CE} event with the previous outflow generates a shock that
contributes to powering the lightcurves of \acp{LRN}. This was
suggested to explain typically observed double peaks.

\cite{pejcha2016b, pejcha2016a} assumed an outflow over an outer
Lagrange point and studied its implications in their simulations, but
they did not model the dynamics of the binary system itself. Although
they focused on the pre-\ac{CE} evolution, they did not attempt to model
how the system enters \ac{CEE}, nor the impact of the simulated phase
on the following inspiral.

This was addressed for the first time by
\cite{macleod2018b}\footnote{Stellar mergers in other contexts have
been studied in numerical simulations before. \cite{motl2017a} give an
overview and discuss differences in the evolution when simulating mass
transfer and mergers between white dwarf stars in different
approaches.}, who studied the evolution from the onset of mass
transfer towards the formation of a \ac{CE} in a system comprising an
evolved primary star and a lower-mass companion. Their choice of
parameters ensured that mass transfer reduces the binary separation
initiating unstable mass transfer.  To simulate the hydrodynamic
evolution in \ac{3D}, \cite{macleod2018b} employed the Eulerian
\ac{MHD} code \textsc{athena++} \citep{stone2020a}, but without
accounting for the evolution of magnetic fields. The spherical polar
mesh was uniformly spaced in angular directions, with the spacing
growing logarithmically with radius. Two nested levels of static grid
refinement were prescribed around the binary orbital plane. The giant
star's envelope was constructed as a polytrope. Both the core of the
primary star and the companion were modeled as point masses. Following
the details of mass transfer required meticulous preparation of the
initial model for the \ac{3D} hydrodynamic simulation. The envelope of
the primary was relaxed damping out spurious velocities to ensure
numerical \ac{HSE}. The gravitational force of the companion was
slowly turned on to allow the structure to adapt to its presence. At
Roche-radius separation, mass transfer commenced in an unstable mode
with an increasing rate so that the orbit of the binary core system
eventually experienced runaway decay. A fit to the loss rates of mass
and angular momentum was provided that allows for an analytic
description of the orbital evolution. The morphology of the gas
outflow was observed to change during the evolution: initially a thin
stream of high-entropy gas was launched from the $L_2$ and $L_3$
Lagrange points, which later developed into a broad fan of gas ejected
on ballistic trajectories. The origin of the transition between the
two modes was identified as desynchronization of the orbital motion
with the rotation of the donor star's envelope. \citet{macleod2018a}
interpreted their results in terms of observational features with
particular emphasis on the formation of bipolar structures observed in
remnants of stellar coalescences. While it is not the only path to
binary coalescence and systems with mass ratios close to unity may
evolve differently, their simulation captured an important case for
the formation of a \ac{CE}.

The case of ``grazing envelope evolution'' impeding or even preventing
the onset of \ac{CEE} (see Sect.~\ref{sect:pre-dynamical_phase}) was
simulated by \citet{shiber2017a} and \citet{shiber2018a} who prepared
a model of a $3.4 \, \msun$ \ac{AGB} star of radius ${\sim} 215 \,
\rsun$ in the \textsc{mesa} code and mapped it into the \textsc{flash}
code \citep{fryxell2000a} on a uniform Cartesian grid and assumed an
ideal gas equation of state. The inner third of the radius was
replaced with a constant density sphere. Gravity was modeled as
constant, temporal evolution of the gravitational field and the effect
of the companion star were ignored. A low-mass main sequence companion
star was initialized on a Keplerian orbit at various radii. This
radius was kept constant by \citet{shiber2017a} while
\citet{shiber2018a} allowed for inspiral to a two thirds of the
initial radios of the primary star at a prescribed rate inspired by
\ac{CE} simulations. At the location of the companion, a bipolar jet
was put in perpendicular to the orbital plane with a mass injection
rate motivated by a \ac{BHL} formalism and a prescribed opening angle
and velocity. \citet{shiber2017a} observed a complicated flow
morphology and an efficient removal of material due to jet
action. \citet{shiber2018a} concluded that the grazing envelope
evolution may substantially delay or even prevent an inspiralling of
the companion.

Simulations trying to model the processes leading to the formation of
a \ac{CE} system have to be distinguished from the conditions in which
multidimensional hydrodynamic simulations are initialized, which are
often chosen for practical and numerical reasons instead of a
stringent physical motivation (see Sect.~\ref{sect:setup_binary}). Is
this a fundamental problem for the credibility of current \ac{CE}
simulations? A look into the currently available simulations and even
semi-analytical descriptions suggests that the ability of the system
to eject the envelope ultimately deciding between the formation of a
compact binary remnant or a merger of the cores is less sensitive than
one might expect to the initial conditions of the simulation, such as
the degree of co-rotation and the initial distance of the companion
from the primary star when setting up global simulations of
\ac{CEE}. Some impact of these parameters, however, is recorded. For
example, the exact value of the final orbital separation
\citep{iaconi2017a} and the rate of orbital decay, do depend on the
degree of co-rotation and the initial orbital angular momentum. The
extent of these dependencies and whether there are large differences
for different systems is not yet established.

Following \citet{iaconi2018a}, \citet{reichardt2019a} carried out full
CE interaction simulations starting at the time of Roche lobe
overflow\footnote{Simulations of the early \ac{CE} phases starting out
with unstable mass transfer were already conducted by
\citet{rasio1996a}, see Sect.~\ref{sect:sim_sph}.}. They tested the
sensitivity of the outcome of the simulations to the conditions at the
start of the interaction. The mass transfer (though Lagrange point
$L_1$) and mass loss (though $L_2$ and $L_3$) rates were found to be
lower for higher resolution and for simulations that were carried out
in the corotating frame, while the time before inspiral was longer. As
a result, the mass lost before the inspiral was relatively insensitive
to numerical conditions. The outcome of the simulations compared to
similar ones that did not model the pre-\ac{CE} inspiral led to the
conclusion that the clearest difference is the shape of the expanded
and ejected \ac{CE}, which was found to be more symmetric in
simulations that model the pre-inspiral thanks to the interaction with
the disk of material formed during the early mass loss. However, it is
likely that these conclusions only pertain to the chosen parameter set
and cannot be generalized.  This leaves us with the question of
Sect.~\ref{sect:setup_binary}: how should we setup the simulation?
Clearly, more work has to be devoted to the onset of \ac{CEE} and its
impact on the processes in the inspiral phase.

\subsection{Simulations of the companion--envelope interaction}
\label{ssec:windtunnel}

One of the fundamental questions of \ac{CEE} is how the companion star
interacts with the envelope gas as it passes through it. This question
was already addressed in the earliest models of \ac{CE} phases in binary
stellar evolution, that were carried out in \ac{1D} in the framework
of classical stellar evolution calculations, see
Sect.~\ref{sec:one_dimensional_implicit_approaches}. The inspiral of
the companion was attributed to the action of a hydrodynamical or
tidal drag force and analytic approximations (see
Sect.~\ref{sssec:drag}) were employed to implement this interaction
into \ac{CE} simulations
(Sect.~\ref{sssec:1d_stellar_structure_models_of_the_dynamicsl_ce_inspiral}).

The simplistic approximate \ac{1D} view on the interaction in
\ac{CEE}, however, diminishes the predictive power of the simulations.
Effects that are not or only approximately captured in the basic
formalism include density gradients in the envelope, asymmetries,
nonlinearities, and the simple fact that not only the companion moves
through the envelope gas but both stellar cores orbit each other
inside it. To test some of these effects and to improve the accuracy
of \ac{1D} models, multidimensional hydrodynamic simulations were used
to study the interaction of objects with the stream of envelope
gas---because of their similarity to technical setups they are
sometimes called \emph{wind tunnel simulations}. In terms of the
physics modeling, they are quite similar to the ``global'' \ac{CE}
simulations in the framework of he basic gravo-hydrodynamic model,
that are discussed in Sect.~\ref{sect:sim_global}. There are, however,
two important differences: the advantage of ``local'' wind tunnel
simulations is that they potentially can resolve the stellar core and
the flow in its immediate vicinity---a region that often remains
under-resolved in global simulations because they have to accommodate
the entire structure of the giant primary star. Their drawback,
however, is that they do not account for the global geometry, the full
gravitational potential of the primary, the eccentricity of the orbit,
large-scale distortions of the flow of envelope gas, instabilities,
and microphysical effects such as the recombination of ions
\citep{macleod2015a}. Nonetheless, wind tunnel simulations may capture
main effects and with their superior spatial resolution they
constitute an important complimentary approach to global \ac{CEE}
simulations.

The astrophysical problem of a compact body moving through gas while
accreting material and being subject to drag forces is rather
ubiquitous. Not all simulations of this type were performed in the
context of \ac{CE} studies. Despite the differences in the setups,
however, their results often carry over to the problem at hand. A
review of such studies can be found in \citet{edgar2004a}. In their
Section 3.1, \citet{macleod2015a} summarize some of the findings.

Early wind tunnel simulations were carried out by \cite{shima1985a}
with a finite-volume discretization of fluid dynamics based on the
Riemann solver of Osher \citep{osher1983a} under the assumption of
axisymmetry. They tested the analytic prescriptions discussed in
Sect.~\ref{sssec:drag} and studied the details of the flow
configurations relaxing the collisionless assumption. Overall, in the
case of high-Mach number flows, the gas was found to converge behind
the object. A bow shock forms slightly upstream of the body and it
encloses the region where gas converges and flows towards the
gravitating object, where it is accreted. This confirms the basic
picture of \ac{BHL} accretion. Consequently, \citet{shima1985a} cast
their results in the corresponding form of the drag force,
\begin{equation}
   F_\mathrm{drag} = C_\mathrm{d} \frac{1}{2} \pi R_\mathrm{a}^2
   \rho_\infty v_\infty^2,
\end{equation}
i.e.\ a variant of Eq.~(\ref{eq:bondi-hoyle}), and determined the drag
coefficients $C_\mathrm{d}$ from their simulations with varying Mach
numbers of the flows and specific heat ratios of the gas.

\ac{3D} hydrodynamic simulations including a density gradient in the
gas were carried out by \citet{soker1986a} and \citet{livio1986a} with
a hybrid particle-grid based scheme, and by \citet{fryxell1988a} and
\citet{ruffert1997a, ruffert1999a} with the piecewise-parabolic
finite-volume method \citep{colella1984a}. These studies were
motivated by other astrophysical scenarios and therefore considered
shallower density gradients than those expected in
\ac{CEE}. Additional simulations focused on flow instabilities that
may occur in the Mach cone behind the moving object
\citep[e.g.][]{ruffert1994b, ruffert1995a, ruffert1996a,
  foglizzo2005a}. \citet{armitage2000a} performed \ac{2D} simulations
with the finite-volume code \textsc{zeus} \citep{stone1992a} to model
the accretion during \ac{CEE} in the context of the formation of black
holes.

The treatment of drag forces discussed in Sect.~\ref{sssec:drag} is
based on linear perturbation theory and applies to low-mass
perturbers. In \ac{CEE}, the companion mass is sufficiently large so
that we expect the density wakes to be in the nonlinear
regime. Following up on their semi-analytical approach
\citep{kim2007a}, \cite{kim2010a} performed \ac{3D} hydrodynamic
simulations with the \textsc{zeus} code \citep{stone1992a} for a
massive perturber moving through gas on circular orbits. They found
that for supersonic motion a bow shock develops upstream of the
perturber through which the flow becomes subsonic. In its immediate
vicinity, the perturber was surrounded by a near-hydrostatic spherical
envelope, which reduced the drag force below the estimates from linear
perturbation analysis, and an extended low-density region trailed the
object.

The \ac{3D} wind tunnel simulations by \cite{macleod2015a} were
specifically aimed at \ac{CE} interaction. They used the
\textsc{flash} code \citep{fryxell2000a} to model a gravitating object
(which one can think of as the companion in a \ac{CE} interaction)
represented by a point particle with a softened gravitational
potential.  The point mass was surrounded by an absorbing boundary so
that gravitational softening did not affect the flow. Block-wise
\ac{AMR} was employed and between six and nine levels of refinement
were used based on the second derivative of pressure, but still
centered at the gravitating object. \cite{macleod2015a} pointed out
that typical Bondi--Hoyle accretion radii in \ac{CE} dynamics are much
larger than the companions and extend over a substantial density
gradient in the envelope. This breaks the symmetry of the flow and
leads to an angular momentum barrier impeding accretion onto the
companion (with consequences for a potential growth of neutron star
masses in \ac{CE} interaction), while the drag force is only mildly
affected. \citet{macleod2015a} pointed out that the drag forces
observed in their simulations are not dominated by hydrodynamic
friction but are of dynamical origin, i.e., they are caused by
envelope gas piling up due to gravitational interaction with the
companion. They fitted their results to the formalism of
\cite{bondi1944a}, see Eq.~(\ref{eq:bondi-hoyle}). Contrary to the
\ac{2D} simulations of \citet{armitage2000a}, the \ac{3D} simulations
of \citet{macleod2015a} showed no formation of a persistent disk
around the companion. The study of \cite{macleod2017b} included the
gravitational potentials of both the primary (assuming a planar
geometry in the wind tunnel setup) and the companion and extended the
parameter range in a systematic way. Different Mach numbers of the
flow, density gradients, and mass ratios between primary and companion
were tested. Again, the accretion rate was found to be reduced by the
angular momentum barrier due to the density gradient.

The universality of the drag forces determined in wind tunnel
simulations remains an open question. Are they sufficient to calibrate
semi-analytic \ac{1D} simulations so that they gain predictive power?
A detailed discussion on the question of which quantities and phases
in \ac{CEE} can be captured this way was presented by
\citet{everson2020a}. \citet{chamandy2019a} compared the drag
calculated using the analytical formalism of \citet{macleod2017b} with
the drag measured in their \ac{3D} global \ac{CE} simulations of a ${\sim} 2 \,
\msun$ \ac{RGB} primary star and a ${\sim} 1 \,\msun$ companion. They
found that the two quantities match well only up to
mid-inspiral. Afterwards, the analytical fit over-predicts the force
experienced by the companion in the simulation by up to an order of
magnitude. Other \ac{3D} simulations compared the measured gravitational
drag with different variants of the analytical drag formalism
(Sect.~\ref{sssec:drag}) to find reasonable agreement
\citep{staff2016a,reichardt2019a}. This said, there has never been a
comprehensive parameter study nor a systematic comparison between drag
estimates and therefore, for the time being, it is impossible to
conclude on the extent to which the drag in \ac{3D} global simulation
matches predictions from analytic formalisms.

In addition to simulations aiming at determining the drag force acting
on the companion, studies have been performed that explore the effects
of accretion and jet formation. In a series of publications,
\citet{lopez2019a}, \citet{lopez2020a}, and \citet{lopez2022a} report
on modeling jets launched from a compact object in a common envelope
environment with the \textsc{Mezcal} code \citep{decolle2012a}. The
compact object is represented by a point particle and a self-regulated
jet is injected through an inner boundary. It is powered by a fraction
of the mass accretion rate onto the compact object as measured at that
boundary. Different parameters (efficiency of jet launching,
inclination angle of the jet) are explored. In some cases, a
neutron-star could drive a jet that breaks through the material piled
up around it \citep{lopez2019a, lopez2020a}. A white dwarf companion
star, however needs a very high efficiency to successfully launch a
jet and for main sequence stars accretion is found to be too weak so
that the jet does not break out of a \ac{RG} star envelope but is
choked \citep{lopez2022a}.

\subsection{Global CE simulations}
\label{sect:sim_global}

Global \ac{CE} simulations have a history dating back to the
1980s. Much of the early work was published in a series of papers on
what was called back then ``double-core evolution'' by different
groups of authors with various numerical methods
\citep{bodenheimer1984a, taam1989a, taam1991a, terman1994a, taam1994a,
  terman1995a, yorke1995a, terman1996a, sandquist1998a}.

The different approaches to numerical fluid dynamics that have been
employed reflect the development of similar simulation techniques in
other branches of computational astrophysics. \ac{CE} simulations have
benefited substantially from the rapid evolution of computational
resources over the past decades, which enabled significant refinement
in the physical and numerical modeling. Each of the approaches has its
strengths and drawbacks and we gave an account of this in
Sect.~\ref{sect:simulations_comparison}.

\subsubsection{Pioneering Eulerian grid-based approaches}
\label{sect:sim_eulerian_pioneering}

At a first glance, Eulerian approaches discretizing the equations of
fluid dynamics on a static spatial grid may seem little appealing for
simulating \ac{CEE}. Missing obvious symmetries, the grid geometry
cannot easily be adapted to the \ac{CE} dynamics. Moreover, as
detailed in Sect.~\ref{sect:cfd}, such schemes are not Galilean
invariant and therefore the representation of physical phenomena, such
as hydrodynamic instabilities, may depend on whether the simulations
are conducted on a static or rotating frame of reference. Notorious
advection errors tend to destabilize orbiting stellar objects and
angular momentum conservation is problematic. Nonetheless, some
Eulerian grid-based methods have been employed to studying \ac{CEE}
early on, motivated as natural extensions of \ac{1D} \ac{CE}
simulations in the drag force formalism. The primary goal was to
alleviate a fundamental problem of the \ac{1D} models, where
gravitational drag luminosity is injected into a spherical layer
instead of a local energy deposition as expected to operate around the
companion in reality.

To give proper credit to early attempts to simulate \ac{CEE} in
multidimensional hydrodynamic approaches, we start out our treatise
with \emph{pioneering} grid-based simulations. Some of their value
lies in the discussion of fundamental processes, that remains valid
until today. Many of the employed methods, however, no longer meet
modern standards. We therefore refrain from discussing them in detail;
they are of historical interest rather than a guide for future
developments.

The first \ac{2D} simulation of \ac{CEE} was conducted by
\citet{bodenheimer1984a}, who used a ``fluid-in-cell'' approach to
discretize the equations of fluid dynamics and assumed
axisymmetry. Their setup system consisted of a $16 \, \msun$
supergiant star, of which only a part of the envelope was modeled, and
a $1 \, \msun$ companion representing a neutron star. Interestingly,
this early multidimensional model already addressed a situation
that---as a progenitor system of a gravitational-wave emitting merger
event---is of utmost interest today, but still challenges numerical
simulations. The transfer of angular momentum and orbital energy from
the companion to the envelope gas was not self-consistently followed,
but instead an analytic expression for frictional heating in the
Bondi--Hoyle--Lyttleton formalism was applied following
\citet{taam1978a}; see also Sect.~\ref{sssec:drag}. The gravitational
interaction of the cores with the envelope gas was not included in
this model. This simulation approach can be seen as a multidimensional
generalization of earlier attempts to follow the inspiral of a
companion into a giant star in classical \ac{1D} stellar evolution
codes, such as discussed in
Sect.~\ref{sssec:1d_stellar_structure_models_of_the_dynamicsl_ce_inspiral}. \cite{bodenheimer1984a}
found that about $1 \, \msun$ of material is ejected in the evolution,
preferentially in the direction of the orbital plane. Based on the
multidimensional structure of the outflow, they argued that modeling
\ac{CE} events based on energy conservation arguments alone is
insufficient. Applying the same numerical method, \citet{taam1989a}
explored \ac{CEE} in a system with a $5\, \msun$ \ac{RG} primary star,
whose core is replaced by a point mass to reduce the associated
time-scale problem, and a $1\, \msun$ companion. They report a
significant mass ejection.

The study of \citet{taam1994a} was performed with the hydrodynamical
code of \citet{rozyczka1989a} and changed the model of the interaction
between the companion and the envelope gas. The orbital decay of the
core binary was explicitly prescribed according to angular momentum
conservation: The angular momentum of the envelope gas was
approximated in a \ac{2D} setup that modeled the dynamics in the
orbital plane. Its temporal change was equated to the change in
angular momentum of the core binary thus providing the evolution of
the orbital separation. The in-fall of the companion was not followed
completely. Instead, the focus was on simulating the gravitational
torques in late stages of \ac{CEE} of a $3\, \msun$ \ac{RG} star,
whose core was replaced by a point mass, and a companion, whose mass
was slowly increased to $1\, \msun$ in a relaxation step. In agreement
with the results of the \ac{SPH} simulation conducted by
\citet{terman1994a} (see Sect.~\ref{sect:sim_sph}), these torques were
found to establish co-rotation of the envelope material close to the
companion.  \citet{taam1994a} suggested this effect to be responsible
for the termination of the inspiral phase.

\citet{yorke1995a} used an axisymmetric setup of a $0.7 \, \msun$ main
sequence star immersed into the envelope of a $3 \, \msun$ \ac{AGB}
primary to follow later stages of \ac{CEE} in this system. In contrast
to previous simulations, both cores were allowed to move and a
nested-grid technique afforded a higher spatial resolution in the
inner regions of the \ac{CE}: eight levels of cylindrical polar grids
were used, each smaller by a factor of two in the radial extend than
the next higher level grid.  The treatment of angular momentum loss of
the cores followed that of \citet{taam1991a}, with some
modifications. \citet{yorke1995a} found that in their simulation the
entire envelope was likely to be ejected without the cores
merging. Orbital decay terminated at a separation of $15 \, \rsun$ due
to a spin-up to near co-rotation of envelope gas in the vicinity of
the cores. Owing to the nested grid approach, the expansion of the
ejected material could be followed to relatively large distances and
\citet{yorke1995a} commented on the shaping the emerging planetary
nebula.

A leap forward in the quality of Eulerian grid-based \ac{CE}
simulations was taken with the \ac{3D} hydrodynamic simulations of
\cite{sandquist1998a}---reported in the last paper of the series on
``double-core evolution''. By abandoning analytic models for the drag
force or the orbital decay, the modeling fidelity of physical
processes drew level with the \ac{SPH} simulations conducted slightly
earlier by \citet{terman1994a, terman1995a}, \citet{terman1996a} and
\citet{rasio1996a}, (see Sect.~\ref{sect:sim_sph}). Therefore, the
simulations of \citet{sandquist1998a} were the first to model \ac{CE}
self-consistently in a \ac{3D} Eulerian grid-based approach. With
their nested-grid approach similar to \citet{yorke1995a}, but extended
to three spatial dimensions in Cartesian geometry,
\citet{sandquist1998a} claimed to reach higher spatial resolution than
the \ac{SPH} simulations. The nested-grid hydrodynamics code was based
on that of \cite{burkert1993a} and employed a second-order
finite-difference method to solve the equations of fluid dynamics. To
determine the gravitational potential of the gas, the Poisson equation
was solved on the coarsest grid with a particle-mesh algorithm. Two
collisionless particles were used to model the core of the primary
star and the companion. Their mutual gravitational interactions and
their interactions with the gas were softened according to
\citet{ruffert1993a}, with an adaptive reduction of the softening
length as the core particles approached each other. The envelope was
modeled with an ideal-gas equation of state in combination with radiation
pressure. The setups followed the interactions of \ac{AGB} primary
stars of $3$ and $5 \, \msun$ with main sequence companions of $0.4$ and
$0.6 \, \msun$.

\citet{sandquist1998a} found that about 20\% to 30\% of the envelope
gas was ejected at the end of the simulations, but most of the mass,
although largely removed from the volume of the original giants,
remained marginally bound at larger distances. Because of limited
spatial resolution, the core orbit did not completely settle to a
final separation, and extrapolating to the further evolution,
\citet{sandquist1998a} predicted full envelope removal for their
setups with the $3 \, \msun$ primary, while for the more massive
\ac{AGB} star they concluded that a merger of the cores may be
possible. The most important result of \citet{sandquist1998a} is
perhaps the first detailed description of the dynamical change in the
envelope structure due to flows that are clearly not
axisymmetric. Their resolution of the coupling between the core
particles and the envelope gas allowed to identify for the first time
spiral density waves that were generated in front of both stellar
cores. \citet{sandquist1998a} concluded that these spiral shock waves
were the primary mechanism for transporting the angular momentum
released in the core inspiral outwards through the envelope \citep[but
  see also][]{rasio1996a}.

With the same numerical approach, \citet{sandquist2000a} performed a
systematic exploration of \ac{CEE} with \ac{RG} primaries of different
total and core masses with companions of various masses. They found
that such low-mass binary systems can survive \ac{CEE} provided that
the total mass of the \ac{RG} primary is below $2 \, \msun$ and its
degenerate helium core exceeds about $0.25 \, \msun$. The remnants of
these interactions are associated with observed helium double
degenerate systems, pre-cataclysmic variables and subdwarf B
stars. With the same code, \cite{demarco2003a, demarco2003b} simulated
the \ac{CE} phase between an \ac{AGB} primary star and companions of
$0.1$ and $0.2 \, \msun$ as a potential origin of Wolf-Rayet central
stars of planetary nebulae.

\subsubsection{Modern Eulerian grid-based approaches}
\label{sect:sim_eulerian_modern}

Up to this point in the history of Eulerian grid-based \ac{CE}
simulations, the treatment of fluid mechanics followed fluid-in-cell
or finite-difference methods that are not commonly in use in modern
astrophysical simulations any longer. The advantages of finite-volume
techniques and high-resolution shock capturing schemes (see
Sect.~\ref{sect:eulerian}) were introduced into the field of \ac{CE}
simulations only relatively late compared with other astrophysical
applications. This is surprising given the fact that conservation
properties and resolution of shock waves are essential ingredients for
capturing the dynamics of such events. Simulations performed after
\citet{sandquist2000a} can be considered \emph{modern efforts}:
Eulerian approaches stepped up using techniques such as finite-volume
methods, Godunov schemes and \ac{AMR}. As such, they started to flank \ac{SPH}
techniques in addressing the complications of the \ac{CE} interaction.

The work of \citet[][already hinted at in \citealt{taam2006a}, who
  discussed some of the technical advantages]{ricker2008a} introduced
the first modern Eulerian grid-based framework for numerical fluid
dynamics to \ac{CE} simulations. The \textsc{flash} code
\citep{fryxell2000a} features the piecewise-parabolic reconstruction
method of \cite{colella1984a}. The use of \ac{AMR} techniques
significantly improves flexibility in spatial resolution over fixed
nested grids. This allows one to refine at shock fronts and in the
vicinity of the stellar cores. As discussed in
Sect.~\ref{sect:eulerian}, it is this technique, that ultimately
renders Eulerian grid-based approaches to simulating \ac{CEE}
competitive to alternative \ac{SPH} and moving mesh techniques that we
discuss in Sects.~\ref{sect:sim_sph} and
\ref{sect:sim_movingmesh}. However, even with \ac{AMR} it is
impossible to resolve the $0.36 \, \msun$ core of the $1.05 \, \msun$
\ac{RG} primary star modeled by \citet{ricker2008a}---in particular
in the view of the small time steps required by the \ac{CFL} condition (see
Sect.~\ref{sect:comp_feasible}). They therefore represented the core
of the primary as well as the $0.6 \, \msun$ companion by clouds of
\num{2e5} particles each occupying a volume given by a length scale
that exceeds the finest grid spacing by a factor of three. The
particles are moved rigidly with their centers of mass. Their
gravitational coupling is modeled with a cloud-in-cell interpolation
and \citet{ricker2008a} argued that particle clouds avoid force
anisotropy problems that would arise from using single particles in
this approach.

With their 8-level \ac{AMR} setup, \cite{ricker2008a} reached an
effective resolution of $2048^3$ cells over the domain size. Transient
motions in the setup of the primary star were damped for one dynamical
timescale before the companion was added and the primary was set
spinning with an angular velocity of 95\% of co-rotation with the
orbit (that, if not previously stabilized, would have induced
expansion). The simulation of \citet{ricker2008a} covered only a very
short initial phase of \ac{CE} interaction---less than one orbit of
the binary system is followed---but it confirmed the earlier finding
of \citet{sandquist1998a} that the orbital decay was dominated by
non-axisymmetric gravitational drag, which is one to two orders of
magnitude larger than hydrodynamic drag. Moreover, \citet{ricker2008a}
found a much faster orbital decay than predicted from the \ac{BHL}
formalism. This was explained by the fact that within the \ac{BHL}
framework, the predicted gravitational drag is similar to the
hydrodynamic counterpart because it is only affected by the local
formation of a wake. In \acp{CE}, however, this drag is
increased by the gravitational forces over a larger volume of the
entire envelope. Although \citet{ricker2008a} emphasized the
shortcomings of their model in capturing accretion onto the companion
caused by the lack of an inner boundary around it, they argued that
the accretion rate onto the companion should be significantly smaller
than expected from \ac{BHL} accretion arguments.

\citet{ricker2012a} presented an update of their previous work, again
employing the \textsc{flash} code, following the same system for
somewhat longer times of the evolution, now covering five orbits. They
explicitly mentioned the treatment of self-gravity in this simulation:
the Poisson equation was solved with a multigrid method (see
Sect.~\ref{sect:grav_alt}). The number of particles used to represent
the stellar cores was increased to \num{2e5} for each cloud and 9
levels of \ac{AMR} were used. \cite{ricker2012a} reported an ejection
of 25\% of the envelope's mass over the simulated time, higher than
reported by \citet{sandquist2000a}, and possibly caused by a larger
mass ratio between primary star and companion in their system.  They
further noted a rapid decrease of the eccentricity of the orbit and
identified imprints of spiral shocks on the morphology of the envelope
gas that was expelled preferentially close to the orbital
plane. Again, they confirmed that gravitational drag dominates over
hydrodynamic drag and emphasized that the \ac{BHL}
prescription overestimates the accretion rate onto the companion.

Recently, the same group, still using \textsc{flash}, has been
attempting the simulation of \ac{CEE} in more massive systems. There is so
far only one brief publication \citep{ricker2019b} presenting a
simulation of \ac{CE} interaction with an $82\, \msun$ red supergiant
primary star of radius of $2891 \, \rsun$ at low metallicity, deriving
from an $88 \, \msun$ main sequence star. The companion was a $35 \,
\msun$ point particle, supposed to represent a black hole. This
simulation included for the first time both the effects of partial
ionization, via a suitable equation of state (see
Sect.~\ref{sect:eos}) and the effects of radiation, via a
single-group, flux-limited diffusion approximation. Interestingly,
\citet{ricker2019b} concluded that envelope unbinding is not
benefiting from recombination energy. Clearly, these simulations are
critical to answer the question of whether double black hole binaries
that merge under the emission of detectable gravitational waves can be
formed via an isolated binary channel \citep{belczynski2016a}, but
more work is needed to confirm the still very preliminary results of
\cite{ricker2019b}.

The next grid-based code to be exercised on the \ac{CE} problem was
\textsc{enzo} \citep{bryan1995a}. This code was originally developed
for cosmological simulations and \citet{passy2012a} adapted it to the
stellar problem. Core and companion were modelled as point masses with
the usual softening of the gravitational potentials (see
Sect.~\ref{sect:gravsoft}). Self-gravity was implemented with a fast
Fourier transform technique (see Sect.~\ref{sect:grav_alt}), while the
gravity of the two point masses was treated analytically by adding the
(softened) potential to that calculated for the gas. Starting out from
a \ac{1D} stellar model, \citet{passy2012a} mapped a $0.88 \,\rsun$,
$100 \,\rsun$ \ac{RGB} star with a core of $0.392 \, \msun$ onto a grid
without any additional levels of refinement. Two resolutions with
$[128]^3$ and $[256]^3$ cells were used. They carried out five \ac{CE}
simulations for each resolution with companion masses of $0.1$,
$0.15$, $0.3$, $0.6$ and $0.9 \,\msun$ and two additional
low-resolution simulations to test the effects of the initial orbital
separation and eccentricity. The companions were initialized on the
stellar surface\footnote{For one simulation, an initial eccentricity
was imposed by increasing slightly the initial separation of the two
core particles.}. One finding of these simulations was the---perhaps
expected---realization, that in none of them the entire \ac{CE} became
unbound\footnote{\citet{passy2012a} do not list the fraction of
envelope that is unbound but these values are very similar to those of
their comparison simulations carried out with an \ac{SPH} code and
presented in Sect.~\ref{sect:sim_sph}}. The range of final separations
was ${\sim}\, 6$ to $30 \, \rsun$, with wider final separations for
more massive companions. The most important achievement of this study,
however, was a direct comparison between Eulerian grid-based and
\ac{SPH} (see Sect.~\ref{sect:sim_sph}) approaches to \ac{CE}
simulations, which used the \textsc{snsph} code of
\citet{fryer2006a}. This paper also raised some questions which remain
unanswered to this day. The most pressing of them is whether the final
separation of the stellar cores measured in the simulations is the
actual post-\ac{CE} separation. The values determined from the simulations
appeared to be on the high side of observations and it was therefore
concluded that in post-dynamic inspiral phase the companion would
migrate inwards.

The \textsc{enzo} code was the basis for a number of follow-up papers,
which, as an upgrade, used its \ac{AMR} capability and the adaptive
particle-mesh solver of
\citet{passy2014a}. \citet{iaconi2017a,iaconi2018a} modelled the same
primary as \citet{passy2012a}, but this time fully investigated the
effect of initial orbital separation. They concluded that starting the
simulation at a wider separation (approximately that of the start of
Roche lobe overflow) does result in a somewhat wider orbital
separation due to the giant expanding before the fast inspiral rather
than spinning up. They too carried out a comparison with an \ac{SPH}
simulation (see Sect.~\ref{sect:sim_sph}), this time with the \textsc{phantom} code
\citep{price2018a}. In a subsequent paper,
\citet{iaconi2018a} compared results from setups identical to those of
\citet{passy2012a} with simulations in which the primary was a more
massive (${\sim} 2 \,\msun$) \ac{AGB} star. The reasoning was that if
the envelope was more massive it would expand more reluctantly and
allow the companion to sink deeper. This was expected to cause an
increased transfer of orbital energy to the envelope. While companions
did spiral in deeper, no substantial increase in unbound mass was
observed: in both the original setup and in the simulations involving
an \ac{AGB} primary star, the unbound mass fractions ranged between a
few and about 15 percent.

\citet{staff2016a} carried out \textsc{enzo}-based \ac{CE} simulations
involving a ${\sim}\, 3 \, \msun$ \ac{AGB} star and companions between
$0.6$ and $3.0 \, \msun$ with a range of eccentricities. Their aim was
to understand systems such as the protoplanetary nebula
OH~231.8+4.2. These had been suggested to result from a \ac{CE}
interaction in an eccentric binary. On the one hand, these simulations
were used primarily as platforms for a series of analytical
discussions on the ability of the companion to launch a jet
\emph{before} the actual \ac{CE} interaction, that would result in the
observed nebular energetics.  On the other hand, these simulations
were among the few that started out with high eccentricities in the
initial system, \ac{AGB} primary stars, and stellar masses above the
common ${\sim}\, 1 \, \msun$. With the relatively low resolution
achieved in these simulations, however, it remained difficult to put
the results appropriately into context. Most of the envelope mass
remained bound (again, between a few and about 15 percent of the
material was unbound) and the separation still ranged on the wide side
compared to observations. Moreover, the study concluded that if mass
transfer takes place at periastron passage, the orbital elements are
altered and it is likely that these orbital changes lead to a \ac{CE}
within a short period of time. This means that it is unlikely that
objects like OH231.8+4.2 could be formed by repeated jet episodes that
take place during periastron passage.

In a companion paper, \citep{staff2016b} used the same $3 \, \msun$
\ac{AGB} primary star as well as an \ac{RGB} star of similar mass to
determine the outcome of planetary intrusions into the envelopes of
giants. Somewhat unsurprisingly the conclusion was that planets would
inspiral and merge with the core within a decade (\ac{RGB} star) to a
century (\ac{AGB} star). Considerable angular momentum would be added
the the envelopes of the giants which would expand somewhat and spin
up to values commensurate with the fastest rotating giant stars
observed by \citet{demedeiros1996a}.
\begin{figure}
    \centering
    \includegraphics[width = \textwidth]{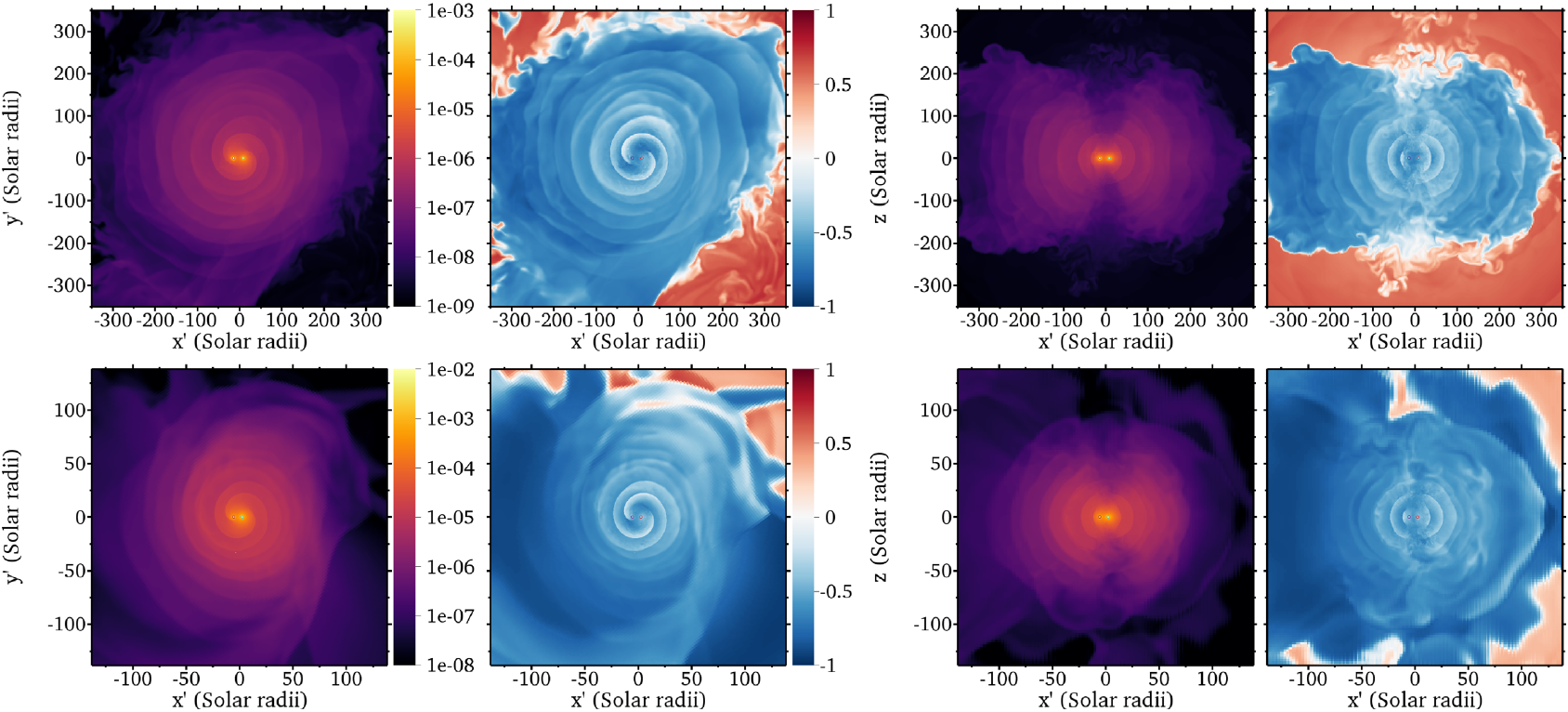}
    \caption{\ac{CEE} simulations with an \ac{AGB}
      (top panels) and an \ac{RGB} (bottom panels) primary star. Shown
      are slices after ten orbits of the core binary system through
      the orbital plane(left) and perpendicular to it in density (left
      panels) and total energy (right panels, where blue is bound gas
      while red is unbound). Figure from \citet{chamandy2020a}}
    \label{fig:Chamandy+20-Fig6b}
\end{figure}

Also using the version of \textsc{enzo} adapted by \citet{passy2012a},
\citet{shiber2019a} carried out a simulation using the same star used
by \citet{passy2012a}. This time, however, the effect of jets was
explored, that are potentially launched by the companion throughout
the inspiral. In particular, the focus was on whether such jets
enhance envelope unbinding. The jets were not self-consistently
simulated, but were instead launched according to analytically derived
quantities such as mass loss rate, launch speed and opening angle. The
outcome of the simulation was not dramatically different from the
equivalent simulation without jets: the final separation was somewhat
wider and additional envelope gas was unbound (the unbound mass
increased from $\sim\,$10\% without jets to $\sim\,$30\% with
jets). This indicated that while jets may aid envelope ejection, it
seems unlikely that they can fully explain it.

\citet{schreier2019a} explore the effect of jets in \ac{CEE} of a
triple system: a tight binary system is placed into the envelope of a
$4 \, \msun$ \ac{AGB} star with $100 \, \rsun$. The inner $5 \, \rsun$
are replaced with a sphere of constant density, pressure, and
temperature. The binary system is not resolved; one of its stars is
assumed to accrete material and to launch two jets in opposite
directions with an inclination relative to the orbital plane of the
binary and the core of the \ac{AGB} star. In the simulation, the jets
are injected with an inclination angle of $45^\circ$ at the location
of the binary system. The simulation is carried out with the
\textsc{Pluto} code \citep{mignone2012a} on a cubical domain with $400
\, \rsun$ side length covered with 48 cells and three to four levels
of \ac{AMR}. Self-gravity is not accounted for in the setup:
gravitational effects of the tight binary system and of the deformed
envelope are ignored. \citet{schreier2019a} observe vortices and
associated density fluctuations perturbing the envelope structure and
discuss implications for dust formation and the shaping of ``messy''
\acp{PN}. The follow-up studies of \citet{schreier2021a} and
\citet{schreier2023a} were carried out with the \textsc{Flash} code
\citep{fryxell2000a} in similar setups but with a more massive primary
star of a zero-age main sequence mass of $15 \, \msun$ evolved to a
red supergiant stage ($M = 12.5 \, \msun$, $R = 881 \, \rsun$). To
avoid prohibitively small time steps in the simulation, the inner 20\%
of the stellar radius are replaced by an inert core. The \ac{AMR}
capabilities of the \textsc{Flash} code are used.
\citet{schreier2021a} assume a neutron star companion that passes
through the envelope of the giant on a highly eccentric orbit---an
event \cite{gilkis2019a} call a ``common-envelope jets supernova impostor''. The
neutron star is modeled as a point particle and the jets are launched
artificially ``by hand''. Again, neither gravity of the neutron star
nor self-gravity of the envelope are included in the model such that
no gravitational drag acts on the companion. Only one orbit of it is
simulated. A complicated clumpy outflow morphology is detected and the
luminosity of the event is estimated to place it into the regime of
``gap transients'' between classical novae and
supernovae. \cite{schreier2023a} explore the deposition of angular
momentum into the envelope due to jets that are launched perpendicular
to the orbit and due to tilted jets (motivated by the triple scenario
where the companion neutron star or black hole is part of a tight
binary system).

Another set of grid-based \ac{CE} simulations was carried out with the
\ac{AMR} code \textsc{AstroBEAR} \citep{carroll2013a}. A few 
studies have been published so far, concentrating on the analysis of
two primary stellar structures and a range of companions. The two
structures are an \ac{RGB} and an \ac{AGB} star deriving from a $2 \, \msun$
main sequence star, with radii of $49 \,\rsun$ and $122\,\rsun$,
respectively \citep{chamandy2018a, chamandy2020a}. All simulations
initialized the companion approximately at the stellar surfaces.

\citet{chamandy2018a} benchmarked \textsc{AstroBEAR} against a similar
simulation carried out with the moving mesh code \textsc{arepo}
\citep[][see Sect.~\ref{sect:sim_movingmesh}]{ohlmann2016a}. They also
investigated the amount of accretion that would likely take place onto
the companion---a very important topic, particularly in the
massive-star regime when massive companions may generate large amounts
of feedback energy. They concluded that the amount of accretion cannot
be estimated just using control surfaces around the companion because
a high-pressure pseudo-atmosphere that settles around the point mass
companion and reduces the accretion. \citet{chamandy2018a} suggested
that the introduction of mass ``disappearance'' likely results in a
more realistic outcome and that high accretion rates (${\sim} \, 1
\,\msun \, \mathrm{yr}^{-1}$) could exist during the short-lived
inspiral phase. In the simulation, this was implemented with a
subgrid-scale accretion model. Following \citet{krumholz2004a}, the
companion was represented by a sink particle in one of the runs. The
simulations were also used to draw a comparison between the
gravitational drag force measured in the simulation and what would be
derived using the analytical approximations of
\citet{macleod2017b}. \citet{chamandy2018a} concluded that the
simulated force was similar to the analytical prescription only early
on and until mid inspiral, but diverges later with the simulated force
being much smaller than the analytical approximation.

\citet{chamandy2020a} were first to model \ac{CEE} of a low-mass
\ac{AGB} star with a ${\sim} \, 1 \, \msun$ companion. The conclusions
that they could draw from a comparison of their \ac{RGB} and \ac{AGB}
simulations was that the two were remarkably similar (see
Fig.~\ref{fig:Chamandy+20-Fig6b}). However, both simulations were
started with the companion at the surface, not accounting for the
redistribution of gas that takes place when the simulation is started
at the time of Roche lobe overflow. Only a fraction of the envelope
mass was unbound during the simulated time, but \cite{chamandy2020a}
estimated a full envelope ejection over 10 years if the rate of mass
unbinding measured in the simulations stayed constant.

\citet{zou2022a} modified the setup of \citet{chamandy2018a} that,
with a $2\, \msun$ \ac{RG} primary star and a $1\, \msun$ or $0.5 \,
\msun$ companion, was similar in the parameters and the treatment of
the \ac{RG} core to that of \citet{ohlmann2016a} but used the
\textsc{AstroBEAR} code. They included a jet launched via a subgrid
model from the companion. This model injects mass at a chosen constant
rate with a certain velocity perpendicular to the orbital plane into a
prescribed opening angle.  Comparing this model to simulations without
jets, the influence of a potential jet on orbital evolution and the
morphology and unbinding of the envelope is explored and the
parameters of the jet model are tested. The jets were found to be
choked in the envelope and transferred additional energy onto it so
that the unbound mass was increased by 10\% over the ten simulated
orbits compared to simulations without jets. Based on this result,
\citet{zou2022a} expect that jets launched from main sequence or white
dwarf companions are unlikely to dominate envelope unbinding.

Except for the study of \citet{ricker2019b}, none of ``modern''
Eulerian grid-based simulations of \ac{CEE} accounted for ionization
effects in the envelope material. None of them achieved a successful
envelope ejection over the simulated time; the unbound mass fraction
typically stayed below $\sim$20\%.

\subsubsection{SPH approaches}
\label{sect:sim_sph}

In logical continuation of previous \ac{1D} models, early Eulerian
grid-based approaches to simulating \ac{CEE} (see
Sect.~\ref{sect:sim_eulerian_pioneering}) modeled the interaction of
the companion with the envelope by explicitly imposing a drag force on
it. \ac{SPH} simulations, in contrast, from the start relied on a
self-consistent treatment of this interaction by implementing
self-gravity. Although with their limited resolution simulations
carried out in the 1980s and 1990s failed to converge in reproducing
the effect, with increasing computer power this approach became
gradually more realistic. In this sense, the development of \ac{SPH}
methods for \ac{CE} simulations was not subject to breaks and
breakthroughs in the modeling technique itself---although an efficient
parallelization of the \ac{SPH} framework to distributed memory
machines is not trivial and was achieved only in the early-to-mid
2000s \citep[e.g.][]{springel2001a, springel2005a, fryer2006a}. It
paved the way to increasing the resolution of the simulations, but
still today not all \ac{SPH}-based \ac{CE} simulations are carried out with
MPI parallelization.

The application of \ac{SPH} for simulating the onset of \ac{CEE} was
pioneered by \cite{dekool1987a}. He modelled the interaction of a $11
\, \msun$ giant with a $4 \, \msun$ companion star. Only \num{600}
particles were used to represent the envelope of the primary star
assuming a polytropic equation of state while its core and the
companion were modeled by single particles. The setup was obtained
from relaxing the stellar envelope to hydrostatic equilibrium
and---for the initial binary setup---by slowly reducing the separation
between the interacting stars.  This setup procedure is largely
followed in \ac{SPH}-based \ac{CE} simulations to this day.
\citet{livio1988a} simulated \ac{CEE} in a hybrid
``particles in cell'' (PIC) discretization of the equations of fluid
dynamics, which combines a pseudo-particle approach with an Eulerian
grid and thus differs from the \ac{SPH} formalism.

The increase of computational power and improvements in numerical
methods---most notably the introduction of a hierarchical tree
algorithm for calculating self-gravity---allowed to conduct more
detailed \ac{3D} \ac{CE} simulations with \ac{SPH}. For the first
time, and in contrast to contemporary grid-based Eulerian simulations
(see Sect.~\ref{sect:sim_eulerian_pioneering}), such simulations were
able to follow the inspiral of the companion into a giant primary with
the gravitational interaction between the cores and the envelope gas
captured by the model instead of using an analytic prescription for a
drag force.

\citet{terman1994a} discretized the co-rotating envelope of a $4.67
\,\msun$ \ac{RG} primary mapped from a polytropic model using
\num{10000} \ac{SPH}-particles. After random initial placement of the
particles, frictional damping established \ac{HSE}. The core of the
primary star as well as the $0.94 \, \msun$ companion were represented
by single massive collisionless particles. A fictitious drag force was
applied only initially to bring the binary system into the \ac{CE}
stage---the actual \ac{CEE} was followed self-consistently with the
drag force resulting from the modeled interaction. With this setup,
\cite{terman1994a} demonstrated that gravitational tidal torques cause
rapid orbital decay of the core binary after the onset of
\ac{CEE}. The evolution of the system was followed for
$\SI{0.6}{\year}$ when the orbital separation settles to about $6 \,
\rsun$ because of a spin-up of the gas in the vicinity of the cores
approaching co-rotation.  This simulation exploited the capability of
\ac{SPH}-based approaches to follow the inspiral phase over many
orbits. It was eventually terminated because the gravitational
softening of the core particles became larger than their separation
rendering the model nonphysical. Only about 13\% of \ac{RG} envelope
mass was ejected to this point. Based on their simulations,
\cite{terman1994a} discussed the importance of the \ac{3D} response of
the envelope to the motion of the binary composed of the giant's core
and the companion, pointing out deficiencies of models that assumed an
inspiral of a companion star into a fixed core--envelope primary
structure.

\citet{terman1995a} extended their numerical approach to simulating
the inspiral of a $1.4 \, \msun$ neutron star into massive primaries
of $16$ and $24 \, \msun$ in various evolutionary stages. These
simulations exploited the efficiency of the \ac{SPH} approach to
simulate \ac{CEE}, although the achieved resolution was insufficient
to accurately separate the core--envelope structure of the primary
stars, which, with the given parameters, remains a challenge even for
current simulations. For some of the considered systems, a mass
ejection efficiency of $\sim$30\%--40\% was reported.
\citet{terman1995a} pointed out that the survival of the core binary,
as opposed to a merger of the neutron star with the primary's core
potentially forming a Thorne--{\.Z}ytkov object \citep{thorne1975a},
depends on the structure and evolutionary stage of the non-compact
star.

\citet{terman1996a} relaxed their assumption of a polytropic equation
of state to include both gas and radiation pressure in simulations of
systems with a $5 \, \msun$ \ac{RG} and companions with $0.5$ and $1.0
\, \msun$. Again, \num{10000} \ac{SPH}-particles were used for
modeling the \ac{RG} envelope, which, in this suite of simulations,
was set up as spherical and non-rotating. The size of the cores was
reduced with respect to the simulations of \cite{terman1994a},
allowing the simulations to follow the inspiral for longer, which
resulted in the ejection of up to 70\% of the envelope gas.

\citet{rasio1996a} used up to \num{50000} particles to follow the
onset of the unstable mass transfer and initial inspiral of a $0.7 \,
\msun$ main sequence star into a \ac{RG} of $4 \, \msun$ until a
quasi-static \ac{CE} had formed. They pioneered the construction of a
rotationally synchronized binary system in \ac{HSE} by means of a
relaxation and damping procedure---in contrast to \cite{terman1996a}
and without the need of applying an initial fictitious drag force as
in \cite{terman1994a}. The core of the primary star and the secondary
were modeled as point particles and their gravitational potentials
were softened over a length comparable to the \ac{SPH} smoothing
length of the innermost particles, typically a hundredth of the radius
of the \ac{RG} primary \citep[which is similar to
  what][used]{terman1996a}. Under this assumption, the core of the
primary star is much larger than in reality.  In their simulation,
\citet{rasio1996a} identified a co-rotating region of gas around the
central binary and found indication for the onset of convection in the
envelope in agreement with the predictions of \citet{meyer1979a}. Only
about 10\% of the envelope material was unbound---most likely because
of the establishment of co-rotation in late stages of the evolution as
opposed to the findings of \citet{terman1996a} whose models eject a
substantial fraction of the envelope. \citet{rasio1996a} attributed
this discrepancy to differences in the spatial resolution, but as
neither the numbers of particles nor the sizes of the core models
differ significantly, this interpretation appears questionable.

In their paper comparing \ac{CE} simulations carried out with a
grid-based Eulerian code (see Sect.~\ref{sect:sim_eulerian_modern})
and \ac{SPH}, \citet{passy2012a} employed the \textsc{snsph} code of
\citet{fryer2006a}. It adopts a distributed-memory parallelization
\citep{fryer2006a} and therefore allowed to conduct simulations with
enhanced resolution. \textsc{snsph} uses a special tree-based method
 \citep[a parallel hashed oct-tree,][]{warren1993a} to calculate
gravitational accelerations. As in their grid-based counterpart
simulations, \citet{passy2012a} followed the interaction of a
non-rotating $0.88 \, \msun$ \ac{RG} primary star of $83 \, \rsun$
with companions of masses between $0.9$ and $0.1 \, \msun$ that were
initially placed at the primary's surface in a circular orbit and were
modeled as collisionless particles. These simulations were set up as
close as possible to the counterpart grid-based \textsc{enzo}
simulations (see Sect.~\ref{sect:sim_eulerian_modern}). The envelope,
modeled with an ideal-gas equation of state, was represented by an
unprecedented number of \num{500000} \ac{SPH}-particles, that were
initially placed with a weighted Voronoi tessellation.  In contrast to
other setups, the core of the giant primary star was treated as a
massive \ac{SPH} particle of $0.392 \, \msun$ and not as a
collisionless particle, such that its pressure stabilized the envelope
against gravitational collapse.  It was placed at the center, embedded
in a region with \ac{SPH}-particles, such that the profile of density
connected smoothly at the core-envelope boundary, which was located at
twice the smoothing length of the core particle, i.e.\ $0.2 \,
\rsun$. Therefore, the interaction of the core with the envelope gas
was better resolved than in any of the previous \ac{SPH} simulations
of \ac{CEE}. The simulations of \citet{passy2012a} followed the
evolution until the orbital separation and the mass of the unbound
envelope gas saturated. The amount of envelope gas ejected during the
simulation varied with the mass ratio and was expelled mostly close to
the orbital plane. It remained below 10\% in all cases and the final
orbital separations between the core particles were systematically
larger than those found in observations of post-\ac{CE}
systems. \citet{passy2012a} discussed additional physical effects that
may increase the envelope ejection efficiency. They emphasized the
capability of the \ac{SPH} method to model the accumulation of gas
around the companion star---a direct implication of the Lagrangian
nature of the scheme in which the spatial resolution automatically
adapts to the mass distribution.

The next group to simulate \ac{CEE} with \ac{SPH} codes were
\citet{nandez2014a}, who modelled the interaction in system V1309~Sco
(a $1.5 \, \msun$ subgiant and a ${\sim}\, 0.15 \, \msun$ main
sequence star; \citealp{tylenda2011a}), using the \ac{SPH} code
\textsc{starsmasher} \citep{lombardi2011a}. In this simulation, the
two stars were spatially resolved because of their relatively similar
sizes (with particle numbers ranging from \num{20000} to \num{200000}
for the primary star and from \num{2000} to \num{19000} for the
companion), and the interaction lead to a merger.

V1309~Sco is a particular system and perhaps not a clear \ac{CE} case
(see Sect.~\ref{ssec:terminology}). A more traditional \ac{CE}
interaction was later simulated by the same group \citep{nandez2015a}
with the goal of modeling the double white dwarf system WD
1001+364. For this purpose, the core of the \ac{RG} primary was
selected to be $0.32 \, \msun$ (slightly larger than the value of the
original \ac{1D} stellar profile in order to match the envelope
structure) and represented by a point mass interacting only
gravitationally. The $0.36 \, \msun$ companion was also modeled as a
point particle. Various masses between $0.668$ and $1.481 \,\msun$
were assumed for the \ac{RG} envelope and resolved with \num{100000}
particles; one run was carried out with twice this
resolution. \citet{nandez2015a} were the first to model the effect of
recombination processes in a \ac{3D} \ac{CE} simulation. For that
purpose, they adapted the \textsc{mesa} equation of state
\citep{paxton2011a}, that accounts for changes in the ionization
structure by implementing the \textsc{opal} tables, see
Sect.~\ref{sect:eos}. They found that while simulations of the basic
gravo-hydrodynamic model removed only about 50\% of the envelope mass,
complete envelope ejection could be achieved when assuming the
released recombination energy to thermalize locally.

The same approach was used by \citet{nandez2016a} for a range of
primary \ac{RG} stars with masses of $1.2$ to $1.8 \, \msun$ and
companions with masses of $0.32$ to $0.40 \,\msun$ meant to be white
dwarfs and modelled as point particles that interact only
gravitationally, as did the cores of the \ac{RG} primaries. The
resolution of \num{100000} \ac{SPH} particles was similar to the
production runs in \citet{nandez2014a}. Almost complete mass ejection
was obtained thanks to their use of the tabulated equation of state
that includes recombination energy.  Their simulations were then used
by \citet{ivanova2016a} to infer a number of parametrizations to be
used, e.g., in population synthesis studies.

\citet{iaconi2017a} carried out a simulation of \ac{CE} interaction
using the same setup as \citet{passy2012a}, but with a wider initial
separation between the primary star and the $0.6 \, \msun$
companion. They used the shared-memory (OpenMP) parallelized \ac{SPH}
code \textsc{phantom} of \citet{price2018a}. This time, both the core
of the primary and the companion were modeled as collisionless massive
single particles. For gravitational softening, the prescription of
\citet{ruffert1993a} was applied instead of the previously used
formulation of \citet{monaghan1992a}. The envelope was represented by
one million particles. Both the obtained post-\ac{CE} orbital separation
and the amount of unbound gas ($\sim$16\% of the initial envelope
material) were found to be slightly larger than those of
\cite{passy2012a}.

\citet{reichardt2019a} followed up on these studies and simulated
\ac{CEE} with the \textsc{phantom} code again in a system similar to
that of \citet{passy2012a}. Their resolution varied between \num{8e4}
and \num{1.1e6} \ac{SPH} particles representing the envelope of the
\ac{RG} primary star, which for one of their simulations was set up to
co-rotate with the binary system. In contrast to previous simulations,
the initial binary separation was chosen such that the \ac{RG} primary
star just filled its Roche lobe and mass transfer started
immediately. This way, \citet{reichardt2019a} extended the simulation
to the time before the inspiral, which was possible thanks to the
excellent angular momentum conservation of \ac{SPH}-based simulations
(see Sect.~\ref{sect:conservation}).  While the duration of the
pre-inspiral evolution was found to be resolution-dependent, this
simulation paves the way for the study of the impact of the initial
conditions on the \ac{CEE}. Moreover, mass lost over the $L_2$ and
$L_3$ Lagrange points of the system before the inspiral was found to
remain bound and to potentially form a circumbinary disk. This disk
was destroyed in the subsequent \ac{CE} ejection, but the authors
speculated about its survival for other system parameters. The
measured drag force in the \ac{CE} inspiral phase was consistent with
analytical models, but only a small fraction of the envelope material
(about 10\% to 30\%) was ejected---confirming previous simulations
that indicated less unbound mass for increasing numerical resolution.

\begin{figure}
    \centering
    \includegraphics[width = \textwidth]{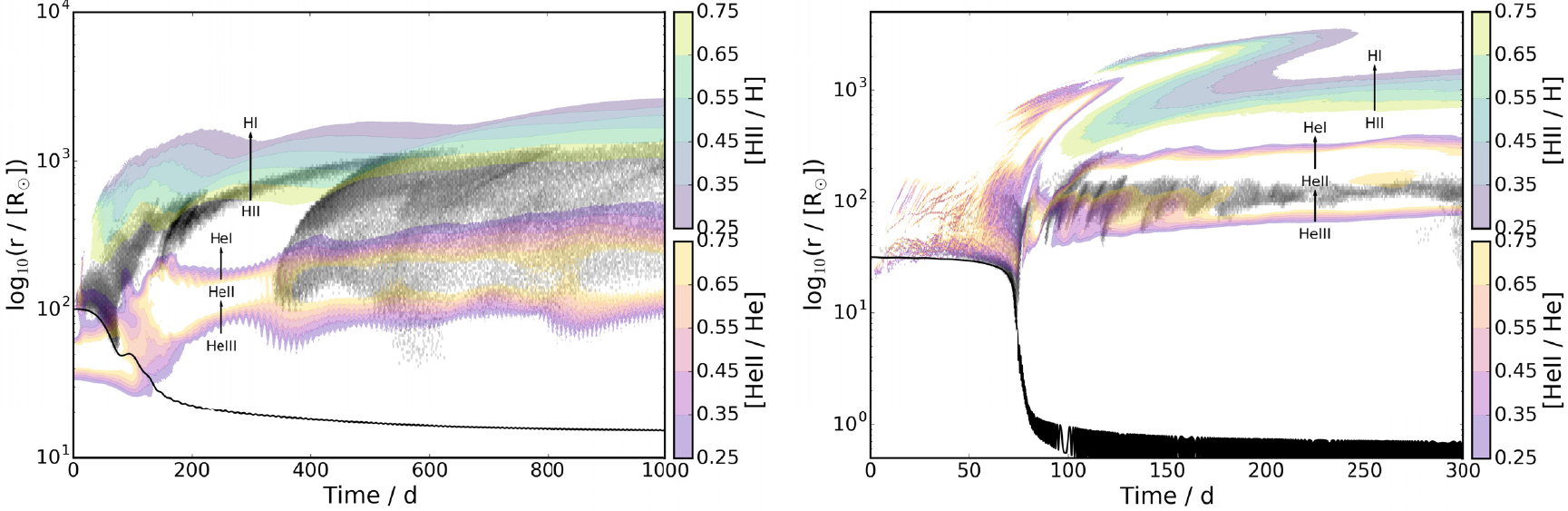}
    \caption{Correlation between unbinding of gas and recombination
      zones in the simulations of \citet{reichardt2020a} with a $0.88
      \, \msun$ (left) and a $1.8 \,\msun$ \ac{RGB} (right). The
      locations and times at which \ac{SPH} particles became
      permanently unbound are plotted in black. The coloured zones
      represent the partially ionized regions of H and He gas. The
      black line indicates the distance between the stellar
      cores. Figure from \cite{reichardt2020a}}
    \label{fig:Reichardt+20_Fig5}
\end{figure}

\citet{reichardt2020a} equipped the \textsc{phantom} \ac{SPH} code
with an equation of state that can account for changes of the
ionization structure of the envelope material. With this modification
and assuming local thermalization of the released recombination
energy, they carried out two simulations with \ac{RG} primary stars of
$0.88 \, \msun$ and $1.8 \, \msun$ and companions of $0.6 \, \msun$
and $0.36 \, \msun$, respectively. The latter configuration was the
same as one of \citet{nandez2016a}, who claim complete envelope
ejection for this case. In agreement with \citet{nandez2016a},
\citet{reichardt2020a} found that including recombination effects
leads to a dramatic increase in envelope ejection. The system with the
less massive primary star achieved complete envelope ejection,
promoted by the release of both helium and hydrogen recombination
energy. The model with the more massive primary, unbound more mass,
seemingly at the hand of helium recombination energy (see
Fig.~\ref{fig:Reichardt+20_Fig5}), but the system did not achieve
complete envelope ejection, contrary to \citet{nandez2016a}. The cause
for this discrepancy is the definition of unbound gas (see also
Sect.~\ref{sect:ionization}): When including recombination energy in
the definition, effectively the entire envelope becomes unbound. If
instead only the thermal energy is accounted for, the conclusion is
that no more than 37\% of the envelope are unbound.  Moreover,
\citet{reichardt2020a} discussed that while the hydrogen recombination
energy may be more readily lost into space, helium recombination
energy is much less prone to be radiated away, because it takes place
deeper in the envelope, and it is therefore more likely to partake in
the envelope ejection. Finally, they pointed out that the formation of
dust in the expanded envelope is extremely likely. This could change
the game-plan for the interaction. \citet{reichardt2020a} compared
their simulations accounting for recombination energy release and
assuming its local thermalization to counterparts carried out with an
ideal-gas equation of state (that do not account for these
effects). They found that the final orbital separation was almost
unaffected by the equation of state and by the success of envelope
ejection.

An efficient \ac{SPH} code with parallelization on distributed memory
architectures is \textsc{gadget2} of \citet{springel2005a}. It was
originally developed for cosmological simulations and has been applied
to simulating \ac{CE} interaction. \cite{glanz2018a} pointed out the
potential importance of dust formation for \ac{CEE} based on a
simulation with \textsc{gadget2}. They set up a solar-like star and
evolved it to the \ac{RG} stage ($R_\star = 83 \, \rsun$, core mass
$M_\mathrm{core} = 0.39 \, \msun$) with the \textsc{mesa} code. The
\ac{CEE} with a main sequence companion was then simulated with
\textsc{gadget2}, but \citet{glanz2018a} give little details about the
setup and the numerical treatment. The structure resulting from the
\ac{CE} simulation was compared to the structure of an \ac{AGB} star
and it was concluded that large parts of the envelope meet the
conditions for dust condensation.

A similar approach was taken by \citet{glanz2021a} to model \ac{CEE}
in binaries with eccentricities of $0 \le e \le 0.95$. Primary stars
of $1\, \msun$, $4\, \msun$, and $8\, \msun$ were evolved to the
\ac{RGB} stage with \textsc{mesa} (core masses of $0.39\, \msun$,
$0.48\, \msun$ and $1.03\, \msun$, respectively), the
intermediate-mass model was in addition followed to the \ac{AGB}
stage. The stellar envelopes were mapped to distributions of
\num{250000} \ac{SPH}-particles assuming an ideal-gas equation of
state, while the cores were represented by point particles. For the
mapping, the \textsc{amuse} framework of \citet{portegies2009a} was
used and the obtained configuration was relaxed in a damping procedure
\citep{glanz2021b}. Companions of $0.6 \, \msun$ were added with
varying eccentricities to the two lower mass primary star models,
while for the high mass model a companion of $2 \, \msun$ was
chosen. These were modeled as point particles interacting only via
gravity. In their simulations, \citet{glanz2021a} found that that the
initial eccentricities only partially circularize. Because of the
closer initial approach of the cores, more mass was unbound in highly
eccentric systems compared to circular setups. However, no matter what
setup was used the unbound mass was between $\sim$5\% to $\sim$25\%,
typical values for all simulations that use an ideal-gas equation of
state.

\citet{glanz2021b} later used similar tools to simulate \ac{CEE} in a
triple system in the ``circumstellar'' case, where a compact binary
system (corresponding to the inner binary in a hierarchical triple)
orbits in the envelope of a giant. In addition to the construction of
the primary star models with \textsc{mesa} and the mapping and
relaxation in the \textsc{gadget2} computational domain, however, they
modeled the dynamics of the binary companion system inside the
potential of the primary giant star with the N-body code
\textsc{huayno} \citep{pelupessy2012a}, likewise integrated in the
\textsc{amuse} framework of \citet{portegies2009a}. As models for
primary stars, a $2 \, \msun$ and an $8\, \msun$ star were evolved to
the \ac{RG} stage (core masses of $0.36 \, \msun$ and $1.03 \, \msun$,
respectively). For mapping these models into the \ac{SPH} code
\textsc{gadget2}, resolution tests were carried out with \num{100000},
\num{250000}, and \num{500000} \ac{SPH}-particles representing the
envelope of the giant primary star. For the production runs,
\cite{glanz2021b} settled to a resolution of \num{250000}
particles. The masses of the objects in the close binary companion
system were set to $0.6 \, \msun$ and $0.4 \, \msun$ for the lower
mass primary star and to $1\, \msun$ for both objects when interacting
with the more massive primary star. For the fate of this compact
binary system, two cases were observed: a merger inside the envelope
of the primary star and a disruption which---together with the core of
the primary star---leads to chaotic triple dynamics. In the latter
case, typically one of the former close-binary components was found to
eventually merge with the core of the primary star, while the second
component is ejected or continues ``classical'' binary \ac{CEE}. The
effect of a compact binary companion system in the case of
triple-\ac{CE} interaction was a slower inspiral and an increased mass
ejection---attributed to the extraction of energy and momentum from
the compact binary companion. Whereas comparison simulations with
classical binary \ac{CE} setups resulted in ejecting about 10\% of the
envelope mass (a typical value for simulations assuming an ideal-gas
equation of state and not considering ionization effects), the triple
systems unbound 11\% to 28\% of the envelope mass. \citet{glanz2021b}
commented about the imprints of the altered triple dynamics on the
morphology of the ejected material.

Recently, \citet{lau2022a} carried out an \ac{SPH} simulation using
\textsc{phantom} of a $12 \,\msun$ star and companions of $3.0 \,
\msun$ and $1.26 \,\msun$, located at an orbital separation such that
the primary just filled its Roche lobe (for their lower-resolved
simulations with 50\,000 and 500\,000 \ac{SPH} particles) or overfilled the
Roche lobe by 25\% (for their higher-resolution simulations, two
million particles, which would have been unfeasibly long if started
farther apart). These simulations were among the first modern attempts
to model \ac{CE} interaction with a massive primary star. Comparisons
were carried out using three equations of state: an ideal gas, an
ideal gas plus radiation, and a ``full'' equation of state that
included the effects of recombination. The final separation of the
binary was of the order of $30$--$44 \,\rsun$ (for the $3.0 \,\msun$
companion; the other companion sinks below the sum of the softening
lengths of the core and companion, where the interaction is not
reliable). Compared to simulations with a simple ideal-gas equation of
state, including radiation pressure resulted in separations larger by
about 10\% and accounting for recombination energy enlarged the final
separations by another $\sim$20\%. The highest-resolution simulations
with the $3.0 \,\msun$ companion unbound 18\%, 28\% and 60\% of the
envelope for the ideal gas, ideal gas plus radiation, and full
equations of state, respectively (using a criterion that accounts for
mechanical and thermal energy, see
Sects.~\ref{ssec:why_is_CE_ejection_hard_to_achieve} and
\ref{sect:ionization}). The measurement of the unbound mass was taken
at a particular point after the inspiral, not at the end of the
simulation. The additional unbound gas in the case of the a full
equation of state derived from the thermalization of the helium
recombination energy, which, even if radiation transport were
included, was released in the deep layers and is unlikely to
escape. For the full equation of state, three quarters of the envelope
were unbound by the end of the simulation and mass unbinding still
continued at 0.2~\msun~yr$^{-1}$, demonstrating that the envelope was
likely to become completely unbound.

Finally, \citet{lau2022b} studied the effects of hydrogen, helium and
molecular hydrogen recombination energy on the same structure as that
investigated by \citet{lau2022a}. They concluded that helium
recombination energy alone, already unbinds 30\% more gas than an
ideal-gas equation of state and that the delivery of that energy is so
deep in the envelope that it is highly likely that the entire payload
can be thermalized and eventually converted to work. Somewhat more
surprisingly, this is also so for hydrogen recombination energy. Only
a fraction of the hydrogen recombination energy is needed to unbind
the envelope and the delivery depth is such that, once again, we would
conclude that it would be mostly thermalized locally and not
escape. For a 12~\msun\ common envelope it is therefore to be
concluded that recombination energy may well help unbind effectively
the entire envelope.

\subsubsection{Moving-mesh simulations of the common-envelope interaction}
\label{sect:sim_movingmesh}

\begin{figure}
  \includegraphics[width=\textwidth]{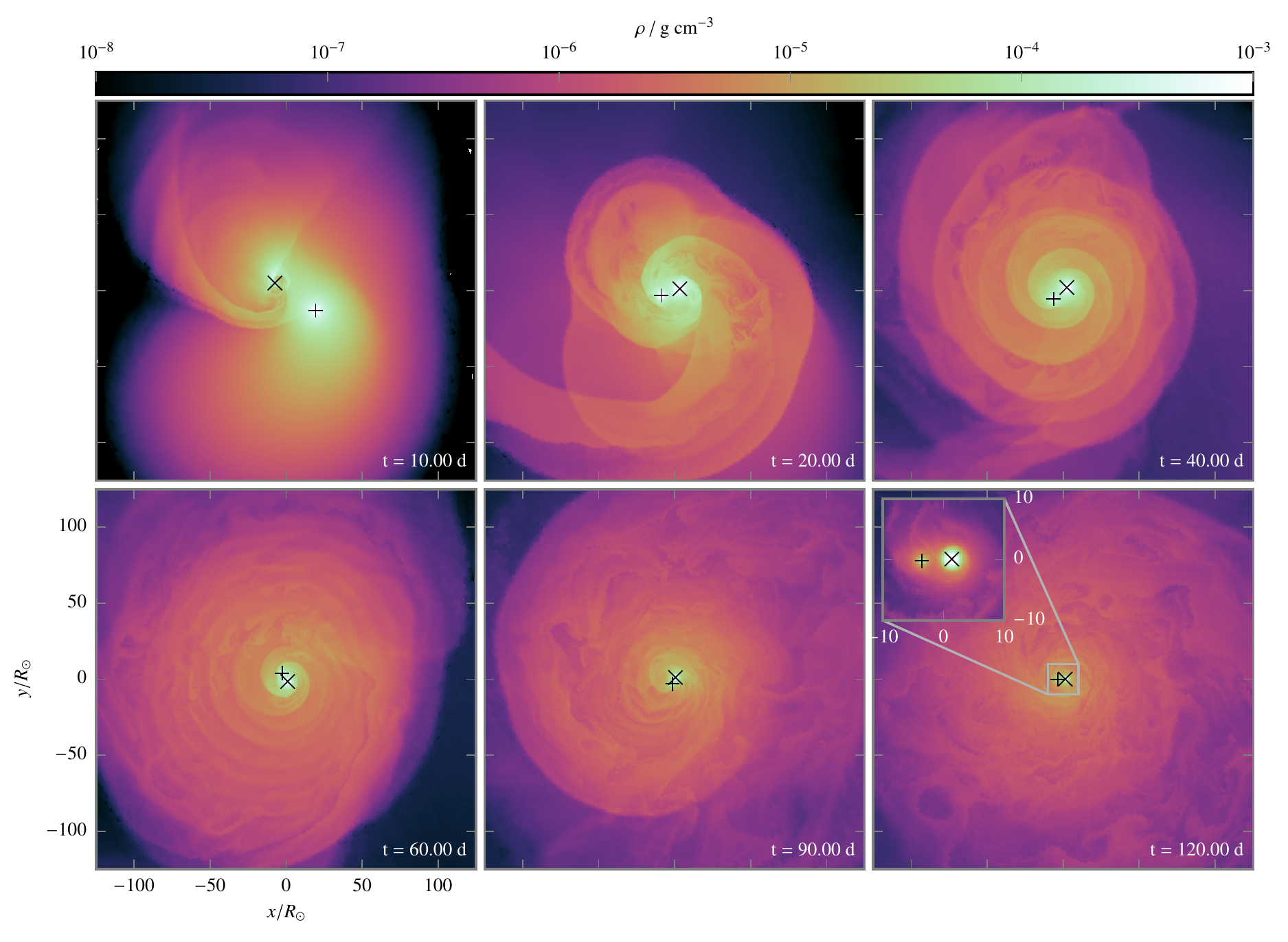}
  \caption{Time series of density cuts through the orbital plane
    during the inspiral simulated with the moving-mesh code
    \textsc{arepo}. The positions of the core of the \ac{RG} primary
    star and the companion are marked by a ``$+$'' and a ``$\times$'',
    respectively. The inset in the last panel shows the central region
    with a diameter of about $20 \, \rsun$. Again, density is
    color-coded and ranges from $10^{-6}$ to $10^{-3}\,
    \mathrm{g}\,\mathrm{cm}^{-3}$. Figure from \citet{ohlmann2016a}.}
  \label{fig:ohlmann2016}
\end{figure}

In Sect.~\ref{sssec:moving_mesh_approaches} we have described the
moving-mesh approach to hydrodynamical simulations. Based on its
implementation in the \textsc{arepo} code of \citet{springel2010a},
\cite{ohlmann2016a} simulated the dynamical inspiral of a $1 \, \msun$
companion into a \ac{RGB} primary star of $2 \, \msun$ with a $0.4 \,
\msun$ core. This rather standard setup was chosen because the scale
problems discussed in Sect.~\ref{sect:challenges_physical} are less
pronounced with a \ac{RG} primary than, e.g., with an \ac{AGB}
star. The simulation is illustrated in Fig.~\ref{fig:ohlmann2016}. It
demonstrated the advantages of using a moving mesh approach. The
method allowed to follow the evolution for many orbits extending over
the initial plunge-in phase into a regime where the orbital separation
of the cores changed only very slightly. This was not only enabled by
the efficiency of the method but also by its excellent conservation of
angular momentum and energy lending credibility to the results even in
late stages of the evolution. As discussed in see
Sect.~\ref{sect:conservation}, this advantage derived from the almost
Lagrangian nature of the scheme. At the same time, the Godunov
approach to compute fluxes in the underlying finite-volume
discretization facilitated high spatial resolution and captured
hydrodynamical instabilities better than \ac{SPH} methods. This
allowed the authors to detect large-scale flow instabilities between
adjacent layers of the spiral shock after about $20$ orbits that were
not noticed in earlier simulations. This effect ultimately lead to
turbulent convection in the envelope. The simulation of
\citet{ohlmann2016a}, however, failed to unbind the envelope. Only
about 8\% of the material was ejected. This failure of the
gravo-hydrodynamic model triggered the inclusion of additional effects
into moving-mesh \ac{CE} simulations.

\begin{figure}
  \includegraphics[width=\textwidth]{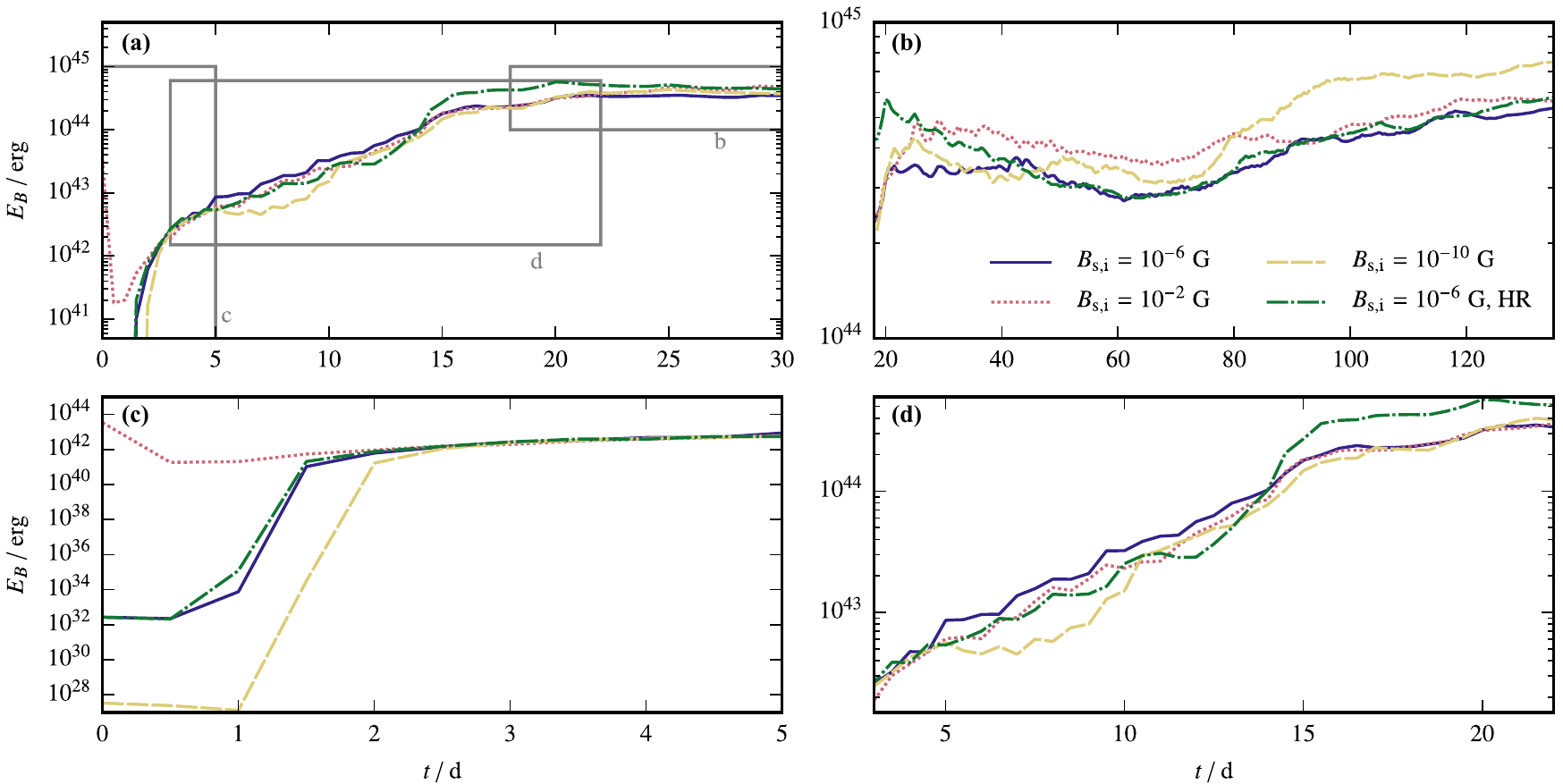}
  \caption{Magnetic field energy in the \ac{CE} simulation of
    \citet{ohlmann2016b}. The three stages of the amplification
    process are marked in panel (a). Panel (c) zooms in on the fast
    amplification phase (i), the slow amplification phase (ii) is
    shown in panel (d) and the phase of saturation (iii) is shown in
    panel (b), where the initial surface fields are given. Figure
    from \cite{ohlmann2016b}}
  \label{fig:magamp}
\end{figure}

The first \ac{MHD} simulation of \ac{CEE} was carried out by
\citet{ohlmann2016a}. It showed that tiny initial magnetic fields are
strongly amplified in the evolution. While previous studies speculated
about dynamo processes in the differentially rotating envelope
\citep{regos1995a}, magnetic field generation in an accretion disk
formed from a tidally disrupted companion \citep{nordhaus2011a}, or in
the outer layers of the degenerate core \citep{wickramasinghe2014a},
the simulation of \cite{ohlmann2016a} allowed to settle the issue by a
direct simulation of the entire \ac{CE} structure.  Three stages of
magnetic field amplification could be distinguished (see
Fig.~\ref{fig:magamp}):
\begin{enumerate}[(i)]
\item In the initial plunge-in of the companion
into the primary's envelope, a fast amplification of the magnetic
field in the forming accretion stream around the companion was
observed.
\item Once this accretion stream had been established, the
  amplification slowed down.
\item Eventually, the magnetic field
saturated and was dispersed over the envelope.
\end{enumerate}
The magnetic field amplification in the accretion flow was found to be
consistent with the action of the magnetorotational instability (MRI,
\citealp{balbus1991a}, \citealp{balbus1995a}). The observed generation
of strong magnetic fields during the dynamical \ac{CE} phase is an
interesting effect and may contribute to shaping the expected
formation of a planetary nebula. However, the resulting fields are too
weak to have a dynamical impact on the \ac{CEE} itself or to alter the
transport of angular momentum significantly. In the setup of
\citet{ohlmann2016b}, the envelope-mass loss exceeded that in non-MHD
simulations with identical parameters only by a few percent.

As discussed before, additional physical processes have to be
considered to achieve envelope ejection. One of the most promising effects for
unbinding envelope material that is ignored in the basic
gravo-(magneto)hydrodynamic \ac{CE} model is the release of ionization
energy when the envelope gas expands and recombines
\citep{nandez2015a, nandez2016a}. This effect has been implemented in
\ac{CE} simulations with moving mesh codes \citep{ohlmann_phd,
  prust2019a}.

\citet{prust2019a} used the moving-mesh hydrodynamics code
\textsc{manga} to simulate \ac{CEE} in setups similar to that of
\citet{ohlmann2016a}. In one of their models, 95\% initial co-rotation
of the primary star's envelope with the companion was assumed while a
second simulation was started with a non-rotating primary
star. Instead of assuming an ideal-gas equation of state, they used
the \textsc{mesa} equation of state \citep{paxton2011a} to include
ionization effects (see Sect.~\ref{sect:eos}). Compared with the
simulations of \citet{ohlmann2016a}, a significant increase in unbound
envelope material ($\gtrsim$ 60\%) was observed.

An interesting new feature in modeling methods was introduced by
\citet{prust2020a}. Moving-mesh hydrodynamics codes allow for moving
boundaries inside the simulated domain \citep{springel2010a}. As a
proof-of-concept, \citet{prust2020a} embedded the companion in a
\ac{CE} simulation in a sphere of $4 \, \rsun$ surrounded by a
reflective moving boundary condition. This paves the way to an
improved modeling of accretion or mass outflow around the companion
object.

\citet{sand2020a} followed \ac{CEE} in systems with a $1.0 \, \msun$
early-\ac{AGB} primary star and companions of different masses. This
was a step towards a computationally more challenging scenario,
because the spatial scale range widens compared with systems involving
\ac{RG} primaries. Moreover, stabilizing the more loosely bound
\ac{AGB} envelope required an increased spatial resolution of the
region around the core of the star. Despite the less tightly bound
\ac{AGB} envelope, \cite{sand2020a} found that envelope ejection in
the basic gravo-hydrodynamic model fails, but the inclusion of
recombination effects render a complete envelope ejection likely---at
least under the assumption of local thermalization of the
recombination energy.

The efficiency of moving mesh approaches to simulate \ac{CEE} has been
exploited to explore the parameter space of models and we give two
examples here. \citet{sand2020a} tested the effect of different ratios
between the mass of an \ac{AGB} primary star and that of the companion
on the mass loss rate and the final orbital separation. Less massive
companions were found to spiral deeper into the \ac{CE} and
\citet{sand2020a} proposed a linear relation between the mass ratio
and the orbital separation at the end of dynamical inspiral. Envelope
ejection was found likely to be complete under the above-mentioned
assumption of local thermalization of recombination
energy. \cite{kramer2020a} addressed the question of whether hot
sub-luminous B-type (sdB) stars can form from \ac{CE} interaction of
primaries close to the tip of the \ac{RG} branch with light companions
as suggested by observations \citep[e.g.][]{geier2011a,
  schaffenroth2014a, schaffenroth2015a}. They conjectured that down to
the regime of brown dwarfs such a formation scenario is likely to
work; whether or not envelope ejection can be triggered by a much less
massive companion, such as a planet, remains to be explored in more
detail although the simulations of \citet{kramer2020a} indicated that
this may be hard.

\begin{figure}
    \centering
    \includegraphics[width=0.75\textwidth]{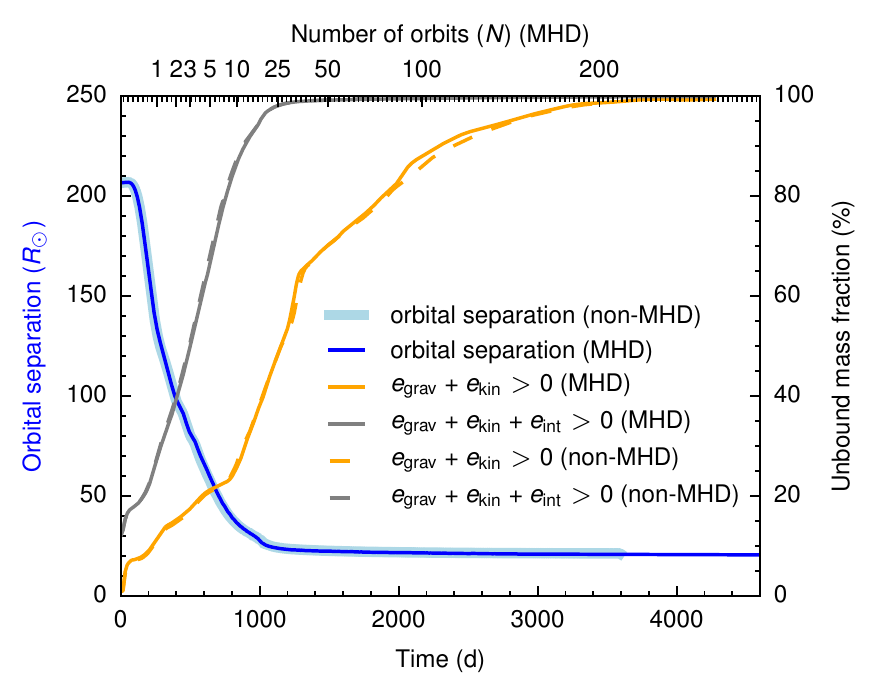}
    \caption{Orbital separation and unbound mass fraction measured
      according to the criteria given in the legend for \ac{CE}
      simulations with a $0.970 \, \msun$ \ac{AGB} primary star and a
      $0.243 \, \msun$ companion. Magnetohydrodynamic simulations are
      labeled with ``MHD'' while ``non-MHD'' indicates pure
      hydrodynamical models. Figure from \cite{ondratschek2022a}.}
    \label{fig:ondratschek_fig2}
\end{figure}

A shortcoming of all recent \ac{CE} simulations that include
ionization effects---whatever method was used for modeling the fluid
dynamics---was that the evolution could only be followed for a rather
short period of time. At termination, not all of the ionization energy
was released by recombination effects. Therefore, while including the
ionization energy in the energy budget to determine unbound material
indicated successful envelope ejection in the case studied by
\citet{sand2020a}, it remained to be proven that recombination indeed
releases this energy so that it is converted into kinetic energy. This
was achieved by \citet{ondratschek2022a}, who were able to follow one
of the systems modeled by \citet{sand2020a} to a stage where the
energy conversion was completed. They found that---under the
assumption of local thermalization of recombination energy and no
radiative losses---complete envelope ejection was indeed reached, see
Fig.~\ref{fig:ondratschek_fig2}. In contrast to \citet{sand2020a},
however, the simulations of \citet{ondratschek2022a} included magnetic
fields. Confirming the results of \citet{ohlmann2016a}, and as
illustrated in Fig.~\ref{fig:ondratschek_fig2}, they found that
although a weak initial field is strongly amplified during plunge-in
of the companion, the magnetic field strength remains dynamically
unimportant for envelope ejection. In the very late stages of \ac{CEE}
a new phenomenon was observed. As the stellar cores came very close,
mass transfer between them initiated and the resulting structure
resembled a contact binary system. This led to a second phase of
strong magnetic field amplification. In the resulting structure, a
high-velocity jet-like outflow was launched. This magnetically-driven
outflow was directed in polar direction of the orbital system and
low-density, high-velocity material propagated along the chimney-like
hole in the ejected \ac{CE} material perpendicular to the orbital
plane. While this material had velocities of ${\sim}\, 10 \,
\mathrm{km} \, \mathrm{s}^{-1}$, the polar outflow reached ${\sim}\,
100 \, \mathrm{km} \, \mathrm{s}^{-1}$. The resulting structures and
the two velocity components of the gas resemble observations of
bipolar \acp{PN} (see Sect.~\ref{sect:pne}).

The first moving-mesh \ac{3D} magneto-hydrodynamic simulations of
\ac{CEE} of objects with masses typical for neutron stars or black
holes with a massive star were presented by \cite{moreno2022a}. For
the primary star model, a $10 \, \msun$ star was evolved with
\textsc{mesa} to the red supergiant stage. The choice of this mass was
motivated by the numerical challenges introduced by massive stars when
represented in \ac{CE} simulations. Therefore, a massive star at the
lower end of the range of masses still capable of producing a neutron
star at the end of its evolution was chosen. \cite{moreno2022a}
demonstrated that with the setup procedure of \cite{ohlmann2017a} it
is possible to construct a sufficiently stable model for a primary
star in \ac{CE} simulations. \ac{CEE} was followed with the
\textsc{arepo} code and the stellar cores were represented by point
masses. Assuming local thermalization of the released recombination
energy, nearly complete envelope ejection seemed possible for the
considered systems. The orbital separation at the end of the
simulations, however, was too large to allow for a gravitational-wave
induced merger of the eventually resulting double compact
system. \citet{moreno2022a} argued that for a favorable alignment of
the expected supernova kicks, some of the systems may still qualify as
progenitors of gravitational-wave emitting mergers of double neutron
stars.

\section{Conclusions and future perspectives}
\label{sec:future_perspective}

\subsection{Where do we stand?}
\label{sect:current}

\ac{3D} hydrodynamic simulations of \ac{CEE} have seen a rapid
improvement over the past decade. But still, these simulations are
insufficient to answer fundamental questions and thus they only mark
the beginning of a deeper physical understanding of \ac{CE}
interaction.

What can potentially be learned from \ac{3D} hydrodynamic \ac{CE}
simulations? For given initial parameters of the pre-\ac{CE} binary
systems, the ultimate questions to \ac{CE} modeling are:
\begin{enumerate}[(1)]
  \item How much of the envelope gas
    is ejected?
  \item What is the final separation of the core binary in
    case it has not merged?
\end{enumerate}
If the stellar cores merge or the companion is destroyed, the
long-term evolution of the remnant object poses an additional
question. Certainly, \ac{3D} hydrodynamic approaches are a
prerequisite to finding answers, but can we settle the questions with
the currently available simulations?

Regarding Question (1) the situation seems rather optimistic. As
discussed in Sects.~\ref{ssec:why_is_CE_ejection_hard_to_achieve} and
\ref{sect:simulations}, when including the ionization energy released
in recombination processes and assuming that it thermalizes locally,
envelope ejection seems plausible for many of the studied systems. It
also seems likely, however, that in certain regions of the parameter
space, envelope ejection still fails. Even if all of the recombination
energy can be used to support envelope ejection, there may be cases
where this energy is still insufficient to unbind the entire
\ac{CE}---especially for high-mass primary stars
\citep{kruckow2016a}. Moreover, \citet{sabach2017a},
\citet{grichener2018a} and \citet{soker2018a} questioned the validity
of the assumption of local thermalization and argued that instead
recombination energy may be transported away by convection
\citep{wilson2019a, wilson2020a} or radiation. However,
\cite{ivanova2018a} estimated that the amount of hydrogen
recombination energy lost by these processes was negligible.
Convection is in principle part of the basic gravo-hydrodynamic model
of \ac{CEE}. The question is whether it is sufficiently resolved in
the simulations. At least moving-mesh hydrodynamics schemes
\citep{ohlmann2016a, ohlmann_phd, prust2019a, sand2020a} seem capable
of reproducing convective flows. However, radiation transport is
usually not accounted for. \citet{sand2020a} estimated that in their
setup, most of the ionized material remains well inside the
photosphere for most parts of the \ac{CEE}, but ultimately the
question of how much of the recombination energy can be used to eject
the envelope material has to be answered by simulations that model
radiation transfer. In summary, the answer to Question (1) is
challenged by potentially missing physical effects and not primarily
by numerical problems or deficiencies in the model setups, although
\citet{reichardt2020a} show that a smaller softening length of the
potentials around the stellar cores alters the rate at which mass is
unbound.

The effects determining the orbital evolution of the system during the
rapid inspiral phase---Phase (ii) of our classification, but not in
the later \ac{CEE} in Phase (iii)---are also implemented in current
models. Some part of the pre-\ac{CE} evolution, the actual plunge-in phase
of \ac{CEE} and several hundred orbits of the inner core binary
following it are accessible to \ac{3D} hydrodynamic
simulations. However, we suspect that the period of time immediately
after these \ac{3D} simulations---perhaps up to the thermal
timescale---may be critical to the parameters characterizing the
binary (or the merged object) later. \citet{iaconi2017a} summarized
final orbital separations obtained global hydrodynamic simulations of
\ac{CEE} and found that they are commonly too wide compared with
observations of post-\ac{CE} binary systems. This points to additional
physical processes (see
Sect.~\ref{ssec:why_is_CE_ejection_hard_to_achieve}) that follow the
simulated dynamical inspiral of the companion into the primary star
and the first few hundred orbits, i.e.\ in Phase (iii) of our
classification.  Bound gas in the outer parts of the envelope may
re-contract on thermal timescales leading to an increase of the drag
force, further inspiral, and more unbound envelope material. This
could proceed in cycles or as a slow process that is regulated by the
energy transport rate through the remaining envelope material (see
Sect.~\ref{ssec:why_is_CE_ejection_hard_to_achieve}). On the other
hand, returning envelope gas has high specific angular momentum, which
could change the orbital separations in Phase (iii) drastically
\citep{kuruwita2016a}.  It is also possible that once a certain
orbital separation is reached, one or both of the cores start to
overflow their Roche lobes. These additional mass transfer episodes
would again alter the orbital separations. It therefore seems likely
that the final orbital separation is not only set by the dynamics in
Phase (ii) but also affected by Phase (iii) of our classification in
Sect.~\ref{sect:phases}. But--\emph{hic sunt dracones}---this is
territory that is difficult or impossible to access with
multidimensional hydrodynamic simulations (see
Sect.~\ref{ssec:why_is_CE_ejection_hard_to_achieve}).  In
Sect.~\ref{ssec:interfacing_3d_global_simulations_with_1d_simulations_of_the_ce_remnant},
we discuss how \ac{3D} hydrodynamic simulations need to be interfaced with
simulations able to model longer timescales, including the future
evolution of the star(s). As for answering Question (1), we again
encounter the problem that not all physics deciding about Question (2)
may be included in \ac{3D} hydrodynamic \ac{CE} simulation.

Before extending our models, however, we need to have confidence in
the parameters at the end of the \ac{3D} hydrodynamic simulations.  It
has to be determined why the plunge-in phase ceases in \ac{3D} simulations,
giving rise the the subsequent phase that takes place on longer
timescales. Regardless of whether or not the separation at the end of
Phase (ii) is the actual final separation, we need to understand
whether the slow-down of the inspiral, temporary as it might be, is in
fact physical. A simplified view is to assume complete envelope
ejection so that there is no matter left behind and the drag forces
acting on the cores cease. But the \ac{3D} hydrodynamic simulations
discussed in Sect.~\ref{sect:sim_global} show a different picture
\citep[e.g.][]{iaconi2018a}: Envelope ejection---if
successful---happens rather late in the evolution, when the expansion
of the envelope leads to the release of recombination energy that
ultimately unbinds the material \citep{nandez2015a}. At this stage,
however, the orbital evolution has already slowed down. But still, the
orbital volume is usually not completely devoid of gas, though the
density has decreased. It is therefore not the evacuation of the orbit
that slackens the inspiral. As discussed in
Sect.~\ref{ssec:why_is_CE_ejection_hard_to_achieve}, other effects may
contribute to reduce the drag: towards the end of the inspiral, the
relative motion between cores and gas becomes very subsonic, for which
the formalisms of \citet{bondi1952a}, \citet{ostriker1999a}, and
\citet{kim2007a} predict a drastic reduction of the drag force.
Ultimately, co-rotation may be re-established between the core binary
and the surrounding envelope gas, particularly for the more massive
companions. This would decrease the drag force dramatically.  But for
tighter orbits, stronger drag is needed to change the orbit
appreciably: As the cores approach each other, their mutual
gravitational attraction becomes stronger. This force is what sets the
orbital separation and only a drag force on the same order of
magnitude would lead to a change in orbital separation. This becomes
ever more unlikely in the late stages of the evolution.

While these reasons plausibly justify the end of the inspiral, there
is still the concern that the representation of the stellar cores as
point particles and the softening of the gravitational potentials
impact the simulated orbital separation. We could assume that, while
the core region is undisturbed, the simulation of the central part of
the star is reasonable. Even so, we do worry that once the envelope
has expanded and the core and companion are near one another, the
structure of the flows may be non-physical. The mass adopted for the
particle representing the core of the primary star is somewhat
arbitrary. Even with the rather controlled approach suggested by
\citet{ohlmann2017a}, a certain cut radius has to be defined, that is
often chosen to be equal to the gravitational softening length of the
core particle. Towards the end of the simulations, the stellar cores
may have come so close that the softened regions overlap and this may
compound any specific problem. Moreover, the physical modeling of
cores as point masses raises other concerns. For example, they
accumulate material in their vicinity which travels with the point
masses. This material could in reality be accreted onto the cores
under the release of energy.  Certainly, the effects that these
choices have on the final orbital separation call for further study.

\subsection{Missing physical effects}
\label{ssec:missing_phisical_effects}

As discussed in Sect.~\ref{sect:gravohydro}, simulations carried out
thus far show that the \ac{CE} can only be ejected if recombination energy
is utilized to do mechanical work, i.e. expand the envelope as opposed
to being convected and radiated away.  Thus, one of the most urgent
questions to \ac{CE} modeling is to test whether this assumption is
justified. Estimates based on \ac{1D} stellar structure models
\citep{ivanova2018a} and on the optical depth at which recombination
energy is released in \ac{3D} hydrodynamic \ac{CEE} simulations
\citep{reichardt2020a, sand2020a, lau2022a} seem to support this
mechanism to aid \ac{CE} ejection. But for a solid conclusion,
radiation transport in the envelope gas has to be modeled. The other
mechanism to drain energy from its location of release by
recombination---convection---is part of the fundamental
gravo-hydrodynamic model and therefore readily included provided a
careful preparation of the setup. While its proper representation is a
challenge to the numerical methods and the spatical resolution reached
in the simulations, radiation is an additional physical process that
has to be modeled explicitly. Introducing a radiation transfer model
into \ac{CE} simulations would not only help to determine the
contribution to envelope ejection by recombination effects, but it
would also allow a more accurate localization of the photosphere and
therefore improve predictions of optical observables from \ac{CE}
events. Moreover, for simulations of \ac{CEE} with very massive
primaries, the setup requires to include radiation in order to
properly represent the structure of these stars \citep{ricker2019a}.

The energy transport by radiation depends critically on the opacity of
the material it passes through. Here, effects can play a role that
extend beyond the ionization structure which is usually followed in
simulations accounting for recombination energy release. Dust
potentially forming in outer layers of the ejecta
(\citealp{clayton2017a}, \citealp{glanz2018a},
\citealp{reichardt2020a}, \citealp{iaconi2019a}, and
\citealp{iaconi2020a}) would increase the opacities
dramatically. Therefore, models of dust formation are another
important ingredient to reaching a comprehensive physical modeling
basis for \ac{CE} simulations.

Magnetic fields have been introduced into \ac{CE} simulations, where
strong field amplification is observed \citep{ohlmann2016a}. Their
dynamical effect remains negligible, but they potentially contribute
to the shaping of planetary nebulae \citep[e.g.][]{garcia2020a,
  ondratschek2022a}. It is also possible that magnetic activity favors
dust formation as has been suggested for \ac{AGB} stars
\citep{soker1998b, rapoport2021a}.

Additional physical effects concern the vicinity of the companion and
the primary core, as well as their structure.  Related processes
include accretion to or perhaps mass loss from the bound cores,
accumulation of material in their vicinity, and mass transfer episodes
between the remaining cores following the initial dynamical \ac{CE}
phase. Some low-mass companions such as planets may dissolve in the
envelope gas \citep[e.g.,][]{staff2016a}. If the companion star is
very compact, for example a neutron star, densities close to it may
reach values where nuclear burning and neutrino processes become
important; in the case of black holes potentially even effects of
general relativity.

\subsection{Unresolved numerical problems}
\label{sect:unresolved_num}

The future challenge to improvements of the numerical approaches is
the same as today: dealing with the scale problems in both time and
space will persist as the main difficulty for \ac{CE} simulations. The
timescale challenge applies in particular to Phases (i) and (iii) of
our classification. As discussed in Sect.~\ref{sect:multiscale},
processes in these phases may take place on longer timescales than
those accessible to \ac{3D} hydrodynamic simulations. This may pertain
to some restructuring of gas remaining bound to the cores that
involves radiation processes and acts on the thermal timescale, but it
may also be caused by minute changes on dynamical timescales whose
cumulative effects after very many orbits may determine, e.g., the
onset of \ac{CEE} or the final orbital separation of the remnant core
binary.

The spatial scale problem mainly concerns the region around the cores
and the cores themselves.  While for special progenitor systems a
(marginal) resolution of companions may be possible, this problem
prevents a direct modeling of the detailed effects in the vicinity of
the cores.  Yet, for reaching the next level of numerical modeling,
the details of the evolution of the cores and their interaction with
the immediate surroundings are of primary interest, see
Sect.~\ref{ssec:missing_phisical_effects}. These processes, however,
are out of reach for common numerical techniques and available
computational resources (see
Sect.~\ref{sect:comp_feasible}). Resolving the core of the primary
spatially may become possible if the timescale problem is
alleviated. For this, a parallelization in the time domain, i.e.\ in
the time level hierarchy instead of a spatial domain decomposition, is
a potential future direction. For the time being, however, the barrier
towards resolving the cores cannot be overcome, and thus the advantage
gained from a modest increase in spatial resolution of the envelope
remains limited. The focus of current simulations lies instead on
determining the main ``global'' quantities, such as orbital separation
between core and companion and mass ejection. For these, numerical
convergence can typically be established at a moderate number of
discretization elements (some $10^6$ grid cells or particles). There
is, however, the prospect of using subgrid-scale models to capture at
least some of the missing effects. Accretion, for instance, can be
represented by sink particles similar to those used in star formation
simulations \citep[e.g.][]{krumholz2004a, chamandy2018a}.

In current and in future models, establishing the convergence of the
simulation results is an important issue. While for the setup,
hydrostatic equilibrium provides a well-understood reference to
compare with, it is less clear what resolution is needed in later
phases of the evolution at different locations. Due to the exploding
computational costs, a global increase of the spatial resolution seems
little promising and may not be necessary. Adaptive mesh refinement
techniques can concentrate resolution to regions where it is needed
for reaching convergence. The task in the near and mid-term future is
to figure out what regions matter and what resolution needs to be
reached in them for convergence.

Another potential problem for \ac{CE} simulations are conservation
properties. In principle, the applied finite-volume and \ac{SPH}
schemes conserve energy. In addition, \ac{SPH} conserves angular
momentum by construction. Moving-mesh codes are nearly-Lagrangian and
inherit this property to good accuracy For Eulerian codes, \ac{AMR}
provides the required resolution for reaching acceptable conservation
of angular momentum. Still, simulations of global \ac{CEE} require to
follow hundreds of orbits and errors may accumulate.

A fundamental problem for all numerical approaches arises from the
coupling between hydrodynamics and gravity as an external force which
enters the equations as a source term. Even with the approximate
schemes discussed in Sect.~\ref{sect:gravity} gravity itself can be
calculated to any desired precision, so it does not by itself
constitute an unavoidable error source. But it is discretized in a
different approach than the hydrodynamic quantities and tiny
mismatches lead to accumulating energy errors. After a large number of
orbits this error may be on the same order of magnitude as the binding
energy of the non-expelled gas and it is therefore difficult to judge
the success of envelope ejection. As mentioned in
Sect.~\ref{sect:challenges_setups}, in setups close to hydrostatic
equilibrium, well-balancing methods improve on this issue
\citep[e.g.][]{edelmann2021a}, but the lack of a clear reference state
renders such approaches less promising in the context of the dynamical
core--envelop interaction.

\subsection{Interfacing 3D global simulations with 1D simulations of the common-envelope remnant}
\label{ssec:interfacing_3d_global_simulations_with_1d_simulations_of_the_ce_remnant}

A promising avenue to understand what happens \emph{after} the
dynamical \ac{CE} inspiral phase is interfacing \ac{3D} hydrodynamic
simulations to \ac{1D} hydrostatic or hydrodynamic time-implicit codes
that can evolve the remnant much further in time.  This is not an easy
project for a number of reasons.  The first is that the post-\ac{CE} object
may be non-spherical and deciding how to map it into a \ac{1D}
configuration is non-trivial. The second problem is that the dynamic
and thermodynamic structure of the object may not be a good fit for
the \ac{1D} code. Nonetheless, the work of \citet{munson2021a}
provides some guidance in this direction.

\begin{figure}
    \centering
    \includegraphics[width=0.65\textwidth]{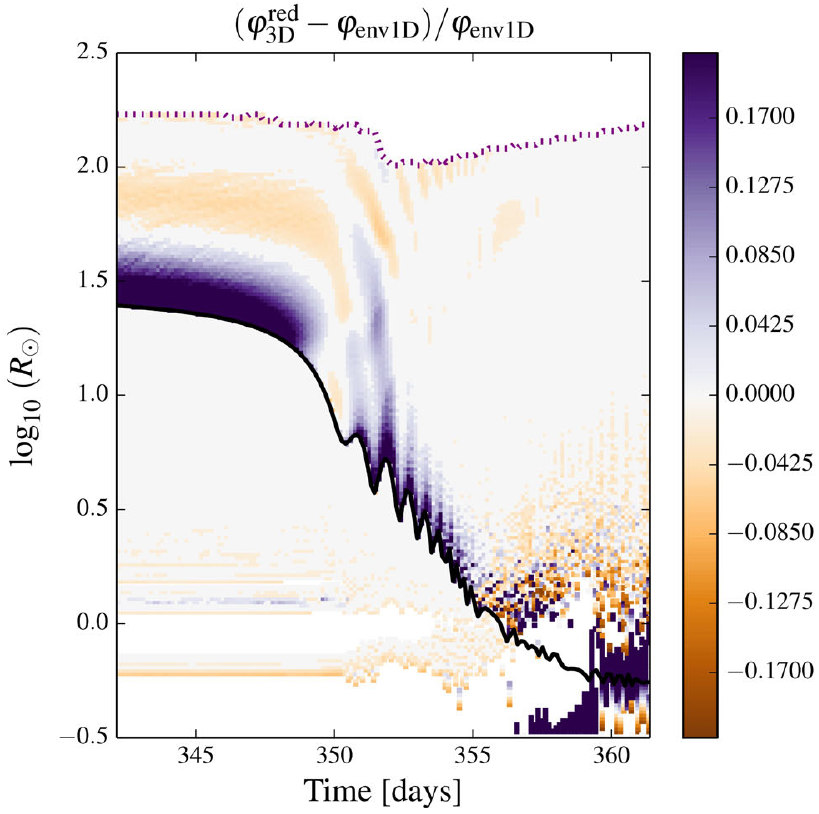}
    \caption{Comparison of the gravitational potentials in the \ac{1D}
      ($\varphi_{\mathrm{env1D}}$) and \ac{3D}
      ($\varphi^{\mathrm{red}}_{\mathrm{3D}}$) approaches. The
      difference is color-coded and shown as a function of radius and
      time. The solid black line indicates the location of the
      companion and the dashed purple line marks the surface of the
      envelope intended as the outer boundary of bound gas. Figure
      adapted from \citet{ivanova2016a}.}
    \label{fig:IvanovaNandez16-Fig4b}
\end{figure}

In a thoughtful paper, \citet{ivanova2016a} compared a \ac{1D} model of the
inspiral with a \ac{3D} model. While the goal was to determine how to
interface a \ac{3D} inspiral model to a \ac{1D}, longer-timescale post-inspiral
model, the paper actually also discusses \ac{1D} simulations of the {\it
  entire} common envelope. They encountered a number of problems which
they catalogued, together with recommendations for further efforts of
this type. For example, the structure of the envelope during inspiral
is quite different, as shown in Fig.~\ref{fig:IvanovaNandez16-Fig4b},
where the difference between the potential of the \ac{1D} and \ac{3D} inspirals
is plotted as a function of time and location. They recommend that if
the fast inspiral phase is to be modelled in \ac{1D}, kinetic energy
should be injected instead of heat into the \ac{1D} model envelope to
simulate the deposition of energy and angular momentum due to orbital
decay. This is preferable because injecting released orbital energy
into heat alters which layers of material release ionization energy in
recombination processes, which has a strong effect on the envelope
dynamics. An important point made by \citet{ivanova2016a} is that
while modelling the inspiral in \ac{1D} is difficult at best,
modelling the inspiral in \ac{3D} and then mapping to \ac{1D} for the
post-inspiral phase also presents some serious problems when trying to
interface the two simulations. This said, interfacing \ac{3D} and
\ac{1D} models is an active research area with some success for the
cases of main-sequence star mergers
\citep{schneider2019a,schneider2020a} and mergers between white dwarfs
\citep{munson2021a}. While the future will no doubt see sophisticated
\ac{3D} hydro simulations, it is likely that short- and
intermediate-term progress will have to rely on the ingenuity of the
researchers in combining different computational tools.

If the binary is definitely going to survive and the envelope is
mostly or fully ejected by the end of the simulation, then mapping the
ejected envelope gas into a \ac{1D} code may poses additional
challenges.  Disks may form with certain properties that may
contribute to change the orbital parameters towards the ``final''
orbital separation. It would be difficult to have a thorough
self-consistent simulation of this phase, but it is possible that its
parts may be addressed separately. For example if the parameters of
the disk could be extracted, then it might be possible to determine
its impact on the inner binary \citep{kashi2011a}.

\subsection{Deriving parametric CE prescriptions from 3D simulations}
\label{ssec:extracting_1d_parametrization_from_multiD_simulations}

A practical goal of simulating the \ac{CE} interaction is to derive simple
analytic parametrizations that can be used in population synthesis
studies.  Classical approaches are the $\alpha_\mathrm{CE}$ and the
$\gamma$ formalisms discussed in
Sect.~\ref{sec:parametrized_models}. The values of the involved
parameters, or---in more sophisticated approaches---the functional
dependence of their values on stellar and system characteristics, has
been investigated extensively with a range of techniques. How can
\ac{3D} hydrodynamic simulations improve on this?

If we are to use global \ac{3D} hydrodynamic simulations to derive
parametric descriptions of successful \ac{CEE}, three requirements must be met:
\begin{enumerate}[(1)]
  \item The \ac{3D} hydrodynamic simulations must be converged and
    include all effects that determine the quantities to be
    parametrized.
  \item Arguably, the envelope must be mostly or even fully unbound.
  \item A suitable parametric model has to be identified that is able
    to capture the functional dependencies that determine the
    quantities.
\end{enumerate}
None of these issues is resolved yet. Determining whether convergence
has been reached is difficult. Even if the envelope was fully unbound
and requirement (2) is met, one wonders whether the recombination
energy, which ultimately unbound the envelope should be fully
thermalized, or whether some of it should have escaped into
space. Fulfilling requirement (1) is even more challenging. Changes to
the orbital parameters take place in the post-\ac{CE} phase on longer
timescales and modeling such effects is outside of the capabilities of
3D time-explicit hydrodynamic approaches.  Also requirement (3) poses
questions: In the context of the parametrization given by
Eqs.~(\ref{eq:alpha_CE})--(\ref{eq:ebind_para}), should the orbital
energy term on the right hand side of Eq.~(\ref{eq:alpha_CE}) be
accompanied by a recombination energy term, and should there be more
than one efficiency parameter to accompany these two energy sources?
Or should the recombination energy, together with thermal energy be
included in the envelope's internal energy, and used on the other side
of Eq.~(\ref{eq:alpha_CE}), that quantifies the binding energy of the
envelope? In any choice and consideration, there are many
approximations and caveats.

An early attempt to derive the parameter $\alpha_\mathrm{CE}$ defined
in Eq.~(\ref{eq:alpha_CE}) from \ac{3D} hydrodynamic simulations of
\ac{CEE} was made by \citet{sandquist1998a, sandquist2000a}, despite
the fact that their models failed to unbind the envelope and therefore
violated requirement (2). In order to circumvent this problem, they
only considered the \emph{change} in binding energy defined as the
initial binding energy of the giant minus the value at the end of
their simulation. In doing so, they presumably included potential and
thermal energy in their binding energy calculation.

\citet{ivanova2016a} carried out a similar calculation using some of
the first \ac{3D} global simulations to successfully eject the
envelope and meeting requirement (2). They derived values for $\alpha
\lambda$ according to Eqs.~(\ref{eq:alpha_CE}) and
(\ref{eq:ebind_para}) from their simulations of \ac{CEE} with \ac{RGB}
primary stars. \cite{sand2020a} carried out a similar analysis, but
they differentiated the two cases where only potential energy is
included in the value of the binding energy, and the case where the
full internal energy (including recombination energy) is accounted
for. This way, the chose an agnostic approach with respect to
requirement (3) and calculated the values of $\lambda$ and $\alpha$
separately for the two cases. When only considering gravitational
potential energy in the binding energy of the envelope, they found
$\ace >1$ which---rather unsurprisingly---implies more than the
orbital energy has been used to unbind the envelope.

\citet{lau2022a} carried out an additional exercise. They took
advantage of the fact that they had carried out a \ac{CE} simulation
between a $12 \,\msun$ red supergiant and a companion using three
different equations of state: ideal gas, ideal gas plus radiation and
an equation of state that additionally accounts for ionization
energy. From these they could calculate three different $\alpha$, one
to quantify the fraction of orbital energy used to unbind the
envelope, the second to quantify the fraction of thermal energy used
to unbind the envelope and the last to quantify the fraction of
recombination energy used to unbind the envelope. It is still to be
seen whether these considerations have any universal significance.

\citet{iaconi2019a} attempted to reconcile information from all
simulations known at the time with observation of single-degenerate
binaries, just to demonstrate how very complex the topic is.  As
discussed in Sect.~\ref{sect:current}, the uncertainty of the final
orbital separation determined from \ac{3D} hydrodynamic \ac{CE}
simulations hampers these efforts.

\subsection{Constraining CE simulations with observations}
\label{sect:observational_constraints}

How can the results of \ac{3D} hydrodynamic simulations be connected
to astronomical observations? What can we learn from such a
connection? Is it possible to constrain models of \ac{CEE} from
observations? To discuss these questions, we distinguish between
observations of stars, binary systems, and \acp{PN}, whose parameters
can directly or indirectly be compared with \ac{CE} simulations, and
observations of transient events that are associated with systems in
the act of undergoing a \ac{CE} interactions.

\subsubsection{Observations of stars, binaries and PNe that constrain CE simulations}
\label{sect:pne}

We know of a number of close binary classes that must be post-\ac{CE}
systems because one or both of the objects in the binary must have
been much larger in the past than the orbit is today. The simplest
objects of this type are those where the companion is a low mass main
sequence star, while the primary is usually a white dwarf remnant of
the giant star or a helium-burning, horizontal branch star. There are
two types of giant producing the observed remnants: \ac{RGB} and \ac{AGB}
stars. In the first case, the post-\ac{CE} remnant tends to be a
core-helium burning, horizontal-branch star, almost always classified
as a subdwarf B or O (sdB or sdO) or it can be a helium white dwarf,
often of lower mass. In the second case the remnant is a carbon-oxygen
white dwarf. In most cases, it is not known whether the primary of the
CE event was an \ac{RGB} or an \ac{AGB} star; however, if the binary
is in the middle of a planetary nebula, we can be \emph{almost}
certain that the giant was recently on the \ac{AGB} \citep[there is
  only a handful of planetary nebulae for which the central star can
  be shown to be a post-\ac{RGB} star; see][]{jones2016c}.
 
These post-\ac{CE} binaries are used to constrain the post-\ac{CE} separation of
the remnant core system if we assume that today's separation is the
same as that at the end of the \ac{CE} interaction. In some cases one
can compensate for the fact that the binary may have suffered
additional orbital reduction due to magnetic breaking or gravitational
wave radiation \citep[e.g.,][]{schreiber2003a}. If the mass of the
primary is known, one can derive the mass and the radius of the giant
at the beginning of the \ac{CE} phase, with some assumptions and
moderate uncertainties \citep[e.g.,][]{zorotovic2010a,
  demarco2011a}. This reconstruction allows one to estimate the
parameters of the pre-\ac{CE} system, which, together with the observed
post-\ac{CE} systems parameters, gives a value for $\alpha_\mathrm{CE}$ in
the energy formalism described in
Sect.~\ref{sec:parametrized_models}. In practice, this method is
fraught with uncertainties and assumptions, and there is a rich
literature of what can and what cannot be derived from it
\citep[see][for a review]{iaconi2019a}.

Another type of post-\ac{CE} binary that recently gained increased
attention are sdB or white dwarf stars orbited by very low mass
companions, most often brown dwarfs, but possibly even massive
planets. There are approximately a dozen systems with companions in the
mass range of $0.027$--$0.07 \,\msun$
\citep{schaffenroth2014a,casewell2018a}. The primary question arising
from these observations is how such low mass companions can have
escaped merging with the primary in the \ac{CE} interaction. As
discussed in Sect.~\ref{sect:sim_movingmesh}, \citet{kramer2020a}
performed \ac{3D} hydrodynamic \ac{CE} simulations of such
systems. They concluded that if the companion is more massive than
${\sim} 0.05 \,\msun$, it can successfully eject the envelope of a
primary star with a main sequence mass of $1 \,\msun$ evolved to the
tip of its \ac{RGB}. If the mass of the companion is between $0.03$
and $0.05 ,\msun$, partial ejection may take place, while below $0.03
\, \msun$, only a small fraction of the envelope is ejected. This is
consistent with the observations of sdB stars, but the final orbital
separations measured at the end of the simulations of
\citet{kramer2020a} are too large compared with the observations.
  
\begin{figure}
  \centering
    \includegraphics[width=0.99\textwidth]{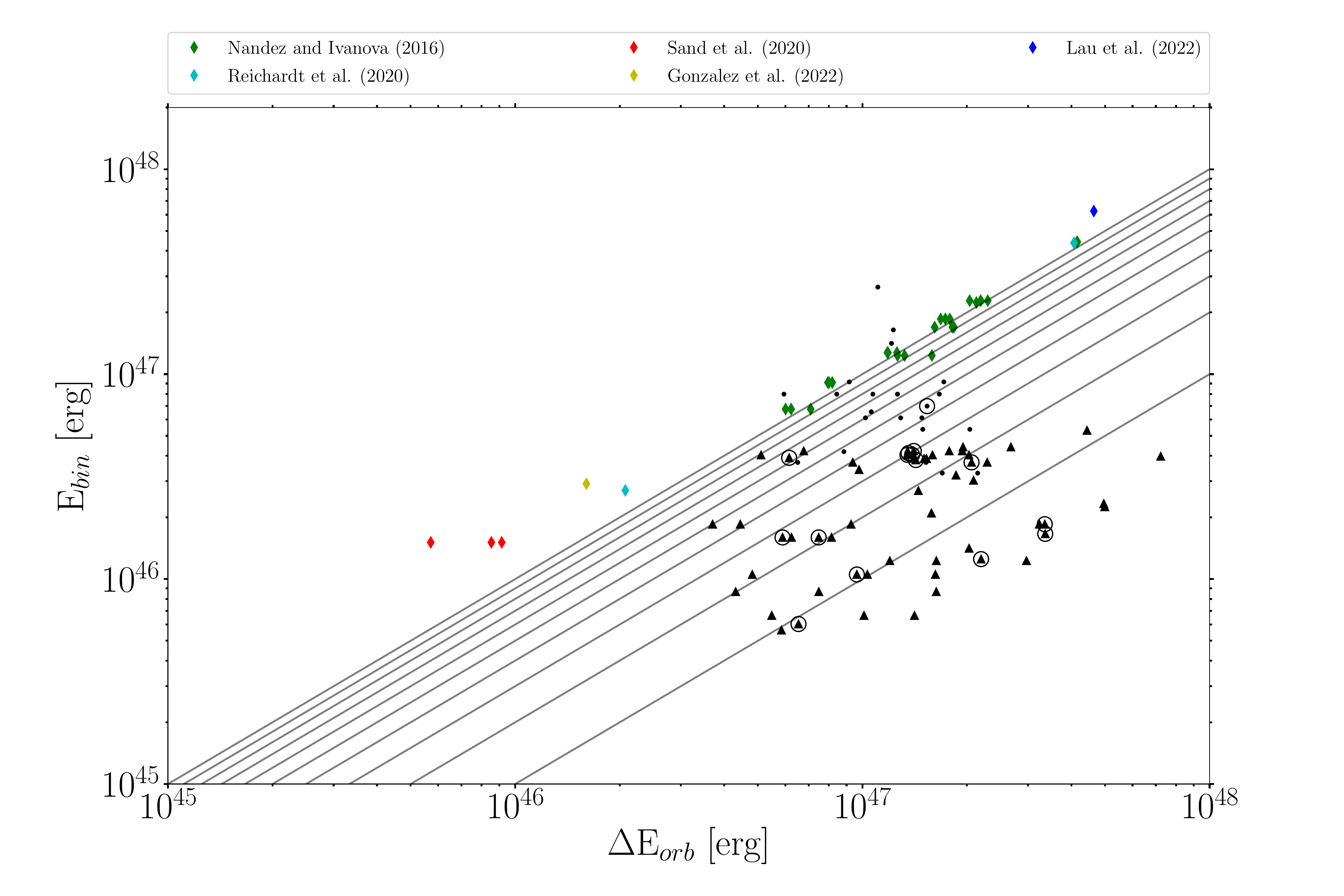}
    \caption{A comparison of simulations (coloured symbols) and
      observations (black symbols; dots: post-\ac{RGB} post-\ac{CE} binaries;
      triangles: post-\ac{AGB} post-\ac{CE} binaries; with circles: a planetary
      nebula surrounds the binary), in the delivered orbital energy
      vs. binding energy plane. The binding energy is that of the
      giant star at the time of the \ac{CE} interaction and includes
      potential and thermal energy. The solid lines denote $\alpha$
      values from 1 to 0.1 (top to bottom).}
    \label{fig:alpha}
\end{figure}

\citet{iaconi2019a} compared a number of \ac{CE} simulations to
observations of post-\ac{CE} binaries such as the ones just described,
where the companion is a main sequence star and the primary is either
an sdO, sdB (O and B, post-\ac{RGB}, core helium burning stars) or a white
dwarf. They analyzed post-\ac{CE} separations as a function of pre-\ac{CE}
system parameters, such as the mass ratio between primary star and
companion and the binding energy of the envelope of the pre-\ac{CE}
giant. Unfortunately most of the simulations available at that time
did not unbind the envelope, making the simulated final separations
upper limits. In Fig.~\ref{fig:alpha} we show an updated figure
mimicking what was presented by \citet{iaconi2019a} in their
Figure~3. Here we only use simulations that eject the entire \ac{CE}. The simulations by \citet{sand2020a} and by
\citet{gonzalez2022a} used an \ac{AGB} primary, those of \citet{lau2022a}
used a massive red supergiant, while the other adopted low mass \ac{RGB}
stars.

Figure~\ref{fig:alpha} shows that the $\alpha$ parameters derived from simulations are approximately
unity (most simulations are even in the $\alpha > 1$ region,
throwing some doubt on the virtue of the formalism itself; for a
discussion see \citealp{lau2022a}).  Moreover, as also observed by
\citet{iaconi2019a}, there seems to be a systematic difference between
post-\ac{RGB} and post-\ac{AGB} observations, but the simulations, in particular
those with \ac{AGB} primaries, have far wider separations than the
observations. With the exception of the \ac{RGB} simulations of
\citet{nandez2016a} that fall on the same region as the post-\ac{RGB}
observations, the post-\ac{AGB}, post-\ac{CE} binaries with their extremely
compact orbits, are not reproduced by simulations.
 
Other post-\ac{CE} systems, such as close white dwarf binaries---sometimes
referred to as \emph{double degenerate systems}, or high- and low-mass
X-ray binaries are too complex to provide useful constraints: Double
degenerate systems went through two episodes of mass transfer, that
could have been two common envelopes. \citet{nelemans2005b} wrote
extensively about the ambiguity of the evolutionary path in these
cases.  X-ray binaries consist of a low- or high-mass giant that
transfers mass to a neutron star or a black hole, hence an object that
derives from a a core-collapse supernova whose effect on the binary is
very difficult to predict. These systems must have gone through a \ac{CE}
but the uncertainties in their evolution leave us with constraints that
are even weaker than those obtained from lower-mass systems discussed
above.

An alternative approach to constraining \ac{CE} simulations is given by the
analysis of the \acp{PN} around post-\ac{CE} binaries. At least one in five
\acp{PN} is formed in \ac{CE} interaction; possibly quite a bit more
\citep{miszalski2009a, jacoby2021a}. The shape of the nebula is likely
not only carrying the blue print of the interaction, but also
guarantees that the \ac{CE} interaction took place in the immediate
past of the object, or else the nebula would have dispersed (\acp{PN}
cannot be seen for much longer than \num{50000} years after ejection,
and usually are quite a bit younger than that). The shapes of \acp{PN}
around post-\ac{CE} systems were already studied in the 1990s
\citep{bond1992a}, and \citet{demarco2009a} showed that there is a
diversity of morphologies even around relatively similar binaries. The
immediate conclusion is that there must be a diversity of
interactions. Elongated or bipolar morphologies are often observed in
remnants of interacting systems such as V838~Mon or Nova 1670
\citep[CK~Vul;][]{kaminski2021a} and can possibly interpreted as a
result of jets or collimated outflows. Such mechanisms potentially
contribute to the shaping of \acp{PN} \citep[e.g.][]{corradi1995a,
  soker1998c, soker2004b, soker2004c}. It is suggestive that shapes of
post-\ac{CE} remnants are a sensitive probe of the details of the
interaction.

Pre-CE interaction and potentially also jets around the companion in
the early inspiral phase may be important for the morphology of
\acp{PN} \citep[e.g.][]{akashi2021a}.  The spaping of biploar \acp{PN}
and proto-\acp{PN} by jets or collimated outflow from a companion in a
bianry system was simulated by, e.g., \citet{lee2003a, lee2004a},
\citet{garcia2004a}, \citet{akashi2008a}, \citet{akashi2008b},
\citet{dennis2008a}, \cite{huarte2012a}, \citet{velazquez2012a},
\citet{akashi2016a, akashi2017a, akashi2018a}, and \cite{akashi2018b}.
These simulations assumed the interaction to take place in an
isotropic wind issued from an \ac{AGB} star or inside the envelope of
such a star but did not explicitly account for \ac{CE} interaction.

\citet{garcia2018a}, \citet{frank2018a}, \citet{zou2020a},
\citet{garcia2020a}, and \citet{garcia2021a} modelled the formation of
\acp{PN} using, as starting point, the circumstellar gas distribution
generated by a \ac{CE} simulation. \citet{garcia2018a} employed the
Eulerian grid-based \ac{CE} simulation of \citet{ricker2012a} under
the assumption that the \ac{CE} gas represents the circumbinary
environment into which the post-\ac{AGB} primary blows its low-density,
spherical, fast wind, generating the visible \ac{PN} \citep[the origin
  of fast wind is not under debate, although there are scenarios in
  which this wind is not spherical at the origin, but already
  collimated;][]{balick2002a}. \citet{garcia2018a} carried out
axisymmetric simulations of the long-term evolution for (\num{10000}
years---typical \acp{PN} have maximum lifetimes of \num{50000} to
\num{100000} years---accounting for ionization effects. The symmetry
of their simulated \acp{PN} is not a surprise given the dimensionality
of their simulation, but it already gives an idea that the nebula has
a bipolar shape because of the typical equatorial concentration of the
expanding \ac{CE}.

\citet{frank2018a} and \citet{zou2020a} carried out \ac{3D}
simulations of \ac{PN} formation using as starting point the \ac{SPH}
\ac{CE} interaction model of \citet{reichardt2019a} taken several
years after the end of the inspiral. As was the case for the work of
\citet{garcia2018a}, they launched a spherical wind of a certain
strength. However, the simulation covered only a shorter timescale of
three years. Interestingly, they observed pronounced inertial
collimation, but noticed that strong \ac{3D} effects do not guarantee
a symmetric \ac{PN}.

\citet{garcia2020a} continued the work started by \citet{garcia2018a}
investigating different types of magnetized fast winds blown into the
\ac{CE} ejecta. They also commented that the collimation observed by
\citet{zou2020a} was due to the pronounced evacuated polar funnels in
the \ac{CE} ejecta which, they remarked, must be due to the adiabatic
nature of the \ac{CE} simulation. Only additional simulations will be
able to tell just how much collimation is expected.

\begin{figure}
  \centering
    \includegraphics[width=0.8\textwidth]{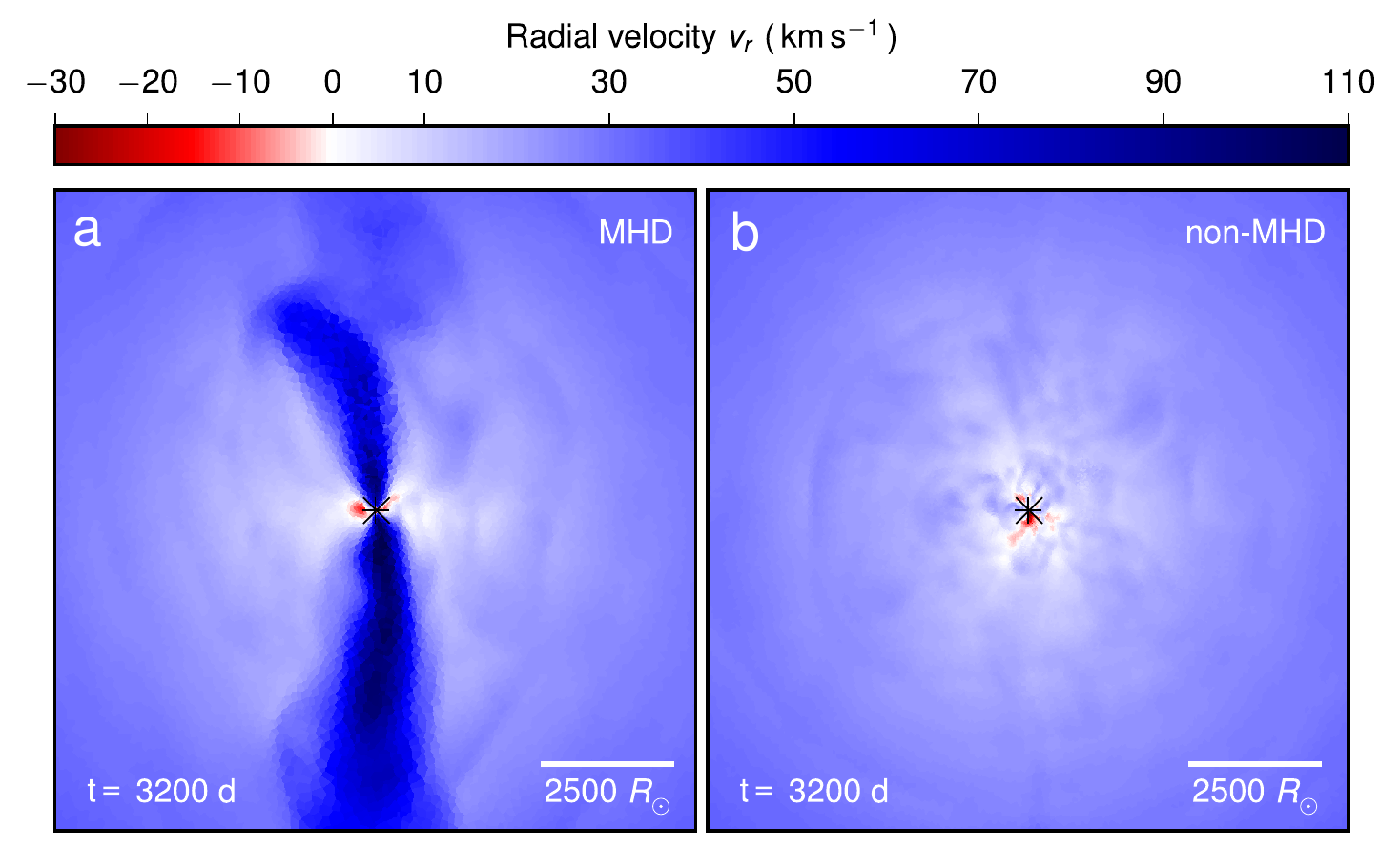}
    \caption{Radial velocity (a) in the \ac{MHD} \ac{CE} simulation of
      \citet{ondratschek2022a} compared with that in a pure
      hydrodynamics model (b). Figure from
      \cite{ondratschek2022a}.}
    \label{fig:ondratschek2022_fig3}
\end{figure}

\citet{ondratschek2022a} took a different approach to the problem of
\ac{PN} formation. Their \ac{CE} simulation, which included magnetic fields
and extended to a later phase of \ac{CE} interaction, where the
envelope is completely unbound even according to the conservative
estimate involving the kinetic energy of the gas only, results in a
collimated jet, magneto-centrifugally launched towards the
end of the inspiral. Similar jet-like outfows were also reported by \cite{garcia2021a}. The resulting bipolar morphology is certainly
different from that of simulated ejected \acp{CE} that do not include
magnetic fields, see Fig.~\ref{fig:ondratschek2022_fig3} for a
comparison. The shaping of the \ac{PN} deriving from such configuration
will still be affected by whatever fast wind is launched next due to
the heating central star and, as is the case for the other post-\ac{CE} \ac{PN}
simulations, by the ionization front propagating through the
material. Clearly this line of comparison between \acp{PN} and \ac{CE}
ejecta is in its early days, but it promises to yield very convincing
results.

\subsubsection{Transient events caused by CE interaction}
\label{sssec:transient_events_likely_due_to_a_ce_interaction}
 
The dynamical part of the \ac{CE} interaction is short-lived and it is
therefore expected to be observed only rarely. With the advent of
time-domain astrophysical observations, e.g., Palomar Transient
Factory \citep{law2009a}, the Catalina Real Time Transient Survey
\citep{drake2009a}, SkyMapper \citep{murphy2009a}, and The Zwicky
Transient Factory \citep{bellm2019a}, we have now realized that a
number of luminous outbursts, or transients, were actually \ac{CE}
interactions, demonstrated beyond doubt for at least some objects
\citep[such as V~1309~Sco,][]{tylenda2011a}. In the near future, the
Legacy Survey of Space and Time at the Rubin Observatory
\citep{ivezic2019a} will undoubtedly make great progress thanks to the
discovery of a large amount of the fainter events \citep[according
  to][about 20 to 750 detections are expected]{howitt2020a}.

The current thinking is that the class of objects known as gap
transients---which we have mentioned in
Sect.~\ref{sect:obs_constraints}---are actually deriving from a number
of physical phenomena, only some of which are likely \ac{CE}
interactions. While the terminology has been slightly confusing in the
earlier literature, with the benefit of more observations,
\citet{pastorello2019a} have proposed that the terms \acfi{RN} and \acfi{LRN} are reserved for
binary coalescence events of a \ac{CE} nature for lower and higher
mass primary stars, respectively.
  
The best examples of \acp{RN} and \acp{LRN} to date are V1309~Sco,
with an inferred primary star mass of ${\sim} 1.5\, \msun$ and a mass
ratio (companion star mass over primary star mass) of the interacting
stars of $q=0.1$ \citep{tylenda2011a}, V838~Mon with a progenitor mass
of $5$--$10 \,\msun$ \citep{tylenda2005a}, M31-LRN2015 with a
progenitor mass of $3$--$5.5 \, \msun$ \citep{macleod2017a} and
M101~OT2015-1 with a progenitor mass ${\sim}\, 18 \, \msun$
\citep{blagorodnova2017a}, with more massive examples being NGC4490-OT
\citep[$20$--$30 \,\msun$][]{smith2016a} and possibly, $\eta$~Carinae
\citep{smith2016a, rest2012a}. For most of these objects,
which---again confusing the terminology---may not all be standard
\ac{CE} events but belong to the broader class of stellar
interactions, we have only partial lightcurves. Occasionally,
progenitor photometry is identified with the best case being the
V1309~Sco system which had complete photometry for three years prior
to outburst \citep{tylenda2011a}.

Due to the rarity of \ac{CE} events, we tend to observe the brightest
cases, usually deriving from more massive objects and/or less evolved
donor stars \citep{kochanek2014a, demarco2017a}. Another observational
bias is that we currently have no compelling evidence that in any of
the observed cases a close binary has survived. This may indicate that
the observed systems failed \ac{CE} ejection and resulted in a merger
of the cores. On the other hand, \citet{howitt2020a} argued that the
brightest events may betray the presence of a surviving binary. In
truth, we do not know. Not having observed the post-\ac{CE} binary is no
proof of the \ac{CE} having resulted in a merger, as in some cases a binary
could have survived that either has not been detected yet due to a
lack of data or simply because it is still embedded in considerable
circumstellar material. As the number of observed \acp{RN} and \acp{LRN}
increases, the interpretation will become more robust. Further
monitoring of past transients will also, hopefully, determine whether
there is a \emph{binary} remnant, in which case we would be able to
connect the event with post-\ac{CE} binary classes. The connection between
\ac{RN} and \ac{LRN} transients and post-\ac{CE} binaries needs to be
established if we are to use these observations as constraints of
simulations. They may also hold a an answer to why, as discussed in
Sect.~\ref{ssec:why_is_CE_ejection_hard_to_achieve}, simulations
struggle to unbind the \ac{CE} material---a prerequisite to the
survival of the binary.
 
Lightcurves of \ac{RN} and \ac{LRN} \citep[for a one-dimensional model
  see, e.g.,][]{matsumoto2022a}, whether they eventually prove to be
\ac{CE} events where the core binary has survived or where it has
merged, should be able to provide some constraints on simulations, if
only simulated lightcurves could be calculated. These, however are not
yet available \citep{galaviz2017a}. For their ``common-envelope jets
supernova impostor'' scenario, \citet{schreier2021a} give an estimate
for optical observations. For the ``grazing envelope evolution''
scenario, \citet{soker2015a, soker2016c} argue that a fraction of the
energy carried away by the jets may be converted to
radiation. Consequently, a bright object may be observable as an
intermediate-luminosity optical transent (perhaps a \ac{LRN}).
Besides the lightcurves, other observables of mergers that can
constrain simulations are parameters such as the photospheric size
\citet{macleod2017a} or the initial ejection velocities, which could
be measured if early spectroscopy of the event were available.

For \ac{CE} mergers with massive primary stars and neutron stars or
black holes as companions, supernova-like events have been suggested
in the scenario of ``common-envelope jets supernovae''
\citep[e.g.][]{chevalier2012a, soker2018b, schroder2020a}. It is still
to be determined how to distinguish this scenario from other
explanations for Type IIn or Type II-P supernovae. Interaction with
\ac{CE} material may lead to a different shaping of the light curve
than what is expected from other circumstellar material
\cite{soker2019b}. \ac{CE} mergers of low-mass stars with white dwarf
companions may give rise to thermonuclear explosions. In this
``core-degenerate'' scenario \cite[e.g.]{kashi2011a, ilkov2012a}, the
merger dynamics is essentially that of the usual double
degenerate scenario for Type Ia supernova explosions. The difference is
in the rate and delay time of the expected events, because inspiral of
the degenerate cores in a \ac{CE} is driven efficiently by drag
whereas without the envelope material it is determined by
gravitational wave emission. For the appearance of the event, however,
the \ac{CE} material is important: hydrogen lines may be present in
the spectra in contrast to the typical Type Ia supernovae
\citep{livio2003a, soker2013a}.

While comparing \ac{CE} models with optical observations seems
promising, there are alternatives. Gravitational waves have been
detected from mergers of compact objects and this naturally implies
the question of whether \ac{CE} events would be detectable with
upcoming observatories, such as the Laser Interferometer Space Antenna
\citep[LISA,][]{baker2019a}. Analytic and semi-analytic studies
\citep{holgado2018a, ginat2020a, renzo2021a} have looked into this question, but
settling it requires a treatment in the framework of full \ac{3D}
hydrodynamic simulations of \ac{CEE}.

\subsection{Epilogue}

Shortly after the \ac{CE} phase in binary stellar systems had been
proposed by \citet{paczynski1976a}, the first numerical simulations of
this binary interaction were presented. However, the spatial and
temporal timescale challenges in combination with lacking spatial
symmetries required coarse approximations and parametrizations of
effects that could not be resolved by the simulations in the 1980a and
1990s. For this reason, early simulations failed to be predictive
about the success of envelope ejection and the final orbital
separation of the remnant core binary system.

The key to understanding \ac{CE} phases are \ac{3D} hydrodynamic
simulations of the inspiral of the companion into the envelope of the
giant primary star. Thanks to increasing computational power and
sophisticated numerical techniques, modeling these processes with
sufficient accuracy is now within reach. This has motivated different
groups using different hydrodynamics codes to join the search for a
full description of the interaction and substantial progress was made
recently.

In our review, we discussed the basic physical modeling approaches and
numerical methods used in \ac{3D} hydrodynamic simulations of
\ac{CEE}. On this basis, we provided an overview of the simulations
performed to date, which has allowed us to emphasize where the
computational frontier currently stands. We hope that this provides a
guide to concentrate efforts on some of the critical problems of
modelling approaches, such as for example the need to interface
different modelling methodologies.  By doing so we will be able to
carry out increasingly sophisticated, direct comparisons of \ac{CE}
simulations with astronomical observables, which are themselves
proliferating with the detection of large numbers of transient
phenomena.

The rapid development in modeling \ac{CE} interaction has the
potential to solve one of the last remaining fundamental problems of
stellar astrophysics. It can complete our understanding of the
evolution of the progenitors of gravitational-wave emitting mergers of
double-compact objects, supernovae, planetary nebulae, gamma-ray
bursts, X-ray binaries, radio-millisecond pulsars in binaries, and
cataclysmic variables, among others.

\begin{acknowledgements}
This review would have been impossible without the help of many of our
colleagues. We are grateful for their input and their permissions to
reproduce figures from their publications.  FKR acknowledges support
from the Klaus Tschira Foundation. OD acknowledges support from the
Australian Research Council's Discovery Project award DP210101094.
\end{acknowledgements}

%
{\small \noindent
\textbf{Conflict of interest} 
The authors declare that they have no conflict of interest.
}

\phantomsection
\addcontentsline{toc}{section}{References}

\bibliographystyle{spbasic-FS-etal}      

\begin{thebibliography}{357}
\expandafter\ifx\csname url\endcsname\relax
 \def\url#1{\burl{#1}}\fi
\expandafter\ifx\csname urlprefix\endcsname\relax\def\urlprefix{URL }\fi
\providecommand{\bibinfo}[2]{#2}
\providecommand{\eprint}[2][]{\url{#2}}
\providecommand{\doi}[1]{\urlstyle{rm}\url{https://doi.org/#1}}

\bibitem[{{Abbott} et~al.(2016){Abbott}, {Abbott}, {Abbott}, {Abernathy},
  {Acernese}, {Ackley}, {Adams}, {Adams}, {Addesso}, {Adhikari}, {Adya},
  {Affeldt}, {Agathos}, {Agatsuma}, {Aggarwal}, {Aguiar}, {Aiello}, {Ain},
  {Ajith}, {Allen}, {Allocca}, {Altin}, {Anderson}, {Anderson}, {Arai},
  {Arain}, {Araya}, {Arceneaux}, {Areeda}, {Arnaud}, {Arun}, {Ascenzi},
  {Ashton}, {Ast}, {Aston}, {Astone}, {Aufmuth}, {Aulbert}, {Babak}, {Bacon},
  {Bader}, {Baker}, {Baldaccini}, {Ballardin}, {Ballmer}, {Barayoga},
  {Barclay}, {Barish}, {Barker}, {Barone}, {Barr}, {Barsotti}, {Barsuglia},
  {Barta}, {Bartlett}, {Barton}, {Bartos}, {Bassiri}, {Basti}, {Batch},
  {Baune}, {Bavigadda}, {Bazzan}, {Behnke}, {Bejger}, {Belczynski}, {Bell},
  {Bell}, {Berger}, {Bergman}, {Bergmann}, {Berry}, {Bersanetti}, {Bertolini},
  {Betzwieser}, {Bhagwat}, {Bhandare}, {Bilenko}, {Billingsley}, {Birch},
  {Birney}, {Birnholtz}, {Biscans}, {Bisht}, {Bitossi}, {Biwer}, {Bizouard},
  {Blackburn}, {Blair}, {Blair}, {Blair}, {Bloemen}, {Bock}, {Bodiya}, {Boer},
  {Bogaert}, {Bogan}, {Bohe}, {Bojtos}, {Bond}, {Bondu}, {Bonnand}, {Boom},
  {Bork}, {Boschi}, {Bose}, {Bouffanais}, {Bozzi}, {Bradaschia}, {Brady},
  {Braginsky}, {Branchesi}, {Brau}, {Briant}, {Brillet}, {Brinkmann},
  {Brisson}, {Brockill}, {Brooks}, {Brown}, {Brown}, {Brown}, {Buchanan},
  {Buikema}, {Bulik}, {Bulten}, {Buonanno}, {Buskulic}, {Buy}, {Byer},
  {Cabero}, {Cadonati}, {Cagnoli}, {Cahillane}, {Bustillo}, {Callister},
  {Calloni}, {Camp}, {Cannon}, {Cao}, {Capano}, {Capocasa}, {Carbognani},
  {Caride}, {Casanueva Diaz}, {Casentini}, {Caudill}, {Cavagli{\`a}},
  {Cavalier}, {Cavalieri}, {Cella}, {Cepeda}, {Baiardi}, {Cerretani},
  {Cesarini}, {Chakraborty}, {Chalermsongsak}, {Chamberlin}, {Chan}, {Chao},
  {Charlton}, {Chassande-Mottin}, {Chen}, {Chen}, {Cheng}, {Chincarini},
  {Chiummo}, {Cho}, {Cho}, {Chow}, {Christensen}, {Chu}, {Chua}, {Chung},
  {Ciani}, {Clara}, {Clark}, {Cleva}, {Coccia}, {Cohadon}, {Colla}, {Collette},
  {Cominsky}, {Constancio}, {Conte}, {Conti}, {Cook}, {Corbitt}, {Cornish},
  {Corsi}, {Cortese}, {Costa}, {Coughlin}, {Coughlin}, {Coulon}, {Countryman},
  {Couvares}, {Cowan}, {Coward}, {Cowart}, {Coyne}, {Coyne}, {Craig},
  {Creighton}, {Creighton}, {Cripe}, {Crowder}, {Cruise}, {Cumming},
  {Cunningham}, {Cuoco}, {Dal Canton}, {Danilishin}, {D'Antonio}, {Danzmann},
  {Darman}, {Da Silva Costa}, {Dattilo}, {Dave}, {Daveloza}, {Davier},
  {Davies}, {Daw}, {Day}, {De}, {DeBra}, {Debreczeni}, {Degallaix}, {De
  Laurentis}, {Del{\'e}glise}, {Del Pozzo}, {Denker}, {Dent}, {Dereli},
  {Dergachev}, {DeRosa}, {De Rosa}, {DeSalvo}, {Dhurandhar}, {D{\'\i}az}, {Di
  Fiore}, {Di Giovanni}, {Di Lieto}, {Di Pace}, {Di Palma}, {Di Virgilio},
  {Dojcinoski}, {Dolique}, {Donovan}, {Dooley}, {Doravari}, {Douglas},
  {Downes}, {Drago}, {Drever}, {Driggers}, {Du}, {Ducrot}, {Dwyer}, {Edo},
  {Edwards}, {Effler}, {Eggenstein}, {Ehrens}, {Eichholz}, {Eikenberry},
  {Engels}, {Essick}, {Etzel}, {Evans}, {Evans}, {Everett}, {Factourovich},
  {Fafone}, {Fair}, {Fairhurst}, {Fan}, {Fang}, {Farinon}, {Farr}, {Farr},
  {Favata}, {Fays}, {Fehrmann}, {Fejer}, {Feldbaum}, {Ferrante}, {Ferreira},
  {Ferrini}, {Fidecaro}, {Finn}, {Fiori}, {Fiorucci}, {Fisher}, {Flaminio},
  {Fletcher}, {Fong}, {Fournier}, {Franco}, {Frasca}, {Frasconi}, {Frede},
  {Frei}, {Freise}, {Frey}, {Frey}, {Fricke}, {Fritschel}, {Frolov}, {Fulda},
  {Fyffe}, {Gabbard}, {Gair}, {Gammaitoni}, {Gaonkar}, {Garufi}, {Gatto},
  {Gaur}, {Gehrels}, {Gemme}, {Gendre}, {Genin}, {Gennai}, {George}, {Gergely},
  {Germain}, {Ghosh}, {Ghosh}, {Ghosh}, {Giaime}, {Giardina}, {Giazotto},
  {Gill}, {Glaefke}, {Gleason}, {Goetz}, {Goetz}, {Gondan}, {Gonz{\'a}lez},
  {Castro}, {Gopakumar}, {Gordon}, {Gorodetsky}, {Gossan}, {Gosselin},
  {Gouaty}, {Graef}, {Graff}, {Granata}, {Grant}, {Gras}, {Gray}, {Greco},
  {Green}, {Greenhalgh}, {Groot}, {Grote}, {Grunewald}, {Guidi}, {Guo},
  {Gupta}, {Gupta}, {Gushwa}, {Gustafson}, {Gustafson}, {Hacker}, {Hall},
  {Hall}, {Hammond}, {Haney}, {Hanke}, {Hanks}, {Hanna}, {Hannam}, {Hanson},
  {Hardwick}, {Harms}, {Harry}, {Harry}, {Hart}, {Hartman}, {Haster},
  {Haughian}, {Healy}, {Heefner}, {Heidmann}, {Heintze}, {Heinzel}, {Heitmann},
  {Hello}, {Hemming}, {Hendry}, {Heng}, {Hennig}, {Heptonstall}, {Heurs},
  {Hild}, {Hoak}, {Hodge}, {Hofman}, {Hollitt}, {Holt}, {Holz}, {Hopkins},
  {Hosken}, {Hough}, {Houston}, {Howell}, {Hu}, {Huang}, {Huerta}, {Huet},
  {Hughey}, {Husa}, {Huttner}, {Huynh-Dinh}, {Idrisy}, {Indik}, {Ingram},
  {Inta}, {Isa}, {Isac}, {Isi}, {Islas}, {Isogai}, {Iyer}, {Izumi}, {Jacobson},
  {Jacqmin}, {Jang}, {Jani}, {Jaranowski}, {Jawahar}, {Jim{\'e}nez-Forteza},
  {Johnson}, {Johnson-McDaniel}, {Jones}, {Jones}, {Jonker}, {Ju}, {Haris},
  {Kalaghatgi}, {Kalogera}, {Kandhasamy}, {Kang}, {Kanner}, {Karki},
  {Kasprzack}, {Katsavounidis}, {Katzman}, {Kaufer}, {Kaur}, {Kawabe},
  {Kawazoe}, {K{\'e}f{\'e}lian}, {Kehl}, {Keitel}, {Kelley}, {Kells},
  {Kennedy}, {Keppel}, {Key}, {Khalaidovski}, {Khalili}, {Khan}, {Khan},
  {Khan}, {Khazanov}, {Kijbunchoo}, {Kim}, {Kim}, {Kim}, {Kim}, {Kim}, {Kim},
  {King}, {King}, {Kinzel}, {Kissel}, {Kleybolte}, {Klimenko}, {Koehlenbeck},
  {Kokeyama}, {Koley}, {Kondrashov}, {Kontos}, {Koranda}, {Korobko}, {Korth},
  {Kowalska}, {Kozak}, {Kringel}, {Krishnan}, {Kr{\'o}lak}, {Krueger}, {Kuehn},
  {Kumar}, {Kumar}, {Kuo}, {Kutynia}, {Kwee}, {Lackey}, {Landry}, {Lange},
  {Lantz}, {Lasky}, {Lazzarini}, {Lazzaro}, {Leaci}, {Leavey}, {Lebigot},
  {Lee}, {Lee}, {Lee}, {Lee}, {Lenon}, {Leonardi}, {Leong}, {Leroy},
  {Letendre}, {Levin}, {Levine}, {Li}, {Libson}, {Littenberg}, {Lockerbie},
  {Logue}, {Lombardi}, {London}, {Lord}, {Lorenzini}, {Loriette}, {Lormand},
  {Losurdo}, {Lough}, {Lousto}, {Lovelace}, {L{\"u}ck}, {Lundgren}, {Luo},
  {Lynch}, {Ma}, {MacDonald}, {Machenschalk}, {MacInnis}, {Macleod},
  {Maga{\~n}a-Sandoval}, {Magee}, {Mageswaran}, {Majorana}, {Maksimovic},
  {Malvezzi}, {Man}, {Mandel}, {Mandic}, {Mangano}, {Mansell}, {Manske},
  {Mantovani}, {Marchesoni}, {Marion}, {M{\'a}rka}, {M{\'a}rka}, {Markosyan},
  {Maros}, {Martelli}, {Martellini}, {Martin}, {Martin}, {Martynov}, {Marx},
  {Mason}, {Masserot}, {Massinger}, {Masso-Reid}, {Matichard}, {Matone},
  {Mavalvala}, {Mazumder}, {Mazzolo}, {McCarthy}, {McClelland}, {McCormick},
  {McGuire}, {McIntyre}, {McIver}, {McManus}, {McWilliams}, {Meacher},
  {Meadors}, {Meidam}, {Melatos}, {Mendell}, {Mendoza-Gandara}, {Mercer},
  {Merilh}, {Merzougui}, {Meshkov}, {Messenger}, {Messick}, {Meyers},
  {Mezzani}, {Miao}, {Michel}, {Middleton}, {Mikhailov}, {Milano}, {Miller},
  {Millhouse}, {Minenkov}, {Ming}, {Mirshekari}, {Mishra}, {Mitra},
  {Mitrofanov}, {Mitselmakher}, {Mittleman}, {Moggi}, {Mohan}, {Mohapatra},
  {Montani}, {Moore}, {Moore}, {Moraru}, {Moreno}, {Morriss}, {Mossavi},
  {Mours}, {Mow-Lowry}, {Mueller}, {Mueller}, {Muir}, {Mukherjee}, {Mukherjee},
  {Mukherjee}, {Mukund}, {Mullavey}, {Munch}, {Murphy}, {Murray}, {Mytidis},
  {Nardecchia}, {Naticchioni}, {Nayak}, {Necula}, {Nedkova}, {Nelemans},
  {Neri}, {Neunzert}, {Newton}, {Nguyen}, {Nielsen}, {Nissanke}, {Nitz},
  {Nocera}, {Nolting}, {Normandin}, {Nuttall}, {Oberling}, {Ochsner}, {O'Dell},
  {Oelker}, {Ogin}, {Oh}, {Oh}, {Ohme}, {Oliver}, {Oppermann}, {Oram},
  {O'Reilly}, {O'Shaughnessy}, {Ott}, {Ottaway}, {Ottens}, {Overmier}, {Owen},
  {Pai}, {Pai}, {Palamos}, {Palashov}, {Palomba}, {Pal-Singh}, {Pan}, {Pan},
  {Pankow}, {Pannarale}, {Pant}, {Paoletti}, {Paoli}, {Papa}, {Paris},
  {Parker}, {Pascucci}, {Pasqualetti}, {Passaquieti}, {Passuello},
  {Patricelli}, {Patrick}, {Pearlstone}, {Pedraza}, {Pedurand}, {Pekowsky},
  {Pele}, {Penn}, {Perreca}, {Pfeiffer}, {Phelps}, {Piccinni}, {Pichot},
  {Pickenpack}, {Piergiovanni}, {Pierro}, {Pillant}, {Pinard}, {Pinto},
  {Pitkin}, {Poeld}, {Poggiani}, {Popolizio}, {Post}, {Powell}, {Prasad},
  {Predoi}, {Premachandra}, {Prestegard}, {Price}, {Prijatelj}, {Principe},
  {Privitera}, {Prix}, {Prodi}, {Prokhorov}, {Puncken}, {Punturo}, {Puppo},
  {P{\"u}rrer}, {Qi}, {Qin}, {Quetschke}, {Quintero}, {Quitzow-James}, {Raab},
  {Rabeling}, {Radkins}, {Raffai}, {Raja}, {Rakhmanov}, {Ramet}, {Rapagnani},
  {Raymond}, {Razzano}, {Re}, {Read}, {Reed}, {Regimbau}, {Rei}, {Reid},
  {Reitze}, {Rew}, {Reyes}, {Ricci}, {Riles}, {Robertson}, {Robie}, {Robinet},
  {Rocchi}, {Rolland}, {Rollins}, {Roma}, {Romano}, {Romano}, {Romanov},
  {Romie}, {Rosi{\'n}ska}, {Rowan}, {R{\"u}diger}, {Ruggi}, {Ryan}, {Sachdev},
  {Sadecki}, {Sadeghian}, {Salconi}, {Saleem}, {Salemi}, {Samajdar}, {Sammut},
  {Sampson}, {Sanchez}, {Sandberg}, {Sandeen}, {Sanders}, {Sanders},
  {Sassolas}, {Sathyaprakash}, {Saulson}, {Sauter}, {Savage}, {Sawadsky},
  {Schale}, {Schilling}, {Schmidt}, {Schmidt}, {Schnabel}, {Schofield},
  {Sch{\"o}nbeck}, {Schreiber}, {Schuette}, {Schutz}, {Scott}, {Scott},
  {Sellers}, {Sengupta}, {Sentenac}, {Sequino}, {Sergeev}, {Serna},
  {Setyawati}, {Sevigny}, {Shaddock}, {Shaffer}, {Shah}, {Shahriar}, {Shaltev},
  {Shao}, {Shapiro}, {Shawhan}, {Sheperd}, {Shoemaker}, {Shoemaker}, {Siellez},
  {Siemens}, {Sigg}, {Silva}, {Simakov}, {Singer}, {Singer}, {Singh}, {Singh},
  {Singhal}, {Sintes}, {Slagmolen}, {Smith}, {Smith}, {Smith}, {Smith}, {Son},
  {Sorazu}, {Sorrentino}, {Souradeep}, {Srivastava}, {Staley}, {Steinke},
  {Steinlechner}, {Steinlechner}, {Steinmeyer}, {Stephens}, {Stevenson},
  {Stone}, {Strain}, {Straniero}, {Stratta}, {Strauss}, {Strigin}, {Sturani},
  {Stuver}, {Summerscales}, {Sun}, {Sutton}, {Swinkels}, {Szczepa{\'n}czyk},
  {Tacca}, {Talukder}, {Tanner}, {T{\'a}pai}, {Tarabrin}, {Taracchini},
  {Taylor}, {Theeg}, {Thirugnanasambandam}, {Thomas}, {Thomas}, {Thomas},
  {Thorne}, {Thorne}, {Thrane}, {Tiwari}, {Tiwari}, {Tokmakov}, {Tomlinson},
  {Tonelli}, {Torres}, {Torrie}, {T{\"o}yr{\"a}}, {Travasso}, {Traylor},
  {Trifir{\`o}}, {Tringali}, {Trozzo}, {Tse}, {Turconi}, {Tuyenbayev},
  {Ugolini}, {Unnikrishnan}, {Urban}, {Usman}, {Vahlbruch}, {Vajente},
  {Valdes}, {Vallisneri}, {van Bakel}, {van Beuzekom}, {van den Brand}, {Van
  Den Broeck}, {Vander-Hyde}, {van der Schaaf}, {van Heijningen}, {van Veggel},
  {Vardaro}, {Vass}, {Vas{\'u}th}, {Vaulin}, {Vecchio}, {Vedovato}, {Veitch},
  {Veitch}, {Venkateswara}, {Verkindt}, {Vetrano}, {Vicer{\'e}}, {Vinciguerra},
  {Vine}, {Vinet}, {Vitale}, {Vo}, {Vocca}, {Vorvick}, {Voss}, {Vousden},
  {Vyatchanin}, {Wade}, {Wade}, {Wade}, {Waldman}, {Walker}, {Wallace},
  {Walsh}, {Wang}, {Wang}, {Wang}, {Wang}, {Wang}, {Ward}, {Ward}, {Warner},
  {Was}, {Weaver}, {Wei}, {Weinert}, {Weinstein}, {Weiss}, {Welborn}, {Wen},
  {We{\ss}els}, {Westphal}, {Wette}, {Whelan}, {Whitcomb}, {White}, {Whiting},
  {Wiesner}, {Wilkinson}, {Willems}, {Williams}, {Williams}, {Williamson},
  {Willis}, {Willke}, {Wimmer}, {Winkelmann}, {Winkler}, {Wipf}, {Wiseman},
  {Wittel}, {Woan}, {Worden}, {Wright}, {Wu}, {Yablon}, {Yakushin}, {Yam},
  {Yamamoto}, {Yancey}, {Yap}, {Yu}, {Yvert}, {Zadrozny}, {Zangrando},
  {Zanolin}, {Zendri}, {Zevin}, {Zhang}, {Zhang}, {Zhang}, {Zhang}, {Zhao},
  {Zhou}, {Zhou}, {Zhu}, {Zucker}, {Zuraw}, {Zweizig}, {LIGO Scientific
  Collaboration}, and {Virgo Collaboration}}]{abbott2016a}
{Abbott} BP, {Abbott} R, {Abbott} TD, et~al. (2016) {Observation of
  Gravitational Waves from a Binary Black Hole Merger}. \prl 116(6):061102.
  \doi{10.1103/PhysRevLett.116.061102}.
  {\href{https://arxiv.org/abs/1602.03837}{{arXiv:1602.03837}}} {[gr-qc]}

\bibitem[{{Abbott} et~al.(2017){Abbott}, {Abbott}, {Abbott}, {Acernese},
  {Ackley}, {Adams}, {Adams}, {Addesso}, {Adhikari}, {Adya}, and
  et~al.}]{abbott2017a}
{Abbott} BP, {Abbott} R, {Abbott} TD, et~al. (2017) {Multi-messenger
  Observations of a Binary Neutron Star Merger}. \apjl 848:L12.
  \doi{10.3847/2041-8213/aa91c9}.
  {\href{https://arxiv.org/abs/1710.05833}{{arXiv:1710.05833}}} {[astro-ph.HE]}

\bibitem[{{Akashi} and {Soker}(2008)}]{akashi2008b}
{Akashi} M, {Soker} N (2008) Shaping planetary nebulae by light jets. \mnras
  391(3):1063--1074. \doi{10.1111/j.1365-2966.2008.13935.x}.
  {\href{https://arxiv.org/abs/0805.2332}{{arXiv:0805.2332}}} {[astro-ph]}

\bibitem[{{Akashi} and {Soker}(2016)}]{akashi2016a}
{Akashi} M, {Soker} N (2016) {Bipolar rings from jet-inflated bubbles around
  evolved binary stars}. \mnras 462(1):206--216. \doi{10.1093/mnras/stw1683}.
  {\href{https://arxiv.org/abs/1605.02574}{{arXiv:1605.02574}}} {[astro-ph.SR]}

\bibitem[{{Akashi} and {Soker}(2017)}]{akashi2017a}
{Akashi} M, {Soker} N (2017) {Shaping planetary nebulae with jets in inclined
  triple stellar systems}. \mnras 469(3):3296--3306.
  \doi{10.1093/mnras/stx1058}.
  {\href{https://arxiv.org/abs/1701.05460}{{arXiv:1701.05460}}} {[astro-ph.SR]}

\bibitem[{{Akashi} and {Soker}(2018)}]{akashi2018a}
{Akashi} M, {Soker} N (2018) The formation of `columns crowns' by jets
  interacting with a circumstellar dense shell. \mnras 481(2):2754--2765.
  \doi{10.1093/mnras/sty2479}.
  {\href{https://arxiv.org/abs/1808.00276}{{arXiv:1808.00276}}} {[astro-ph.SR]}

\bibitem[{{Akashi} and {Soker}(2021)}]{akashi2021a}
{Akashi} M, {Soker} N (2021) {Shaping ``Ears'' in Planetary Nebulae by Early
  Jets}. \apj 913(2):91. \doi{10.3847/1538-4357/abf7bb}.
  {\href{https://arxiv.org/abs/2012.08917}{{arXiv:2012.08917}}} {[astro-ph.GA]}

\bibitem[{{Akashi} et~al.(2008){Akashi}, {Meiron}, and {Soker}}]{akashi2008a}
{Akashi} M, {Meiron} Y, {Soker} N (2008) X-ray emission from jet wind
  interaction in planetary nebulae. \na 13(8):563--568.
  \doi{10.1016/j.newast.2008.03.002}.
  {\href{https://arxiv.org/abs/0711.3265}{{arXiv:0711.3265}}} {[astro-ph]}

\bibitem[{{Akashi} et~al.(2018){Akashi}, {Bear}, and {Soker}}]{akashi2018b}
{Akashi} M, {Bear} E, {Soker} N (2018) Forming h-shaped and barrel-shaped
  nebulae with interacting jets. \mnras 475(4):4794--4808.
  \doi{10.1093/mnras/sty029}.
  {\href{https://arxiv.org/abs/1712.07156}{{arXiv:1712.07156}}} {[astro-ph.SR]}

\bibitem[{{Alexander} et~al.(1976){Alexander}, {Chau}, and
  {Henriksen}}]{alexander1976a}
{Alexander} ME, {Chau} WY, {Henriksen} RN (1976) Orbital evolution of a singly
  condensed, close binary, by mass loss from the primary and by accretion drag
  on the condensed member. \apj 204:879--888. \doi{10.1086/154236}

\bibitem[{Amdahl(1967)}]{amdahl1967a}
Amdahl GM (1967) Validity of the single processor approach to achieving large
  scale computing capabilities. In: Proceedings of the April 18-20, 1967,
  Spring Joint Computer Conference. AFIPS '67 (Spring). Association for
  Computing Machinery, New York, NY, USA, p 483–485.
  \doi{10.1145/1465482.1465560},
  \urlprefix\url{https://doi.org/10.1145/1465482.1465560}

\bibitem[{{Appel}(1985)}]{appel1985a}
{Appel} AW (1985) {An Efficient Program for Many-Body Simulation}. SIAM Journal
  on Scientific and Statistical Computing 6(1):85--103

\bibitem[{{Armitage} and {Livio}(2000)}]{armitage2000a}
{Armitage} PJ, {Livio} M (2000) {Black Hole Formation via Hypercritical
  Accretion during Common-Envelope Evolution}. \apj 532(1):540--547.
  \doi{10.1086/308548}.
  {\href{https://arxiv.org/abs/astro-ph/9906028}{{arXiv:astro-ph/9906028}}}
  {[astro-ph]}

\bibitem[{{Baiotti} and {Rezzolla}(2017)}]{baiotti2017a}
{Baiotti} L, {Rezzolla} L (2017) Binary neutron star mergers: a review of
  einstein's richest laboratory. Reports on Progress in Physics 80(9):096901.
  \doi{10.1088/1361-6633/aa67bb}.
  {\href{https://arxiv.org/abs/1607.03540}{{arXiv:1607.03540}}} {[gr-qc]}

\bibitem[{{Baker} et~al.(2019){Baker}, {Bellovary}, {Bender}, {Berti},
  {Caldwell}, {Camp}, {Conklin}, {Cornish}, {Cutler}, {DeRosa}, {Eracleous},
  {Ferrara}, {Francis}, {Hewitson}, {Holley-Bockelmann}, {Hornschemeier},
  {Hogan}, {Kamai}, {Kelly}, {Shapiro Key}, {Larson}, {Livas},
  {Manthripragada}, {McKenzie}, {McWilliams}, {Mueller}, {Natarajan}, {Numata},
  {Rioux}, {Sankar}, {Schnittman}, {Shoemaker}, {Shoemaker}, {Slutsky},
  {Spero}, {Stebbins}, {Thorpe}, {Vallisneri}, {Ware}, {Wass}, {Yu}, and
  {Ziemer}}]{baker2019a}
{Baker} J, {Bellovary} J, {Bender} PL, et~al. (2019) The laser interferometer
  space antenna: Unveiling the millihertz gravitational wave sky. arXiv
  e-prints arXiv:1907.06482.
  {\href{https://arxiv.org/abs/1907.06482}{{arXiv:1907.06482}}} {[astro-ph.IM]}

\bibitem[{{Balbus}(1995)}]{balbus1995a}
{Balbus} SA (1995) {General Local Stability Criteria for Stratified, Weakly
  Magnetized Rotating Systems}. \apj 453:380. \doi{10.1086/176397}

\bibitem[{{Balbus} and {Hawley}(1991)}]{balbus1991a}
{Balbus} SA, {Hawley} JF (1991) {A powerful local shear instability in weakly
  magnetized disks. I - Linear analysis. II - Nonlinear evolution}. \apj
  376:214--233. \doi{10.1086/170270}

\bibitem[{{Balick} and {Frank}(2002)}]{balick2002a}
{Balick} B, {Frank} A (2002) Shapes and shaping of planetary nebulae. \araa
  40:439--486. \doi{10.1146/annurev.astro.40.060401.093849}

\bibitem[{{Barnes} and {Hut}(1986)}]{barnes1986a}
{Barnes} J, {Hut} P (1986) A hierarchical $\mathcal{O}(n \log n)$
  force-calculation algorithm. \nat 324(6096):446--449. \doi{10.1038/324446a0}

\bibitem[{{Belczynski} et~al.(2014){Belczynski}, {Buonanno}, {Cantiello},
  {Fryer}, {Holz}, {Mandel}, {Miller}, and {Walczak}}]{belczynski2014a}
{Belczynski} K, {Buonanno} A, {Cantiello} M, et~al. (2014) The formation and
  gravitational-wave detection of massive stellar black hole binaries. \apj
  789(2):120. \doi{10.1088/0004-637X/789/2/120}.
  {\href{https://arxiv.org/abs/1403.0677}{{arXiv:1403.0677}}} {[astro-ph.HE]}

\bibitem[{{Belczynski} et~al.(2016){Belczynski}, {Holz}, {Bulik}, and
  {O'Shaughnessy}}]{belczynski2016a}
{Belczynski} K, {Holz} DE, {Bulik} T, {O'Shaughnessy} R (2016) {The first
  gravitational-wave source from the isolated evolution of two stars in the
  40-100 solar mass range}. \nat 534(7608):512--515. \doi{10.1038/nature18322}.
  {\href{https://arxiv.org/abs/1602.04531}{{arXiv:1602.04531}}} {[astro-ph.HE]}

\bibitem[{{Bellm} et~al.(2019){Bellm}, {Kulkarni}, {Graham}, {Dekany}, {Smith},
  {Riddle}, {Masci}, {Helou}, {Prince}, {Adams}, {Barbarino}, {Barlow},
  {Bauer}, {Beck}, {Belicki}, {Biswas}, {Blagorodnova}, {Bodewits}, {Bolin},
  {Brinnel}, {Brooke}, {Bue}, {Bulla}, {Burruss}, {Cenko}, {Chang}, {Connolly},
  {Coughlin}, {Cromer}, {Cunningham}, {De}, {Delacroix}, {Desai}, {Duev},
  {Eadie}, {Farnham}, {Feeney}, {Feindt}, {Flynn}, {Franckowiak}, {Frederick},
  {Fremling}, {Gal-Yam}, {Gezari}, {Giomi}, {Goldstein}, {Golkhou}, {Goobar},
  {Groom}, {Hacopians}, {Hale}, {Henning}, {Ho}, {Hover}, {Howell}, {Hung},
  {Huppenkothen}, {Imel}, {Ip}, {Ivezi{\'c}}, {Jackson}, {Jones}, {Juric},
  {Kasliwal}, {Kaspi}, {Kaye}, {Kelley}, {Kowalski}, {Kramer}, {Kupfer},
  {Landry}, {Laher}, {Lee}, {Lin}, {Lin}, {Lunnan}, {Giomi}, {Mahabal}, {Mao},
  {Miller}, {Monkewitz}, {Murphy}, {Ngeow}, {Nordin}, {Nugent}, {Ofek},
  {Patterson}, {Penprase}, {Porter}, {Rauch}, {Rebbapragada}, {Reiley},
  {Rigault}, {Rodriguez}, {van Roestel}, {Rusholme}, {van Santen}, {Schulze},
  {Shupe}, {Singer}, {Soumagnac}, {Stein}, {Surace}, {Sollerman}, {Szkody},
  {Taddia}, {Terek}, {Van Sistine}, {van Velzen}, {Vestrand}, {Walters},
  {Ward}, {Ye}, {Yu}, {Yan}, and {Zolkower}}]{bellm2019a}
{Bellm} EC, {Kulkarni} SR, {Graham} MJ, et~al. (2019) {The Zwicky Transient
  Facility: System Overview, Performance, and First Results}. \pasp
  131(995):018002. \doi{10.1088/1538-3873/aaecbe}.
  {\href{https://arxiv.org/abs/1902.01932}{{arXiv:1902.01932}}} {[astro-ph.IM]}

\bibitem[{{Benz} and {Hills}(1987)}]{benz1987a}
{Benz} W, {Hills} JG (1987) {Three-dimensional hydrodynamical simulations of
  stellar collisions. I - Equal-mass main-sequence stars}. \apj 323:614--628.
  \doi{10.1086/165857}

\bibitem[{{Benz} et~al.(1990){Benz}, {Bowers}, {Cameron}, and
  {Press}}]{benz1990a}
{Benz} W, {Bowers} RL, {Cameron} AGW, {Press} WH (1990) {Dynamic Mass Exchange
  in Doubly Degenerate Binaries. I. 0.9 and 1.2 M$_{sun}$ Stars}. \apj 348:647.
  \doi{10.1086/168273}

\bibitem[{{Berger} and {Colella}(1989)}]{berger1989a}
{Berger} MJ, {Colella} P (1989) Local adaptive mesh refinement for shock
  hydrodynamics. Journal of Computational Physics 82(1):64--84.
  \doi{10.1016/0021-9991(89)90035-1}

\bibitem[{{Berger} and {Oliger}(1984)}]{berger1984a}
{Berger} MJ, {Oliger} J (1984) {Adaptive Mesh Refinement for Hyperbolic Partial
  Differential Equations}. Journal of Computational Physics 53(3):484--512.
  \doi{10.1016/0021-9991(84)90073-1}

\bibitem[{{Blagorodnova} et~al.(2017){Blagorodnova}, {Kotak}, {Polshaw},
  {Kasliwal}, {Cao}, {Cody}, {Doran}, {Elias-Rosa}, {Fraser}, {Fremling},
  {Gonzalez-Fernand ez}, {Harmanen}, {Jencson}, {Kankare}, {Kudritzki},
  {Kulkarni}, {Magnier}, {Manulis}, {Masci}, {Mattila}, {Nugent}, {Ochner},
  {Pastorello}, {Reynolds}, {Smith}, {Sollerman}, {Taddia}, {Terreran},
  {Tomasella}, {Turatto}, {Vreeswijk}, {Wozniak}, and
  {Zaggia}}]{blagorodnova2017a}
{Blagorodnova} N, {Kotak} R, {Polshaw} J, et~al. (2017) {Common Envelope
  Ejection for a Luminous Red Nova in M101}. \apj 834(2):107.
  \doi{10.3847/1538-4357/834/2/107}.
  {\href{https://arxiv.org/abs/1607.08248}{{arXiv:1607.08248}}} {[astro-ph.SR]}

\bibitem[{{Bodenheimer} and {Taam}(1984)}]{bodenheimer1984a}
{Bodenheimer} P, {Taam} RE (1984) {Double-core evolution. II - Two-dimensional
  hydrodynamic effects}. \apj 280:771--779. \doi{10.1086/162049}

\bibitem[{{Bond} et~al.(1992){Bond}, {Ciardullo}, and {Meakes}}]{bond1992a}
{Bond} HE, {Ciardullo} R, {Meakes} MG (1992) Close binary nuclei of planetary
  nebulae. In: {Kondo} Y, {Sistero} R, {Polidan} RS (eds) Evolutionary
  Processes in Interacting Binary Stars. vol 151. p 517

\bibitem[{{Bond} et~al.(2003){Bond}, {Henden}, {Levay}, {Panagia}, {Sparks},
  {Starrfield}, {Wagner}, {Corradi}, and {Munari}}]{bond2003a}
{Bond} HE, {Henden} A, {Levay} ZG, et~al. (2003) {An energetic stellar outburst
  accompanied by circumstellar light echoes}. \nat 422:405--408.
  \doi{10.1038/nature01508}.
  {\href{https://arxiv.org/abs/astro-ph/0303513}{{astro-ph/0303513}}}

\bibitem[{{Bondi}(1952)}]{bondi1952a}
{Bondi} H (1952) {On spherically symmetrical accretion}. \mnras 112:195.
  \doi{10.1093/mnras/112.2.195}

\bibitem[{{Bondi} and {Hoyle}(1944)}]{bondi1944a}
{Bondi} H, {Hoyle} F (1944) {On the mechanism of accretion by stars}. \mnras
  104:273. \doi{10.1093/mnras/104.5.273}

\bibitem[{{Botticella} et~al.(2009){Botticella}, {Pastorello}, {Smartt},
  {Meikle}, {Benetti}, {Kotak}, {Cappellaro}, {Crockett}, {Mattila}, {Sereno},
  {Patat}, {Tsvetkov}, {van Loon}, {Abraham}, {Agnoletto}, {Arbour}, {Benn},
  {di Rico}, {Elias-Rosa}, {Gorshanov}, {Harutyunyan}, {Hunter}, {Lorenzi},
  {Keenan}, {Maguire}, {Mendez}, {Mobberley}, {Navasardyan}, {Ries},
  {Stanishev}, {Taubenberger}, {Trundle}, {Turatto}, and
  {Volkov}}]{botticella2009a}
{Botticella} MT, {Pastorello} A, {Smartt} SJ, et~al. (2009) {SN 2008S:} an
  electron-capture {SN} from a super-{AGB} progenitor? \mnras
  398(3):1041--1068. \doi{10.1111/j.1365-2966.2009.15082.x}.
  {\href{https://arxiv.org/abs/0903.1286}{{arXiv:0903.1286}}} {[astro-ph.SR]}

\bibitem[{{Brandt}(1977)}]{brandt1977a}
{Brandt} A (1977) Multi-level adaptive solutions to boundary-value problems.
  Mathematics of Computation 39:333--390

\bibitem[{{Brown}(1995)}]{brown1995a}
{Brown} GE (1995) {Neutron star accretion and binary pulsar formation}. \apj
  440:270--279. \doi{10.1086/175268}

\bibitem[{{Bryan} et~al.(1995){Bryan}, {Norman}, {Stone}, {Cen}, and
  {Ostriker}}]{bryan1995a}
{Bryan} GL, {Norman} ML, {Stone} JM, {Cen} R, {Ostriker} JP (1995) A piecewise
  parabolic method for cosmological hydrodynamics. Computer Physics
  Communications 89(1-3):149--168. \doi{10.1016/0010-4655(94)00191-4}

\bibitem[{{Burkert} and {Bodenheimer}(1993)}]{burkert1993a}
{Burkert} A, {Bodenheimer} P (1993) Multiple fragmentation in collapsing
  protostars. \mnras 264:798. \doi{10.1093/mnras/264.4.798}

\bibitem[{{Cai} et~al.(2019){Cai}, {Pastorello}, {Fraser}, {Prentice},
  {Reynolds}, {Cappellaro}, {Benetti}, {Morales-Garoffolo}, {Reguitti},
  {Elias-Rosa}, {Brennan}, {Callis}, {Cannizzaro}, {Fiore}, {Gromadzki},
  {Galindo-Guil}, {Gall}, {Heikkil{\"a}}, {Mason}, {Moran}, {Onori},
  {Sagu{\'e}s Carracedo}, and {Valerin}}]{cai2019a}
{Cai} YZ, {Pastorello} A, {Fraser} M, et~al. (2019) The transitional gap
  transient {AT~2018hso:} new insights into the luminous red nova phenomenon.
  \aap 632:L6. \doi{10.1051/0004-6361/201936749}.
  {\href{https://arxiv.org/abs/1909.13147}{{arXiv:1909.13147}}} {[astro-ph.HE]}

\bibitem[{{Campbell}(2018)}]{campbell2018a}
{Campbell} CG (2018) Magneto-hydrodynamics in binary stars, 2nd edn. Springer
  Nature, Switzerland

\bibitem[{{Carroll-Nellenback} et~al.(2013){Carroll-Nellenback}, {Shroyer},
  {Frank}, and {Ding}}]{carroll2013a}
{Carroll-Nellenback} JJ, {Shroyer} B, {Frank} A, {Ding} C (2013) {Efficient
  parallelization for AMR MHD multiphysics calculations; implementation in
  AstroBEAR}. Journal of Computational Physics 236:461--476.
  \doi{10.1016/j.jcp.2012.10.004}

\bibitem[{{Casewell} et~al.(2018){Casewell}, {Braker}, {Parsons}, {Hermes},
  {Burleigh}, {Belardi}, {Chaushev}, {Finch}, {Roy}, {Littlefair}, {Goad}, and
  {Dennihy}}]{casewell2018a}
{Casewell} SL, {Braker} IP, {Parsons} SG, et~al. (2018) The first sub-70 min
  non-interacting {WD-BD} system: {EPIC212235321}. \mnras 476(1):1405--1411.
  \doi{10.1093/mnras/sty245}.
  {\href{https://arxiv.org/abs/1801.07773}{{arXiv:1801.07773}}} {[astro-ph.SR]}

\bibitem[{{Chamandy} et~al.(2018){Chamandy}, {Frank}, {Blackman},
  {Carroll-Nellenback}, {Liu}, {Tu}, {Nordhaus}, {Chen}, and
  {Peng}}]{chamandy2018a}
{Chamandy} L, {Frank} A, {Blackman} EG, et~al. (2018) {Accretion in common
  envelope evolution}. \mnras 480(2):1898--1911. \doi{10.1093/mnras/sty1950}.
  {\href{https://arxiv.org/abs/1805.03607}{{arXiv:1805.03607}}} {[astro-ph.SR]}

\bibitem[{{Chamandy} et~al.(2019){Chamandy}, {Blackman}, {Frank},
  {Carroll-Nellenback}, {Zou}, and {Tu}}]{chamandy2019a}
{Chamandy} L, {Blackman} EG, {Frank} A, et~al. (2019) {How drag force evolves
  in global common envelope simulations}. \mnras 490(3):3727--3739.
  \doi{10.1093/mnras/stz2813}.
  {\href{https://arxiv.org/abs/1908.06195}{{arXiv:1908.06195}}} {[astro-ph.SR]}

\bibitem[{{Chamandy} et~al.(2020){Chamandy}, {Blackman}, {Frank},
  {Carroll-Nellenback}, and {Tu}}]{chamandy2020a}
{Chamandy} L, {Blackman} EG, {Frank} A, {Carroll-Nellenback} J, {Tu} Y (2020)
  {Common Envelope Evolution on the Asymptotic Giant Branch: Unbinding within a
  Decade?} \mnras \doi{10.1093/mnras/staa1273}

\bibitem[{{Chandrasekhar}(1943)}]{chandrasekhar1943a}
{Chandrasekhar} S (1943) {Dynamical Friction. I. General Considerations: the
  Coefficient of Dynamical Friction.} \apj 97:255. \doi{10.1086/144517}

\bibitem[{{Chen} and {Liu}(2013)}]{chen2014b}
{Chen} WC, {Liu} WM (2013) Evolution of neutron star + he star binaries: an
  alternative evolutionary channel to intermediate-mass binary pulsars. \mnras
  432:L75--L79. \doi{10.1093/mnrasl/slt043}.
  {\href{https://arxiv.org/abs/1303.6155}{{arXiv:1303.6155}}} {[astro-ph.SR]}

\bibitem[{{Chen} et~al.(2019){Chen}, {Coleman}, {Blackman}, and
  {Frank}}]{chen2019a}
{Chen} Z, {Coleman} MSB, {Blackman} EG, {Frank} A (2019) Solving the riemann
  problem for realistic astrophysical fluids. Journal of Computational Physics
  388:490--517. \doi{10.1016/j.jcp.2019.03.016}.
  {\href{https://arxiv.org/abs/1903.04568}{{arXiv:1903.04568}}}
  {[physics.comp-ph]}

\bibitem[{{Chevalier}(2012)}]{chevalier2012a}
{Chevalier} RA (2012) Common envelope evolution leading to supernovae with
  dense interaction. \apjl 752(1):L2. \doi{10.1088/2041-8205/752/1/L2}.
  {\href{https://arxiv.org/abs/1204.3300}{{arXiv:1204.3300}}} {[astro-ph.HE]}

\bibitem[{{Clayton} et~al.(2017){Clayton}, {Podsiadlowski}, {Ivanova}, and
  {Justham}}]{clayton2017a}
{Clayton} M, {Podsiadlowski} P, {Ivanova} N, {Justham} S (2017) {Episodic mass
  ejections from common-envelope objects}. \mnras 470(2):1788--1808.
  \doi{10.1093/mnras/stx1290}.
  {\href{https://arxiv.org/abs/1705.08457}{{arXiv:1705.08457}}} {[astro-ph.SR]}

\bibitem[{{Colella} and {Glaz}(1985)}]{colella1985a}
{Colella} P, {Glaz} HM (1985) Efficient solution algorithms for the riemann
  problem for real gases. Journal of Computational Physics 59:264--289

\bibitem[{{Colella} and {Woodward}(1984)}]{colella1984a}
{Colella} P, {Woodward} PR (1984) The {P}iecewise {P}arabolic {M}ethod {(PPM)}
  for gas-dynamical simulations. Journal of Computational Physics 54:174--201

\bibitem[{{Corradi} and {Schwarz}(1995)}]{corradi1995a}
{Corradi} RLM, {Schwarz} HE (1995) {Morphological populations of planetary
  nebulae: which progenitors? I. Comparative properties of bipolar nebulae.}
  \aap 293:871--888

\bibitem[{{Courant} et~al.(1928){Courant}, {Friedrichs}, and
  {Lewy}}]{courant1928a}
{Courant} R, {Friedrichs} KO, {Lewy} H (1928) {{\"U}ber die partiellen
  Differentialgleichungen der mathematischen Physik}. Math Ann 100:32--74

\bibitem[{{Croft} et~al.(2009){Croft}, {Di Matteo}, {Springel}, and
  {Hernquist}}]{croft2009a}
{Croft} RAC, {Di Matteo} T, {Springel} V, {Hernquist} L (2009) Galaxy
  morphology, kinematics and clustering in a hydrodynamic simulation of a
  {\ensuremath{\Lambda}} cold dark matter universe. \mnras 400(1):43--67.
  \doi{10.1111/j.1365-2966.2009.15446.x}.
  {\href{https://arxiv.org/abs/0803.4003}{{arXiv:0803.4003}}} {[astro-ph]}

\bibitem[{{Dan} et~al.(2009){Dan}, {Rosswog}, and {Br{\"u}ggen}}]{dan2009a}
{Dan} M, {Rosswog} S, {Br{\"u}ggen} M (2009) {Mass transfer dynamics in double
  degenerate binary systems}. Journal of Physics Conference Series
  172(1):012034--+. \doi{10.1088/1742-6596/172/1/012034}.
  {\href{https://arxiv.org/abs/0811.1517}{{arXiv:0811.1517}}}

\bibitem[{{Dan} et~al.(2011){Dan}, {Rosswog}, {Guillochon}, and
  {Ramirez-Ruiz}}]{dan2011a}
{Dan} M, {Rosswog} S, {Guillochon} J, {Ramirez-Ruiz} E (2011) {Prelude to A
  Double Degenerate Merger: The Onset of Mass Transfer and Its Impact on
  Gravitational Waves and Surface Detonations}. \apj 737:89.
  \doi{10.1088/0004-637X/737/2/89}.
  {\href{https://arxiv.org/abs/1101.5132}{{arXiv:1101.5132}}} {[astro-ph.HE]}

\bibitem[{{Dan} et~al.(2012){Dan}, {Rosswog}, {Guillochon}, and
  {Ramirez-Ruiz}}]{dan2012a}
{Dan} M, {Rosswog} S, {Guillochon} J, {Ramirez-Ruiz} E (2012) {How the merger
  of two white dwarfs depends on their mass ratio: orbital stability and
  detonations at contact}. \mnras p 2742.
  \doi{10.1111/j.1365-2966.2012.20794.x}.
  {\href{https://arxiv.org/abs/1201.2406}{{arXiv:1201.2406}}} {[astro-ph.HE]}

\bibitem[{Darwin(1879)}]{darwin1879a}
Darwin GH (1879) The determination of the secular effects of tidal friction by
  a graphical method. Proceedings of the Royal Society of London
  29(196-199):168--181. \doi{10.1098/rspl.1879.0028}

\bibitem[{{De} et~al.(2020){De}, {MacLeod}, {Everson}, {Antoni}, {Mandel}, and
  {Ramirez-Ruiz}}]{de2020a}
{De} S, {MacLeod} M, {Everson} RW, et~al. (2020) {Common Envelope Wind Tunnel:
  The Effects of Binary Mass Ratio and Implications for the Accretion-driven
  Growth of LIGO Binary Black Holes}. \apj 897(2):130.
  \doi{10.3847/1538-4357/ab9ac6}.
  {\href{https://arxiv.org/abs/1910.13333}{{arXiv:1910.13333}}} {[astro-ph.SR]}

\bibitem[{{De Colle} et~al.(2012){De Colle}, {Granot}, {L{\'o}pez-C{\'a}mara},
  and {Ramirez-Ruiz}}]{decolle2012a}
{De Colle} F, {Granot} J, {L{\'o}pez-C{\'a}mara} D, {Ramirez-Ruiz} E (2012)
  Gamma-ray burst dynamics and afterglow radiation from adaptive mesh
  refinement, special relativistic hydrodynamic simulations. \apj 746(2):122.
  \doi{10.1088/0004-637X/746/2/122}.
  {\href{https://arxiv.org/abs/1111.6890}{{arXiv:1111.6890}}} {[astro-ph.HE]}

\bibitem[{{de Kool}(1987)}]{dekool1987a}
{de Kool} M (1987) {Models of interacting binary stars}. PhD thesis, University
  of Amsterdam

\bibitem[{{de Kool} et~al.(1987){de Kool}, {van den Heuvel}, and
  {Pylyser}}]{dekool1987b}
{de Kool} M, {van den Heuvel} EPJ, {Pylyser} E (1987) {An evolutionary scenario
  for the black hole binary A0620-00.} \aap 183:47--52

\bibitem[{{De Marco}(2009)}]{demarco2009a}
{De Marco} O (2009) {The Origin and Shaping of Planetary Nebulae: Putting the
  Binary Hypothesis to the Test}. \pasp 121:316--342. \doi{10.1086/597765}.
  {\href{https://arxiv.org/abs/0902.1137}{{arXiv:0902.1137}}}

\bibitem[{{De Marco} and {Izzard}(2017)}]{demarco2017a}
{De Marco} O, {Izzard} RG (2017) {Dawes Review 6: The Impact of Companions on
  Stellar Evolution}. \pasa 34:e001. \doi{10.1017/pasa.2016.52}.
  {\href{https://arxiv.org/abs/1611.03542}{{arXiv:1611.03542}}} {[astro-ph.SR]}

\bibitem[{{De Marco} et~al.(2003{\natexlab{a}}){De Marco}, {Sandquist}, {Mac
  Low}, {Herwig}, and {Taam}}]{demarco2003a}
{De Marco} O, {Sandquist} EL, {Mac Low} MM, {Herwig} F, {Taam} RE
  (2003{\natexlab{a}}) {Of Wolf-Rayet Central Stars and Common Envelopes}. In:
  {Arthur} J, {Henney} WJ (eds) Revista Mexicana de Astronomia y Astrofisica
  Conference Series. Revista Mexicana de Astronomia y Astrofisica Conference
  Series, vol~15. pp 34--37

\bibitem[{{De Marco} et~al.(2003{\natexlab{b}}){De Marco}, {Sandquist}, {Mac
  Low}, {Herwig}, and {Taam}}]{demarco2003b}
{De Marco} O, {Sandquist} EL, {Mac Low} MM, {Herwig} F, {Taam} RE
  (2003{\natexlab{b}}) {Wolf-Rayet Central Stars and the Binary Evolution
  Channel}. In: {Reyes-Ruiz} M, {V{\'a}zquez-Semadeni} E (eds) Revista Mexicana
  de Astronomia y Astrofisica Conference Series. Revista Mexicana de Astronomia
  y Astrofisica Conference Series, vol~18. pp 24--30

\bibitem[{{De Marco} et~al.(2011){De Marco}, {Passy}, {Moe}, {Herwig}, {Mac
  Low}, and {Paxton}}]{demarco2011a}
{De Marco} O, {Passy} JC, {Moe} M, et~al. (2011) {On the {$\alpha$} formalism
  for the common envelope interaction}. \mnras 411:2277--2292.
  \doi{10.1111/j.1365-2966.2010.17891.x}.
  {\href{https://arxiv.org/abs/1010.4374}{{arXiv:1010.4374}}} {[astro-ph.SR]}

\bibitem[{{de Medeiros} et~al.(1996){de Medeiros}, {Da Rocha}, and
  {Mayor}}]{demedeiros1996a}
{de Medeiros} JR, {Da Rocha} C, {Mayor} M (1996) The distribution of rotational
  velocity for evolved stars. \aap 314:499--502

\bibitem[{{Dehnen}(2001)}]{dehnen2001a}
{Dehnen} W (2001) {Towards optimal softening in three-dimensional N-body codes
  - I. Minimizing the force error}. \mnras 324(2):273--291.
  \doi{10.1046/j.1365-8711.2001.04237.x}.
  {\href{https://arxiv.org/abs/astro-ph/0011568}{{arXiv:astro-ph/0011568}}}
  {[astro-ph]}

\bibitem[{{Dehnen} and {Aly}(2012)}]{dehnen2012a}
{Dehnen} W, {Aly} H (2012) Improving convergence in smoothed particle
  hydrodynamics simulations without pairing instability. \mnras
  425(2):1068--1082. \doi{10.1111/j.1365-2966.2012.21439.x}.
  {\href{https://arxiv.org/abs/1204.2471}{{arXiv:1204.2471}}} {[astro-ph.IM]}

\bibitem[{{Delgado}(1980)}]{delgado1980a}
{Delgado} AJ (1980) {Evolution of a blue supergiant with a neutron star
  companion immersed in its envelope}. \aap 87:343--348

\bibitem[{{Deloye} and {Taam}(2010)}]{deloye2010a}
{Deloye} CJ, {Taam} RE (2010) Adiabatic mass loss and the outcome of the common
  envelope phase of binary evolution. \apjl 719(1):L28--L31.
  \doi{10.1088/2041-8205/719/1/L28}.
  {\href{https://arxiv.org/abs/1007.1036}{{arXiv:1007.1036}}} {[astro-ph.SR]}

\bibitem[{{Dennis} et~al.(2008){Dennis}, {Cunningham}, {Frank}, {Balick},
  {Blackman}, and {Mitran}}]{dennis2008a}
{Dennis} TJ, {Cunningham} AJ, {Frank} A, et~al. (2008) Proto-planetary nebulae
  as explosions: Bullets versus jets and nebular shaping. \apj
  679(2):1327--1337. \doi{10.1086/587730}.
  {\href{https://arxiv.org/abs/0707.1641}{{arXiv:0707.1641}}} {[astro-ph]}

\bibitem[{{Dewi} and {Tauris}(2000)}]{dewi2000a}
{Dewi} JDM, {Tauris} TM (2000) {On the energy equation and efficiency parameter
  of the common envelope evolution}. \aap 360:1043--1051.
  {\href{https://arxiv.org/abs/astro-ph/0007034}{{astro-ph/0007034}}}

\bibitem[{{Dewi} et~al.(2006){Dewi}, {Podsiadlowski}, and {Sena}}]{dewi2006a}
{Dewi} JDM, {Podsiadlowski} P, {Sena} A (2006) {Double-core evolution and the
  formation of neutron star binaries with compact companions}. \mnras
  368:1742--1748. \doi{10.1111/j.1365-2966.2006.10233.x}.
  {\href{https://arxiv.org/abs/astro-ph/0602510}{{astro-ph/0602510}}}

\bibitem[{{Dokuchaev}(1964)}]{dokuchaev1964a}
{Dokuchaev} VP (1964) Emission of magnetoacoustic waves in the motion of stars
  in cosmic space. \sovast 8:23

\bibitem[{{Dominik} et~al.(2012){Dominik}, {Belczynski}, {Fryer}, {Holz},
  {Berti}, {Bulik}, {Mandel}, and {O'Shaughnessy}}]{dominik2012a}
{Dominik} M, {Belczynski} K, {Fryer} C, et~al. (2012) {Double Compact Objects.
  I. The Significance of the Common Envelope on Merger Rates}. \apj 759(1):52.
  \doi{10.1088/0004-637X/759/1/52}.
  {\href{https://arxiv.org/abs/1202.4901}{{arXiv:1202.4901}}} {[astro-ph.HE]}

\bibitem[{{Drake} et~al.(2009){Drake}, {Djorgovski}, {Mahabal}, {Beshore},
  {Larson}, {Graham}, {Williams}, {Christensen}, {Catelan}, {Boattini},
  {Gibbs}, {Hill}, and {Kowalski}}]{drake2009a}
{Drake} AJ, {Djorgovski} SG, {Mahabal} A, et~al. (2009) First results from the
  catalina real-time transient survey. \apj 696(1):870--884.
  \doi{10.1088/0004-637X/696/1/870}.
  {\href{https://arxiv.org/abs/0809.1394}{{arXiv:0809.1394}}} {[astro-ph]}

\bibitem[{{Eckart}(1960)}]{eckart1960a}
{Eckart} C (1960) Variation principles of hydrodynamics. Physics of Fluids
  3(3):421--427. \doi{10.1063/1.1706053}

\bibitem[{{Edelmann} et~al.(2017){Edelmann}, {R{\"o}pke}, {Hirschi}, {Georgy},
  and {Jones}}]{edelmann2017a}
{Edelmann} PVF, {R{\"o}pke} FK, {Hirschi} R, {Georgy} C, {Jones} S (2017)
  {Testing a one-dimensional prescription of dynamical shear mixing with a
  two-dimensional hydrodynamic simulation}. \aap 604:A25.
  \doi{10.1051/0004-6361/201629873}.
  {\href{https://arxiv.org/abs/1704.06261}{{arXiv:1704.06261}}} {[astro-ph.SR]}

\bibitem[{{Edelmann} et~al.(2021){Edelmann}, {Horst}, {Berberich}, {Andrassy},
  {Higl}, {Leidi}, {Klingenberg}, and {R{\"o}pke}}]{edelmann2021a}
{Edelmann} PVF, {Horst} L, {Berberich} JP, et~al. (2021) Well-balanced
  treatment of gravity in astrophysical fluid dynamics simulations at low mach
  numbers. \aap 652:A53. \doi{10.1051/0004-6361/202140653}.
  {\href{https://arxiv.org/abs/2102.13111}{{arXiv:2102.13111}}} {[astro-ph.SR]}

\bibitem[{{Edgar}(2004)}]{edgar2004a}
{Edgar} R (2004) A review of bondi-hoyle-lyttleton accretion. \nar
  48(10):843--859. \doi{10.1016/j.newar.2004.06.001}.
  {\href{https://arxiv.org/abs/astro-ph/0406166}{{arXiv:astro-ph/0406166}}}
  {[astro-ph]}

\bibitem[{{Eggleton}(1983)}]{eggleton1983a}
{Eggleton} PP (1983) {Approximations to the radii of Roche lobes}. \apj
  268:368. \doi{10.1086/160960}

\bibitem[{{Everson} et~al.(2020){Everson}, {MacLeod}, {De}, {Macias}, and
  {Ramirez-Ruiz}}]{everson2020a}
{Everson} RW, {MacLeod} M, {De} S, {Macias} P, {Ramirez-Ruiz} E (2020) Common
  envelope wind tunnel: Range of applicability and self-similarity in realistic
  stellar envelopes. \apj 899(1):77. \doi{10.3847/1538-4357/aba75c}.
  {\href{https://arxiv.org/abs/2006.07471}{{arXiv:2006.07471}}} {[astro-ph.SR]}

\bibitem[{{Faber} and {Rasio}(2012)}]{faber2012a}
{Faber} JA, {Rasio} FA (2012) Binary neutron star mergers. Living Reviews in
  Relativity 15(1):8. \doi{10.12942/lrr-2012-8}.
  {\href{https://arxiv.org/abs/1204.3858}{{arXiv:1204.3858}}} {[gr-qc]}

\bibitem[{{Ferziger} et~al.(2020){Ferziger}, {Peri\'{c}}, and
  {Street}}]{ferziger2020a}
{Ferziger} JH, {Peri\'{c}} M, {Street} RL (2020) {Computational Methods for
  Fluid Dynamics}, 4th edn. Springer Nature, Switzerland

\bibitem[{{Foglizzo} et~al.(2005){Foglizzo}, {Galletti}, and
  {Ruffert}}]{foglizzo2005a}
{Foglizzo} T, {Galletti} P, {Ruffert} M (2005) {A fresh look at the unstable
  simulations of Bondi-Hoyle-Lyttleton accretion}. \aap 435(2):397--411.
  \doi{10.1051/0004-6361:20042201}.
  {\href{https://arxiv.org/abs/astro-ph/0502168}{{arXiv:astro-ph/0502168}}}
  {[astro-ph]}

\bibitem[{{Fragos} et~al.(2019){Fragos}, {Andrews}, {Ramirez-Ruiz}, {Meynet},
  {Kalogera}, {Taam}, and {Zezas}}]{fragos2019a}
{Fragos} T, {Andrews} JJ, {Ramirez-Ruiz} E, et~al. (2019) The complete
  evolution of a neutron-star binary through a common envelope phase using {1D}
  hydrodynamic simulations. \apjl 883(2):L45. \doi{10.3847/2041-8213/ab40d1}.
  {\href{https://arxiv.org/abs/1907.12573}{{arXiv:1907.12573}}} {[astro-ph.HE]}

\bibitem[{{Frank} et~al.(2018){Frank}, {Chen}, {Reichardt}, {De Marco},
  {Blackman}, and {Nordhaus}}]{frank2018a}
{Frank} A, {Chen} Z, {Reichardt} T, et~al. (2018) {Planetary Nebulae Shaped by
  Common Envelope Evolution}. Galaxies 6(4):113. \doi{10.3390/galaxies6040113}.
  {\href{https://arxiv.org/abs/1807.05925}{{arXiv:1807.05925}}} {[astro-ph.SR]}

\bibitem[{{Fryer} and {Woosley}(1998)}]{fryer1998b}
{Fryer} CL, {Woosley} SE (1998) Helium star/black hole mergers: A new gamma-ray
  burst model. \apjl 502(1):L9--L12. \doi{10.1086/311493}.
  {\href{https://arxiv.org/abs/astro-ph/9804167}{{arXiv:astro-ph/9804167}}}
  {[astro-ph]}

\bibitem[{{Fryer} et~al.(2006){Fryer}, {Rockefeller}, and
  {Warren}}]{fryer2006a}
{Fryer} CL, {Rockefeller} G, {Warren} MS (2006) {SNSPH}: A parallel
  three-dimensional smoothed particle radiation hydrodynamics code. \apj
  643(1):292--305. \doi{10.1086/501493}.
  {\href{https://arxiv.org/abs/astro-ph/0512532}{{arXiv:astro-ph/0512532}}}
  {[astro-ph]}

\bibitem[{{Fryxell} et~al.(2000){Fryxell}, {Olson}, {Ricker}, {Timmes},
  {Zingale}, {Lamb}, {MacNeice}, {Rosner}, {Truran}, and {Tufo}}]{fryxell2000a}
{Fryxell} B, {Olson} K, {Ricker} P, et~al. (2000) {FLASH}: {A}n adaptive mesh
  hydrodynamics code for modeling astrophysical thermonuclear flashes. \apjs
  131:273--334. \doi{10.1086/317361}

\bibitem[{{Fryxell} and {Taam}(1988)}]{fryxell1988a}
{Fryxell} BA, {Taam} RE (1988) {Numerical Simulations of Nonaxisymmetric
  Adiabatic Accretion Flow}. \apj 335:862. \doi{10.1086/166973}

\bibitem[{{Galaviz} et~al.(2017){Galaviz}, {De Marco}, {Passy}, {Staff}, and
  {Iaconi}}]{galaviz2017a}
{Galaviz} P, {De Marco} O, {Passy} JC, {Staff} JE, {Iaconi} R (2017) {Common
  Envelope Light Curves. I. Grid-code Module Calibration}. \apjs 229(2):36.
  \doi{10.3847/1538-4365/aa64e1}.
  {\href{https://arxiv.org/abs/1702.07872}{{arXiv:1702.07872}}} {[astro-ph.SR]}

\bibitem[{{Garc{\'\i}a-Arredondo} and {Frank}(2004)}]{garcia2004a}
{Garc{\'\i}a-Arredondo} F, {Frank} A (2004) Collimated outflow formation via
  binary stars: Three-dimensional simulations of asymptotic giant branch wind
  and disk wind interactions. \apj 600(2):992--1003. \doi{10.1086/379821}.
  {\href{https://arxiv.org/abs/astro-ph/0307454}{{arXiv:astro-ph/0307454}}}
  {[astro-ph]}

\bibitem[{{Garc{\'i}a-Segura} et~al.(2018){Garc{\'i}a-Segura}, {Ricker}, and
  {Taam}}]{garcia2018a}
{Garc{\'i}a-Segura} G, {Ricker} PM, {Taam} RE (2018) {Common Envelope Shaping
  of Planetary Nebulae}. \apj 860(1):19. \doi{10.3847/1538-4357/aac08c}.
  {\href{https://arxiv.org/abs/1804.09309}{{arXiv:1804.09309}}} {[astro-ph.SR]}

\bibitem[{{Garc{\'\i}a-Segura} et~al.(2020){Garc{\'\i}a-Segura}, {Taam}, and
  {Ricker}}]{garcia2020a}
{Garc{\'\i}a-Segura} G, {Taam} RE, {Ricker} PM (2020) {Common Envelope Shaping
  of Planetary Nebulae. II. Magnetic Solutions and Self-collimated Outflows}.
  \apj 893(2):150. \doi{10.3847/1538-4357/ab8006}.
  {\href{https://arxiv.org/abs/2003.06073}{{arXiv:2003.06073}}} {[astro-ph.SR]}

\bibitem[{{Garc{\'\i}a-Segura} et~al.(2021){Garc{\'\i}a-Segura}, {Taam}, and
  {Ricker}}]{garcia2021a}
{Garc{\'\i}a-Segura} G, {Taam} RE, {Ricker} PM (2021) {Common Envelope Shaping
  of Planetary Nebulae. III. The Launching of Jets in Proto-Planetary Nebulae}.
  \apj 914(2):111. \doi{10.3847/1538-4357/abfc4e}.
  {\href{https://arxiv.org/abs/2104.12831}{{arXiv:2104.12831}}} {[astro-ph.SR]}

\bibitem[{{Ge} et~al.(2010){Ge}, {Hjellming}, {Webbink}, {Chen}, and
  {Han}}]{ge2010a}
{Ge} H, {Hjellming} MS, {Webbink} RF, {Chen} X, {Han} Z (2010) Adiabatic mass
  loss in binary stars. i. computational method. \apj 717(2):724--738.
  \doi{10.1088/0004-637X/717/2/724}.
  {\href{https://arxiv.org/abs/1005.3099}{{arXiv:1005.3099}}} {[astro-ph.SR]}

\bibitem[{{Geier} et~al.(2011){Geier}, {Classen}, and {Heber}}]{geier2011a}
{Geier} S, {Classen} L, {Heber} U (2011) {The Fast-rotating, Low-gravity
  Subdwarf B Star EC 22081-1916: Remnant of a Common Envelope Merger Event}.
  \apjl 733(1):L13. \doi{10.1088/2041-8205/733/1/L13}.
  {\href{https://arxiv.org/abs/1104.4202}{{arXiv:1104.4202}}} {[astro-ph.SR]}

\bibitem[{{Gilkis} et~al.(2019){Gilkis}, {Soker}, and {Kashi}}]{gilkis2019a}
{Gilkis} A, {Soker} N, {Kashi} A (2019) {Common envelope jets supernova (CEJSN)
  impostors resulting from a neutron star companion}. \mnras 482(3):4233--4242.
  \doi{10.1093/mnras/sty3008}.
  {\href{https://arxiv.org/abs/1802.08669}{{arXiv:1802.08669}}} {[astro-ph.HE]}

\bibitem[{{Ginat} et~al.(2020){Ginat}, {Glanz}, {Perets}, {Grishin}, and
  {Desjacques}}]{ginat2020a}
{Ginat} YB, {Glanz} H, {Perets} HB, {Grishin} E, {Desjacques} V (2020)
  Gravitational waves from in-spirals of compact objects in binary
  common-envelope evolution. \mnras 493(4):4861--4867.
  \doi{10.1093/mnras/staa465}.
  {\href{https://arxiv.org/abs/1903.11072}{{arXiv:1903.11072}}} {[astro-ph.SR]}

\bibitem[{{Gingold} and {Monaghan}(1977)}]{gingold1977a}
{Gingold} RA, {Monaghan} JJ (1977) {Smoothed particle hydrodynamics - Theory
  and application to non-spherical stars}. \mnras 181:375--389

\bibitem[{{Glanz} and {Perets}(2018)}]{glanz2018a}
{Glanz} H, {Perets} HB (2018) {Efficient common-envelope ejection through
  dust-driven winds}. \mnras 478(1):L12--L17. \doi{10.1093/mnrasl/sly065}.
  {\href{https://arxiv.org/abs/1801.08130}{{arXiv:1801.08130}}} {[astro-ph.SR]}

\bibitem[{{Glanz} and {Perets}(2021{\natexlab{a}})}]{glanz2021a}
{Glanz} H, {Perets} HB (2021{\natexlab{a}}) Common envelope evolution of
  eccentric binaries. \mnras 507(2):2659--2670. \doi{10.1093/mnras/stab2291}.
  {\href{https://arxiv.org/abs/2105.02227}{{arXiv:2105.02227}}} {[astro-ph.SR]}

\bibitem[{{Glanz} and {Perets}(2021{\natexlab{b}})}]{glanz2021b}
{Glanz} H, {Perets} HB (2021{\natexlab{b}}) Simulations of common envelope
  evolution in triple systems: circumstellar case. \mnras 500(2):1921--1932.
  \doi{10.1093/mnras/staa3242}.
  {\href{https://arxiv.org/abs/2004.00020}{{arXiv:2004.00020}}} {[astro-ph.SR]}

\bibitem[{{Godunov}(1959)}]{godunov1959a}
{Godunov} SK (1959) Finite difference method for numerical computation of
  discontinous solution of the equations of fluid dynamics. Matematicheskii
  Sbornik 47:271

\bibitem[{{Gonzalez-Bolivar} et~al.(2022){Gonzalez-Bolivar}, {De Marco}, {Lau},
  {Hirai}, and {Price}}]{gonzalez2022a}
{Gonzalez-Bolivar} M, {De Marco} O, {Lau} MYM, {Hirai} R, {Price} DJ (2022)
  Common envelope binary interaction simulations between a thermally-pulsating
  {AGB} star and a low mass companion. arXiv e-prints arXiv:2205.09749.
  {\href{https://arxiv.org/abs/2205.09749}{{arXiv:2205.09749}}} {[astro-ph.SR]}

\bibitem[{{Goranskij} et~al.(2004){Goranskij}, {Shugarov}, {Barsukova}, and
  {Kroll}}]{goranskij2004a}
{Goranskij} VP, {Shugarov} SY, {Barsukova} EA, {Kroll} P (2004) {V838 Mon}
  before and after its outburst. Information Bulletin on Variable Stars 5511:1

\bibitem[{{G{\'o}rski} et~al.(2005){G{\'o}rski}, {Hivon}, {Banday}, {Wandelt},
  {Hansen}, {Reinecke}, and {Bartelmann}}]{gorski2005a}
{G{\'o}rski} KM, {Hivon} E, {Banday} AJ, et~al. (2005) {HEALPix: A Framework
  for High-Resolution Discretization and Fast Analysis of Data Distributed on
  the Sphere}. \apj 622:759--771. \doi{10.1086/427976}.
  {\href{https://arxiv.org/abs/arXiv:astro-ph/0409513}{{arXiv:astro-ph/0409513}}}

\bibitem[{{Grichener} et~al.(2018){Grichener}, {Sabach}, and
  {Soker}}]{grichener2018a}
{Grichener} A, {Sabach} E, {Soker} N (2018) {The limited role of recombination
  energy in common envelope removal}. \mnras 478(2):1818--1824.
  \doi{10.1093/mnras/sty1178}.
  {\href{https://arxiv.org/abs/1803.05864}{{arXiv:1803.05864}}} {[astro-ph.SR]}

\bibitem[{{Han} et~al.(1995){Han}, {Podsiadlowski}, and {Eggleton}}]{han1995a}
{Han} Z, {Podsiadlowski} P, {Eggleton} PP (1995) The formation of bipolar
  planetary nebulae and close white dwarf binaries. \mnras 272:800--820.
  \doi{10.1093/mnras/272.4.800}

\bibitem[{{Hillel} et~al.(2022){Hillel}, {Schreier}, and {Soker}}]{hillel2022a}
{Hillel} S, {Schreier} R, {Soker} N (2022) Three-dimensional simulations of the
  jet feedback mechanism in common envelope jets supernovae. \mnras
  514(3):3212--3221. \doi{10.1093/mnras/stac1341}.
  {\href{https://arxiv.org/abs/2112.01459}{{arXiv:2112.01459}}} {[astro-ph.HE]}

\bibitem[{{Hillwig} et~al.(2016){Hillwig}, {Jones}, {De Marco}, {Bond},
  {Margheim}, and {Frew}}]{hillwig2016a}
{Hillwig} TC, {Jones} D, {De Marco} O, et~al. (2016) {Observational
  Confirmation of a Link Between Common Envelope Binary Interaction and
  Planetary Nebula Shaping}. \apj 832(2):125.
  \doi{10.3847/0004-637X/832/2/125}.
  {\href{https://arxiv.org/abs/1609.02185}{{arXiv:1609.02185}}} {[astro-ph.SR]}

\bibitem[{{Hjellming} and {Webbink}(1987)}]{hjellming1987a}
{Hjellming} MS, {Webbink} RF (1987) Thresholds for rapid mass transfer in
  binary system. {I.} polytropic models. \apj 318:794. \doi{10.1086/165412}

\bibitem[{{Holgado} et~al.(2018){Holgado}, {Ricker}, and
  {Huerta}}]{holgado2018a}
{Holgado} AM, {Ricker} PM, {Huerta} EA (2018) Gravitational waves from
  accreting neutron stars undergoing common-envelope inspiral. \apj 857(1):38.
  \doi{10.3847/1538-4357/aab6a9}.
  {\href{https://arxiv.org/abs/1706.09413}{{arXiv:1706.09413}}} {[astro-ph.HE]}

\bibitem[{{Howitt} et~al.(2020){Howitt}, {Stevenson}, {Vigna-G{\'o}mez},
  {Justham}, {Ivanova}, {Woods}, {Neijssel}, and {Mandel}}]{howitt2020a}
{Howitt} G, {Stevenson} S, {Vigna-G{\'o}mez} Ar, et~al. (2020) Luminous red
  novae: population models and future prospects. \mnras 492(3):3229--3240.
  \doi{10.1093/mnras/stz3542}.
  {\href{https://arxiv.org/abs/1912.07771}{{arXiv:1912.07771}}} {[astro-ph.HE]}

\bibitem[{{Hoyle} and {Lyttleton}(1939)}]{hoyle1939a}
{Hoyle} F, {Lyttleton} RA (1939) {The effect of interstellar matter on climatic
  variation}. Proceedings of the Cambridge Philosophical Society 35(3):405.
  \doi{10.1017/S0305004100021150}

\bibitem[{{Huarte-Espinosa} et~al.(2012){Huarte-Espinosa}, {Frank}, {Balick},
  {Blackman}, {De Marco}, {Kastner}, and {Sahai}}]{huarte2012a}
{Huarte-Espinosa} M, {Frank} A, {Balick} B, et~al. (2012) From bipolar to
  elliptical: simulating the morphological evolution of planetary nebulae.
  \mnras 424(3):2055--2068. \doi{10.1111/j.1365-2966.2012.21348.x}.
  {\href{https://arxiv.org/abs/1107.0415}{{arXiv:1107.0415}}} {[astro-ph.SR]}

\bibitem[{{Iaconi} and {De Marco}(2019)}]{iaconi2019a}
{Iaconi} R, {De Marco} O (2019) {Speaking with one voice: simulations and
  observations discuss the common envelope {\ensuremath{\alpha}} parameter}.
  \mnras 490(2):2550--2566. \doi{10.1093/mnras/stz2756}.
  {\href{https://arxiv.org/abs/1902.02039}{{arXiv:1902.02039}}} {[astro-ph.SR]}

\bibitem[{{Iaconi} et~al.(2017{\natexlab{a}}){Iaconi}, {Reichardt}, {Staff},
  {De Marco}, {Passy}, {Price}, {Wurster}, and Herwig}]{iaconi2016a}
{Iaconi} R, {Reichardt} T, {Staff} J, et~al. (2017{\natexlab{a}}) {Effect of
  initial separation on common envelope simulations: The effect of a wider
  initial separation on common envelope binary interaction simulations}. \mnras
  464(4):4028--4044. \doi{10.1093/mnras/stw2377}.
  {\href{https://arxiv.org/abs/1603.01953}{{arXiv:1603.01953}}} {[astro-ph.SR]}

\bibitem[{{Iaconi} et~al.(2017{\natexlab{b}}){Iaconi}, {Reichardt}, {Staff},
  {De Marco}, {Passy}, {Price}, {Wurster}, and Herwig}]{iaconi2017a}
{Iaconi} R, {Reichardt} T, {Staff} J, et~al. (2017{\natexlab{b}}) {Effect of
  initial separation on common envelope simulations: The effect of a wider
  initial separation on common envelope binary interaction simulations}. \mnras
  464(4):4028--4044. \doi{10.1093/mnras/stw2377}.
  {\href{https://arxiv.org/abs/1603.01953}{{arXiv:1603.01953}}} {[astro-ph.SR]}

\bibitem[{{Iaconi} et~al.(2018){Iaconi}, {De Marco}, {Passy}, and
  {Staff}}]{iaconi2018a}
{Iaconi} R, {De Marco} O, {Passy} JC, {Staff} J (2018) {The effect of binding
  energy and resolution in simulations of the common envelope binary
  interaction}. \mnras 477(2):2349--2365. \doi{10.1093/mnras/sty794}.
  {\href{https://arxiv.org/abs/1706.09786}{{arXiv:1706.09786}}} {[astro-ph.SR]}

\bibitem[{{Iaconi} et~al.(2020){Iaconi}, {Maeda}, {Nozawa}, {De Marco}, and
  {Reichardt}}]{iaconi2020a}
{Iaconi} R, {Maeda} K, {Nozawa} T, {De Marco} O, {Reichardt} T (2020)
  Properties of the post in-spiral common envelope ejecta {II:} dust formation.
  \mnras \doi{10.1093/mnras/staa2169}.
  {\href{https://arxiv.org/abs/2003.06151}{{arXiv:2003.06151}}} {[astro-ph.SR]}

\bibitem[{{Ilkov} and {Soker}(2012)}]{ilkov2012a}
{Ilkov} M, {Soker} N (2012) Type ia supernovae from very long delayed explosion
  of core-white dwarf merger. \mnras 419:1695--1700.
  \doi{10.1111/j.1365-2966.2011.19833.x}.
  {\href{https://arxiv.org/abs/1106.2027}{{arXiv:1106.2027}}} {[astro-ph.SR]}

\bibitem[{{Ivanova}(2018)}]{ivanova2018a}
{Ivanova} N (2018) On the use of hydrogen recombination energy during common
  envelope events. \apj 858(2):L24. \doi{10.3847/2041-8213/aac101}.
  {\href{https://arxiv.org/abs/1804.10681}{{arXiv:1804.10681}}} {[astro-ph.SR]}

\bibitem[{{Ivanova} and {Nandez}(2016)}]{ivanova2016a}
{Ivanova} N, {Nandez} JLA (2016) {Common envelope events with low-mass giants:
  understanding the transition to the slow spiral-in}. \mnras 462:362--381.
  \doi{10.1093/mnras/stw1676}.
  {\href{https://arxiv.org/abs/1606.04923}{{arXiv:1606.04923}}} {[astro-ph.SR]}

\bibitem[{{Ivanova} et~al.(2013){Ivanova}, {Justham}, {Chen}, {De Marco},
  {Fryer}, {Gaburov}, {Ge}, {Glebbeek}, {Han}, {Li}, {Lu}, {Marsh},
  {Podsiadlowski}, {Potter}, {Soker}, {Taam}, {Tauris}, {van den Heuvel}, and
  {Webbink}}]{ivanova2013a}
{Ivanova} N, {Justham} S, {Chen} X, et~al. (2013) {Common envelope evolution:
  where we stand and how we can move forward}. \aapr 21:59.
  \doi{10.1007/s00159-013-0059-2}.
  {\href{https://arxiv.org/abs/1209.4302}{{arXiv:1209.4302}}} {[astro-ph.HE]}

\bibitem[{{Ivanova} et~al.(2015){Ivanova}, {Justham}, and
  {Podsiadlowski}}]{ivanova2015a}
{Ivanova} N, {Justham} S, {Podsiadlowski} P (2015) {On the role of
  recombination in common-envelope ejections}. \mnras 447:2181--2197.
  \doi{10.1093/mnras/stu2582}.
  {\href{https://arxiv.org/abs/1409.3260}{{arXiv:1409.3260}}} {[astro-ph.SR]}

\bibitem[{{Ivezi{\'c}} et~al.(2019){Ivezi{\'c}}, {Kahn}, {Tyson}, {Abel},
  {Acosta}, {Allsman}, {Alonso}, {AlSayyad}, {Anderson}, {Andrew}, {Angel},
  {Angeli}, {Ansari}, {Antilogus}, {Araujo}, {Armstrong}, {Arndt}, {Astier},
  {Aubourg}, {Auza}, {Axelrod}, {Bard}, {Barr}, {Barrau}, {Bartlett}, {Bauer},
  {Bauman}, {Baumont}, {Bechtol}, {Bechtol}, {Becker}, {Becla}, {Beldica},
  {Bellavia}, {Bianco}, {Biswas}, {Blanc}, {Blazek}, {Blandford}, {Bloom},
  {Bogart}, {Bond}, {Booth}, {Borgland}, {Borne}, {Bosch}, {Boutigny},
  {Brackett}, {Bradshaw}, {Brandt}, {Brown}, {Bullock}, {Burchat}, {Burke},
  {Cagnoli}, {Calabrese}, {Callahan}, {Callen}, {Carlin}, {Carlson},
  {Chandrasekharan}, {Charles-Emerson}, {Chesley}, {Cheu}, {Chiang}, {Chiang},
  {Chirino}, {Chow}, {Ciardi}, {Claver}, {Cohen-Tanugi}, {Cockrum}, {Coles},
  {Connolly}, {Cook}, {Cooray}, {Covey}, {Cribbs}, {Cui}, {Cutri}, {Daly},
  {Daniel}, {Daruich}, {Daubard}, {Daues}, {Dawson}, {Delgado}, {Dellapenna},
  {de Peyster}, {de Val-Borro}, {Digel}, {Doherty}, {Dubois},
  {Dubois-Felsmann}, {Durech}, {Economou}, {Eifler}, {Eracleous}, {Emmons},
  {Fausti Neto}, {Ferguson}, {Figueroa}, {Fisher-Levine}, {Focke}, {Foss},
  {Frank}, {Freemon}, {Gangler}, {Gawiser}, {Geary}, {Gee}, {Geha}, {Gessner},
  {Gibson}, {Gilmore}, {Glanzman}, {Glick}, {Goldina}, {Goldstein}, {Goodenow},
  {Graham}, {Gressler}, {Gris}, {Guy}, {Guyonnet}, {Haller}, {Harris},
  {Hascall}, {Haupt}, {Hernandez}, {Herrmann}, {Hileman}, {Hoblitt}, {Hodgson},
  {Hogan}, {Howard}, {Huang}, {Huffer}, {Ingraham}, {Innes}, {Jacoby}, {Jain},
  {Jammes}, {Jee}, {Jenness}, {Jernigan}, {Jevremovi{\'c}}, {Johns}, {Johnson},
  {Johnson}, {Jones}, {Juramy-Gilles}, {Juri{\'c}}, {Kalirai}, {Kallivayalil},
  {Kalmbach}, {Kantor}, {Karst}, {Kasliwal}, {Kelly}, {Kessler}, {Kinnison},
  {Kirkby}, {Knox}, {Kotov}, {Krabbendam}, {Krughoff}, {Kub{\'a}nek},
  {Kuczewski}, {Kulkarni}, {Ku}, {Kurita}, {Lage}, {Lambert}, {Lange},
  {Langton}, {Le Guillou}, {Levine}, {Liang}, {Lim}, {Lintott}, {Long},
  {Lopez}, {Lotz}, {Lupton}, {Lust}, {MacArthur}, {Mahabal}, {Mandelbaum},
  {Markiewicz}, {Marsh}, {Marshall}, {Marshall}, {May}, {McKercher}, {McQueen},
  {Meyers}, {Migliore}, {Miller}, {Mills}, {Miraval}, {Moeyens}, {Moolekamp},
  {Monet}, {Moniez}, {Monkewitz}, {Montgomery}, {Morrison}, {Mueller},
  {Muller}, {Mu{\~n}oz Arancibia}, {Neill}, {Newbry}, {Nief}, {Nomerotski},
  {Nordby}, {O'Connor}, {Oliver}, {Olivier}, {Olsen}, {O'Mullane}, {Ortiz},
  {Osier}, {Owen}, {Pain}, {Palecek}, {Parejko}, {Parsons}, {Pease},
  {Peterson}, {Peterson}, {Petravick}, {Libby Petrick}, {Petry},
  {Pierfederici}, {Pietrowicz}, {Pike}, {Pinto}, {Plante}, {Plate}, {Plutchak},
  {Price}, {Prouza}, {Radeka}, {Rajagopal}, {Rasmussen}, {Regnault}, {Reil},
  {Reiss}, {Reuter}, {Ridgway}, {Riot}, {Ritz}, {Robinson}, {Roby}, {Roodman},
  {Rosing}, {Roucelle}, {Rumore}, {Russo}, {Saha}, {Sassolas}, {Schalk},
  {Schellart}, {Schindler}, {Schmidt}, {Schneider}, {Schneider}, {Schoening},
  {Schumacher}, {Schwamb}, {Sebag}, {Selvy}, {Sembroski}, {Seppala}, {Serio},
  {Serrano}, {Shaw}, {Shipsey}, {Sick}, {Silvestri}, {Slater}, {Smith},
  {Smith}, {Sobhani}, {Soldahl}, {Storrie-Lombardi}, {Stover}, {Strauss},
  {Street}, {Stubbs}, {Sullivan}, {Sweeney}, {Swinbank}, {Szalay}, {Takacs},
  {Tether}, {Thaler}, {Thayer}, {Thomas}, {Thornton}, {Thukral}, {Tice},
  {Trilling}, {Turri}, {Van Berg}, {Vanden Berk}, {Vetter}, {Virieux},
  {Vucina}, {Wahl}, {Walkowicz}, {Walsh}, {Walter}, {Wang}, {Wang}, {Warner},
  {Wiecha}, {Willman}, {Winters}, {Wittman}, {Wolff}, {Wood-Vasey}, {Wu},
  {Xin}, {Yoachim}, and {Zhan}}]{ivezic2019a}
{Ivezi{\'c}} {\v{Z}}, {Kahn} SM, {Tyson} JA, et~al. (2019) {LSST: From Science
  Drivers to Reference Design and Anticipated Data Products}. \apj 873(2):111.
  \doi{10.3847/1538-4357/ab042c}.
  {\href{https://arxiv.org/abs/0805.2366}{{arXiv:0805.2366}}} {[astro-ph]}

\bibitem[{{Jacoby} et~al.(2021){Jacoby}, {Hillwig}, {Jones}, {Martin}, {De
  Marco}, {Kronberger}, {Hurowitz}, {Crocker}, and {Dey}}]{jacoby2021a}
{Jacoby} GH, {Hillwig} TC, {Jones} D, et~al. (2021) Binary central stars of
  planetary nebulae identified with kepler/{K2}. \mnras 506(4):5223--5246.
  \doi{10.1093/mnras/stab2045}.
  {\href{https://arxiv.org/abs/2104.07934}{{arXiv:2104.07934}}} {[astro-ph.SR]}

\bibitem[{{Jia} and {Spruit}(2018)}]{jia2018a}
{Jia} S, {Spruit} HC (2018) Disruption of a planet spiraling into its host
  star. \apj 864(2):169. \doi{10.3847/1538-4357/aad77c}.
  {\href{https://arxiv.org/abs/1808.00467}{{arXiv:1808.00467}}} {[astro-ph.EP]}

\bibitem[{{Jones}(2016)}]{jones2016c}
{Jones} D (2016) {The discovery and characterisation of binary central stars in
  planetary nebulae}. In: Journal of Physics Conference Series. Journal of
  Physics Conference Series, vol 728. p 032014.
  \doi{10.1088/1742-6596/728/3/032014}.
  {\href{https://arxiv.org/abs/1602.00846}{{arXiv:1602.00846}}} {[astro-ph.SR]}

\bibitem[{{Kalogera} et~al.(2007){Kalogera}, {Belczynski}, {Kim},
  {O'Shaughnessy}, and {Willems}}]{kalogera2007a}
{Kalogera} V, {Belczynski} K, {Kim} C, {O'Shaughnessy} R, {Willems} B (2007)
  {Formation of double compact objects}. \physrep 442(1-6):75--108.
  \doi{10.1016/j.physrep.2007.02.008}.
  {\href{https://arxiv.org/abs/astro-ph/0612144}{{arXiv:astro-ph/0612144}}}
  {[astro-ph]}

\bibitem[{{Kami{\'n}ski} et~al.(2021){Kami{\'n}ski}, {Steffen}, {Bujarrabal},
  {Tylenda}, {Menten}, and {Hajduk}}]{kaminski2021a}
{Kami{\'n}ski} T, {Steffen} W, {Bujarrabal} V, et~al. (2021) {Molecular remnant
  of Nova 1670 (CK Vulpeculae). II. A three-dimensional view of the gas
  distribution and velocity field}. \aap 646:A1.
  \doi{10.1051/0004-6361/202039634}.
  {\href{https://arxiv.org/abs/2010.05832}{{arXiv:2010.05832}}} {[astro-ph.SR]}

\bibitem[{{Kashi} and {Soker}(2011)}]{kashi2011a}
{Kashi} A, {Soker} N (2011) A circumbinary disc in the final stages of common
  envelope and the core-degenerate scenario for {T}ype {I}a supernovae. \mnras
  417:1466--1479. \doi{10.1111/j.1365-2966.2011.19361.x}.
  {\href{https://arxiv.org/abs/1105.5698}{{arXiv:1105.5698}}} {[astro-ph.SR]}

\bibitem[{{Kasliwal}(2012)}]{kasliwal2012b}
{Kasliwal} MM (2012) {Systematically Bridging the Gap Between Novae and
  Supernovae}. \pasa 29:482--488. \doi{10.1071/AS11061}

\bibitem[{{Kim} and {Kim}(2007)}]{kim2007a}
{Kim} H, {Kim} WT (2007) Dynamical friction of a circular-orbit perturber in a
  gaseous medium. \apj 665(1):432--444. \doi{10.1086/519302}.
  {\href{https://arxiv.org/abs/0705.0084}{{arXiv:0705.0084}}} {[astro-ph]}

\bibitem[{{Kim}(2010)}]{kim2010a}
{Kim} WT (2010) Nonlinear dynamical friction of a circular-orbit perturber in a
  gaseous medium. \apj 725(1):1069--1081. \doi{10.1088/0004-637X/725/1/1069}.
  {\href{https://arxiv.org/abs/1010.1995}{{arXiv:1010.1995}}} {[astro-ph.GA]}

\bibitem[{{Kippenhahn} et~al.(2012){Kippenhahn}, {Weigert}, and
  {Weiss}}]{kippenhahn2012a}
{Kippenhahn} R, {Weigert} A, {Weiss} A (2012) {Stellar Structure and
  Evolution}. Springer-Verlag, Berlin Heidelberg.
  \doi{10.1007/978-3-642-30304-3}

\bibitem[{{Kochanek} et~al.(2014){Kochanek}, {Adams}, and
  {Belczynski}}]{kochanek2014a}
{Kochanek} CS, {Adams} SM, {Belczynski} K (2014) Stellar mergers are common.
  \mnras 443(2):1319--1328. \doi{10.1093/mnras/stu1226}.
  {\href{https://arxiv.org/abs/1405.1042}{{arXiv:1405.1042}}} {[astro-ph.SR]}

\bibitem[{{Kramer} et~al.(2020){Kramer}, {Schneider}, {Ohlmann}, {Geier},
  {Schaffenroth}, {Pakmor}, and {R{\"o}pke}}]{kramer2020a}
{Kramer} M, {Schneider} FRN, {Ohlmann} ST, et~al. (2020) Formation of
  {sdB}-stars via common envelope ejection by substellar companions. \aap
  642:A97. \doi{10.1051/0004-6361/202038702}.
  {\href{https://arxiv.org/abs/2007.00019}{{arXiv:2007.00019}}} {[astro-ph.SR]}

\bibitem[{{Kruckow} et~al.(2016){Kruckow}, {Tauris}, {Langer}, {Sz{\'e}csi},
  {Marchant}, and {Podsiadlowski}}]{kruckow2016a}
{Kruckow} MU, {Tauris} TM, {Langer} N, et~al. (2016) {Common-envelope ejection
  in massive binary stars. Implications for the progenitors of GW150914 and
  GW151226}. \aap 596:A58. \doi{10.1051/0004-6361/201629420}.
  {\href{https://arxiv.org/abs/1610.04417}{{arXiv:1610.04417}}} {[astro-ph.SR]}

\bibitem[{{Krumholz} et~al.(2004){Krumholz}, {McKee}, and
  {Klein}}]{krumholz2004a}
{Krumholz} MR, {McKee} CF, {Klein} RI (2004) {Embedding Lagrangian Sink
  Particles in Eulerian Grids}. \apj 611(1):399--412. \doi{10.1086/421935}.
  {\href{https://arxiv.org/abs/astro-ph/0312612}{{arXiv:astro-ph/0312612}}}
  {[astro-ph]}

\bibitem[{{Kuruwita} et~al.(2016){Kuruwita}, {Staff}, and {De
  Marco}}]{kuruwita2016a}
{Kuruwita} RL, {Staff} J, {De Marco} O (2016) {Considerations on the role of
  fall-back discs in the final stages of the common envelope binary
  interaction}. \mnras 461(1):486--496. \doi{10.1093/mnras/stw1414}.
  {\href{https://arxiv.org/abs/1606.04635}{{arXiv:1606.04635}}} {[astro-ph.SR]}

\bibitem[{{Kwok} et~al.(1978){Kwok}, {Purton}, and {Fitzgerald}}]{kwok1978a}
{Kwok} S, {Purton} CR, {Fitzgerald} PM (1978) On the origin of planetary
  nebulae. \apjl 219:L125--L127. \doi{10.1086/182621}

\bibitem[{{Lau} et~al.(2022{\natexlab{a}}){Lau}, {Hirai},
  {Gonz{\'a}lez-Bol{\'i}var}, {Price}, {De Marco}, and {Mandel}}]{lau2022a}
{Lau} MYM, {Hirai} R, {Gonz{\'a}lez-Bol{\'i}var} M, et~al. (2022{\natexlab{a}})
  Common envelopes in massive stars: Towards the role of radiation pressure and
  recombination energy in ejecting red supergiant envelopes. \mnras
  \doi{10.1093/mnras/stac049}.
  {\href{https://arxiv.org/abs/2111.00923}{{arXiv:2111.00923}}} {[astro-ph.SR]}

\bibitem[{{Lau} et~al.(2022{\natexlab{b}}){Lau}, {Hirai}, {Price}, and
  {Mandel}}]{lau2022b}
{Lau} MYM, {Hirai} R, {Price} DJ, {Mandel} I (2022{\natexlab{b}}) Common
  envelopes in massive stars {II}: The distinct roles of hydrogen and helium
  recombination. arXiv e-prints arXiv:2206.06411.
  {\href{https://arxiv.org/abs/2206.06411}{{arXiv:2206.06411}}} {[astro-ph.SR]}

\bibitem[{{Law} et~al.(2009){Law}, {Kulkarni}, {Dekany}, {Ofek}, {Quimby},
  {Nugent}, {Surace}, {Grillmair}, {Bloom}, {Kasliwal}, {Bildsten}, {Brown},
  {Cenko}, {Ciardi}, {Croner}, {Djorgovski}, {van Eyken}, {Filippenko}, {Fox},
  {Gal-Yam}, {Hale}, {Hamam}, {Helou}, {Henning}, {Howell}, {Jacobsen},
  {Laher}, {Mattingly}, {McKenna}, {Pickles}, {Poznanski}, {Rahmer}, {Rau},
  {Rosing}, {Shara}, {Smith}, {Starr}, {Sullivan}, {Velur}, {Walters}, and
  {Zolkower}}]{law2009a}
{Law} NM, {Kulkarni} SR, {Dekany} RG, et~al. (2009) The palomar transient
  factory: System overview, performance, and first results. \pasp
  121:1395--1408. \doi{10.1086/648598}.
  {\href{https://arxiv.org/abs/0906.5350}{{arXiv:0906.5350}}}

\bibitem[{{Law-Smith} et~al.(2020){Law-Smith}, {Everson}, {Ramirez-Ruiz}, {de
  Mink}, {van Son}, {G{\"o}tberg}, {Zellmann}, {Vigna-G{\'o}mez}, {Renzo},
  {Wu}, {Schr{\o}der}, {Foley}, and {Hutchinson-Smith}}]{law-smith2020a}
{Law-Smith} JAP, {Everson} RW, {Ramirez-Ruiz} E, et~al. (2020) {Successful
  Common Envelope Ejection and Binary Neutron Star Formation in 3D
  Hydrodynamics}. arXiv e-prints arXiv:2011.06630.
  {\href{https://arxiv.org/abs/2011.06630}{{arXiv:2011.06630}}} {[astro-ph.HE]}

\bibitem[{{Lax} and {Richtmyer}(1956)}]{lax1956a}
{Lax} PD, {Richtmyer} RD (1956) Survey of the stability of linear finite
  difference equations. Communications an Pure and Applied Mathematics
  9:267--293

\bibitem[{{Lee} and {Sahai}(2003)}]{lee2003a}
{Lee} CF, {Sahai} R (2003) Shaping proto-planetary and young planetary nebulae
  with collimated fast winds. \apj 586(1):319--337. \doi{10.1086/346265}.
  {\href{https://arxiv.org/abs/astro-ph/0211510}{{arXiv:astro-ph/0211510}}}
  {[astro-ph]}

\bibitem[{{Lee} and {Sahai}(2004)}]{lee2004a}
{Lee} CF, {Sahai} R (2004) Magnetohydrodynamic models of the bipolar knotty jet
  in henize 2-90. \apj 606(1):483--496. \doi{10.1086/381677}

\bibitem[{{Liu} et~al.(2016){Liu}, {Di Matteo}, and {Feng}}]{liu2016b}
{Liu} M, {Di Matteo} T, {Feng} Y (2016) {The effects of AGN feedback and SPH
  formulation on black hole growth in galaxies}. \mnras 458(2):1402--1416.
  \doi{10.1093/mnras/stw342}

\bibitem[{{Livio} and {Riess}(2003)}]{livio2003a}
{Livio} M, {Riess} AG (2003) Have the elusive progenitors of type ia supernovae
  been discovered? \apjl 594:L93--L94. \doi{10.1086/378765}.
  {\href{https://arxiv.org/abs/arXiv:astro-ph/0308018}{{arXiv:astro-ph/0308018}}}

\bibitem[{{Livio} and {Soker}(1984{\natexlab{a}})}]{livio1984b}
{Livio} M, {Soker} N (1984{\natexlab{a}}) {On the masses of the white dwarfs in
  cataclysmic variables}. \mnras 208:783--797. \doi{10.1093/mnras/208.4.783}

\bibitem[{{Livio} and {Soker}(1984{\natexlab{b}})}]{livio1984a}
{Livio} M, {Soker} N (1984{\natexlab{b}}) {Star-planet systems as possible
  progenitors of cataclysmic binaries}. \mnras 208:763--781.
  \doi{10.1093/mnras/208.4.763}

\bibitem[{{Livio} and {Soker}(1988)}]{livio1988a}
{Livio} M, {Soker} N (1988) {The common envelope phase in the evolution of
  binary stars}. \apj 329:764--779. \doi{10.1086/166419}

\bibitem[{{Livio} et~al.(1986){Livio}, {Soker}, {de Kool}, and
  {Savonije}}]{livio1986a}
{Livio} M, {Soker} N, {de Kool} M, {Savonije} GJ (1986) {Accretion from an
  inhomogeneous medium - III. General case and observational consequences.}
  \mnras 222:235--250. \doi{10.1093/mnras/222.2.235}

\bibitem[{{Lombardi} et~al.(2011){Lombardi}, {Holtzman}, {Dooley}, {Gearity},
  {Kalogera}, and {Rasio}}]{lombardi2011a}
{Lombardi} J J~C, {Holtzman} W, {Dooley} KL, et~al. (2011) Twin binaries:
  Studies of stability, mass transfer, and coalescence. \apj 737(2):49.
  \doi{10.1088/0004-637X/737/2/49}.
  {\href{https://arxiv.org/abs/1009.1300}{{arXiv:1009.1300}}} {[astro-ph.SR]}

\bibitem[{{L{\'o}pez-C{\'a}mara} et~al.(2019){L{\'o}pez-C{\'a}mara}, {De
  Colle}, and {Moreno M{\'e}ndez}}]{lopez2019a}
{L{\'o}pez-C{\'a}mara} D, {De Colle} F, {Moreno M{\'e}ndez} E (2019)
  {Self-regulating jets during the common-envelope phase}. \mnras
  482(3):3646--3655. \doi{10.1093/mnras/sty2959}.
  {\href{https://arxiv.org/abs/1806.11115}{{arXiv:1806.11115}}} {[astro-ph.HE]}

\bibitem[{{L{\'o}pez-C{\'a}mara} et~al.(2020){L{\'o}pez-C{\'a}mara}, {Moreno
  M{\'e}ndez}, and {De Colle}}]{lopez2020a}
{L{\'o}pez-C{\'a}mara} D, {Moreno M{\'e}ndez} E, {De Colle} F (2020) Disc
  formation and jet inclination effects in common envelopes. \mnras
  497(2):2057--2065. \doi{10.1093/mnras/staa1983}.
  {\href{https://arxiv.org/abs/2004.04158}{{arXiv:2004.04158}}} {[astro-ph.HE]}

\bibitem[{{L{\'o}pez-C{\'a}mara} et~al.(2022){L{\'o}pez-C{\'a}mara}, {De
  Colle}, {Moreno M{\'e}ndez}, {Shiber}, and {Iaconi}}]{lopez2022a}
{L{\'o}pez-C{\'a}mara} D, {De Colle} F, {Moreno M{\'e}ndez} E, {Shiber} S,
  {Iaconi} R (2022) {Jets in common envelopes: a low-mass main-sequence star in
  a red giant}. \mnras 513(3):3634--3645. \doi{10.1093/mnras/stac932}.
  {\href{https://arxiv.org/abs/2110.02227}{{arXiv:2110.02227}}} {[astro-ph.HE]}

\bibitem[{{Loveridge} et~al.(2011){Loveridge}, {van der Sluys}, and
  {Kalogera}}]{loveridge2011a}
{Loveridge} AJ, {van der Sluys} MV, {Kalogera} V (2011) Analytical expressions
  for the envelope binding energy of giants as a function of basic stellar
  parameters. \apj 743(1):49. \doi{10.1088/0004-637X/743/1/49}.
  {\href{https://arxiv.org/abs/1009.5400}{{arXiv:1009.5400}}} {[astro-ph.SR]}

\bibitem[{{Lucy}(1977)}]{lucy1977a}
{Lucy} LB (1977) {A numerical approach to the testing of the fission
  hypothesis}. \aj 82:1013--1024. \doi{10.1086/112164}

\bibitem[{{MacLeod} and {Ramirez-Ruiz}(2015{\natexlab{a}})}]{macleod2015a}
{MacLeod} M, {Ramirez-Ruiz} E (2015{\natexlab{a}}) {Asymmetric Accretion Flows
  within a Common Envelope}. \apj 803:41. \doi{10.1088/0004-637X/803/1/41}.
  {\href{https://arxiv.org/abs/1410.3823}{{arXiv:1410.3823}}} {[astro-ph.SR]}

\bibitem[{{MacLeod} and {Ramirez-Ruiz}(2015{\natexlab{b}})}]{macleod2015b}
{MacLeod} M, {Ramirez-Ruiz} E (2015{\natexlab{b}}) {On the Accretion-fed Growth
  of Neutron Stars during Common Envelope}. \apjl 798:L19.
  \doi{10.1088/2041-8205/798/1/L19}.
  {\href{https://arxiv.org/abs/1410.5421}{{arXiv:1410.5421}}} {[astro-ph.SR]}

\bibitem[{{MacLeod} et~al.(2017{\natexlab{a}}){MacLeod}, {Antoni},
  {Murguia-Berthier}, {Macias}, and {Ramirez-Ruiz}}]{macleod2017b}
{MacLeod} M, {Antoni} A, {Murguia-Berthier} A, {Macias} P, {Ramirez-Ruiz} E
  (2017{\natexlab{a}}) Common envelope wind tunnel: Coefficients of drag and
  accretion in a simplified context for studying flows around objects embedded
  within stellar envelopes. \apj 838(1):56. \doi{10.3847/1538-4357/aa6117}.
  {\href{https://arxiv.org/abs/1704.02372}{{arXiv:1704.02372}}} {[astro-ph.SR]}

\bibitem[{{MacLeod} et~al.(2017{\natexlab{b}}){MacLeod}, {Macias},
  {Ramirez-Ruiz}, {Grindlay}, {Batta}, and {Montes}}]{macleod2017a}
{MacLeod} M, {Macias} P, {Ramirez-Ruiz} E, et~al. (2017{\natexlab{b}}) {Lessons
  from the Onset of a Common Envelope Episode: the Remarkable M31 2015 Luminous
  Red Nova Outburst}. \apj 835:282. \doi{10.3847/1538-4357/835/2/282}.
  {\href{https://arxiv.org/abs/1605.01493}{{arXiv:1605.01493}}} {[astro-ph.SR]}

\bibitem[{{MacLeod} et~al.(2018{\natexlab{a}}){MacLeod}, {Ostriker}, and
  {Stone}}]{macleod2018a}
{MacLeod} M, {Ostriker} EC, {Stone} JM (2018{\natexlab{a}}) {Bound Outflows,
  Unbound Ejecta, and the Shaping of Bipolar Remnants during Stellar
  Coalescence}. \apj 868(2):136. \doi{10.3847/1538-4357/aae9eb}.
  {\href{https://arxiv.org/abs/1808.05950}{{arXiv:1808.05950}}} {[astro-ph.SR]}

\bibitem[{{MacLeod} et~al.(2018{\natexlab{b}}){MacLeod}, {Ostriker}, and
  {Stone}}]{macleod2018b}
{MacLeod} M, {Ostriker} EC, {Stone} JM (2018{\natexlab{b}}) {Runaway
  Coalescence at the Onset of Common Envelope Episodes}. \apj 863(1):5.
  \doi{10.3847/1538-4357/aacf08}.
  {\href{https://arxiv.org/abs/1803.03261}{{arXiv:1803.03261}}} {[astro-ph.SR]}

\bibitem[{{MacLeod} et~al.(2022){MacLeod}, {Vick}, and {Loeb}}]{macleod2022a}
{MacLeod} M, {Vick} M, {Loeb} A (2022) Tidal wave breaking in the eccentric
  lead-in to mass transfer and common envelope phases. arXiv e-prints
  arXiv:2203.01947.
  {\href{https://arxiv.org/abs/2203.01947}{{arXiv:2203.01947}}} {[astro-ph.SR]}

\bibitem[{{Madappatt} et~al.(2016){Madappatt}, {De Marco}, and
  {Villaver}}]{madappatt2016a}
{Madappatt} N, {De Marco} O, {Villaver} E (2016) The effect of tides on the
  population of {PN} from interacting binaries. \mnras 463(1):1040--1056.
  \doi{10.1093/mnras/stw2025}.
  {\href{https://arxiv.org/abs/1608.03041}{{arXiv:1608.03041}}} {[astro-ph.SR]}

\bibitem[{{Marchant} et~al.(2021){Marchant}, {Pappas}, {Gallegos-Garcia},
  {Berry}, {Taam}, {Kalogera}, and {Podsiadlowski}}]{marchant2021a}
{Marchant} P, {Pappas} KMW, {Gallegos-Garcia} M, et~al. (2021) The role of mass
  transfer and common envelope evolution in the formation of merging binary
  black holes. \aap 650:A107. \doi{10.1051/0004-6361/202039992}.
  {\href{https://arxiv.org/abs/2103.09243}{{arXiv:2103.09243}}} {[astro-ph.SR]}

\bibitem[{{Matsumoto} and {Metzger}(2022)}]{matsumoto2022a}
{Matsumoto} T, {Metzger} BD (2022) {Light-curve Model for Luminous Red Novae
  and Inferences about the Ejecta of Stellar Mergers}. \apj 938(1):5.
  \doi{10.3847/1538-4357/ac6269}.
  {\href{https://arxiv.org/abs/2202.10478}{{arXiv:2202.10478}}} {[astro-ph.SR]}

\bibitem[{{Menon} et~al.(2015){Menon}, {Wesolowski}, {Zheng}, {Jetley}, {Kale},
  {Quinn}, and {Governato}}]{menon2015a}
{Menon} H, {Wesolowski} L, {Zheng} G, et~al. (2015) Adaptive techniques for
  clustered {N}-body cosmological simulations. Computational Astrophysics and
  Cosmology 2:1. \doi{10.1186/s40668-015-0007-9}.
  {\href{https://arxiv.org/abs/1409.1929}{{arXiv:1409.1929}}} {[astro-ph.IM]}

\bibitem[{{Metzger} and {Pejcha}(2017)}]{metzger2017a}
{Metzger} BD, {Pejcha} O (2017) {Shock-powered light curves of luminous red
  novae as signatures of pre-dynamical mass-loss in stellar mergers}. \mnras
  471:3200--3211. \doi{10.1093/mnras/stx1768}.
  {\href{https://arxiv.org/abs/1705.03895}{{arXiv:1705.03895}}} {[astro-ph.HE]}

\bibitem[{{Meyer} and {Meyer-Hofmeister}(1979)}]{meyer1979a}
{Meyer} F, {Meyer-Hofmeister} E (1979) {Formation of cataclysmic binaries
  through common envelope evolution}. \aap 78:167--176

\bibitem[{{Mignone} et~al.(2012){Mignone}, {Zanni}, {Tzeferacos}, {van
  Straalen}, {Colella}, and {Bodo}}]{mignone2012a}
{Mignone} A, {Zanni} C, {Tzeferacos} P, et~al. (2012) {The PLUTO Code for
  Adaptive Mesh Computations in Astrophysical Fluid Dynamics}. \apjs 198:7.
  \doi{10.1088/0067-0049/198/1/7}.
  {\href{https://arxiv.org/abs/1110.0740}{{arXiv:1110.0740}}} {[astro-ph.HE]}

\bibitem[{{Miszalski} et~al.(2009){Miszalski}, {Acker}, {Moffat}, {Parker}, and
  {Udalski}}]{miszalski2009a}
{Miszalski} B, {Acker} A, {Moffat} AFJ, {Parker} QA, {Udalski} A (2009) {Binary
  planetary nebulae nuclei towards the Galactic bulge. I. Sample discovery,
  period distribution, and binary fraction}. \aap 496:813--825.
  \doi{10.1051/0004-6361/200811380}.
  {\href{https://arxiv.org/abs/0901.4419}{{arXiv:0901.4419}}} {[astro-ph.SR]}

\bibitem[{{Moe} and {Di Stefano}(2017)}]{moe2017a}
{Moe} M, {Di Stefano} R (2017) Mind your {P}s and {Q}s: The interrelation
  between period ({P}) and mass-ratio ({Q}) distributions of binary stars.
  \apjs 230(2):15. \doi{10.3847/1538-4365/aa6fb6}.
  {\href{https://arxiv.org/abs/1606.05347}{{arXiv:1606.05347}}} {[astro-ph.SR]}

\bibitem[{{Monaghan}(1992)}]{monaghan1992a}
{Monaghan} JJ (1992) Smoothed particle hydrodynamics. \araa 30:543--574.
  \doi{10.1146/annurev.aa.30.090192.002551}

\bibitem[{{Monaghan} and {Gingold}(1983)}]{monaghan1983a}
{Monaghan} JJ, {Gingold} RA (1983) Shock simulation by the particle method
  {SPH}. Journal of Computational Physics 52(2):374--389.
  \doi{10.1016/0021-9991(83)90036-0}

\bibitem[{{Monaghan} and {Lattanzio}(1985)}]{monaghan1985a}
{Monaghan} JJ, {Lattanzio} JC (1985) {A refined particle method for
  astrophysical problems}. \aap 149(1):135--143

\bibitem[{{Monaghan} and {Price}(2001)}]{monaghan2001a}
{Monaghan} JJ, {Price} DJ (2001) Variational principles for relativistic
  smoothed particle hydrodynamics. \mnras 328(2):381--392.
  \doi{10.1046/j.1365-8711.2001.04742.x}

\bibitem[{{Moreno} et~al.(2021){Moreno}, {Schneider}, {R{\"o}pke}, {Ohlmann},
  {Pakmor}, {Podsiadlowski}, and {Sand}}]{moreno2022a}
{Moreno} MM, {Schneider} FRN, {R{\"o}pke} FK, et~al. (2021) From {3D}
  hydrodynamic simulations of common-envelope interaction to gravitational-wave
  mergers. submitted to A\&A
  {\href{https://arxiv.org/abs/2111.12112}{{arXiv:2111.12112}}} {[astro-ph.SR]}

\bibitem[{{Moreno M{\'e}ndez} et~al.(2017){Moreno M{\'e}ndez},
  {L{\'o}pez-C{\'a}mara}, and {De Colle}}]{moreno2017a}
{Moreno M{\'e}ndez} E, {L{\'o}pez-C{\'a}mara} D, {De Colle} F (2017) Dynamics
  of jets during the common-envelope phase. \mnras 470(3):2929--2937.
  \doi{10.1093/mnras/stx1385}.
  {\href{https://arxiv.org/abs/1702.03293}{{arXiv:1702.03293}}} {[astro-ph.HE]}

\bibitem[{{Motl} et~al.(2017){Motl}, {Frank}, {Staff}, {Clayton}, {Fryer},
  {Even}, {Diehl}, and {Tohline}}]{motl2017a}
{Motl} PM, {Frank} J, {Staff} J, et~al. (2017) {A Comparison of Grid-based and
  SPH Binary Mass-transfer and Merger Simulations}. \apjs 229(2):27.
  \doi{10.3847/1538-4365/aa5bde}.
  {\href{https://arxiv.org/abs/1702.03562}{{arXiv:1702.03562}}} {[astro-ph.SR]}

\bibitem[{{Munari} et~al.(2002){Munari}, {Henden}, {Kiyota}, {Laney}, {Marang},
  {Zwitter}, {Corradi}, {Desidera}, {Marrese}, {Giro}, {Boschi}, and
  {Schwartz}}]{munari2002a}
{Munari} U, {Henden} A, {Kiyota} S, et~al. (2002) The mysterious eruption of
  {V838 Mon}. \aap 389:L51--L56. \doi{10.1051/0004-6361:20020715}.
  {\href{https://arxiv.org/abs/astro-ph/0205288}{{astro-ph/0205288}}}

\bibitem[{{Munday} et~al.(2020){Munday}, {Jones}, {Garc{\'\i}a-Rojas},
  {Boffin}, {Miszalski}, {Corradi}, {Rodr{\'\i}guez-Gil}, {Rubio-D{\'\i}ez},
  {Santander-Garc{\'\i}a}, and {Sowicka}}]{munday2020a}
{Munday} J, {Jones} D, {Garc{\'\i}a-Rojas} J, et~al. (2020) The
  post-common-envelope binary central star of the planetary nebula {ETHOS} 1.
  \mnras 498(4):6005--6012. \doi{10.1093/mnras/staa2753}.
  {\href{https://arxiv.org/abs/2009.03577}{{arXiv:2009.03577}}} {[astro-ph.SR]}

\bibitem[{{Munson} et~al.(2021){Munson}, {Chatzopoulos}, {Frank}, {Clayton},
  {Crawford}, {Denissenkov}, and {Herwig}}]{munson2021a}
{Munson} B, {Chatzopoulos} E, {Frank} J, et~al. (2021) {R Coronae Borealis Star
  Evolution: Simulating 3D Merger Events to 1D Stellar Evolution Including
  Large-scale Nucleosynthesis}. \apj 911(2):103.
  \doi{10.3847/1538-4357/abeb6c}.
  {\href{https://arxiv.org/abs/2103.01741}{{arXiv:2103.01741}}} {[astro-ph.SR]}

\bibitem[{{Murphy} et~al.(2009){Murphy}, {Keller}, {Schmidt}, {Tisserand},
  {Bessell}, {Francis}, and {Costa}}]{murphy2009a}
{Murphy} S, {Keller} S, {Schmidt} B, et~al. (2009) Skymapper and the southern
  sky survey: A valuable resource for stellar astrophysics. In: {S~J~Murphy \&
  M~S~Bessell} (ed) Astronomical Society of the Pacific Conference Series.
  Astronomical Society of the Pacific Conference Series, vol 404. pp 356--+

\bibitem[{{Mustill} and {Villaver}(2012)}]{mustill2012a}
{Mustill} AJ, {Villaver} E (2012) {Foretellings of Ragnar{\"o}k:
  World-engulfing Asymptotic Giants and the Inheritance of White Dwarfs}. \apj
  761:121. \doi{10.1088/0004-637X/761/2/121}.
  {\href{https://arxiv.org/abs/1210.0328}{{arXiv:1210.0328}}} {[astro-ph.EP]}

\bibitem[{{Nandez} and {Ivanova}(2016)}]{nandez2016a}
{Nandez} JLA, {Ivanova} N (2016) {Common envelope events with low-mass giants:
  understanding the energy budget}. \mnras 460:3992--4002.
  \doi{10.1093/mnras/stw1266}.
  {\href{https://arxiv.org/abs/1606.04922}{{arXiv:1606.04922}}} {[astro-ph.SR]}

\bibitem[{{Nandez} et~al.(2014){Nandez}, {Ivanova}, and
  {Lombardi}}]{nandez2014a}
{Nandez} JLA, {Ivanova} N, {Lombardi} JC Jr (2014) {V1309 Sco} -- understanding
  a merger. \apj 786:39. \doi{10.1088/0004-637X/786/1/39}.
  {\href{https://arxiv.org/abs/1311.6522}{{arXiv:1311.6522}}} {[astro-ph.SR]}

\bibitem[{{Nandez} et~al.(2015){Nandez}, {Ivanova}, and
  {Lombardi}}]{nandez2015a}
{Nandez} JLA, {Ivanova} N, {Lombardi} JC (2015) {Recombination energy in double
  white dwarf formation}. \mnras 450:L39--L43. \doi{10.1093/mnrasl/slv043}.
  {\href{https://arxiv.org/abs/1503.02750}{{arXiv:1503.02750}}} {[astro-ph.SR]}

\bibitem[{{Nelemans} and {Tauris}(1998)}]{nelemans1998a}
{Nelemans} G, {Tauris} TM (1998) Formation of undermassive single white dwarfs
  and the influence of planets on late stellar evolution. \aap 335:L85--L88.
  {\href{https://arxiv.org/abs/astro-ph/9806011}{{arXiv:astro-ph/9806011}}}
  {[astro-ph]}

\bibitem[{{Nelemans} and {Tout}(2005)}]{nelemans2005b}
{Nelemans} G, {Tout} CA (2005) {Reconstructing the evolution of white dwarf
  binaries: further evidence for an alternative algorithm for the outcome of
  the common-envelope phase in close binaries}. \mnras 356:753--764.
  \doi{10.1111/j.1365-2966.2004.08496.x}.
  {\href{https://arxiv.org/abs/astro-ph/0410301}{{astro-ph/0410301}}}

\bibitem[{{Nelemans} et~al.(2000){Nelemans}, {Verbunt}, {Yungelson}, and
  {Portegies Zwart}}]{nelemans2000a}
{Nelemans} G, {Verbunt} F, {Yungelson} LR, {Portegies Zwart} SF (2000)
  {Reconstructing the evolution of double helium white dwarfs: envelope loss
  without spiral-in}. \aap 360:1011--1018.
  {\href{https://arxiv.org/abs/arXiv:astro-ph/0006216}{{arXiv:astro-ph/0006216}}}

\bibitem[{{Nelemans} et~al.(2001{\natexlab{a}}){Nelemans}, {Portegies Zwart},
  {Verbunt}, and {Yungelson}}]{nelemans2001b}
{Nelemans} G, {Portegies Zwart} SF, {Verbunt} F, {Yungelson} LR
  (2001{\natexlab{a}}) {Population synthesis for double white dwarfs. II.
  Semi-detached systems: AM CVn stars}. \aap 368:939--949.
  \doi{10.1051/0004-6361:20010049}.
  {\href{https://arxiv.org/abs/astro-ph/0101123}{{astro-ph/0101123}}}

\bibitem[{{Nelemans} et~al.(2001{\natexlab{b}}){Nelemans}, {Yungelson},
  {Portegies Zwart}, and {Verbunt}}]{nelemans2001a}
{Nelemans} G, {Yungelson} LR, {Portegies Zwart} SF, {Verbunt} F
  (2001{\natexlab{b}}) {Population synthesis for double white dwarfs . I. Close
  detached systems}. \aap 365:491--507. \doi{10.1051/0004-6361:20000147}.
  {\href{https://arxiv.org/abs/astro-ph/0010457}{{astro-ph/0010457}}}

\bibitem[{{Nordhaus} et~al.(2011){Nordhaus}, {Wellons}, {Spiegel}, {Metzger},
  and {Blackman}}]{nordhaus2011a}
{Nordhaus} J, {Wellons} S, {Spiegel} DS, {Metzger} BD, {Blackman} EG (2011)
  {Formation of high-field magnetic white dwarfs from common envelopes}.
  Proceedings of the National Academy of Science 108:3135--3140.
  \doi{10.1073/pnas.1015005108}.
  {\href{https://arxiv.org/abs/1010.1529}{{arXiv:1010.1529}}} {[astro-ph.SR]}

\bibitem[{{Ohlmann}(2016)}]{ohlmann_phd}
{Ohlmann} ST (2016) Hydrodynamics of the common envelope phase in binary
  stellar evolution. Dissertation, Universit\"at Heidelberg.
  \doi{10.11588/heidok.00021513}, available at
  http://www.ub.uni-heidelberg.de/archiv/21513

\bibitem[{{Ohlmann} et~al.(2016{\natexlab{a}}){Ohlmann}, {R{\"o}pke}, {Pakmor},
  and {Springel}}]{ohlmann2016a}
{Ohlmann} ST, {R{\"o}pke} FK, {Pakmor} R, {Springel} V (2016{\natexlab{a}})
  {Hydrodynamic moving-mesh simulations of the common envelope phase in binary
  stellar systems}. \apjl 816(1):L9. \doi{10.3847/2041-8205/816/1/L9},
  \urlprefix\url{http://stacks.iop.org/2041-8205/816/i=1/a=L9}.
  {\href{https://arxiv.org/abs/1512.04529}{{arXiv:1512.04529}}} {[astro-ph.SR]}

\bibitem[{{Ohlmann} et~al.(2016{\natexlab{b}}){Ohlmann}, {R\"opke}, {Pakmor},
  {Springel}, and {M\"uller}}]{ohlmann2016b}
{Ohlmann} ST, {R\"opke} FK, {Pakmor} R, {Springel} V, {M\"uller} E
  (2016{\natexlab{b}}) {Magnetic Field Amplification During the Common Envelope
  Phase}. \mnras 462(1):L121--L125. \doi{10.1093/mnrasl/slw144},
  \urlprefix\url{http://mnrasl.oxfordjournals.org/content/462/1/L121.abstract}.
  {\href{https://arxiv.org/abs/1607.05996}{{arXiv:1607.05996}}} {[astro-ph.SR]}

\bibitem[{{Ohlmann} et~al.(2017){Ohlmann}, {R\"{o}pke}, {Pakmor}, and
  {Springel}}]{ohlmann2017a}
{Ohlmann} ST, {R\"{o}pke} FK, {Pakmor} R, {Springel} V (2017) {Constructing
  stable 3D hydrodynamical models of giant stars}. \aap 599:A5.
  \doi{10.1051/0004-6361/201629692}.
  {\href{https://arxiv.org/abs/1612.00008}{{arXiv:1612.00008}}} {[astro-ph.SR]}

\bibitem[{{Ondratschek} et~al.(2022){Ondratschek}, {R{\"o}pke}, {Schneider},
  {Fendt}, {Sand}, {Ohlmann}, {Pakmor}, and {Springel}}]{ondratschek2022a}
{Ondratschek} PA, {R{\"o}pke} FK, {Schneider} FRN, et~al. (2022) Bipolar
  planetary nebulae from common-envelope evolution of binary stars. \aap
  660:L8. \doi{10.1051/0004-6361/202142478}.
  {\href{https://arxiv.org/abs/2110.13177}{{arXiv:2110.13177}}} {[astro-ph.SR]}

\bibitem[{{Osher} and {Chakravarthy}(1983)}]{osher1983a}
{Osher} S, {Chakravarthy} S (1983) {Upwind Schemes and Boundary Conditions with
  Applications to Euler Equations in General Geometries}. Journal of
  Computational Physics 50(3):447--481. \doi{10.1016/0021-9991(83)90106-7}

\bibitem[{{Ostriker}(1999)}]{ostriker1999a}
{Ostriker} EC (1999) Dynamical friction in a gaseous medium. \apj 513:252--258.
  \doi{10.1086/306858}.
  {\href{https://arxiv.org/abs/astro-ph/9810324}{{astro-ph/9810324}}}

\bibitem[{Paczy{\'n}ski(1976)}]{paczynski1976a}
Paczy{\'n}ski B (1976) {Common Envelope Binaries}. In: {Eggleton} P, {Mitton}
  S, {Whelan} J (eds) Structure and Evolution of Close Binary Systems. IAU
  Symposium, vol~73. p~75

\bibitem[{{Pakmor} and {Springel}(2013)}]{pakmor2013b}
{Pakmor} R, {Springel} V (2013) {Simulations of magnetic fields in isolated
  disc galaxies}. \mnras 432:176--193. \doi{10.1093/mnras/stt428}.
  {\href{https://arxiv.org/abs/1212.1452}{{arXiv:1212.1452}}} {[astro-ph.CO]}

\bibitem[{{Pakmor} et~al.(2011){Pakmor}, {Bauer}, and {Springel}}]{pakmor2011d}
{Pakmor} R, {Bauer} A, {Springel} V (2011) {Magnetohydrodynamics on an
  unstructured moving grid}. \mnras 418:1392--1401.
  \doi{10.1111/j.1365-2966.2011.19591.x}.
  {\href{https://arxiv.org/abs/1108.1792}{{arXiv:1108.1792}}} {[astro-ph.IM]}

\bibitem[{{Pakmor} et~al.(2012){Pakmor}, {Edelmann}, {R{\"o}pke}, and
  {Hillebrandt}}]{pakmor2012b}
{Pakmor} R, {Edelmann} P, {R{\"o}pke} FK, {Hillebrandt} W (2012) {Stellar
  GADGET: a smoothed particle hydrodynamics code for stellar astrophysics and
  its application to Type Ia supernovae from white dwarf mergers}. \mnras
  424:2222--2231. \doi{10.1111/j.1365-2966.2012.21383.x}.
  {\href{https://arxiv.org/abs/1205.5806}{{arXiv:1205.5806}}} {[astro-ph.HE]}

\bibitem[{{Pakmor} et~al.(2013){Pakmor}, {Kromer}, {Taubenberger}, and
  {Springel}}]{pakmor2013a}
{Pakmor} R, {Kromer} M, {Taubenberger} S, {Springel} V (2013) Helium-ignited
  violent mergers as a unified model for normal and rapidly declining type {I}a
  supernovae. \apjl 770:L8. \doi{10.1088/2041-8205/770/1/L8}.
  {\href{https://arxiv.org/abs/1302.2913}{{arXiv:1302.2913}}} {[astro-ph.HE]}

\bibitem[{{Pakmor} et~al.(2016){Pakmor}, {Springel}, {Bauer}, {Mocz}, {Munoz},
  {Ohlmann}, {Schaal}, and {Zhu}}]{pakmor2016a}
{Pakmor} R, {Springel} V, {Bauer} A, et~al. (2016) {Improving the convergence
  properties of the moving-mesh code AREPO}. \mnras 455(1):1134--1143.
  \doi{10.1093/mnras/stv2380}.
  {\href{https://arxiv.org/abs/1503.00562}{{arXiv:1503.00562}}} {[astro-ph.GA]}

\bibitem[{{Passy} and {Bryan}(2014)}]{passy2014a}
{Passy} JC, {Bryan} GL (2014) An adaptive particle-mesh gravity solver for
  {ENZO}. \apjs 215(1):8. \doi{10.1088/0067-0049/215/1/8}.
  {\href{https://arxiv.org/abs/1410.0010}{{arXiv:1410.0010}}} {[astro-ph.IM]}

\bibitem[{{Passy} et~al.(2012){Passy}, {De Marco}, {Fryer}, {Herwig}, {Diehl},
  {Oishi}, {Mac Low}, {Bryan}, and {Rockefeller}}]{passy2012a}
{Passy} JC, {De Marco} O, {Fryer} CL, et~al. (2012) {Simulating the Common
  Envelope Phase of a Red Giant Using Smoothed-particle Hydrodynamics and
  Uniform-grid Codes}. \apj 744:52. \doi{10.1088/0004-637X/744/1/52}.
  {\href{https://arxiv.org/abs/1107.5072}{{arXiv:1107.5072}}} {[astro-ph.SR]}

\bibitem[{{Pastorello} et~al.(2019){Pastorello}, {Chen}, {Cai},
  {Morales-Garoffolo}, {Cano}, {Mason}, {Barsukova}, {Benetti}, {Berton},
  {Bose}, {Bufano}, {Callis}, {Cannizzaro}, {Cartier}, {Chen}, {Dong},
  {Dyrbye}, {Elias-Rosa}, {Fl{\"o}rs}, {Fraser}, {Geier}, {Goranskij}, {Kann},
  {Kuncarayakti}, {Onori}, {Reguitti}, {Reynolds}, {Losada}, {Sagu{\'e}s
  Carracedo}, {Schweyer}, {Smartt}, {Tatarnikov}, {Valeev}, {Vogl}, {Wevers},
  {de Ugarte Postigo}, {Izzo}, {Inserra}, {Kankare}, {Maguire}, {Smith},
  {Stalder}, {Tartaglia}, {Th{\"o}ne}, {Valerin}, and
  {Young}}]{pastorello2019a}
{Pastorello} A, {Chen} TW, {Cai} YZ, et~al. (2019) {The evolution of luminous
  red nova AT 2017jfs in NGC 4470}. \aap 625:L8.
  \doi{10.1051/0004-6361/201935511}.
  {\href{https://arxiv.org/abs/1906.00811}{{arXiv:1906.00811}}} {[astro-ph.SR]}

\bibitem[{{Paxton} et~al.(2011){Paxton}, {Bildsten}, {Dotter}, {Herwig},
  {Lesaffre}, and {Timmes}}]{paxton2011a}
{Paxton} B, {Bildsten} L, {Dotter} A, et~al. (2011) Modules for experiments in
  stellar astrophysics (mesa). \apjs 192:3. \doi{10.1088/0067-0049/192/1/3}.
  {\href{https://arxiv.org/abs/1009.1622}{{arXiv:1009.1622}}} {[astro-ph.SR]}

\bibitem[{{Pejcha}(2014)}]{pejcha2014a}
{Pejcha} O (2014) {Burying a Binary: Dynamical Mass Loss and a Continuous
  Optically thick Outflow Explain the Candidate Stellar Merger V1309 Scorpii}.
  \apj 788(1):22. \doi{10.1088/0004-637X/788/1/22}.
  {\href{https://arxiv.org/abs/1307.4088}{{arXiv:1307.4088}}} {[astro-ph.SR]}

\bibitem[{{Pejcha} et~al.(2016{\natexlab{a}}){Pejcha}, {Metzger}, and
  {Tomida}}]{pejcha2016a}
{Pejcha} O, {Metzger} BD, {Tomida} K (2016{\natexlab{a}}) {Binary stellar
  mergers with marginally bound ejecta: excretion discs, inflated envelopes,
  outflows, and their luminous transients}. \mnras 461:2527--2539.
  \doi{10.1093/mnras/stw1481}.
  {\href{https://arxiv.org/abs/1604.07414}{{arXiv:1604.07414}}} {[astro-ph.SR]}

\bibitem[{{Pejcha} et~al.(2016{\natexlab{b}}){Pejcha}, {Metzger}, and
  {Tomida}}]{pejcha2016b}
{Pejcha} O, {Metzger} BD, {Tomida} K (2016{\natexlab{b}}) {Cool and luminous
  transients from mass-losing binary stars}. \mnras 455:4351--4372.
  \doi{10.1093/mnras/stv2592}.
  {\href{https://arxiv.org/abs/1509.02531}{{arXiv:1509.02531}}} {[astro-ph.SR]}

\bibitem[{{Pelupessy} et~al.(2012){Pelupessy}, {J{\"a}nes}, and {Portegies
  Zwart}}]{pelupessy2012a}
{Pelupessy} FI, {J{\"a}nes} J, {Portegies Zwart} S (2012) N-body integrators
  with individual time steps from hierarchical splitting. \na 17(8):711--719.
  \doi{10.1016/j.newast.2012.05.009}.
  {\href{https://arxiv.org/abs/1205.5668}{{arXiv:1205.5668}}} {[astro-ph.IM]}

\bibitem[{{Podsiadlowski}(2001)}]{podsiadlowski2001a}
{Podsiadlowski} P (2001) {Common-Envelope Evolution and Stellar Mergers}. In:
  {Podsiadlowski} P, {Rappaport} S, {King} AR, {D'Antona} F, {Burderi} L (eds)
  Evolution of Binary and Multiple Star Systems. Astronomical Society of the
  Pacific Conference Series, vol 229. p 239

\bibitem[{{Politano} and {Weiler}(2007)}]{politano2007a}
{Politano} M, {Weiler} KP (2007) Population synthesis studies of close binary
  systems using a variable common envelope efficiency parameter. {I.}
  dependence on secondary mass. \apj 665:663--679. \doi{10.1086/518997}.
  {\href{https://arxiv.org/abs/astro-ph/0702662}{{astro-ph/0702662}}}

\bibitem[{{Portegies Zwart} et~al.(2009){Portegies Zwart}, {McMillan},
  {Harfst}, {Groen}, {Fujii}, {Nuall{\'a}in}, {Glebbeek}, {Heggie}, {Lombardi},
  {Hut}, {Angelou}, {Banerjee}, {Belkus}, {Fragos}, {Fregeau}, {Gaburov},
  {Izzard}, {Juri{\'c}}, {Justham}, {Sottoriva}, {Teuben}, {van Bever},
  {Yaron}, and {Zemp}}]{portegies2009a}
{Portegies Zwart} S, {McMillan} S, {Harfst} S, et~al. (2009) A multiphysics and
  multiscale software environment for modeling astrophysical systems. \na
  14(4):369--378. \doi{10.1016/j.newast.2008.10.006}.
  {\href{https://arxiv.org/abs/0807.1996}{{arXiv:0807.1996}}} {[astro-ph]}

\bibitem[{{Press} et~al.(2007){Press}, {Teukolsky}, {Vetterling}, and
  {Flannery}}]{press2007a}
{Press} WH, {Teukolsky} SA, {Vetterling} WT, {Flannery} BP (2007) Numerical
  Recipes: The Art of Scientific Computing, vol~3. Cambridge University Press

\bibitem[{{Price}(2012)}]{price2012a}
{Price} DJ (2012) Smoothed particle hydrodynamics and magnetohydrodynamics.
  Journal of Computational Physics 231(3):759--794.
  \doi{10.1016/j.jcp.2010.12.011}.
  {\href{https://arxiv.org/abs/1012.1885}{{arXiv:1012.1885}}} {[astro-ph.IM]}

\bibitem[{{Price} and {Monaghan}(2007)}]{price2007a}
{Price} DJ, {Monaghan} JJ (2007) {An energy-conserving formalism for adaptive
  gravitational force softening in smoothed particle hydrodynamics and N-body
  codes}. \mnras 374:1347--1358. \doi{10.1111/j.1365-2966.2006.11241.x}.
  {\href{https://arxiv.org/abs/arXiv:astro-ph/0610872}{{arXiv:astro-ph/0610872}}}

\bibitem[{{Price} et~al.(2018){Price}, {Wurster}, {Tricco}, {Nixon}, {Toupin},
  {Pettitt}, {Chan}, {Mentiplay}, {Laibe}, {Glover}, {Dobbs}, {Nealon},
  {Liptai}, {Worpel}, {Bonnerot}, {Dipierro}, {Ballabio}, {Ragusa},
  {Federrath}, {Iaconi}, {Reichardt}, {Forgan}, {Hutchison}, {Constantino},
  {Ayliffe}, {Hirsh}, and {Lodato}}]{price2018a}
{Price} DJ, {Wurster} J, {Tricco} TS, et~al. (2018) Phantom: {A} smoothed
  particle hydrodynamics and magnetohydrodynamics code for astrophysics. \pasa
  35:e031. \doi{10.1017/pasa.2018.25}.
  {\href{https://arxiv.org/abs/1702.03930}{{arXiv:1702.03930}}} {[astro-ph.IM]}

\bibitem[{{Prust}(2020)}]{prust2020a}
{Prust} LJ (2020) Moving and reactive boundary conditions in moving-mesh
  hydrodynamics. \mnras 494(4):4616--4626. \doi{10.1093/mnras/staa1031}.
  {\href{https://arxiv.org/abs/2002.04287}{{arXiv:2002.04287}}}
  {[physics.comp-ph]}

\bibitem[{{Prust} and {Chang}(2019)}]{prust2019a}
{Prust} LJ, {Chang} P (2019) {Common envelope evolution on a moving mesh}.
  \mnras 486(4):5809--5818. \doi{10.1093/mnras/stz1219}.
  {\href{https://arxiv.org/abs/1904.09256}{{arXiv:1904.09256}}} {[astro-ph.SR]}

\bibitem[{{Radice} et~al.(2020){Radice}, {Bernuzzi}, and
  {Perego}}]{radice2020a}
{Radice} D, {Bernuzzi} S, {Perego} A (2020) The dynamics of binary neutron star
  mergers and {GW170817}. Annual Review of Nuclear and Particle Science
  70:95--119. \doi{10.1146/annurev-nucl-013120-114541}.
  {\href{https://arxiv.org/abs/2002.03863}{{arXiv:2002.03863}}} {[astro-ph.HE]}

\bibitem[{{Rapoport} et~al.(2021){Rapoport}, {Bear}, and
  {Soker}}]{rapoport2021a}
{Rapoport} I, {Bear} E, {Soker} N (2021) The future influence of six exoplanets
  on the envelope properties of their parent stars on the giant branches.
  \mnras 506(1):468--472. \doi{10.1093/mnras/stab1774}.
  {\href{https://arxiv.org/abs/2103.14335}{{arXiv:2103.14335}}} {[astro-ph.SR]}

\bibitem[{{Rasio} and {Livio}(1996)}]{rasio1996a}
{Rasio} FA, {Livio} M (1996) {On the Formation and Evolution of Common Envelope
  Systems}. \apj 471:366. \doi{10.1086/177975}.
  {\href{https://arxiv.org/abs/astro-ph/9511054}{{astro-ph/9511054}}}

\bibitem[{{Rasio} and {Shapiro}(1992)}]{rasio1992a}
{Rasio} FA, {Shapiro} SL (1992) Hydrodynamical evolution of coalescing binary
  neutron stars. \apj 401:226. \doi{10.1086/172055}

\bibitem[{{Reg{\H o}s} and {Tout}(1995)}]{regos1995a}
{Reg{\H o}s} E, {Tout} CA (1995) {The effect of magnetic fields in
  common-envelope evolution on the formation of cataclysmic variables}. \mnras
  273:146--156. \doi{10.1093/mnras/273.1.146}

\bibitem[{{Reichardt} et~al.(2019){Reichardt}, {De Marco}, {Iaconi}, {Tout},
  and {Price}}]{reichardt2019a}
{Reichardt} TA, {De Marco} O, {Iaconi} R, {Tout} CA, {Price} DJ (2019)
  Extending common envelope simulations from roche lobe overflow to the nebular
  phase. \mnras 484(1):631--647. \doi{10.1093/mnras/sty3485}.
  {\href{https://arxiv.org/abs/1809.02297}{{arXiv:1809.02297}}} {[astro-ph.SR]}

\bibitem[{{Reichardt} et~al.(2020){Reichardt}, {De Marco}, {Iaconi},
  {Chamandy}, and {Price}}]{reichardt2020a}
{Reichardt} TA, {De Marco} O, {Iaconi} R, {Chamandy} L, {Price} DJ (2020) {The
  impact of recombination energy on simulations of the common-envelope binary
  interaction}. \mnras 494(4):5333--5349. \doi{10.1093/mnras/staa937}.
  {\href{https://arxiv.org/abs/1911.02759}{{arXiv:1911.02759}}} {[astro-ph.SR]}

\bibitem[{{Renzo} et~al.(2021){Renzo}, {Callister}, {Chatziioannou}, {van Son},
  {Mingarelli}, {Cantiello}, {Ford}, {McKernan}, and {Ashton}}]{renzo2021a}
{Renzo} M, {Callister} T, {Chatziioannou} K, et~al. (2021) Prospects of
  gravitational wave detections from common envelope evolution with {LISA}.
  \apj 919(2):128. \doi{10.3847/1538-4357/ac1110}.
  {\href{https://arxiv.org/abs/2102.00078}{{arXiv:2102.00078}}} {[astro-ph.SR]}

\bibitem[{{Rephaeli} and {Salpeter}(1980)}]{rephaeli1980a}
{Rephaeli} Y, {Salpeter} EE (1980) {Flow past a massive object and the
  gravitational drag}. \apj 240:20--24. \doi{10.1086/158202}

\bibitem[{{Rest} et~al.(2012){Rest}, {Prieto}, {Walborn}, {Smith}, {Bianco},
  {Chornock}, {Welch}, {Howell}, {Huber}, {Foley}, {Fong}, {Sinnott}, {Bond},
  {Smith}, {Toledo}, {Minniti}, and {Mandel}}]{rest2012a}
{Rest} A, {Prieto} JL, {Walborn} NR, et~al. (2012) {Light echoes reveal an
  unexpectedly cool {\ensuremath{\eta}} Carinae during its nineteenth-century
  Great Eruption}. \nat 482(7385):375--378. \doi{10.1038/nature10775}.
  {\href{https://arxiv.org/abs/1112.2210}{{arXiv:1112.2210}}} {[astro-ph.GA]}

\bibitem[{{Ricker} and {Taam}(2008)}]{ricker2008a}
{Ricker} PM, {Taam} RE (2008) {The Interaction of Stellar Objects within a
  Common Envelope}. \apjl 672:L41--L44. \doi{10.1086/526343}.
  {\href{https://arxiv.org/abs/0710.3631}{{arXiv:0710.3631}}}

\bibitem[{{Ricker} and {Taam}(2012)}]{ricker2012a}
{Ricker} PM, {Taam} RE (2012) {An AMR Study of the Common-envelope Phase of
  Binary Evolution}. \apj 746:74. \doi{10.1088/0004-637X/746/1/74}.
  {\href{https://arxiv.org/abs/1107.3889}{{arXiv:1107.3889}}} {[astro-ph.SR]}

\bibitem[{{Ricker} et~al.(2019{\natexlab{a}}){Ricker}, {Taam}, {Webbink},
  {Timmes}, and {Holgado}}]{ricker2019a}
{Ricker} PM, {Taam} RE, {Webbink} RF, {Timmes} FX, {Holgado} AM
  (2019{\natexlab{a}}) {Common Envelope Evolution of Massive Binaries}. In:
  American Astronomical Society Meeting Abstracts \#233. American Astronomical
  Society Meeting Abstracts, vol 233. p 348.11

\bibitem[{{Ricker} et~al.(2019{\natexlab{b}}){Ricker}, {Timmes}, {Taam}, and
  {Webbink}}]{ricker2019b}
{Ricker} PM, {Timmes} FX, {Taam} RE, {Webbink} RF (2019{\natexlab{b}}) Common
  envelope evolution of massive stars. In: {Oskinova} LM, {Bozzo} E, {Bulik} T,
  {Gies} DR (eds) IAU Symposium. IAU Symposium, vol 346. pp 449--454.
  \doi{10.1017/S1743921318007433}.
  {\href{https://arxiv.org/abs/1811.03656}{{arXiv:1811.03656}}} {[astro-ph.SR]}

\bibitem[{{Robertson} et~al.(2010){Robertson}, {Kravtsov}, {Gnedin}, {Abel},
  and {Rudd}}]{robertson2010a}
{Robertson} BE, {Kravtsov} AV, {Gnedin} NY, {Abel} T, {Rudd} DH (2010)
  Computational eulerian hydrodynamics and {G}alilean invariance. \mnras
  401(4):2463--2476. \doi{10.1111/j.1365-2966.2009.15823.x}.
  {\href{https://arxiv.org/abs/0909.0513}{{arXiv:0909.0513}}} {[astro-ph.CO]}

\bibitem[{{Rogers} and {Nayfonov}(2002)}]{rogers2002a}
{Rogers} FJ, {Nayfonov} A (2002) {Updated and Expanded OPAL Equation-of-State
  Tables: Implications for Helioseismology}. \apj 576:1064--1074.
  \doi{10.1086/341894}

\bibitem[{{Rogers} et~al.(1996){Rogers}, {Swenson}, and
  {Iglesias}}]{rogers1996a}
{Rogers} FJ, {Swenson} FJ, {Iglesias} CA (1996) {OPAL Equation-of-State Tables
  for Astrophysical Applications}. \apj 456:902. \doi{10.1086/176705}

\bibitem[{{Rosswog}(2015)}]{rosswog2015a}
{Rosswog} S (2015) {SPH Methods in the Modelling of Compact Objects}. Living
  Reviews in Computational Astrophysics 1. \doi{10.1007/lrca-2015-1}.
  {\href{https://arxiv.org/abs/1406.4224}{{arXiv:1406.4224}}} {[astro-ph.IM]}

\bibitem[{{Rosswog} et~al.(1999){Rosswog}, {Liebend{\"o}rfer}, {Thielemann},
  {Davies}, {Benz}, and {Piran}}]{rosswog1999a}
{Rosswog} S, {Liebend{\"o}rfer} M, {Thielemann} FK, et~al. (1999) {Mass
  ejection in neutron star mergers}. \aap 341:499--526.
  {\href{https://arxiv.org/abs/astro-ph/9811367}{{arXiv:astro-ph/9811367}}}
  {[astro-ph]}

\bibitem[{{Rosswog} et~al.(2004){Rosswog}, {Speith}, and {Wynn}}]{rosswog2004a}
{Rosswog} S, {Speith} R, {Wynn} GA (2004) {Accretion dynamics in neutron
  star-black hole binaries}. \mnras 351:1121--1133.
  \doi{10.1111/j.1365-2966.2004.07865.x}.
  {\href{https://arxiv.org/abs/arXiv:astro-ph/0403500}{{arXiv:astro-ph/0403500}}}

\bibitem[{{R\'{o}{ż}yczka} and {Spruit}(1989)}]{rozyczka1989a}
{R\'{o}{ż}yczka} M, {Spruit} HC (1989) {Spiral shocks in accretion disks: a
  preliminary numerical study}. In: {Meyer} F (ed) Theory of Accretion Disks.
  NATO Advanced Study Institute (ASI) Series C, vol 290. pp 341--354

\bibitem[{{Ruderman} and {Spiegel}(1971)}]{ruderman1971a}
{Ruderman} MA, {Spiegel} EA (1971) Galactic wakes. \apj 165:1.
  \doi{10.1086/150870}

\bibitem[{{Ruffert}(1993)}]{ruffert1993a}
{Ruffert} M (1993) {Collisions between a white dwarf and a main-sequence star.
  3: Simulations including the white dwarf surface}. \aap 280(1):141--156

\bibitem[{{Ruffert}(1994)}]{ruffert1994b}
{Ruffert} M (1994) {Three-dimensional hydrodynamic Bondi-Hoyle accretion. III.
  Mach 0.6, 1.4 and 10; {\ensuremath{\gamma}}=5/3.} \aaps 106:505--522

\bibitem[{{Ruffert}(1995)}]{ruffert1995a}
{Ruffert} M (1995) {Three-dimensional hydrodynamic Bondi-Hoyle accretion. IV.
  Specific heat ratio 4/3.} \aaps 113:133.
  {\href{https://arxiv.org/abs/astro-ph/9503026}{{arXiv:astro-ph/9503026}}}
  {[astro-ph]}

\bibitem[{{Ruffert}(1996)}]{ruffert1996a}
{Ruffert} M (1996) {Three-dimensional hydrodynamic Bondi-Hoyle accretion. V.
  Specific heat ratio 1.01, nearly isothermal flow.} \aap 311:817--832.
  {\href{https://arxiv.org/abs/astro-ph/9510021}{{arXiv:astro-ph/9510021}}}
  {[astro-ph]}

\bibitem[{{Ruffert}(1997)}]{ruffert1997a}
{Ruffert} M (1997) {Non-axisymmetric wind-accretion simulations. I. Velocity
  gradients of 3\% and 20\% over one accretion radius.} \aap 317:793--814.
  {\href{https://arxiv.org/abs/astro-ph/9605072}{{arXiv:astro-ph/9605072}}}
  {[astro-ph]}

\bibitem[{{Ruffert}(1999)}]{ruffert1999a}
{Ruffert} M (1999) {Non-axisymmetric wind-accretion simulations. II. Density
  gradients}. \aap 346:861--877.
  {\href{https://arxiv.org/abs/astro-ph/9903304}{{arXiv:astro-ph/9903304}}}
  {[astro-ph]}

\bibitem[{{Ruffert} et~al.(1996){Ruffert}, {Janka}, and
  {Schaefer}}]{ruffert1996b}
{Ruffert} M, {Janka} HT, {Schaefer} G (1996) Coalescing neutron stars - a step
  towards physical models. {I.} hydrodynamic evolution and gravitational-wave
  emission. \aap 311:532--566.
  {\href{https://arxiv.org/abs/astro-ph/9509006}{{arXiv:astro-ph/9509006}}}
  {[astro-ph]}

\bibitem[{{Ruiter} et~al.(2009){Ruiter}, {Belczynski}, and
  {Fryer}}]{ruiter2009a}
{Ruiter} AJ, {Belczynski} K, {Fryer} C (2009) Rates and delay times of type ia
  supernovae. \apj 699:2026--2036. \doi{10.1088/0004-637X/699/2/2026}.
  {\href{https://arxiv.org/abs/0904.3108}{{arXiv:0904.3108}}}

\bibitem[{{Sabach} and {Soker}(2015)}]{sabach2015a}
{Sabach} E, {Soker} N (2015) {A formation scenario for the triple pulsar PSR
  J0337+1715: breaking a binary system inside a common envelope}. \mnras
  450(2):1716--1723. \doi{10.1093/mnras/stv717}.
  {\href{https://arxiv.org/abs/1501.06787}{{arXiv:1501.06787}}} {[astro-ph.SR]}

\bibitem[{{Sabach} et~al.(2017){Sabach}, {Hillel}, {Schreier}, and
  {Soker}}]{sabach2017a}
{Sabach} E, {Hillel} S, {Schreier} R, {Soker} N (2017) {Energy transport by
  convection in the common envelope evolution}. \mnras 472(4):4361--4367.
  \doi{10.1093/mnras/stx2272}.
  {\href{https://arxiv.org/abs/1706.05838}{{arXiv:1706.05838}}} {[astro-ph.SR]}

\bibitem[{{Sana} et~al.(2012){Sana}, {de Mink}, {de Koter}, {Langer}, {Evans},
  {Gieles}, {Gosset}, {Izzard}, {Le Bouquin}, and {Schneider}}]{sana2012a}
{Sana} H, {de Mink} SE, {de Koter} A, et~al. (2012) {Binary Interaction
  Dominates the Evolution of Massive Stars}. Science 337:444.
  \doi{10.1126/science.1223344}.
  {\href{https://arxiv.org/abs/1207.6397}{{arXiv:1207.6397}}} {[astro-ph.SR]}

\bibitem[{{Sand} et~al.(2020){Sand}, {Ohlmann}, {Schneider}, {Pakmor}, and
  {R{\"o}pke}}]{sand2020a}
{Sand} C, {Ohlmann} ST, {Schneider} FRN, {Pakmor} R, {R{\"o}pke} FK (2020)
  {Common-envelope evolution with an asymptotic giant branch star}. \aap
  644:A60. \doi{10.1051/0004-6361/202038992}.
  {\href{https://arxiv.org/abs/2007.11000}{{arXiv:2007.11000}}} {[astro-ph.SR]}

\bibitem[{{Sandquist} et~al.(1998){Sandquist}, {Taam}, {Chen}, {Bodenheimer},
  and {Burkert}}]{sandquist1998a}
{Sandquist} EL, {Taam} RE, {Chen} X, {Bodenheimer} P, {Burkert} A (1998)
  {Double Core Evolution. X. Through the Envelope Ejection Phase}. \apj
  500:909. \doi{10.1086/305778}.
  {\href{https://arxiv.org/abs/arXiv:astro-ph/9801230}{{arXiv:astro-ph/9801230}}}

\bibitem[{{Sandquist} et~al.(2000){Sandquist}, {Taam}, and
  {Burkert}}]{sandquist2000a}
{Sandquist} EL, {Taam} RE, {Burkert} A (2000) {On the Formation of Helium
  Double Degenerate Stars and Pre-Cataclysmic Variables}. \apj 533:984--997.
  \doi{10.1086/308687}.
  {\href{https://arxiv.org/abs/astro-ph/9912243}{{astro-ph/9912243}}}

\bibitem[{{Schaffenroth} et~al.(2014){Schaffenroth}, {Classen}, {Nagel},
  {Geier}, {Koen}, {Heber}, and {Edelmann}}]{schaffenroth2014a}
{Schaffenroth} V, {Classen} L, {Nagel} K, et~al. (2014) {Two candidate brown
  dwarf companions around core helium-burning stars}. \aap 570:A70.
  \doi{10.1051/0004-6361/201424616}.
  {\href{https://arxiv.org/abs/1409.4357}{{arXiv:1409.4357}}} {[astro-ph.SR]}

\bibitem[{{Schaffenroth} et~al.(2015){Schaffenroth}, {Barlow}, {Drechsel}, and
  {Dunlap}}]{schaffenroth2015a}
{Schaffenroth} V, {Barlow} BN, {Drechsel} H, {Dunlap} BH (2015) An eclipsing
  post common-envelope system consisting of a pulsating hot subdwarf {B} star
  and a brown dwarf companion. \aap 576:A123.
  \doi{10.1051/0004-6361/201525701}.
  {\href{https://arxiv.org/abs/1502.04459}{{arXiv:1502.04459}}} {[astro-ph.SR]}

\bibitem[{{Schaffenroth} et~al.(2019){Schaffenroth}, {Barlow}, {Geier},
  {Vu{\v{c}}kovi{\'c}}, {Kilkenny}, {Wolz}, {Kupfer}, {Heber}, {Drechsel},
  {Kimeswenger}, {Marsh}, {Wolf}, {Pelisoli}, {Freudenthal}, {Dreizler},
  {Kreuzer}, and {Ziegerer}}]{schaffenroth2019a}
{Schaffenroth} V, {Barlow} BN, {Geier} S, et~al. (2019) The {EREBOS} project:
  Investigating the effect of substellar and low-mass stellar companions on
  late stellar evolution. survey, target selection, and atmospheric parameters.
  \aap 630:A80. \doi{10.1051/0004-6361/201936019}.
  {\href{https://arxiv.org/abs/1907.09892}{{arXiv:1907.09892}}} {[astro-ph.SR]}

\bibitem[{{Schneider} et~al.(2019){Schneider}, {Ohlmann}, {Podsiadlowski},
  {R{\"o}pke}, {Balbus}, {Pakmor}, and {Springel}}]{schneider2019a}
{Schneider} FRN, {Ohlmann} ST, {Podsiadlowski} P, et~al. (2019) {Stellar
  mergers as the origin of magnetic massive stars}. \nat 574(7777):211--214.
  \doi{10.1038/s41586-019-1621-5}.
  {\href{https://arxiv.org/abs/1910.14058}{{arXiv:1910.14058}}} {[astro-ph.SR]}

\bibitem[{{Schneider} et~al.(2020){Schneider}, {Ohlmann}, {Podsiadlowski},
  {R{\"o}pke}, {Balbus}, and {Pakmor}}]{schneider2020a}
{Schneider} FRN, {Ohlmann} ST, {Podsiadlowski} P, et~al. (2020) Long-term
  evolution of a magnetic massive merger product. \mnras 495(3):2796--2812.
  \doi{10.1093/mnras/staa1326}.
  {\href{https://arxiv.org/abs/2005.05335}{{arXiv:2005.05335}}} {[astro-ph.SR]}

\bibitem[{{Schreiber} and {G{\"a}nsicke}(2003)}]{schreiber2003a}
{Schreiber} MR, {G{\"a}nsicke} BT (2003) {The age, life expectancy, and space
  density of Post Common Envelope Binaries}. \aap 406:305--321.
  \doi{10.1051/0004-6361:20030801}.
  {\href{https://arxiv.org/abs/astro-ph/0305531}{{astro-ph/0305531}}}

\bibitem[{{Schreier} et~al.(2019){Schreier}, {Hillel}, and
  {Soker}}]{schreier2019a}
{Schreier} R, {Hillel} S, {Soker} N (2019) Inclined jets inside a common
  envelope of a triple stellar system. \mnras 490(4):4748--4755.
  \doi{10.1093/mnras/stz2914}.
  {\href{https://arxiv.org/abs/1907.13175}{{arXiv:1907.13175}}} {[astro-ph.SR]}

\bibitem[{{Schreier} et~al.(2021){Schreier}, {Hillel}, {Shiber}, and
  {Soker}}]{schreier2021a}
{Schreier} R, {Hillel} S, {Shiber} S, {Soker} N (2021) Simulating highly
  eccentric common envelope jet supernova impostors. \mnras 508(2):2386--2398.
  \doi{10.1093/mnras/stab2687}.
  {\href{https://arxiv.org/abs/2106.11601}{{arXiv:2106.11601}}} {[astro-ph.HE]}

\bibitem[{{Schreier} et~al.(2022){Schreier}, {Hillel}, and
  {Soker}}]{schreier2023a}
{Schreier} R, {Hillel} S, {Soker} N (2022) Simulating the deposition of angular
  momentum by jets in common envelope evolution. arXiv e-prints
  arXiv:2209.13573.
  {\href{https://arxiv.org/abs/2209.13573}{{arXiv:2209.13573}}} {[astro-ph.HE]}

\bibitem[{{Schr{\o}der} et~al.(2020){Schr{\o}der}, {MacLeod}, {Loeb},
  {Vigna-G{\'o}mez}, and {Mandel}}]{schroder2020a}
{Schr{\o}der} SL, {MacLeod} M, {Loeb} A, {Vigna-G{\'o}mez} A, {Mandel} I (2020)
  {Explosions Driven by the Coalescence of a Compact Object with the Core of a
  Massive-star Companion inside a Common Envelope: Circumstellar Properties,
  Light Curves, and Population Statistics}. \apj 892(1):13.
  \doi{10.3847/1538-4357/ab7014}.
  {\href{https://arxiv.org/abs/1906.04189}{{arXiv:1906.04189}}} {[astro-ph.HE]}

\bibitem[{{Shibata} and {Hotokezaka}(2019)}]{shibata2019a}
{Shibata} M, {Hotokezaka} K (2019) Merger and mass ejection of neutron star
  binaries. Annual Review of Nuclear and Particle Science 69:41--64.
  \doi{10.1146/annurev-nucl-101918-023625}.
  {\href{https://arxiv.org/abs/1908.02350}{{arXiv:1908.02350}}} {[astro-ph.HE]}

\bibitem[{{Shiber} and {Soker}(2018)}]{shiber2018a}
{Shiber} S, {Soker} N (2018) Simulating a binary system that experiences the
  grazing envelope evolution. \mnras 477(2):2584--2598.
  \doi{10.1093/mnras/sty843}.
  {\href{https://arxiv.org/abs/1706.00398}{{arXiv:1706.00398}}} {[astro-ph.SR]}

\bibitem[{{Shiber} et~al.(2017){Shiber}, {Kashi}, and {Soker}}]{shiber2017a}
{Shiber} S, {Kashi} A, {Soker} N (2017) {Simulating the onset of grazing
  envelope evolution of binary stars}. \mnras 465(1):L54--L58.
  \doi{10.1093/mnrasl/slw208}.
  {\href{https://arxiv.org/abs/1607.00839}{{arXiv:1607.00839}}} {[astro-ph.SR]}

\bibitem[{{Shiber} et~al.(2019){Shiber}, {Iaconi}, {De Marco}, and
  {Soker}}]{shiber2019a}
{Shiber} S, {Iaconi} R, {De Marco} O, {Soker} N (2019) Companion-launched jets
  and their effect on the dynamics of common envelope interaction simulations.
  \mnras 488(4):5615--5632. \doi{10.1093/mnras/stz2013}.
  {\href{https://arxiv.org/abs/1902.03931}{{arXiv:1902.03931}}} {[astro-ph.SR]}

\bibitem[{{Shima} et~al.(1985){Shima}, {Matsuda}, {Takeda}, and
  {Sawada}}]{shima1985a}
{Shima} E, {Matsuda} T, {Takeda} H, {Sawada} K (1985) {Hydrodynamic
  calculations of axisymmetric accretion flow}. \mnras 217:367--386.
  \doi{10.1093/mnras/217.2.367}

\bibitem[{{Smith} et~al.(2016){Smith}, {Andrews}, {Van Dyk}, {Mauerhan},
  {Kasliwal}, {Bond}, {Filippenko}, {Clubb}, {Graham}, {Perley}, {Jencson},
  {Bally}, {Ubeda}, and {Sabbi}}]{smith2016a}
{Smith} N, {Andrews} JE, {Van Dyk} SD, et~al. (2016) {Massive star mergers and
  the recent transient in NGC 4490: a more massive cousin of V838 Mon and V1309
  Sco}. \mnras 458:950--962. \doi{10.1093/mnras/stw219}.
  {\href{https://arxiv.org/abs/1602.05203}{{arXiv:1602.05203}}} {[astro-ph.SR]}

\bibitem[{{Soker}(1994)}]{soker1994a}
{Soker} N (1994) Influences of wide binaries on the structures of planetary
  nebulae. \mnras 270:774

\bibitem[{{Soker}(1998{\natexlab{a}})}]{soker1998c}
{Soker} N (1998{\natexlab{a}}) {Binary Progenitor Models for Bipolar Planetary
  Nebulae}. \apj 496(2):833--841. \doi{10.1086/305407}

\bibitem[{{Soker}(1998{\natexlab{b}})}]{soker1998a}
{Soker} N (1998{\natexlab{b}}) Can planets influence the horizontal branch
  morphology? \aj 116(3):1308--1313. \doi{10.1086/300503}.
  {\href{https://arxiv.org/abs/astro-ph/9803223}{{arXiv:astro-ph/9803223}}}
  {[astro-ph]}

\bibitem[{{Soker}(1998{\natexlab{c}})}]{soker1998b}
{Soker} N (1998{\natexlab{c}}) Magnetic field, dust and axisymmetrical mass
  loss on the asymptotic giant branch. \mnras 299(4):1242--1248.
  \doi{10.1046/j.1365-8711.1998.01884.x}.
  {\href{https://arxiv.org/abs/astro-ph/9808289}{{arXiv:astro-ph/9808289}}}
  {[astro-ph]}

\bibitem[{{Soker}(2004{\natexlab{a}})}]{soker2004c}
{Soker} N (2004{\natexlab{a}}) Bubbles in planetary nebulae and clusters of
  galaxies: Jet properties. \aap 414:943--947.
  \doi{10.1051/0004-6361:20034120}.
  {\href{https://arxiv.org/abs/astro-ph/0309095}{{arXiv:astro-ph/0309095}}}
  {[astro-ph]}

\bibitem[{{Soker}(2004{\natexlab{b}})}]{soker2004b}
{Soker} N (2004{\natexlab{b}}) Shaping planetary nebulae and related objects.
  In: {Meixner} M, {Kastner} JH, {Balick} B, {Soker} N (eds) Asymmetrical
  Planetary Nebulae III: Winds, Structure and the Thunderbird. Astronomical
  Society of the Pacific Conference Series, vol 313. p 562.
  {\href{https://arxiv.org/abs/astro-ph/0309228}{{arXiv:astro-ph/0309228}}}
  {[astro-ph]}

\bibitem[{{Soker}(2015)}]{soker2015a}
{Soker} N (2015) {Close Stellar Binary Systems by Grazing Envelope Evolution}.
  \apj 800(2):114. \doi{10.1088/0004-637X/800/2/114}.
  {\href{https://arxiv.org/abs/1410.5363}{{arXiv:1410.5363}}} {[astro-ph.SR]}

\bibitem[{{Soker}(2016{\natexlab{a}})}]{soker2016c}
{Soker} N (2016{\natexlab{a}}) {Intermediate luminosity optical transients
  during the grazing envelope evolution (GEE)}. \na 47:16--18.
  \doi{10.1016/j.newast.2016.01.005}.
  {\href{https://arxiv.org/abs/1601.05913}{{arXiv:1601.05913}}} {[astro-ph.SR]}

\bibitem[{{Soker}(2016{\natexlab{b}})}]{soker2016b}
{Soker} N (2016{\natexlab{b}}) The jet feedback mechanism ({JFM}) in stars,
  galaxies and clusters. \nar 75:1--23. \doi{10.1016/j.newar.2016.08.002}.
  {\href{https://arxiv.org/abs/1605.02672}{{arXiv:1605.02672}}} {[astro-ph.HE]}

\bibitem[{{Soker}(2017)}]{soker2017a}
{Soker} N (2017) {Energizing the last phase of common-envelope removal}. \mnras
  471(4):4839--4843. \doi{10.1093/mnras/stx1978}.
  {\href{https://arxiv.org/abs/1706.03720}{{arXiv:1706.03720}}} {[astro-ph.SR]}

\bibitem[{{Soker} and {Gilkis}(2018)}]{soker2018b}
{Soker} N, {Gilkis} A (2018) {Explaining iPTF14hls as a common-envelope jets
  supernova}. \mnras 475(1):1198--1202. \doi{10.1093/mnras/stx3287}.
  {\href{https://arxiv.org/abs/1711.05180}{{arXiv:1711.05180}}} {[astro-ph.HE]}

\bibitem[{{Soker} and {Tylenda}(2006)}]{soker2006a}
{Soker} N, {Tylenda} R (2006) {Violent stellar merger model for transient
  events}. \mnras 373:733--738. \doi{10.1111/j.1365-2966.2006.11056.x}.
  {\href{https://arxiv.org/abs/astro-ph/0606467}{{astro-ph/0606467}}}

\bibitem[{{Soker} et~al.(1984){Soker}, {Livio}, and {Harpaz}}]{soker1984a}
{Soker} N, {Livio} M, {Harpaz} A (1984) {The evolution of a star-'planet'
  system in the double core phase}. \mnras 210:189--195.
  \doi{10.1093/mnras/210.2.189}

\bibitem[{{Soker} et~al.(1986){Soker}, {Livio}, {de Kool}, and
  {Savonije}}]{soker1986a}
{Soker} N, {Livio} M, {de Kool} M, {Savonije} GJ (1986) {Accretion of angular
  momentum from an inhomogeneous medium. II - Isothermal flow}. \mnras
  221:445--452. \doi{10.1093/mnras/221.2.445}

\bibitem[{{Soker} et~al.(2013){Soker}, {Kashi}, {Garc{\'{\i}}a-Berro},
  {Torres}, and {Camacho}}]{soker2013a}
{Soker} N, {Kashi} A, {Garc{\'{\i}}a-Berro} E, {Torres} S, {Camacho} J (2013)
  Explaining the type ia supernova ptf 11kx with a violent prompt merger
  scenario. \mnras 431:1541--1546. \doi{10.1093/mnras/stt271}.
  {\href{https://arxiv.org/abs/1207.5770}{{arXiv:1207.5770}}} {[astro-ph.SR]}

\bibitem[{{Soker} et~al.(2018){Soker}, {Grichener}, and {Sabach}}]{soker2018a}
{Soker} N, {Grichener} A, {Sabach} E (2018) {Radiating the Hydrogen
  Recombination Energy during Common Envelope Evolution}. \apjl 863(1):L14.
  \doi{10.3847/2041-8213/aad736}.
  {\href{https://arxiv.org/abs/1805.08543}{{arXiv:1805.08543}}} {[astro-ph.SR]}

\bibitem[{{Soker} et~al.(2019){Soker}, {Grichener}, and {Gilkis}}]{soker2019b}
{Soker} N, {Grichener} A, {Gilkis} A (2019) {Diversity of common envelope jets
  supernovae and the fast transient AT2018cow}. \mnras 484(4):4972--4979.
  \doi{10.1093/mnras/stz364}.
  {\href{https://arxiv.org/abs/1811.11106}{{arXiv:1811.11106}}} {[astro-ph.HE]}

\bibitem[{{Sparks} and {Stecher}(1974)}]{sparks1974a}
{Sparks} WM, {Stecher} TP (1974) Supernova: The result of the death spiral of a
  white dwarf into a red giant. \apj 188:149. \doi{10.1086/152697}

\bibitem[{{Springel}(2005)}]{springel2005a}
{Springel} V (2005) The cosmological simulation code {GADGET-2}. \mnras
  364:1105--1134. \doi{10.1111/j.1365-2966.2005.09655.x}.
  {\href{https://arxiv.org/abs/arXiv:astro-ph/0505010}{{arXiv:astro-ph/0505010}}}

\bibitem[{{Springel}(2010{\natexlab{a}})}]{springel2010a}
{Springel} V (2010{\natexlab{a}}) E pur si muove: Galilean-invariant
  cosmological hydrodynamical simulations on a moving mesh. \mnras
  401:791--851. \doi{10.1111/j.1365-2966.2009.15715.x}.
  {\href{https://arxiv.org/abs/0901.4107}{{arXiv:0901.4107}}} {[astro-ph.CO]}

\bibitem[{{Springel}(2010{\natexlab{b}})}]{springel2010c}
{Springel} V (2010{\natexlab{b}}) {Smoothed Particle Hydrodynamics in
  Astrophysics}. \araa 48:391--430. \doi{10.1146/annurev-astro-081309-130914}

\bibitem[{{Springel} and {Hernquist}(2002)}]{springel2002a}
{Springel} V, {Hernquist} L (2002) {Cosmological smoothed particle
  hydrodynamics simulations: the entropy equation}. \mnras 333:649--664.
  \doi{10.1046/j.1365-8711.2002.05445.x}.
  {\href{https://arxiv.org/abs/arXiv:astro-ph/0111016}{{arXiv:astro-ph/0111016}}}

\bibitem[{{Springel} et~al.(2001){Springel}, {Yoshida}, and
  {White}}]{springel2001a}
{Springel} V, {Yoshida} N, {White} SDM (2001) {GADGET: a code for collisionless
  and gasdynamical cosmological simulations}. \na 6:79--117.
  \doi{10.1016/S1384-1076(01)00042-2}.
  {\href{https://arxiv.org/abs/arXiv:astro-ph/0003162}{{arXiv:astro-ph/0003162}}}

\bibitem[{{Springel} et~al.(2021){Springel}, {Pakmor}, {Zier}, and
  {Reinecke}}]{springel2021a}
{Springel} V, {Pakmor} R, {Zier} O, {Reinecke} M (2021) Simulating cosmic
  structure formation with the {GADGET-4} code. \mnras 506(2):2871--2949.
  \doi{10.1093/mnras/stab1855}.
  {\href{https://arxiv.org/abs/2010.03567}{{arXiv:2010.03567}}} {[astro-ph.IM]}

\bibitem[{{Staff} et~al.(2016{\natexlab{a}}){Staff}, {De Marco}, {Macdonald},
  {Galaviz}, {Passy}, {Iaconi}, and {Low}}]{staff2016b}
{Staff} JE, {De Marco} O, {Macdonald} D, et~al. (2016{\natexlab{a}})
  {Hydrodynamic simulations of the interaction between an AGB star and a
  main-sequence companion in eccentric orbits}. \mnras 455:3511--3525.
  \doi{10.1093/mnras/stv2548}.
  {\href{https://arxiv.org/abs/1510.08429}{{arXiv:1510.08429}}} {[astro-ph.SR]}

\bibitem[{{Staff} et~al.(2016{\natexlab{b}}){Staff}, {De Marco}, {Wood},
  {Galaviz}, and {Passy}}]{staff2016a}
{Staff} JE, {De Marco} O, {Wood} P, {Galaviz} P, {Passy} JC
  (2016{\natexlab{b}}) {Hydrodynamic simulations of the interaction between
  giant stars and planets}. \mnras 458:832--844. \doi{10.1093/mnras/stw331}.
  {\href{https://arxiv.org/abs/1602.03130}{{arXiv:1602.03130}}} {[astro-ph.SR]}

\bibitem[{{Stinson} et~al.(2013){Stinson}, {Brook}, {Macci{\`o}}, {Wadsley},
  {Quinn}, and {Couchman}}]{stinson2013a}
{Stinson} GS, {Brook} C, {Macci{\`o}} AV, et~al. (2013) Making galaxies in a
  cosmological context: the need for early stellar feedback. \mnras
  428(1):129--140. \doi{10.1093/mnras/sts028}.
  {\href{https://arxiv.org/abs/1208.0002}{{arXiv:1208.0002}}} {[astro-ph.CO]}

\bibitem[{{Stone} and {Norman}(1992)}]{stone1992a}
{Stone} JM, {Norman} ML (1992) {ZEUS-2D: A Radiation Magnetohydrodynamics Code
  for Astrophysical Flows in Two Space Dimensions. II. The Magnetohydrodynamic
  Algorithms and Tests}. \apjs 80:791--+. \doi{10.1086/191681}

\bibitem[{{Stone} et~al.(2020){Stone}, {Tomida}, {White}, and
  {Felker}}]{stone2020a}
{Stone} JM, {Tomida} K, {White} CJ, {Felker} KG (2020) {The Athena++ Adaptive
  Mesh Refinement Framework: Design and Magnetohydrodynamic Solvers}. \apjs
  249(1):4. \doi{10.3847/1538-4365/ab929b}.
  {\href{https://arxiv.org/abs/2005.06651}{{arXiv:2005.06651}}} {[astro-ph.IM]}

\bibitem[{{Sz{\"o}lgy{\'e}n} et~al.(2022){Sz{\"o}lgy{\'e}n}, {MacLeod}, and
  {Loeb}}]{szolgyen2022a}
{Sz{\"o}lgy{\'e}n} {\'A}, {MacLeod} M, {Loeb} A (2022) Eccentricity evolution
  in gaseous dynamical friction. arXiv e-prints arXiv:2203.01334.
  {\href{https://arxiv.org/abs/2203.01334}{{arXiv:2203.01334}}} {[astro-ph.EP]}

\bibitem[{{Taam}(1979)}]{taam1979a}
{Taam} RE (1979) Double core evolution and {X}-ray binaries. \aplett 20:29

\bibitem[{{Taam} and {Bodenheimer}(1989)}]{taam1989a}
{Taam} RE, {Bodenheimer} P (1989) {Double-core evolution. VIII - The evolution
  of a 5 solar mass red giant with a 1 solar mass companion}. \apj
  337:849--857. \doi{10.1086/167155}

\bibitem[{{Taam} and {Bodenheimer}(1991)}]{taam1991a}
{Taam} RE, {Bodenheimer} P (1991) {Double core evolution. IV - The late stages
  of evolution of a 2-solar mass red giant with a 1-solar mass companion}. \apj
  373:246--249. \doi{10.1086/170043}

\bibitem[{{Taam} and {Ricker}(2006)}]{taam2006a}
{Taam} RE, {Ricker} PM (2006) {Common Envelope Evolution}. arXiv e-prints
  astro-ph/0611043.
  {\href{https://arxiv.org/abs/astro-ph/0611043}{{arXiv:astro-ph/0611043}}}
  {[astro-ph]}

\bibitem[{{Taam} et~al.(1978){Taam}, {Bodenheimer}, and {Ostriker}}]{taam1978a}
{Taam} RE, {Bodenheimer} P, {Ostriker} JP (1978) {Double core evolution. I - A
  16 solar mass star with a 1 solar mass neutron-star companion}. \apj
  222:269--280. \doi{10.1086/156142}

\bibitem[{{Taam} et~al.(1994){Taam}, {Bodenheimer}, and
  {R\'{o}{ż}yczka}}]{taam1994a}
{Taam} RE, {Bodenheimer} P, {R\'{o}{ż}yczka} M (1994) {Double Core Evolution.
  VI. Effects of Gravitational Torques}. \apj 431:247. \doi{10.1086/174482}

\bibitem[{{Tauris} and {Dewi}(2001)}]{tauris2001a}
{Tauris} TM, {Dewi} JDM (2001) Research note on the binding energy parameter of
  common envelope evolution. dependency on the definition of the stellar core
  boundary during spiral-in. \aap 369:170--173.
  \doi{10.1051/0004-6361:20010099}.
  {\href{https://arxiv.org/abs/astro-ph/0101530}{{arXiv:astro-ph/0101530}}}
  {[astro-ph]}

\bibitem[{{Tauris} and {van den Heuvel}(2014)}]{tauris2014a}
{Tauris} TM, {van den Heuvel} EPJ (2014) {Formation of the Galactic Millisecond
  Pulsar Triple System PSR J0337+1715{\textemdash}A Neutron Star with Two
  Orbiting White Dwarfs}. \apjl 781(1):L13. \doi{10.1088/2041-8205/781/1/L13}.
  {\href{https://arxiv.org/abs/1401.0941}{{arXiv:1401.0941}}} {[astro-ph.SR]}

\bibitem[{{Terman} and {Taam}(1996)}]{terman1996a}
{Terman} JL, {Taam} RE (1996) {Double-Core Evolution. IX. The Infall of a
  Main-Sequence Star through the Envelope of Its Intermediate-Mass Red Giant
  Companion}. \apj 458:692. \doi{10.1086/176850}

\bibitem[{{Terman} et~al.(1994){Terman}, {Taam}, and {Hernquist}}]{terman1994a}
{Terman} JL, {Taam} RE, {Hernquist} L (1994) {Double-core evolution. V.
  Three-dimensional effects in the merger of a red giant with a dwarf
  companion}. \apj 422:729--736. \doi{10.1086/173765}

\bibitem[{{Terman} et~al.(1995){Terman}, {Taam}, and {Hernquist}}]{terman1995a}
{Terman} JL, {Taam} RE, {Hernquist} L (1995) Double core evolution. vii. the
  infall of a neutron star through the envelope of its massive star companion.
  \apj 445:367--376. \doi{10.1086/175702}

\bibitem[{{Thorne} and {Zytkow}(1975)}]{thorne1975a}
{Thorne} KS, {Zytkow} AN (1975) Red giants and supergiants with degenerate
  neutron cores. \apjl 199:L19--L24. \doi{10.1086/181839}

\bibitem[{{Tocknell} et~al.(2014){Tocknell}, {De Marco}, and
  {Wardle}}]{tocknell2014a}
{Tocknell} J, {De Marco} O, {Wardle} M (2014) {Constraints on common envelope
  magnetic fields from observations of jets in planetary nebulae}. \mnras
  439:2014--2024. \doi{10.1093/mnras/stu079}.
  {\href{https://arxiv.org/abs/1308.5027}{{arXiv:1308.5027}}} {[astro-ph.SR]}

\bibitem[{{Toonen} et~al.(2012){Toonen}, {Nelemans}, and {Portegies
  Zwart}}]{toonen2012a}
{Toonen} S, {Nelemans} G, {Portegies Zwart} S (2012) Supernova type ia
  progenitors from merging double white dwarfs. using a new population
  synthesis model. \aap 546:A70. \doi{10.1051/0004-6361/201218966}.
  {\href{https://arxiv.org/abs/1208.6446}{{arXiv:1208.6446}}} {[astro-ph.HE]}

\bibitem[{{Toonen} et~al.(2018){Toonen}, {Perets}, {Igoshev}, {Michaely}, and
  {Zenati}}]{toonen2018a}
{Toonen} S, {Perets} HB, {Igoshev} AP, {Michaely} E, {Zenati} Y (2018) {The
  demographics of neutron star - white dwarf mergers. Rates, delay-time
  distributions, and progenitors}. \aap 619:A53.
  \doi{10.1051/0004-6361/201833164}.
  {\href{https://arxiv.org/abs/1804.01538}{{arXiv:1804.01538}}} {[astro-ph.HE]}

\bibitem[{Toro(2009)}]{toro2009a}
Toro EF (2009) Riemann Solvers and Numerical Methods for Fluid Dynamics: A
  Practical Introduction. Springer, Berlin Heidelberg.
  \urlprefix\url{http://books.google.de/books?id=SqEjX0um8o0C}

\bibitem[{{Tout}(1991)}]{tout1991a}
{Tout} CA (1991) On the relation between the mass-ratio distribution in binary
  stars and the mass function for single stars. \mnras 250:701--706

\bibitem[{{Tylenda} and {Soker}(2006)}]{tylenda2006a}
{Tylenda} R, {Soker} N (2006) {Eruptions of the V838 Mon type: stellar merger
  versus nuclear outburst models}. \aap 451(1):223--236.
  \doi{10.1051/0004-6361:20054201}.
  {\href{https://arxiv.org/abs/astro-ph/0509379}{{arXiv:astro-ph/0509379}}}
  {[astro-ph]}

\bibitem[{{Tylenda} et~al.(2005){Tylenda}, {Soker}, and
  {Szczerba}}]{tylenda2005a}
{Tylenda} R, {Soker} N, {Szczerba} R (2005) {On the progenitor of V838
  Monocerotis}. \aap 441(3):1099--1109. \doi{10.1051/0004-6361:20042485}.
  {\href{https://arxiv.org/abs/astro-ph/0412183}{{arXiv:astro-ph/0412183}}}
  {[astro-ph]}

\bibitem[{{Tylenda} et~al.(2011){Tylenda}, {Hajduk}, {Kami{\'n}ski}, {Udalski},
  {Soszy{\'n}ski}, {Szyma{\'n}ski}, {Kubiak}, {Pietrzy{\'n}ski}, {Poleski},
  {Wyrzykowski}, and {Ulaczyk}}]{tylenda2011a}
{Tylenda} R, {Hajduk} M, {Kami{\'n}ski} T, et~al. (2011) {V1309 Scorpii: merger
  of a contact binary}. \aap 528:A114. \doi{10.1051/0004-6361/201016221}.
  {\href{https://arxiv.org/abs/1012.0163}{{arXiv:1012.0163}}} {[astro-ph.SR]}

\bibitem[{{van den Heuvel}(1976)}]{vandenheuvel1976a}
{van den Heuvel} EPJ (1976) Late stages of close binary systems. In: {Eggleton}
  P, {Mitton} S, {Whelan} J (eds) Structure and Evolution of Close Binary
  Systems. vol~73. p~35

\bibitem[{{Vanderburg} et~al.(2020){Vanderburg}, {Rappaport}, {Xu},
  {Crossfield}, {Becker}, {Gary}, {Murgas}, {Blouin}, {Kaye}, {Palle}, {Melis},
  {Morris}, {Kreidberg}, {Gorjian}, {Morley}, {Mann}, {Parviainen}, {Pearce},
  {Newton}, {Carrillo}, {Zuckerman}, {Nelson}, {Zeimann}, {Brown},
  {Tronsgaard}, {Klein}, {Ricker}, {Vanderspek}, {Latham}, {Seager}, {Winn},
  {Jenkins}, {Adams}, {Benneke}, {Berardo}, {Buchhave}, {Caldwell},
  {Christiansen}, {Collins}, {Col{\'o}n}, {Daylan}, {Doty}, {Doyle},
  {Dragomir}, {Dressing}, {Dufour}, {Fukui}, {Glidden}, {Guerrero}, {Guo},
  {Heng}, {Henriksen}, {Huang}, {Kaltenegger}, {Kane}, {Lewis}, {Lissauer},
  {Morales}, {Narita}, {Pepper}, {Rose}, {Smith}, {Stassun}, and
  {Yu}}]{vanderburg2020a}
{Vanderburg} A, {Rappaport} SA, {Xu} S, et~al. (2020) A giant planet candidate
  transiting a white dwarf. \nat 585(7825):363--367.
  \doi{10.1038/s41586-020-2713-y}.
  {\href{https://arxiv.org/abs/2009.07282}{{arXiv:2009.07282}}} {[astro-ph.EP]}

\bibitem[{{Vel{\'a}zquez} et~al.(2012){Vel{\'a}zquez}, {Raga}, {Riera},
  {Steffen}, {Esquivel}, {Cant{\'o}}, and {Haro-Corzo}}]{velazquez2012a}
{Vel{\'a}zquez} PF, {Raga} AC, {Riera} A, et~al. (2012) {Multipolar young
  planetary nebulae modelled as a precessing and orbiting jet with
  time-dependent ejection velocity}. \mnras 419(4):3529--3536.
  \doi{10.1111/j.1365-2966.2011.19991.x}

\bibitem[{{Vigna-G{\'o}mez} et~al.(2018){Vigna-G{\'o}mez}, {Neijssel},
  {Stevenson}, {Barrett}, {Belczynski}, {Justham}, {de Mink}, {M{\"u}ller},
  {Podsiadlowski}, {Renzo}, {Sz{\'e}csi}, and {Mandel}}]{vigna2018a}
{Vigna-G{\'o}mez} A, {Neijssel} CJ, {Stevenson} S, et~al. (2018) {On the
  formation history of Galactic double neutron stars}. \mnras
  481(3):4009--4029. \doi{10.1093/mnras/sty2463}.
  {\href{https://arxiv.org/abs/1805.07974}{{arXiv:1805.07974}}} {[astro-ph.SR]}

\bibitem[{{Vigna-G{\'o}mez} et~al.(2022){Vigna-G{\'o}mez}, {Wassink},
  {Klencki}, {Istrate}, {Nelemans}, and {Mandel}}]{vigna2022a}
{Vigna-G{\'o}mez} A, {Wassink} M, {Klencki} J, et~al. (2022) Stellar response
  after stripping as a model for common-envelope outcomes. \mnras
  511(2):2326--2338. \doi{10.1093/mnras/stac237}.
  {\href{https://arxiv.org/abs/2107.14526}{{arXiv:2107.14526}}} {[astro-ph.HE]}

\bibitem[{{Vogelsberger} et~al.(2013){Vogelsberger}, {Genel}, {Sijacki},
  {Torrey}, {Springel}, and {Hernquist}}]{vogelsberger2013a}
{Vogelsberger} M, {Genel} S, {Sijacki} D, et~al. (2013) A model for
  cosmological simulations of galaxy formation physics. \mnras 436:3031--3067.
  \doi{10.1093/mnras/stt1789}.
  {\href{https://arxiv.org/abs/1305.2913}{{arXiv:1305.2913}}}

\bibitem[{{Vogelsberger} et~al.(2014){Vogelsberger}, {Genel}, {Springel},
  {Torrey}, {Sijacki}, {Xu}, {Snyder}, {Nelson}, and
  {Hernquist}}]{vogelsberger2014a}
{Vogelsberger} M, {Genel} S, {Springel} V, et~al. (2014) {Introducing the
  Illustris Project: simulating the coevolution of dark and visible matter in
  the Universe}. \mnras 444:1518--1547. \doi{10.1093/mnras/stu1536}.
  {\href{https://arxiv.org/abs/1405.2921}{{arXiv:1405.2921}}}

\bibitem[{{Wang}(2018)}]{wang2018a}
{Wang} B (2018) {Mass-accreting white dwarfs and type Ia supernovae}. Research
  in Astronomy and Astrophysics 18(5):049. \doi{10.1088/1674-4527/18/5/49}.
  {\href{https://arxiv.org/abs/1801.04031}{{arXiv:1801.04031}}} {[astro-ph.SR]}

\bibitem[{Warren and Salmon(1993)}]{warren1993a}
Warren MS, Salmon JK (1993) A parallel hashed oct-tree {N}-body algorithm. In:
  Proceedings of the 1993 ACM/IEEE Conference on Supercomputing. Supercomputing
  '93. Association for Computing Machinery, New York, NY, USA, pp 12--21.
  \doi{10.1145/169627.169640},
  \urlprefix\url{https://doi.org/10.1145/169627.169640}

\bibitem[{{Webbink}(1984)}]{webbink1984a}
{Webbink} RF (1984) Double white dwarfs as progenitors of {R} {C}oronae
  {B}orealis stars and {T}ype {I} supernovae. \apj 277:355--360.
  \doi{10.1086/161701}

\bibitem[{{Webbink}(2008)}]{webbink2008a}
{Webbink} RF (2008) {Common Envelope Evolution Redux}. In: {Milone} EF, {Leahy}
  DA, {Hobill} DW (eds) Astrophysics and Space Science Library. Astrophysics
  and Space Science Library, vol 352. p 233.
  {\href{https://arxiv.org/abs/0704.0280}{{arXiv:0704.0280}}}

\bibitem[{{Wickramasinghe} et~al.(2014){Wickramasinghe}, {Tout}, and
  {Ferrario}}]{wickramasinghe2014a}
{Wickramasinghe} DT, {Tout} CA, {Ferrario} L (2014) {The most magnetic stars}.
  \mnras 437:675--681. \doi{10.1093/mnras/stt1910}.
  {\href{https://arxiv.org/abs/1310.2696}{{arXiv:1310.2696}}} {[astro-ph.SR]}

\bibitem[{{Wilson} and {Nordhaus}(2019)}]{wilson2019a}
{Wilson} EC, {Nordhaus} J (2019) {The role of convection in determining the
  ejection efficiency of common envelope interactions}. \mnras
  485(4):4492--4501. \doi{10.1093/mnras/stz601}.
  {\href{https://arxiv.org/abs/1811.03161}{{arXiv:1811.03161}}} {[astro-ph.SR]}

\bibitem[{{Wilson} and {Nordhaus}(2020)}]{wilson2020a}
{Wilson} EC, {Nordhaus} J (2020) Convection and spin-up during common envelope
  evolution: the formation of short-period double white dwarfs. \mnras
  497(2):1895--1903. \doi{10.1093/mnras/staa2088}.
  {\href{https://arxiv.org/abs/2006.09360}{{arXiv:2006.09360}}} {[astro-ph.SR]}

\bibitem[{{Xiong} et~al.(2017){Xiong}, {Chen}, {Podsiadlowski}, {Li}, and
  {Han}}]{xiong2017a}
{Xiong} H, {Chen} X, {Podsiadlowski} P, {Li} Y, {Han} Z (2017) Subdwarf {B}
  stars from the common envelope ejection channel. \aap 599:A54.
  \doi{10.1051/0004-6361/201629622}.
  {\href{https://arxiv.org/abs/1608.08739}{{arXiv:1608.08739}}} {[astro-ph.SR]}

\bibitem[{{Yorke} et~al.(1995){Yorke}, {Bodenheimer}, and {Taam}}]{yorke1995a}
{Yorke} HW, {Bodenheimer} P, {Taam} RE (1995) {Double Core Evolution. VIII. The
  Spiral-in of a Main-Sequence Star through the Envelope of an Asymptotic Giant
  Branch Companion}. \apj 451:308. \doi{10.1086/176220}

\bibitem[{{Zahn}(1977)}]{zahn1977a}
{Zahn} JP (1977) Tidal friction in close binary stars. \aap 57:383--394

\bibitem[{{Zhang} and {Fryer}(2001)}]{zhang2001a}
{Zhang} W, {Fryer} CL (2001) The merger of a helium star and a black hole:
  Gamma-ray bursts. \apj 550(1):357--367. \doi{10.1086/319734}.
  {\href{https://arxiv.org/abs/astro-ph/0011236}{{arXiv:astro-ph/0011236}}}
  {[astro-ph]}

\bibitem[{{Zhu} et~al.(2015){Zhu}, {Pakmor}, {van Kerkwijk}, and
  {Chang}}]{zhu2015a}
{Zhu} C, {Pakmor} R, {van Kerkwijk} MH, {Chang} P (2015) {Magnetized Moving
  Mesh Merger of a Carbon--Oxygen White Dwarf Binary}. \apjl 806:L1.
  \doi{10.1088/2041-8205/806/1/L1}.
  {\href{https://arxiv.org/abs/1504.01732}{{arXiv:1504.01732}}} {[astro-ph.SR]}

\bibitem[{{Zingale} et~al.(2002){Zingale}, {Dursi}, {ZuHone}, {Calder},
  {Fryxell}, {Plewa}, {Truran}, {Caceres}, {Olson}, {Ricker}, {Riley},
  {Rosner}, {Siegel}, {Timmes}, and {Vladimirova}}]{zingale2002a}
{Zingale} M, {Dursi} LJ, {ZuHone} J, et~al. (2002) Mapping initial hydrostatic
  models in {G}odunov codes. \apjs 143(2):539--565. \doi{10.1086/342754}.
  {\href{https://arxiv.org/abs/astro-ph/0208031}{{arXiv:astro-ph/0208031}}}
  {[astro-ph]}

\bibitem[{{Zorotovic} et~al.(2010){Zorotovic}, {Schreiber}, {G{\"a}nsicke}, and
  {Nebot G{\'o}mez-Mor{\'a}n}}]{zorotovic2010a}
{Zorotovic} M, {Schreiber} MR, {G{\"a}nsicke} BT, {Nebot G{\'o}mez-Mor{\'a}n} A
  (2010) {Post-common-envelope binaries from SDSS. IX: Constraining the
  common-envelope efficiency}. \aap 520:A86. \doi{10.1051/0004-6361/200913658}.
  {\href{https://arxiv.org/abs/1006.1621}{{arXiv:1006.1621}}} {[astro-ph.SR]}

\bibitem[{{Zou} et~al.(2020){Zou}, {Frank}, {Chen}, {Reichardt}, {De Marco},
  {Blackman}, {Nordhaus}, {Balick}, {Carroll-Nellenback}, {Chamandy}, and
  {Liu}}]{zou2020a}
{Zou} Y, {Frank} A, {Chen} Z, et~al. (2020) Bipolar planetary nebulae from
  outflow collimation by common envelope evolution. \mnras 497(3):2855--2869.
  \doi{10.1093/mnras/staa2145}.
  {\href{https://arxiv.org/abs/1912.01647}{{arXiv:1912.01647}}} {[astro-ph.SR]}

\bibitem[{{Zou} et~al.(2022){Zou}, {Chamandy}, {Carroll-Nellenback},
  {Blackman}, and {Frank}}]{zou2022a}
{Zou} Y, {Chamandy} L, {Carroll-Nellenback} J, {Blackman} EG, {Frank} A (2022)
  Jets from main sequence and white dwarf companions during common envelope
  evolution. \mnras 514(2):3041--3057. \doi{10.1093/mnras/stac1529}.
  {\href{https://arxiv.org/abs/2202.05715}{{arXiv:2202.05715}}} {[astro-ph.SR]}

\end{thebibliography}

\section*{Acronyms}
\begin{acronym}

\acro{1D}{one-dimensional}
\acro{2D}{two-dimensional}
\acro{3D}{three-dimensional}
\acro{AGB}{asymptotic giant branch}
\acro{AMR}{adaptive mesh refinement}
\acro{BHL}{Bondi--Hoyle--Lyttleton}
\acro{CE}{common envelope}
\acro{CEE}{common-envelope evolution}
\acro{CFL}{Courant--Friedrichs--Lewy}
\acro{HSE}{hydrostatic equilibrium}
\acro{LRN}[LRN]{luminous red nova}
\acroplural{LRN}[LRNe]{luminous red novae}
\acro{MHD}{magneto-hydrodynamics}
\acro{PN}[PN]{planetary nebula}
\acroplural{PN}[PNe]{planetary nebulae}
\acro{RG}{red giant}
\acro{RN}[RN]{red nova}
\acroplural{RN}[RNe]{red novae}
\acro{ILRT}[ILRT]{intermediate-luminosity red transient}
\acro{RLOF}{Roche-lobe overflow}
\acro{RGB}{red giant branch}
\acro{SPH}{smoothed particle hydrodynamics}

\end{acronym}

\end{document}